\documentclass{madoc}
\usepackage{graphics}
\usepackage{ngerman}
\usepackage{helvet}
\usepackage{cite}
\usepackage{a4} 
\usepackage{epsf}
\usepackage{amsmath}
\usepackage{amssymb}
\usepackage{bbm} 
\usepackage{textcomp}
\usepackage[NoDate]{currvita}
\renewcommand{\v}[1]{{\mathbf #1}} 
 
\begin{document}
\thispagestyle{empty}

\begin{center}
\large

\mbox{}\\[1cm]

\textbf{\Huge
Modeling and Simulation of Strained Heteroepitaxial Growth}\\

\vspace{2cm}

\vfill

\begin{centering}
Dissertation zur Erlangung des \\
naturwissenschaftlichen Doktorgrades \\
der Bayerischen Julius-Maximilians-Universit\"{a}t\\
W\"{u}rzburg
\end{centering}

\vspace{2cm}

vorgelegt von\\[2ex]

{\Large Florian Much} \\[1ex]

aus  Straubing \\

\vfill

Institut f\"{u}r Theoretische Physik und Astrophysik\\
Bayerische Julius-Maximilians-Universit\"{a}t\\
W\"{u}rzburg\\[1ex]

W\"{u}rzburg 2003
 \vspace{1cm}
\end{center}
\newpage 

\mbox{ }

\thispagestyle{empty}

\newpage 

\thispagestyle{empty}

\begin{center}
\large

\mbox{}\\[1cm]

\textbf{\Huge
Modeling and Simulation of Strained Heteroepitaxial Growth}\\

\vspace{2cm}

\vfill

\begin{centering}
Dissertation zur Erlangung des \\
naturwissenschaftlichen Doktorgrades \\
der Bayerischen Julius-Maximilians-Universit\"{a}t\\
W\"{u}rzburg
\end{centering}

\vspace{2cm}

vorgelegt von\\[2ex]

{\Large Florian Much} \\[1ex]

aus  Straubing \\

\vfill

Institut f\"{u}r Theoretische Physik und Astrophysik\\
Bayerische Julius-Maximilians-Universit\"{a}t\\
W\"{u}rzburg\\[1ex]

W\"{u}rzburg 2003
 \vspace{1cm}
\end{center}
\newpage
\thispagestyle{empty}

\large
\ \\
\vspace{6cm}

\noindent
Eingereicht am ................................... 26.11.2003\\
bei der Fakult\"{a}t f\"{u}r Physik und Astronomie\\
\vspace{0.5cm}
\noindent \\
Beurteilung der Dissertation:\\
\noindent \\
1. Gutachter: ......... Priv.-Doz. Dr. Michael Biehl\\
2. Gutachter: ............. Prof. Dr. Wolfgang Kinzel\\
\ \\
\ \\
M\"{u}ndliche Pr\"{u}fung:\\
\ \\
1. Pr\"{u}fer: ............... Priv.-Doz. Dr. Michael Biehl\\
2. Pr\"{u}fer: .......................... Prof. Dr. Jean Geurts\\
\ \\
Tag der m\"{u}ndlichen \\
Pr\"{u}fung: ............................................. 12.12.2003\\
\noindent \\
Doktorurkunde \\
ausgeh\"{a}ndigt am: ................................................\\

\newpage
\thispagestyle{empty}
\normalsize
\selectlanguage{ngerman}
\section*{Zusammenfassung}
Diese Doktorarbeit behandelt die Modellierung und Simulation von Heteroepitaxie--Wachstum. Dabei wird 
insbesondere der Gitterunterschied der am Wachstumsprozess beteiligten Materialien miteinbezogen.

Im Einleitungskapitel wird ein \"Uberblick \"uber die wichtigsten Ober\-fl\"a\-chen\-pro\-zesse 
und die wesentlichen Mechanismen des Verspannungsabbaus beim 
heteroepitaktischen Wachstum gegeben. Es folgt eine Zusammenstellung g\"angiger Methoden der Modellierung und 
Simulation von Heteroepitaxie wie sie z.B. in der {\it Molekularstrahlepitaxie} realisiert ist.

In Kapitel \ref{KAP-2} wird die so genannte {\it Molekular--Statik} Methode zur Berechnung von 
Diffusionsbarrieren vorgestellt. Unter der Annahme, dass die Teilchen des Kristalls \"uber Paarpotentiale 
miteinander wechselwirken 
berechnen wir H\"upf-- und Austauschbarrieren f\"ur Diffusionsereignisse am Rand von 
Adsorbatinseln.
Hierbei interessiert vor allem die Abh\"angigkeit der Barrieren von der Inselgr\"o{\ss}e, 
vom Gitterunterschied und dem verwendeten Potential. Es zeigt sich, dass die Barriere f\"ur Austauschdiffusion 
besonders stark von diesen Faktoren beeinflusst wird. Im Bereich gro{\ss}er Gitterunterschiede zum Beispiel 
nimmt diese Barriere deutlich ab. Damit wird hier die Austauschdiffusion zum dominanten Mechanismus der 
Interlagendiffusion.
 
Im folgenden Kapitel \ref{KAP-3} wird ein von uns weiterentwickelter Algorithmus zur Simulation von 
Heteroepitaxie--Wachstum vorgestellt. Um die Verspannungen und ihren Abbau realistisch modellieren zu k\"onnen, bedarf 
die Beschreibung von Heteroepitaxie der M\"oglichkeit {\it kontinuierlicher} Abst\"ande zwischen 
den Teilchen des Kristalls.
Der Leit\-gedanke dieser {\it gitterfreien} Methode ist es daher, die Diffusionsbarrieren f\"ur jedes 
Teilchen auf der Kristalloberfl{\"a}che unter Verwendung von Paarpotentialen als Funktion der 
Teilchenabst\"ande zu berechnen. 
Die so gewonnenen Aktivierungsenergien werden dann in einer {\it Kinetischen Monte Carlo} (KMC) 
Simulation zur Modellierung der Wachstumsdynamik verwendet. Als entscheidender Vorteil von 
KMC gegen\"uber anderen Methoden (wie z.B. der Molekular--Dynamik) k\"onnen so die relevanten Zeitskalen des 
Kristallwachstums abgedeckt werden. 
Wir beschreiben in diesem Kapitel ausf\"uhrlich die Berechnung der 
Diffusionsbarrieren, die Relaxation des Kristalls sowie Strategien f\"ur eine effiziente Umsetzung des 
Algorithmus. 

Die folgenden Kapitel besch\"aftigen sich mit der Analyse von drei wichtigen Mechanismen 
des Verspannungsabbaus unter Verwendung der gitterfreien KMC Methode.
In Kapitel \ref{KAP-4} untersuchen wir den Verspannungsabbau aufgrund der Bildung von {\it Versetzungen }
im Adsorbatfilm. Im ersten Teil des Kapitels werden Mechanismen der Versetzungsbildung ausf\"uhrlich 
anhand von Simulationsergebnissen diskutiert. Weiterhin wird die Abh\"angigkeit der so genannten 
{\it kritischen} Filmdicke f\"ur das Auftreten von Versetzungen in Abh\"angigkeit vom Gitterunterschied 
untersucht. Wir finden dabei in \"Ubereinstimmung mit zahlreichen experimentellen Ergebnissen, dass sich 
die kritische Filmdicke mit Hilfe eines Potenzgesetzes als Funktion des 
Gitterunterschieds beschreiben l\"asst. Der zweite Teil des Kapitels behandelt das pseudomorphe Wachstum mit 
anschlie{\ss}ender gradueller Relaxation des Adsorbatfilms f\"ur den Bereich relativ kleiner Gitterunterschiede.
Diese Untersuchung ist durch neuartige in--situ Messungen der vertikalen Gitterkonstante am System ZnSe/GaAs 
motiviert. Es zeigt sich 
eine sehr gute qualitative \"Ubereinstimmung der Simulationsergebnisse 
mit dem Ex\-peri\-ment. Au{\ss}erdem kann gezeigt werden, dass der Bereich pseudomorphen  
Wachstums n\"aherungsweise nach einem Potenzgesetz mit dem Gitterunterschied
skaliert.

Im folgenden Kapitel \ref{KAP-5} gehen wir auf die Entstehung {\it selbstbildender} Inseln 
({\it Stranski--Krastanov Wachstum}) 
als einen weiteren m\"oglichen Relaxationsmechanismus der Heteroepitaxie ein. 
Wir f\"uhren hier die Bildung einer benetzenden Adsorbatschicht auf verlangsamte Adsorbat--Diffusion auf dem Substrat 
zur\"uck. Die anschlie{\ss}ende Inselbildung finden wir durch zwei {\it kinetische} Faktoren beg\"unstigt: eine 
verlangsamte Diffusion auf der Inseloberfl\"ache und eine gerichtete Diffusion zur Inselmitte hin. Beide Ph\"anomene 
haben ihren Ursprung in der teilweisen Relaxation von Adsorbatmaterial in den Inseloberfl\"achen. Weiter bestimmen 
wir die Abh\"angigkeit der Inselgr\"o{\ss}e und Inseldichte vom Gitterunterschied, der Temperatur und dem Teilchenfluss.
Dabei werden sehr gute qualitative \"Ubereinstimmungen mit {\it MOVPE} (metal--organic vapor phase epitaxy)
Experimenten erzielt.

Im abschlie{\ss}enden Kapitel \ref{KAP-6} untersuchen wir anhand eines Dreikomponentensystems 
die Bildung von Oberfl\"achen--{\it Legierungen} als m\"oglichen Relaxationsmechanismus. Anhand von 
{\it Gleichgewichtssimulationen} gelingt es nachzuweisen, dass die Konkurrenz von Teilchenbindungen und 
Gitterunterschied zu einem regelm\"a{\ss}igen  Streifenmuster der beteiligten Materialien f\"uhrt.
Wir untersuchen anschlie{\ss}end inwieweit sich diese Musterbildung in ein kinetisches Modell 
\"ubertragen l\"asst. Abh\"angig von Temperatur, Gitterunterschied und Potentialtyp finden wir sowohl 
die Streifenbildung als auch die experimentell berichtete Ver\"astelung der Oberl\"achenstrukturen.
Der abschlie{\ss}ende Vergleich mit einem Gittergasmodell zeigt, dass Musterbildung zwar 
allein aufgrund kinetischer Ursachen m\"oglich ist, der Gitterunterschied zwischen den 
beteiligten Materialien aber die Ver\"astelung und Stabilisierung der Strukturen bewirkt.

Schlie{\ss}lich werden im Anhang der Arbeit Ursprung und Eigenschaften der verwendeten Potentiale f\"ur die 
Teilchenwechselwirkungen besprochen.
\selectlanguage{USenglish}
\newpage
\thispagestyle{empty}
\normalsize
\section*{Abstract}
In this PhD thesis,we develop models of heteroepitaxial growth, where the lattice misfit of the 
involved particle species is of special interest.
In the introductory chapter \ref{KAP-1} we introduce important physical processes which occur on 
the crystal surface. We give an overview on relevant strain relaxation mechanisms and discuss 
different methods for the simulation of heteroepitaxial growth.

In chapter \ref{KAP-2}, we introduce the so--called {\it Molecular Static} method for the 
calculation of diffusion barriers. Provided that the particles of the crystal interact with 
each other via a {\it pair--potential} we calculate barriers for hopping and exchange diffusion moves
over island step edges. In this investigation the dependence of the barriers on the island size, 
the misfit and the potential--type is of special interest. We show that the exchange barriers are 
particularly sensitive to these parameters. For example, in the case of large misfits the exchange diffusion 
barrier decreases dramatically and the exchange process becomes the dominant diffusion mechanism at island edges.

In chapter \ref{KAP-3}, we introduce an algorithm for the simulation of heteroepitaxial growth. 
In order to account for strain effects caused by the atomic mismatch in the crystal it is 
essential to allow for continuous particle positions in our simulations. The main idea of our 
off--lattice method is to compute the barriers for each diffusion event from a pair--potential 
and to use the barriers in a rejection--free Kinetic Monte Carlo (KMC) simulation.
The main advantage of the KMC method is the feasibility of the relevant time scales necessary for the simulation 
of crystal growth, as realized, e.g., in molecular beam epitaxy (MBE). We discuss in this
chapter the calculation of activation barriers, the relaxation of the crystal and an efficient 
implementation of the algorithm. 

In the following chapters we discuss the application of
our off--lattice KMC method to the simulation of three important strain relaxation mechanisms.
In chapter \ref{KAP-4} we investigate the strain relaxation by introduction of misfit dislocations 
in the adsorbate film. In the first part of the chapter we discuss in detail different formation mechanisms 
of dislocations and investigate the so--called critical thickness for the first appearance of misfit dislocations.
In agreement with various experimental studies the dependency of the critical thickness on the misfit is 
given by a power--law.
In the second part of the chapter we treat the pseudomorphic region of heteroepitaxial growth and the subsequent  
gradual relaxation of the adsorbate film. These studies are motivated by a new in--situ method for 
the determination of the vertical lattice constant during MBE growth. Our simulation results show an  
excellent qualitative agreement with experimental results for ZnSe/GaAs heteroepitaxy.
We are able to demonstrate that the region of pseudomorphic growth scales with the misfit. 

In chapter \ref{KAP-5} follows the investigation of another prominent relaxation mechanism in heteroepitaxial growth:
the so--called {\it Stranski--Krastanov} growth mode, where $3d$ islands self--assemble on a thin adsorbate 
wetting--layer. We are able to show, that an increased barrier for diffusion of adsorbate particles on the 
substrate is a possible kinetic reason for the formation of a stable wetting--layer. The formation 
of adsorbate islands is due to a partial relaxation of adsorbate material, which causes two kinetic effects:
diffusion on the relaxed island surface is slower than on the pseudomorphic strained wetting--layer and 
a diffusion bias drives particles on the island surface to the island center. We further measure 
the dependence of island size and island density on the misfit, the temperature and the particle flux and find 
good qualitative agreement with metal--organic vapor phase epitaxy (MOVPE) experiments.

Finally, in chapter \ref{KAP-6} we consider surface alloying as a possible strain relaxation mechanism in 
multi--component systems.
By means of equilibrium simulations we are able to show that the competition between binding and 
strain energy yields the formation of regular stripe patterns on the crystal surface. 
The pattern formation is then investigated in a kinetic model: depending on temperature, misfit and 
potential--type we find as well the pattern formation as the experimental reported ramification of 
the structures. The comparison with a lattice gas model shows that the observed pattern formation 
can solely be due to kinetic effects but the misfit is essential for the ramification and stabilization 
of the surface structures. 

In the appendix of the work we discuss origin and properties of the used pair--potentials.


       \cleardoublepage
\setcounter{page}{1}
\tableofcontents    \cleardoublepage
\chapter{Theoretical descriptions and models of heteroepitaxial growth }
\label{KAP-1} 
In recent years heteroepitaxial growth has been a field of intense study.
This is mainly due to the fact that countless technical applications - including laser diodes, 
solar cells and magnetic or optic data storage devices - are heterosystems. 
In this work we focus in particular on atomic size mismatched systems where the participating
materials crystallize in the same lattice structure but have different lattice constants.
 
In technical application this so--called misfit can lead to both wanted and unwanted
effects.
On the one hand one often wishes to deposit a smooth adsorbate layer of 
certain thickness on a given substrate with a different lattice constant. 
This is for example needed in the fabrication of computer memory chips. 
Here, the relaxation 
of strain due to the misfit can cause perturbations of the lattice structure (e.g. dislocations)
or modulations of the surface which may affect the characteristics of the device in a negative way. \\
On the other hand a moderate misfit can lead to self--assembly of
islands which is an interesting process for the fabrication of so--called quantum dots. These 
quantum dots are solid state
structures typically made of metals or semiconductors which confine a small number of electrons to
a small space used for optoelectronic devices or single electron transistors.
However, in both cases the understanding and control of misfit caused phenomena is essential.

\section{Microscopic processes on crystal surfaces}
\begin{figure}[hbt]
\centerline{
\begin{minipage}{0.70 \textwidth}
\epsfxsize= 0.99\textwidth
  \epsffile{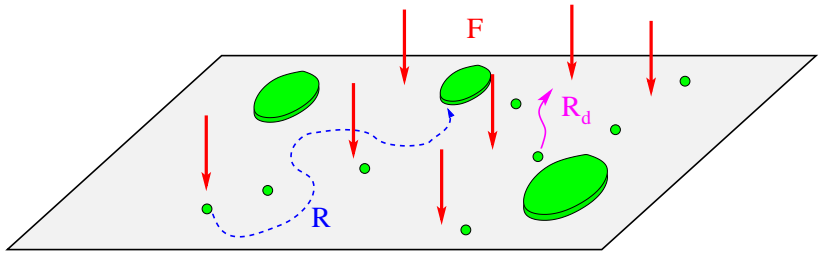}
\end{minipage}
\hfill
\begin{minipage}{0.25 \textwidth}
\caption{Sche\-ma\-tic depiction of relevant processes during MBE growth.} 
\label{MBE_FIG}
\end{minipage}}
\end{figure}

A particularly important technique for the fabrication of heterostructures is the 
molecular beam epitaxy (MBE) and related techniques:
in an ultra high vacuum (UHV) environment the substrate surface is exposed to a uniform flux $F$ of
adsorbate particles, which are evaporated in a thermal effusion cell.
The interplay of three relevant microscopic processes on the crystal surface 
determines the morphology of the growing adsorbate film: {\it deposition}, {\it desorption} and
{\it surface diffusion} (see fig. \ref{MBE_FIG}) of adsorbate particles 
\cite{Barabasi:1995:FCS,Zangwill:1992:PS,Pimpinelli:1998:PCG}.
\subsection{Deposition}
From the vapor beam adsorbate particles arrive with a flux $F$ - normally measured in monolayers per second ($ML/s$) - 
at a random position on the surface of the crystal. Since the
temperature of the effusion cell is typically higher than the substrate temperature, 
the energy of the newly deposited 
particle is higher than that of 
particles already in thermal contact with the surface. This may result in a increased mobility
of the arriving particles, which is accommodated in some simulation models by effective rules like {\it incorporation} or
{\it downhill funneling} (see e.g. \cite{Ahr:2002:SPE}). In the latter case 
the initial motion of the particle is biased to energetically 
favorable 
sites with a high coordination number where the particle then stick to the surface.
After the deposition process has ended 
the particle is considered to be in a chemisorbed state at the bottom of a potential well - the so--called 
{\it binding state} -
thermalized at the crystal temperature.

In the following we consider only the chemisorbed state and 
neglect physisorbed states, in which the adsorbed particle is held at the surface by much 
weaker Van der Waals forces. 
This can lead to an enhanced mobility of the adatoms and is addressed in detail in recent publications 
\cite{Ahr:2002:SPE,Volkmann:Phd}. 
\subsection{Desorption}
A competing effect of deposition is the desorption of a bound adsorbate particle. The probability of 
desorption depends on the
depth of the potential well - the so--called {\it binding energy} $E_b$ - and the temperature $T$ of the crystal surface. 
Thermal fluctuations tend to drive the adsorbate particle back to the gas phase with a rate 
proportional to $\exp(-E_b/kT)$. 
The desorption rate $R_{d}$ is given according to an {\it Arrhenius} law \cite{Zangwill:1992:PS}:
\begin{equation}
\label{desorption}
 R_{d}=\nu_0 e^{-\frac{E_b}{kT}}.
\end{equation}
Here $\nu_0$ is the attempt frequency which is on the order of magnitude of the {\it Debye} frequency of the crystal. 
The binding energy $E_b$ depends
both on the specific particle types and the local geometry of the surface {\it i.e.} the coordination number of 
the binding site. In heteroepitaxial growth $E_b$ depends also on the misfit between adsorbate and substrate particles 
and varies locally depending on the local strain.

Due to
the high binding energies (typically $E_b \approx 2.5 eV$) the desorption rate is small compared to the rates 
of deposition 
and diffusion. Thus for many material systems under typical MBE conditions 
the desorption of chemisorbed adsorbate particles is negligible. 
  
\subsection{Surface diffusion}
\label{KAP-1:diffusion} 
\begin{figure}[hbt]
\centerline{
\begin{minipage}{0.59 \textwidth}
  \epsfxsize= 0.97\textwidth
  \epsffile{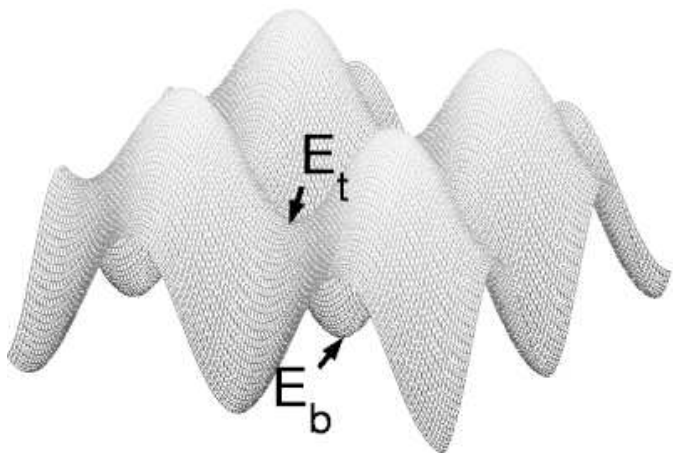}
\end{minipage}
\begin{minipage}{0.39 \textwidth}
\caption{PES for a test particle on a plain surface. The particles of the crystal interact via a $3d$ cubic 
Lennard--Jones potential (cf. \ref{AP-1}).}
\label{PES_3d}
\end{minipage}
}
\end{figure}

The surface diffusion of adsorbate particles is described by an activated process where particles {\it jump} 
laterally along the 
surface from one
binding place to another. In this process they have to overcome an energy barrier, the so--called activation energy $E_a$. 
At the transition state - the transition energy $E_t$ - the particle is less well bound
than at the binding site but still far from zero of energy (corresponding to desorption).
The activation energy is therefore given by 
\begin{equation}
\label{E_A}
 E_a=E_t-E_b
\end{equation}
and again according to an {\it Arrhenius} law the rate for surface diffusion $R$ results to
\begin{equation}
\label{diffusion}
  R=\nu_0 e^{-\frac{E_a}{kT}}.
\end{equation}
In the multi--dimensional case ($d>2$) the transition state corresponds to a first order 
saddle point in the {\it potential energy surface} (PES) \cite{Spjut:1994:CSS,Kratzer:2001:SKT} with a maximum
in the direction of diffusion and minimum along all other coordinates 
of the crystal surface \cite{Jensen:1999:ICC}.
The binding site is represented by a minimum in the energy landscape (see fig. \ref{PES_3d}). 
\subsubsection{Potential energy surface}
\label{KAP-1_PES}
Consider a crystal containing of $n-1$ particles interacting with each other via a pair--potential.
In order to compute the PES of the crystal surface a test particle $n$ is moved in small steps across the surface.
After each step the total potential energy of the $n$--particle system is minimized by variation of all particle 
coordinates plus the test particle's coordinate perpendicular to the surface. 
Potential energy and coordinates of particle $n$ result then in the PES. This so--called {\it Molecular Static} 
method for the calculation of the energy surface is explained in detail in chapter \ref{KAP-2}.
The energy 
surface represents the {\it stable} (minima) and {\it metastable} (saddle points) sites for an adatom at
crystal temperature $T=0$
(disregarding the vibrations of the lattice). Figure \ref{PES_3d} shows such a PES for a particle on a plain $2d$ surface.
\subsubsection{Diffusion bias}
\label{KAP-1_DB}
\begin{figure}[hbt]
\centerline{
\begin{minipage}{0.60 \textwidth}
  \epsfxsize= 0.99\textwidth
  \epsffile{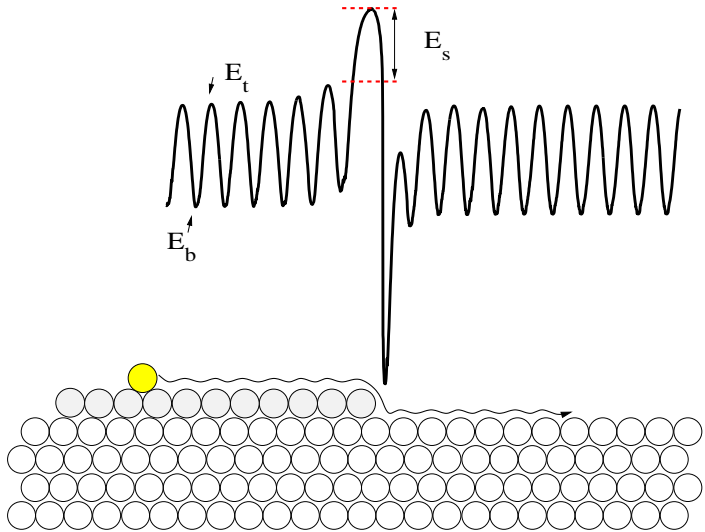}
\end{minipage}
\hfill
\begin{minipage}{0.35 \textwidth}
\caption{PES for a test particle on the surface of a $2d$ Lennard--Jones crystal. The misfit is $\varepsilon =5\%$.
Note that the the diffusion barriers rise from the island center to the edge.}
\label{PES_2d}
\end{minipage}}
\end{figure}

Figure \ref{PES_2d} shows the PES of a $2d$ {\it Lennard--Jones} crystal, where due to the isotropy of the pair--potential 
the particles arrange into a triangular lattice. 
On the substrate with lattice constant $a_s$ a  monolayer island of adsorbate particles
(lattice constant $a_a$) is placed. The test particle moves from the middle of the island towards the right side of 
the crystal.
As one would expect for a triangular lattice the minima of the PES correspond to the sites in the middle between 
two particles in the underlying
layer (bridge sites). 
As a $2d$ crystal is considered here the transition states are represented by maxima in the PES and coincide 
with the vertices of underlying
particles (top sites). 
\\
When the test particle crosses the edge of the island it has to overcome an increased energy barrier $E_s$. 
This so--called {\it Schwoebel} barrier \cite{Schwoebel:1966:SMC,Schwoebel:1969:SMC} results from the 
reduced number of binding partners at the rim of the island. The probability 
of being reflected at the edges is therefore higher than the one for jumping from the island.
On the other hand a test particle moving towards the bottom of a monolayer islands has to 
overcome a decreased barrier in order to reach the island step. 
Within the range of the interaction potential between the particles of the system 
islands act attractive to surrounding particles. Because of the increased 
number of binding partners a particle  
located at the step edge is much stronger bound than on a flat surface.
\par
Another phenomenon of biased diffusion on top of islands is especially pronounced in heteroepitaxial growth, 
when the misfit  
\begin{equation}
\label{Misfit}
 \varepsilon =\ \frac{a_a-a_s}{a_s}. 
\end{equation}
between substrate and adsorbate becomes nonzero. In order to minimize the potential energy of the system,
adsorbate islands 
always try to achieve
their favored lattice constant $a_a$. Because of the higher mobility this works most efficiently for particles at 
the edges of islands, whereas
the center of an island is still strained and the lattice constant there is closer to $a_s$.
As a consequence of the inhomogeneous relaxation the binding and transition energies for diffusion depend both on 
the position of the adatom on the island and the misfit between adsorbate and substrate. The diffusion close to the 
edge is different from 
that near the island center \cite{Schroeder:1997:DSS}. 
For negative misfit ($\varepsilon <0$) this leads to a diffusion current from the island center towards the edges.
This can e.g. result in a reduction of
the next layer nucleation rate. On the other hand in the case of positive misfit ($\varepsilon >0$) the diffusion 
is biased to the island center and
next layer nucleation can be enhanced. Biased diffusion from the island edge to the center can also
be concluded from figure 
\ref{PES_2d}, where the misfit between adsorbate
and substrate is $\varepsilon =5\%$.
\par
In systems with pair--potentials such a behavior of the diffusion barriers is a general phenomenon \cite{Schroeder:1997:DSS} 
and is observed for all types of pair--potentials used in this work: if the diffusing particle {\it feels} 
a smaller underlying lattice constant than its natural one (i.e. diffusion on a {\it compressive} strained crystal) 
the diffusion barrier is decreased and diffusion becomes faster. 
For diffusion on a {\it tensile} strained crystal the situation is 
contrariwise and the activation energy is increased. This can be understood in an intuitive way: compression 
of the lattice moves the diffusing particle a bit away from the surface. 
For that reason the particle experiences a less undulated PES.
For the extreme situation of a very large compressive strain the underlying crystal can be viewed as a continuous film with 
no discrete binding sites and the diffusion barriers therefore vanish. In the limit of large tensile strain diffusion is
equivalent to breaking a pair of atoms and build a new one resulting in a diffusion barrier equal to the pair binding 
energy. As shown in \cite{Schroeder:1997:DSS}  for the Lennard--Jones potential the dependence between activation energy and
strain is linear, at least for moderate values of $\varepsilon$. Given real materials metallic systems show
the same trend of the strain dependence. In case of semiconductors the strain dependence of the diffusion barriers 
cannot be explained that easily \cite{Cherepanov:2002:ISD}.
For example first--principles calculations for the In/GaAs(001) surface showed that the diffusion barrier has a 
non--monotonic strain dependence with a maximum at compressive strain values \cite{Penev:2001:ESS}.
\subsection{Exchange diffusion}
Although the above described hopping diffusion is the intuitive way one would think of atoms moving over an
surface, there is another important diffusion mechanism.  
Especially in the case of  face--centered--cubic (fcc) metals the so--called exchange diffusion 
becomes relevant. In the latter, the
adatom takes the place of a lattice atom whereas the displaced atom becomes the new adatom and 
continues the diffusion \cite{Wrigley:1980:SDA}.
This mechanism is most likely for diffusion over descending steps where the smaller number of neighboring atoms
makes the displacement of an atom from the step rather easy (see chapter \ref{KAP-2}).
\section{Strain relaxation mechanisms}
 \label{KAP-1-RELAX}
In the following the bulk lattice constant of the substrate and the adsorbate will be denoted as $a_s$ and $a_a$.
The misfit between the substrate and
the adsorbate film is given according to equation (\ref{Misfit}).
\par
If the misfit is not too high ($\left| \varepsilon \right| << 1$) the adsorbate is
coherent with the substrate during the early stages of growth. 
In this state the crystal topology is
that of a perfect crystal, {\it i.e.} each particle has the same coordination
number and its nearest and next--nearest neighbors form the same 
geometrical figure with only slightly modified distances \cite{Politi:2000:ICG,Pimpinelli:1998:PCG}.
\par
As the thickness $h$ of the adsorbate film increases the elastic
energy of the film rises and there are two possible relaxation mechanisms: the  
introduction of {\it misfit dislocations} in the adsorbate layer and the formation 
of three--dimensional adsorbate island.
\subsection{Misfit dislocations}
\label{KAP-1-Dislocations} 
An important strain relaxation mechanism which is often ineluctable for large adsorbate layers in large misfit 
heteroepitaxy is
the formation of dislocations, where the strain energy is released by plastic deformation.
In this case, if the absolute value of the misfit between substrate and adsorbate 
$\left| \varepsilon \right|$ 
is sufficiently small \cite{Politi:2000:ICG}, the crystal topology is only weakly perturbed in large
domains separated by lines.
Along these lines, called {\it misfit dislocations} the perturbation is large.
The adsorbate film thickness
at which dislocations first occur is known as critical thickness $h_c^d$
and qualitatively increases with decreasing values of $\left| \varepsilon \right|$.
\par
The existence of a dislocation is indicated by a non--vanishing dislocation--dis\-place\-ment vector, the so--called
{\it Burgers vector} $\v{b}$ \cite{Hirth:1968:TD}. As a result of {\it continuous elasticity theory} 
\cite{Matthews:1974:DEM,Matthews:1975:DAA,Politi:2000:ICG} only 
the component of $\v{b}$ parallel to the substrate/adsorbate interface contributes to the relaxation of strain.
With the help of the Burgers vector dislocations can be categorized in {\it climb} and {\it glide} dislocations.
\begin{figure}[h]
\centerline{
\begin{minipage}{0.33 \textwidth}
\epsfxsize= 0.95\textwidth
  \epsffile{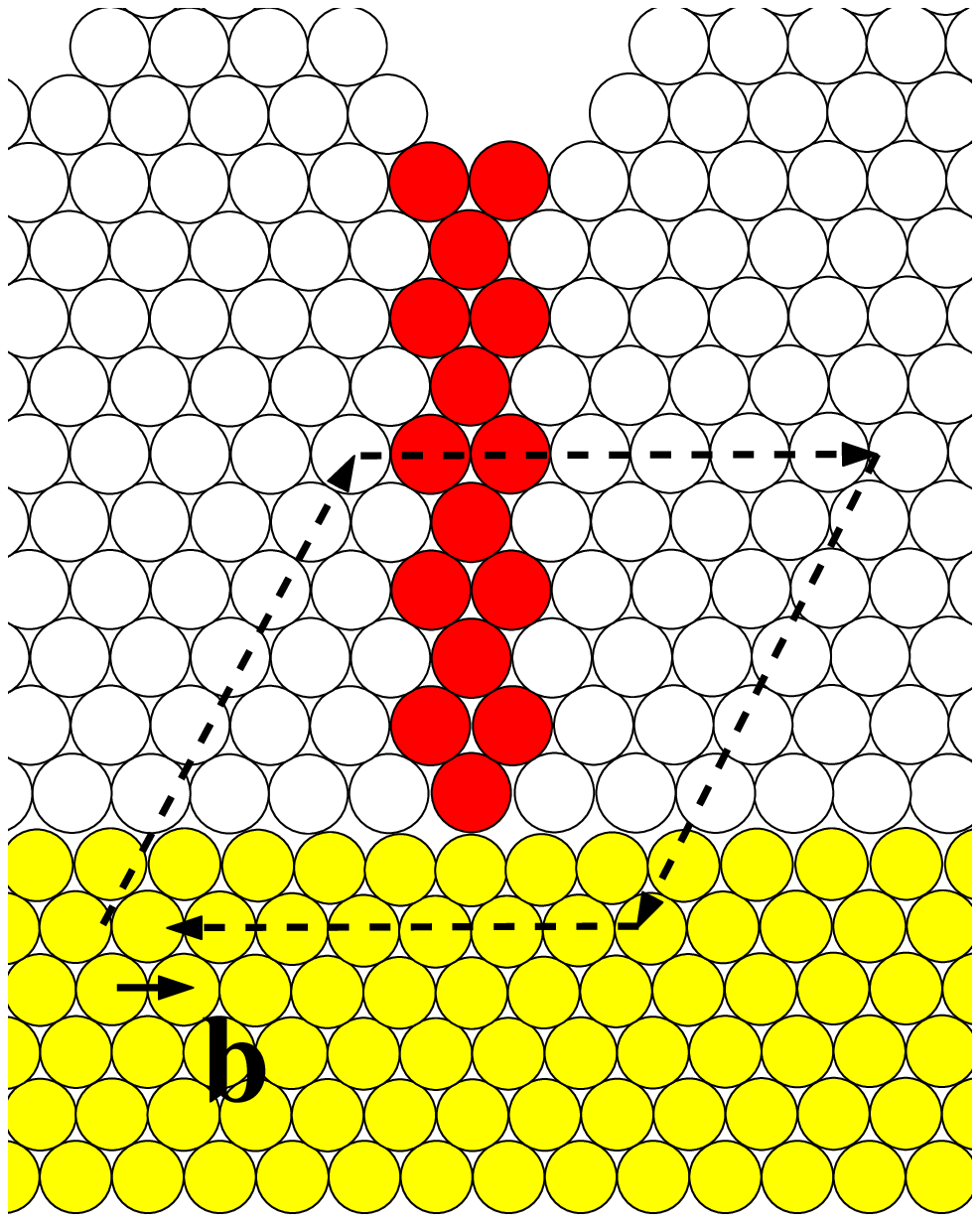}
\centerline{(a) $\varepsilon = 10\%$}
\end{minipage}
\begin{minipage}{0.33 \textwidth}
\epsfxsize= 0.95\textwidth
  \epsffile{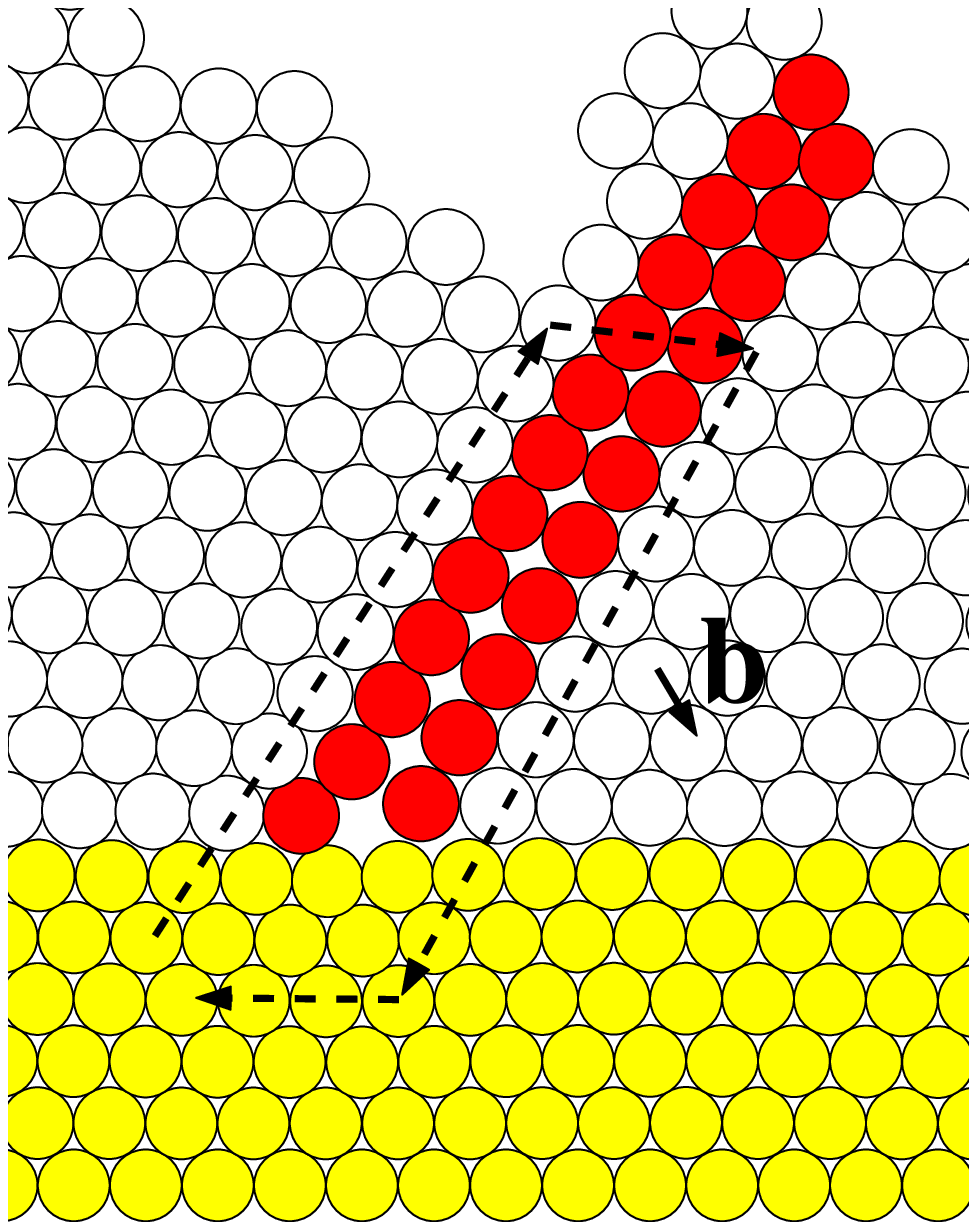}
\centerline{(b) $\varepsilon = 5\%$}
\end{minipage}
\begin{minipage}{0.33 \textwidth}
\epsfxsize= 0.95\textwidth
  \epsffile{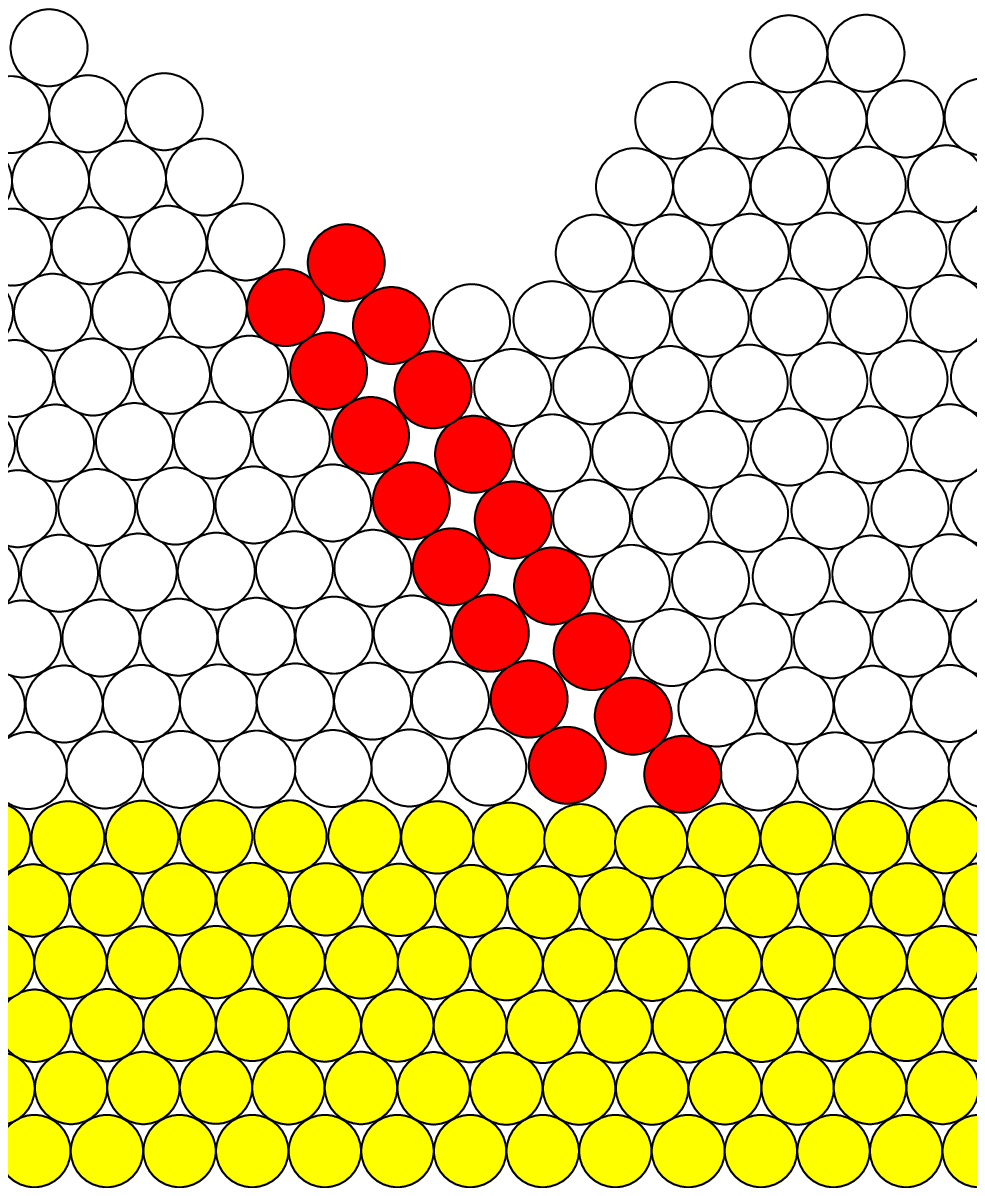}
\centerline{(c) $\varepsilon = 6\%$}
\end{minipage}}
  \caption{Sections of crystals obtained in simulations with perfect climb (a) and glide (b) dislocations.
        The dashed arrows show the Burgers {\it circuit}, which is drawn to determine the dislocation 
        Burgers vector $\v{b}$.
        Panel (c) shows a partial glide dislocation with an partial atomic step at the surface.}
  \label{Climb_Glide}
\end{figure}
Climb dislocations (fig. \ref{Climb_Glide}(a)) - with a Burgers vector parallel to the interface - relax the 
elastic energy best but the system has 
to overcome a high activation energy to create them (see e.g. \cite{Trushin:2003:EAM}). 
The easier to form glide dislocations (fig. \ref{Climb_Glide}(b))
have a component of $\v{b}$ vertical to 
the interface, which does not contribute to the relaxation.
\par
If the crystal topology is only perturbed near the dislocation line and
far from the interface the topology of the crystal is the same as in the coherent state
the Burgers vector is a lattice vector and the dislocation is called {\it perfect} (fig. \ref{Climb_Glide}(a),(b)). 
Otherwise if $h$ is not large enough {\it partial}
dislocations are formed (fig. \ref{Climb_Glide}(c)), characterized by a Burgers vector which is a rational 
fraction of the lattice vector.
In this case lattice planes have a discontinuity when crossing the glide plane and a partial atomic step 
appears on the crystal surface.
\par
\subsection{Island formation}
Mismatched epitaxial films relax their strain not only by the introduction of misfit dislocations, 
but also in an elastic way by deformation of
the surface or the so--called $2d-3d$ transition.
Historically, three growth modes are distinguished in heteroepitaxial growth: the {\it Frank--Van der Merwe}
(FM), the {\it Volmer--Weber} (VW) and the {\it Stranski--Krastanov} (SK) type of growth.
Which of these growth modes in thermal equilibrium occurs depends upon the relative magnitudes of the surface energies
$\gamma_s$, $\gamma_a$ and $\gamma_i \left( h \right)$ of the substrate, the adsorbate film and the interfacial 
energy, respectively.
$\gamma_s$ and $\gamma_a$ are the values for the semi--infinite crystals. 
The strain energy depending on the thickness of the adsorbate film $h$ has been absorbed in $\gamma_i$ here 
\cite{Bauer:1986:SGC,Chambers:2000:EGP}. 
\subsubsection{Frank--Van der Merwe growth mode}
In the FM growth mode the adsorbate forms a flat film on the substrate in a layer--by--layer way.
It occurs when
\begin{equation}
\label{FM_EQU}
 \Delta\gamma =\gamma_a + \gamma_i \left(h \right) - \gamma_s \leq 0
\end{equation}
for all film thicknesses $h$.
It is observed for several material systems with small misfits $\left| \varepsilon \right| < 2\%$ at low deposition 
fluxes and
high temperatures, where $3d$ nucleation is suppressed and adsorbate particles are mobile enough to reach surface steps. 
Since the strain in the adsorbate
film rises with increasing thickness $h$ the FM growth mode is metastable \cite{Seifert:1996:SGQ}. It can be relaxed
mainly by the introduction of {\it misfit dislocations} or the deformation of the surface,
known as {\it Asaro--Tiller--Grinfeld} instability \cite{Politi:2000:ICG}. 
\par
In the latter case a weak perturbation of the adsorbate surface leads to a quasi--periodic modulation:
For instance, in a heteroepitaxial system with a small positive misfit the stress is relaxed at the peaks of a
weak surface modulation,  because the film is tensed there and the adsorbate can get closer to 
its own lattice constant $a_a$.
On the other hand stress is concentrated at the valley, where the adsorbate film is under compression.
\begin{figure}[hbt]
\centerline{
\begin{minipage}{0.49 \textwidth}
\epsfxsize= 0.97\textwidth
  \epsffile{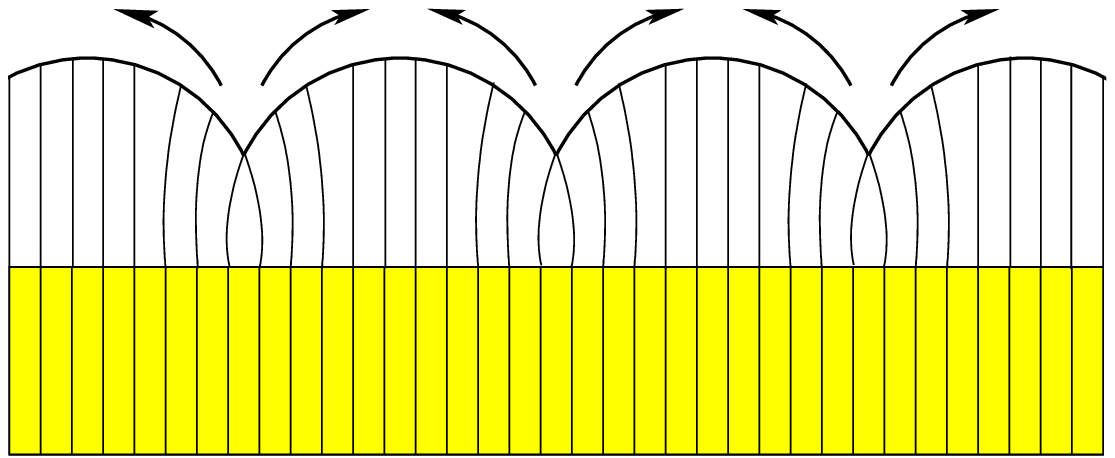}
\end{minipage}
\begin{minipage}{0.49 \textwidth}
  \caption{Strain--induced epilayer roughening: strain is reduced at the peak of the modulation and increased at the 
troughs.
        The arrows symbolize the direction of surface diffusion.}
\label{FM_FIG}
\end{minipage}
}
\end{figure}
This difference in strain energy density causes mass transport by surface diffusion from high to low strained 
regions (see fig. 
\ref{FM_FIG}) and 
leads to a quasi--periodic three--dimensional modulation of the surface of a rather thick adsorbate 
film \cite{Hull:2000:TFH}.
For instance, this thickness is several hundreds
of monolayers for Si$_{0.84}$Ge$_{0.16}$ on Si(001) with a misfit of  $\varepsilon \approx 0.6\%$ \cite{Dutartre:1994:DFS}.
\subsubsection{Stranski--Krastanov growth mode}
\begin{figure}[hbt]
\centerline{
\begin{minipage}{0.49 \textwidth}
\epsfxsize= 0.97\textwidth
  \epsffile{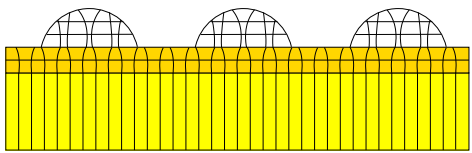}
\end{minipage}
\begin{minipage}{0.49 \textwidth}
  \caption{Schematic representation of the Stranski--Krastanov growth mode, where $3d$ adsorbate islands form
 on an adsorbate 
wetting--layer.}
\label{SK_FIG}
\end{minipage}
}
\end{figure}

If the FM condition (\ref{FM_EQU}) is only satisfied for a small number of adsorbate layers SK growth is energetically 
possible.
In this growth mode {\it coherent} three--dimensional ($3d$) islands form on a so--called {\it wetting--layer} 
of coherently growing
adsorbate (see fig. \ref{SK_FIG}). Typically the thickness of this wetting--layer is between one and four monolayers.
The formation of SK islands can be understood as a phase transition: the $2d$ wetting--layer is metastable 
and grows up to
a supercritical thickness $h_c^*$, when more stable islands begin to form. Further growth of these islands is 
fed both by capturing 
of newly deposited adsorbate particles and the decomposition of the 
supercritical layer. After the transition, the thickness of the wetting--layer decreases to a
stationary value, commonly denoted by $h_c$.
It is important to stress, that this process even takes place without any further deposition of particles 
\cite{Landin:1999:OSI}.
\par
The exact mechanism of island formation is still under discussion.
It seems clear, that monolayer ($2d$) islands - located on the wetting--layer - play an important role 
as precursors for the formation of $3d$ islands. One theoretical model is that at a critical size
these $2d$ islands become unstable. Because of the strain the more weakly bound edge atoms jump to the next 
island level and build up
the second island layer. This process is repeated in the transformation from bi-- to three--layer islands, and so on 
\cite{Stoyanov:1982:TET,Korutcheva:2000:CSK}. It is known as the {\it spontaneous} $2d-3d$ transition which results in the 
formation of {\it self--assembled} quantum dots.
\par
\begin{figure}[hbt]
\centerline{
\begin{minipage}{0.55 \textwidth}
\epsfxsize= 0.97\textwidth
  \epsffile{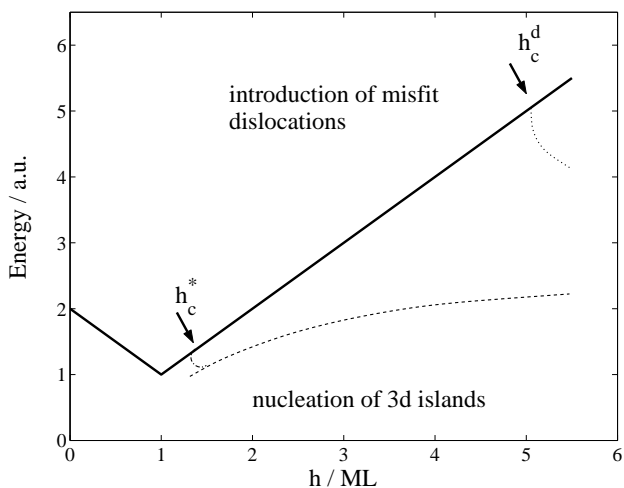}
\end{minipage}
\begin{minipage}{0.43 \textwidth}
  \caption{Total energy of a strained heteroepitaxial system versus adsorbate layer thickness $h$. $3d$ island nucleation
starts at $h_c^*$, dislocations nucleate at $h_c^d$ \cite{Seifert:1996:SGQ}.}
\label{SK_vs_MD}
\end{minipage}
}
\end{figure}

This transition involves an activation barrier $E_{2d-3d}$ which the system has to overcome. Otherwise the film
thickness may increase until the introduction of dislocations becomes favorable.
Figure \ref{SK_vs_MD} gives schematically the energetic situation for the two cases of SK and dislocation--relaxed FM 
growth.
First the total energy of the system 
decreases due to the energy contribution from the substrate/adsorbate interface 
until the substrate is covered by a complete monolayer of adsorbate. Then the elastic energy in the 
strained adsorbate film increases linearly with each film layer. If the systems does not manage to overcome the 
barrier $E_{2d-3d}$ FM growth is maintained until the introduction of misfit dislocations at a film thickness $h_c^d$. 
If on the other hand growth conditions favor the SK mode $3d$ island nucleation
starts at $h_c^*$.
This SK growth is observed in various strained heteroepitaxial systems of the IV--IV, III--V and II--VI families 
of semiconductors
(for a review see \cite{Seifert:1996:SGQ}). In all these cases the misfit is positive and quite large 
($2\% \leq \varepsilon \leq 7\%$).

\subsubsection{Volmer--Weber growth mode}
\begin{figure}[hbt]
\centerline{
\begin{minipage}{0.49 \textwidth}
\epsfxsize= 0.97\textwidth
  \epsffile{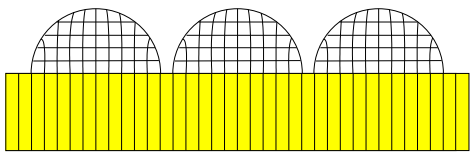}
\end{minipage}
\begin{minipage}{0.49 \textwidth}
  \caption{Schematic representation of the Volmer--Weber growth mode, where $3d$ adsorbate islands form 
directly on the substrate.}
\label{VW_FIG}
\end{minipage}
}
\end{figure}

In the case of VM growth equation (\ref{FM_EQU}) is not fulfilled and $3d$ adsorbate islands form directly on the 
substrate (see fig. \ref{VW_FIG}).
Due to the fact that there is no technological application of incoherent islands - where misfit dislocations are 
introduced to relieve the
strain energy - coherent VM growth is of special interest.
In this case the base of the islands is still constrained by the substrate but the adsorbate can reach its own 
lattice constant $a_a$ 
on the top, the sides and to some degree in the island center. The mechanism of island formation should be 
similar to the process of
$2d-3d$ transition described above.
\par
Coherent VW growth is observed mainly in heteroepitaxial systems with large positive misfit,
e.g. for ZnTe/ZnSe \cite{Kuo:2002:FSA}, Mn/Si(111) \cite{Evans:1995:EGM} or Si/Ge(111) \cite{Raviswaran:2001:ECI}.
\subsection{Surface confined alloying}
\label{KAP-1_ALLOY}
A further strain relaxation mechanism, which arises generically in systems dominated by atomic size mismatch
is surface confined alloying of the materials \cite{Tersoff:1995:SCA}. Surface alloying is observed for 
both cases: two component systems with mixing of adsorbate and substrate particles (e.g. Na/Al(111), K/Al(111) 
\cite{Neugebauer:1993:MIF,Stampfl:1992:ISM}, Au/Ni(110) \cite{Nielsen:1993:IGA}, Ag/Pt(111) \cite{Roder:1993:MCM}, 
Sb/Ag(111) \cite{Oppo:1993:TAS}) and three component growth with alloying of two adsorbate species on a different substrate
(e.g. CoAg/Ru(0001) \cite{Hwang:1996:CIS,Hwang:1997:STM,Thayer:2001:RST,Thayer:2002:LSS},  
CoAg/Mo(110) \cite{Tober:1998:SAL}, 
FeAg/Mo(110) \cite{Tober:1998:SAL},  
AgCu/Ru(0001) \cite{Stevens:1995:SSA},   
PdAu/Ru(0001) \cite{Sadigh:1999:SRO}).

In principle, e.g. for the case of a binary system, surface alloying is again understood in terms of surface and
interface energies \cite{Roder:1993:MCM}. If the interfacial energy $\gamma_i \leq 0$ adsorbate and substrate can lower 
their energy by intermixing. Otherwise the adsorbate material will segregate.
\begin{figure}[hbt]
\centerline{
\begin{minipage}{0.49 \textwidth}
\epsfxsize= 0.97\textwidth
  \epsffile{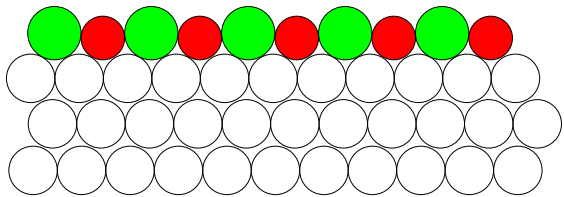}
\end{minipage}
\begin{minipage}{0.49 \textwidth}
  \caption{Schematic representation of surface confined alloying of two adsorbate species (dark and light gray)
on a different substrate.}
\label{ALLOY}
\end{minipage}
}
  
\end{figure}

Also for the three component system the strain relaxation due to  surface alloying can be understood intuitively.
Consider the situation of figure \ref{ALLOY}, with two adsorbate species A, B which imply a misfit of the same 
absolute value but opposite sign with the substrate. As long as no further difference between the adsorbate 
particles exists and in particular the binding $E^{AB}$ between the two species is the same as for 
the A--A and the B--B interaction, an alternating arrangement of both adsorbate types 
is likely to be the energetically most favorable state.
A weaker A--B interaction can complicate the situation 
and causes a competition between 
alloying and the introduction of misfit dislocations as the preferred relaxation mechanism. 

\section{Models of heteroepitaxial growth}
Various methods have been proposed recently to model strain relaxation and growth of atomic 
mismatched systems by means of computer calculations and simulations. Due to the steady growth of computer 
power they became an important tool for analyzing and understanding microscopic processes and their effects. 
In this section we give a short overview of methods for the computational treatment of, in particular,  
heteroepitaxial growth.
\subsection{Density--functional--theory}
For calculations and simulations of growth processes it is desirable to describe elementary processes, like 
the interaction between atoms or molecules, on a high quality level i.e. without making empirical assumptions.
In contrast to classical approaches describing the atom--atom interaction using empirical potentials or simple 
bond--counting rules, density--functional--theory (DFT) (see e.g. \cite{Kratzer:2001:SKT})
accounts for the quantum mechanical nature of the electrons.

For DFT calculations only very few assumptions about the 
electronic structure of a poly--atomic system have to be made: an approximate density functional is used to solve 
separate, Schr{\"o}dinger--like equations for all electrons. Electron many particle (exchange or correlation) 
effects are added in terms of an additional potential.

With respect to the high computational effort of the method it is not possible to use DFT for in--situ calculation
of e.g. diffusion barriers in Monte Carlo simulations. Only the calculation of single events like e.g. the adsorption
of an $As_2$ molecule on a $GaAs$ surface at zero temperature and pressure is within the scope of the method 
right now \cite{Kratzer:2001:SKT}.

However, DFT methods were used with some success for energy calculations of exemplary situations like found in 
heteroepitaxial growth. For example in \cite{Scheffler:2000:ITS,Moll:1998:ISS} the total energy of a system 
consisting of a pseudomorphic
InAs layer located on a GaAs substrate was compared to the energy of a system with an InAs island of certain 
shape and size placed on a thinner InAs film. DFT methods were used for the calculation of surface energies, 
whereas the elastic energy was treated by means of classical elastic continuum equations.
The investigation yielded results about energetically most favorable island shapes and sizes and predicted a 
non--vanishing wetting--layer due to energetic reasons for the InAs/GaAs system.

In conclusion, though DFT is at present 
not suitable for large scale simulations of heteroepitaxial growth processes, it is an appropriate 
method for material specific calculations regarding prototype situation in atomic mismatch systems.

\subsection{Molecular Dynamics simulations}
Since the more or less exact calculations of chemical bonds between atoms or molecules are computationally far too 
demanding for the simulation of many particle systems and relevant time scales often empirical potentials are chosen.
These potentials can give the interaction strength between particles of the systems as a function of e.g. the particle 
distance and the bond directions. Material specific potentials are fitted in order to reproduce bulk properties 
like elastic constants, the vacancy formation energy, the stacking fault energy, surface energy, 
and phonon frequencies \cite{Abraham:1998:MOA}. One prominent example for empirical material specific potentials 
are the Embedded Atom Method (EAM) potentials used for the modeling of metallic systems. EAM potentials 
are calculated as a sum of pairwise interactions between the particles of the system and a many body term. 

Empirical potentials for the simulation of semiconductor properties are more complicated and 
computationally more demanding. 
An important set of potentials for semiconductor systems are e.g. the so--called Tersoff--potentials 
\cite{Abraham:1998:MOA,Tersoff:1988:NEA}. Tersoff--potentials are based on the concept of bond order: the strength of a bond 
between two atoms is not constant, but depends on the local environment. 
The basic idea is that the bond between two atoms $i$ and $j$ 
is weakened by the presence of other bonds $i-k$ also involving the atom $i$. 
The amount of weakening is determined by 
where these other bonds are placed. Hence angular terms become 
necessary for the construction of a realistic model.  

Given at least an empirical potential for the interaction between particles of the system, Molecular Dynamics (MD) 
simulations are clearly the most realistic and desirable method for the simulation of heteroepitaxial growth.
In MD simulations the time evolution of a particle system is described by integrating 
the equations of motion using the interaction potential for the computation of the forces on each particle.
MD techniques are applied with success to e.g. the investigation of dislocation formation and motion 
\cite{Dong:1998:SRM,Patriarca:2002:MMD,Bailey:2000:DMT} 
or the formation of grain boundaries \cite{Schiotz:1998:SMS} for rather large system sizes (up to $10^6$ particles).

However, MD simulation suffer generally from the restriction to short physical times on the order of 
$10^{-6}s$ or less. The simulation of crystal growth - like e.g. in the MBE environment - requires the coverage of 
seconds up to minutes.
That is because the properties of a growing crystal are ruled by rather rare thermally activated processes like 
surface diffusion jumps from one local minimum to a neighboring one. These events normally occur with an 
exponentially decreasing probability (see eq. (\ref{diffusion})) \cite{Kratzer:2001:SKT}.

MD simulations are altogether surely an adequate method for the investigation of concerted moves involving many 
particles like dislocation motions or interdiffusion processes at the interface between two material types. 
But MD methods are not suitable for the simulation of growth processes since relevant time scales are currently 
not feasible. The simulation of heteroepitaxial {\it growth} requires a further simplification with concentration 
on the relevant rare events.
\subsection{Off--lattice Monte Carlo simulations}
Monte Carlo simulations have been proved to be a powerful tool for the simulation of homoepitaxial growth
(see e.g. \cite{Ahr:2002:SPE}). Here rare events (e.g. diffusion events) are performed according to their 
probability, given by an Arrhenius law (see eq. (\ref{diffusion})).
However, for the simulation of heteroepitaxial growth it is essential to allow for continuous particle distances 
in order to account for strain effects in the crystal.

\begin{figure}[hbt]
\begin{minipage}{0.49 \textwidth}
\epsfxsize= 0.97\textwidth
  \epsffile{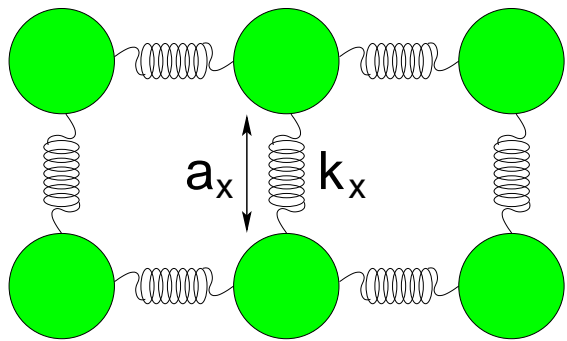}
\end{minipage}
\hfill
\begin{minipage}{0.48 \textwidth}
  \caption{Schematic representation of the ball and spring model. The elastic energy is determined by 
the spring constant $k_x$ and the natural length $a_x$.}
\label{BAS}
\end{minipage}
\end{figure}

The computationally least demanding simulation technique which fulfills this requirement is the so--called ball and
spring model \cite{Madhukar:1983:FEV,Ghaisas:1986:RSM,Orr:1992:MSI,Barabasi:1997:SAI,Khor:2000:QDS,Lam:2002:CRM}(see fig. \ref{BAS}).
Here the activation energy for a diffusion jump of a surface particle is split into bond and 
strain energy $E_a=E_{bond}-E_{strain}$. The bond energy is determined by the exact number of nearest (nn) and 
next--nearest neighbors (nnn) e.g. by a simple bond--counting rule. The strain energy for a site $i$ is
obtained by the difference of the system's elastic energies with site $i$ occupied and unoccupied.  The elastic 
energy is given by harmonic interactions ({\it springs}) between an atom and its nn and nnn. Interaction between 
substrate particles is represented by the spring constant $k_s$ and the natural length $a_s$. The adsorbate--adsorbate
interaction is given by $k_a$, $a_a=(1+\varepsilon)a_s$, accordingly.

Since the calculation of $E_{bond}$ and $E_{strain}$ requires the number and positions of 
surrounding particles (nn and nnn) of each particle this 
method does not allow for the simulation of misfit dislocations. The method also suffers form the rather rough 
description of the elastic energy by simple harmonic interactions. However, ball and spring models yield some 
interesting results on the growth of coherent islands and the different growth modes.

A step towards a more realistic treatment of the binding between particles are continuous--space Monte Carlo 
simulations \cite{Plotz:1992:MCS,Sitter:1995:MGM,Kew:1993:CSM}. The interaction between the particles of the system is here given 
by pair--potentials. A surface move of a particle is e.g. accepted with the probability $\exp{(-E_b/kT)}$, where $E_b$ 
is the binding energy of the particle due to the interaction with surrounding particles of the crystal within a 
certain range. Since - unlike in the ball and spring simulations - number and positions of surrounding particles are 
not required for the calculation of $E_b$ this method allows for the simulation of dislocations \cite{Kew:1993:CSM}.

In this work we go a different way and put forward an off--lattice Kinetic Monte Carlo (KMC) method introduced in
\cite{Schindler:1999:TAG}. In the KMC method the activation energy for a hopping diffusion jump from one local minimum 
to another is calculated with respect to a pair--potential that mediates the interaction between the particles 
of the system. The method will be explained in detail in chapter \ref{KAP-3}. 
In chapter \ref{KAP-4} we analyze the formation of misfit dislocations with this method. Chapter \ref{KAP-5} gives
investigations of the SK growth mode and in chapter \ref{KAP-6} we use the off--lattice KMC method for the simulation 
of multicomponent growth.

  \cleardoublepage
\chapter{Diffusion barriers on step edges}
\label{KAP-2} 
In this chapter we focus on the influence of the misfit between substrate and
adsorbate on the diffusion barriers across steps.

The barrier for the downward movement from the top of an island plays an 
important role for the growth mode: high barriers for downward movement can
lead to rough growth, whereas a low barrier for descending jumps favors
smoother layers at a given temperature. 

Large effort has been spent on the calculation of diffusion barriers 
for various material systems by the means of empirical potential methods 
\cite{Trushin:1997:EBS,Maca:1999:EBD,Maca:2000:EBD,Liu:1993:DBS,Villarba:1994:DMR} 
or density functional theory (DFT) \cite{Feibelman:1998:ISD,Feibelman:1999:SDA}, mostly for the case
of self--diffusion. 
But to our knowledge there is no systematical study on the influence of 
the lattice misfit between substrate and the adsorbate island on the 
downward barriers. Since the control of the surface morphology is a 
major goal in MBE techniques the influence of the misfit on barriers
for descending diffusion moves is of special interest.

By means of a {\it Molecular Static} method we study these barriers for different
misfits, island sizes and interaction potentials for the $2d$ and $3d$ case.
The difference of the barriers for hopping and exchange diffusion are of
particular interest. 
The computation of these barriers is also of some relevance for the following parts of 
this work: in the case of off--lattice simulations 
it is important to know whether the rather complicated 
concerted moves - like e.g. exchange diffusion - should be taken into account.
\section{Hopping and exchange diffusion}
\begin{figure}[hbt]
\centerline{
\begin{minipage}{0.42 \textwidth}
  \epsfxsize= 0.90\textwidth
  \epsffile{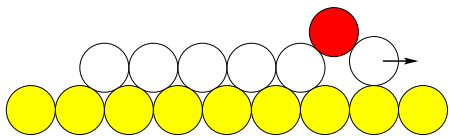}
\end{minipage}
\hfill
\begin{minipage}{0.58 \textwidth}
\caption{Schematic representation of a exchange diffusion move at a descending step 
edge for the $2d$ case. The adatom takes the place of the surface particle at
the island's edge.}
\label{EX_FIG}
\end{minipage}
}
\end{figure}

\begin{figure}[hbt]
\centerline{
\begin{minipage}{0.42 \textwidth}
  \epsfxsize= 0.90\textwidth
  \epsffile{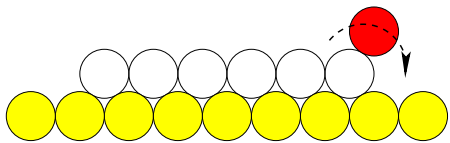}
\end{minipage}
\hfill
\begin{minipage}{0.58 \textwidth}
\caption{Schematic representation of a hopping diffusion move over the rim of an island.}
\label{HOP_FIG}
\end{minipage}
}
\end{figure}

As mentioned in chapter \ref{KAP-1} surface diffusion of an adatom can proceed by 
two different processes:
\begin{itemize}
\item {\it Hopping} diffusion: The adatom {\it jumps} from one minimum in the potential energy surface 
(PES) to another by overcoming the activation barrier $E_a=E_t-E_b$, where $E_b$ is the energy of 
the binding state and $E_t$ is the energy at the transition state.
\item {\it Exchange} diffusion: The adatom {\it replaces} a surface atom and this surface atom resumes
further diffusion \cite{Wrigley:1980:SDA}. The activation energy of the exchange process can also 
be written as $E_a=E_t-E_b$.
\end{itemize}
The latter diffusion mechanism may become particularly important at descending steps. Due to their lower
coordination number atoms at the step edge can be more easily {\it pushed} away by the adatom 
(see fig. \ref{EX_FIG}),
whereas for the case of hopping over the rim of an island in many material systems 
the adatom has to overcome an energetically unfavorable situation due to the
weaker binding during the diffusion step. This is especially pronounced in the $2d$ 
case (see fig. \ref{HOP_FIG}).
\section{Calculation of diffusion barriers}
The technique we apply here in order to compute the barriers for downward hopping and exchange diffusion at
a descending step edge is the so--called Molecular Static method (see {\it e.g.} 
\cite{Trushin:1997:EBS,Maca:1999:EBD,Maca:2000:EBD,Liu:1993:DBS,Villarba:1994:DMR,Schindler:1999:TAG}).

Consider a crystal with $n-1$ particles interacting via a pair--potential $U_{ij}$ which is 
a function of the distance $|\vec{r}_{ij}|$ between two particles $i$ and $j$ of the crystal.
An additional {\it test} particle $n$ is placed on the crystal's surface and the system's total potential
energy
\begin{equation}
\label{E_TOT}
 E_{tot}= \sum_{i=1}^{n-1} \sum_{j=i+1}^{n} U_{ij}
\end{equation}{
is minimized by the means of a {\it conjugate gradient} method \cite{Press:1992:NRC}.
The coordinates of the $n-1$ crystal particles and the test particle's coordinate 
perpendicular to the surface are varied. After reaching the minimum 
energy configuration of the system $E_{tot}$ together with the position of the test 
particle is noted and the test particle is moved by a small step $\delta$ in a given direction.
This procedure is repeated until the particle $n$ has reached a pre--determined 
position. The recorded particle coordinates together with $E_{tot}$ then result 
in the PES of the crystal (see figures \ref{PES_2d} and  
\ref{PES_3d} for the $2d$ and $3d$ case, respectively).

\section{Calculations for the two--dimensional case}
For the calculation of the downward hopping barrier in the $2d$ case the test particle is set on top 
of the island, near the step edge (see fig. \ref{HOP_FIG}) and moved towards the step edge. 
The activation energy $E_{a,hop}$ is given according to equation (\ref{E_A}) (see fig. \ref{HOP_EX_M5}).  
For the exchange diffusion move the test particle
is {\it drawn} out of the step edge and the particle on top of the island relaxes into the
resulting gap (see fig. \ref{EX_FIG}). Figure \ref{HOP_EX_M5} shows the resulting exchange barrier
$E_{a,ex}$.
\begin{figure}[hbt]
\centerline{
\begin{minipage}{0.55 \textwidth}
  \epsfxsize= 0.95\textwidth
  \epsffile{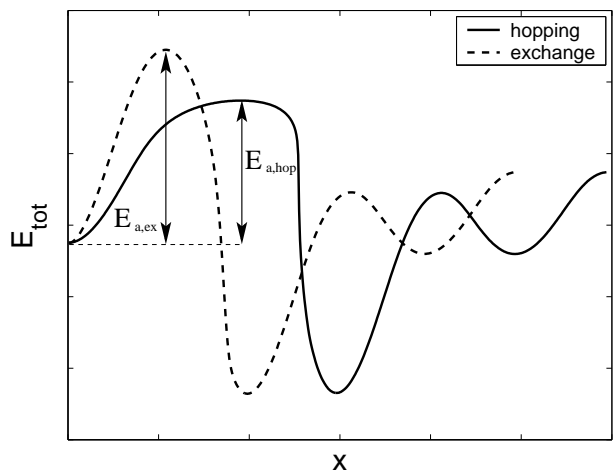}
\end{minipage}
\hfill
\begin{minipage}{0.40 \textwidth}
\caption{PES for hopping and exchange diffusion from the top of an island in two dimensions. The particles of the system  
interact via a Morse potential ($a=5.0$ and $\varepsilon=0$). Note that in case of exchange diffusion $x$ gives 
the position of the edge particle, whereas in case of hopping diffusion $x$ denotes the position of the hopping 
particle.}
\label{HOP_EX_M5}
\end{minipage}
}
\end{figure} 
\subsection{Parameters}
For the calculation of the PES we take a 15 layers thick substrate (h=15), which contains $L=70$ particles in each layer.
Periodic boundary conditions are applied in the lateral $x$--direction. To stabilize the crystal during the relaxations, 
the bottom layer
is frozen, that means particles in this layer are not allowed to relax.

On the substrate an adsorbate monolayer island is placed consisting of $3 \leq l \leq 20$ particles.
As interaction potential $U_{ij}$ for the particles of the crystal the Lennard--Jones $12,6$ potential 
\begin{equation}
\label{LJ_2}
 {U}_{ij} =\ 4 E_{ij} \left[\left(\frac{{\sigma}_{ij}}{r_{ij}}\right)^{12}
-\left(\frac{{\sigma}_{ij}}{r_{ij}}\right)^{6}\right] 
\end{equation}
and Morse potentials 
 \begin{equation}
\label{MORSE_2}
 {U}_{ij} =\ E_{ij} e^{a\left({\sigma}_{ij}-r_{ij}\right)} \left(e^{a\left({\sigma}_{ij}-r_{ij}\right)}-2 \right)
\end{equation}
for values of $a$ between $a=4.5$ and $a=7.0$ are chosen (also see appendix \ref{AP-1}).

In the following the potential depth is set to 
$E_{ij}=1.0eV$ for all types of interactions. The potential is cut off at a distance $r_{cut}=6.0 r_{0}$, which
is perfectly justified since all used pair--potentials converge fast to $U_{ij}=0$ for the particle distance 
$r_{ij} \to +\infty$. The results were also compared to preliminary calculations with $r_{cut}=4.0 r_{0}$
and $r_{cut}=12.0 r_{0}$ which yield no significant differences in the obtained barriers. 

The misfit between the substrate and the island is varied between 
$\varepsilon = -14\%$ and $\varepsilon = 14\%$ in steps of $1\%$.
For each value of the island size $l$ and the misfit 
$\varepsilon$ the PES near the step edge is calculated for both possible diffusion
mechanisms. The barriers $E_{a,hop}$ and $E_{a,ex}$ then result from the obtained energy landscapes.
\subsection{Exchange vs. hopping}
In the following we examine the difference between the barriers for exchange and hopping $E_{a,hop}-E_{a,ex}$
as functions of island size and misfit.
Figures \ref{M6_EH}(a) and \ref{M6_EH}(b) show this difference for a Morse potential ($a=6.0$) as a
function of $l$. Each curve represents a value of $\varepsilon$.
\begin{figure}[hbt]
\centerline{
\begin{minipage}{0.49 \textwidth}
  \epsfxsize= 0.95\textwidth
  \epsffile{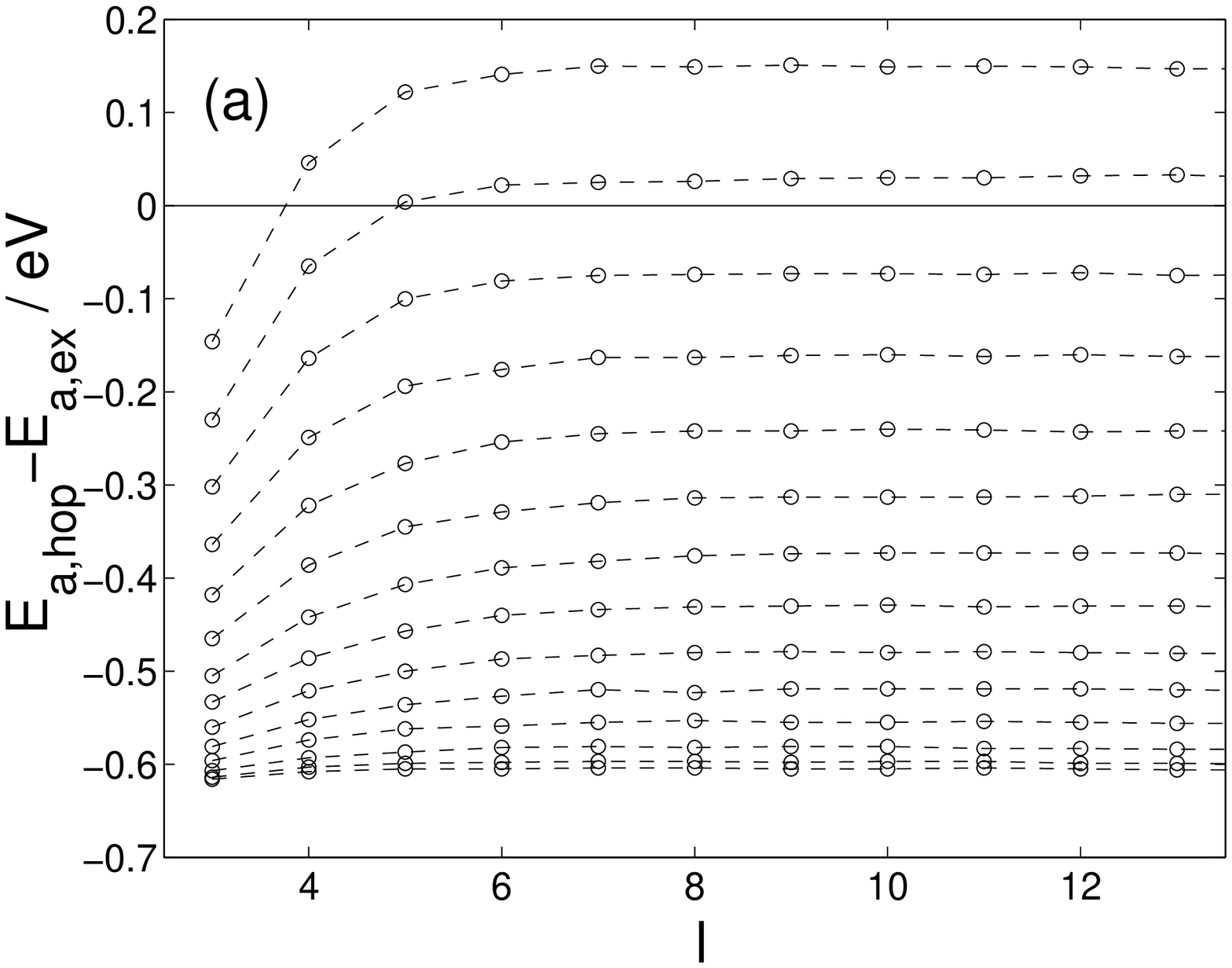}
\end{minipage}
\hfill
\begin{minipage}{0.49 \textwidth}
 \epsfxsize= 0.95\textwidth
  \epsffile{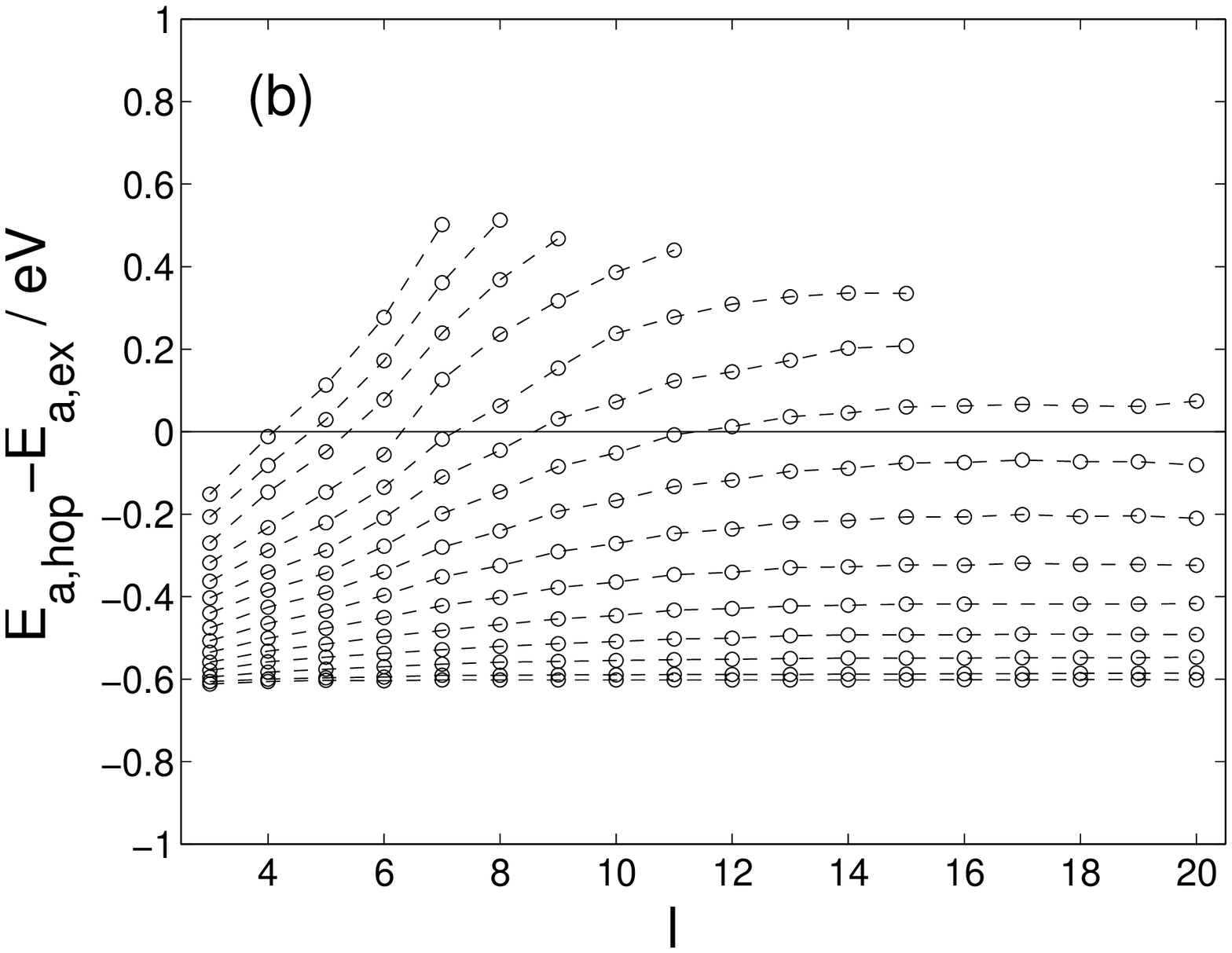}
\end{minipage}}
\caption{$E_{a,hop}-E_{a,ex}$ as a function of island size $l$ for different misfits $\varepsilon$.
(a) from down to top: $\varepsilon=-0.01$ to $\varepsilon=-0.14$ in steps of $0.01$, 
(b) from down to top: $\varepsilon=0.0$ to $\varepsilon=0.14$ in steps of $0.01$.}
\label{M6_EH}
\end{figure}  
The characteristics of these curves are typical for all used potentials. First $E_{a,hop}-E_{a,ex}$ increases
with $l$ until the difference becomes more or less constant. If $|\varepsilon|$ exceeds a critical value
$\varepsilon_c$  the difference becomes positive, which means $E_{a,ex}<E_{a,hop}$ and thus exchange diffusion
becomes likely. The absolute value of the critical misfit varies 
for positive and negative misfit. For each $|\varepsilon| \geq 
|\varepsilon_c|$ there is a critical island size $l_c$:
for island sizes $l\geq l_c$ the barrier for exchange is smaller than
the barrier for hopping.

This behavior is easy to understand if one considers the position of the particle which is drawn out of the island 
edge.
For the homoepitaxy situation ($\varepsilon=0$) all particles of an island - and therefore also the edge particles -
match perfectly the lattice structure of the substrate. But at a given island size with increasing 
misfit the edge particles move toward the top site of an underlying substrate particle and become less well bound.
The same is true for a given misfit and increasing island size:
the larger the island the less favorable is the position of the edge particles.

This affects the exchange diffusion more than the hopping diffusion: in the hopping diffusion mode the active particle
always has to overcome an energetically unfavorable top position, whereas in exchange diffusion a weakly bound
edge particle is much easier {\it kicked} out of its place.
In case of e.g. the $a=6.0$ Morse potential shown in figure \ref{M6_EH}  
at $\varepsilon=0.12$ and $l=9$ the hopping barrier has
$84\%$ of its value at $l=3$, whereas the exchange barrier dropped to $28\%$ of its value at $l=3$.

To explain the mechanism of exchange diffusion in more detail figure 
\ref{LEN12_12_8} shows a PES for a high positive misfit ($\varepsilon=0.12$) at island size $l=8$ 
in case of the Lennard--Jones $12,6$ interaction. 
In figure \ref{LEN12_12_8_ISLANDS} corresponding snapshots of the particle positions during the exchange process
are shown:
\begin{figure}[hbt]
\centerline{
\begin{minipage}{0.49 \textwidth}
  \epsfxsize= 0.95\textwidth
  \epsffile{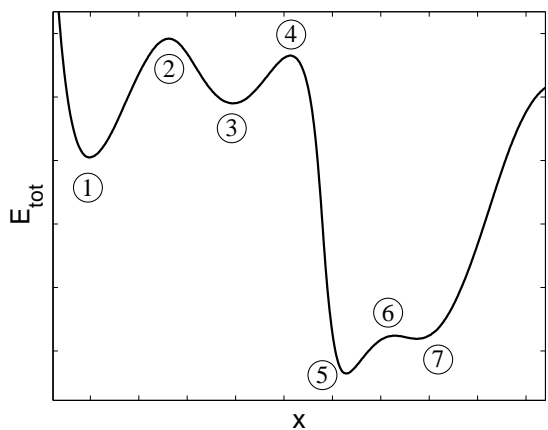}
\end{minipage}
\hfill
\begin{minipage}{0.49 \textwidth}
\caption{PES for exchange diffusion from the top of an island. The particles of the system  
interact via the Lennard--Jones $12,6$ potential. The misfit is $\varepsilon=0.12$ und island size is $l=8$.}
\label{LEN12_12_8}
\end{minipage}
}
\end{figure} 
\begin{figure}[h]
\centerline{
\begin{minipage}{0.49 \textwidth}
  \epsfxsize= 0.75\textwidth
  \epsffile{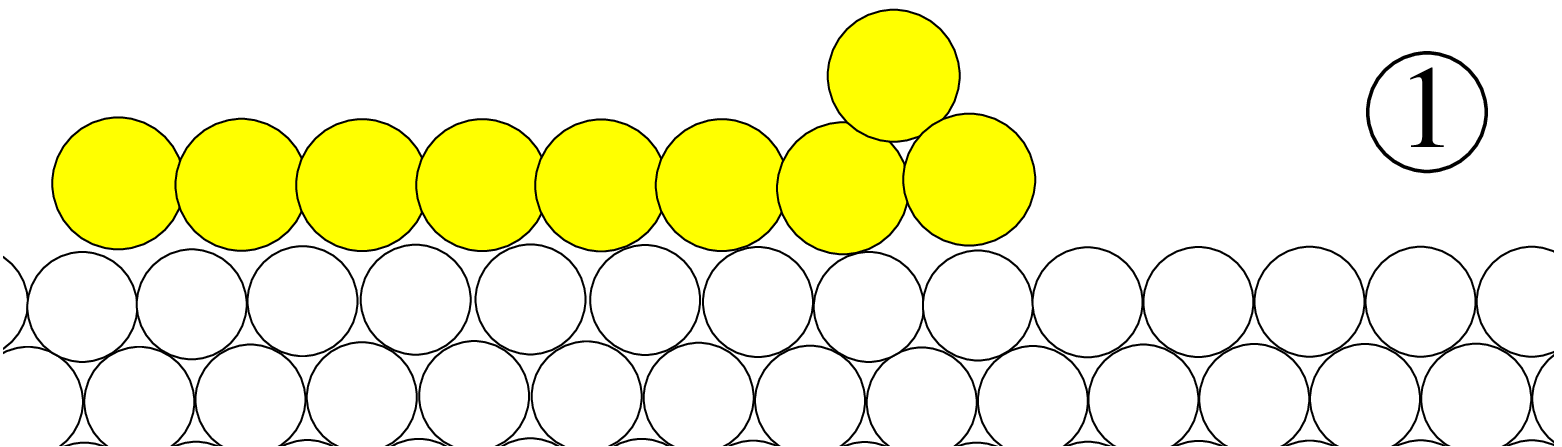}
  \epsfxsize= 0.75\textwidth
  \epsffile{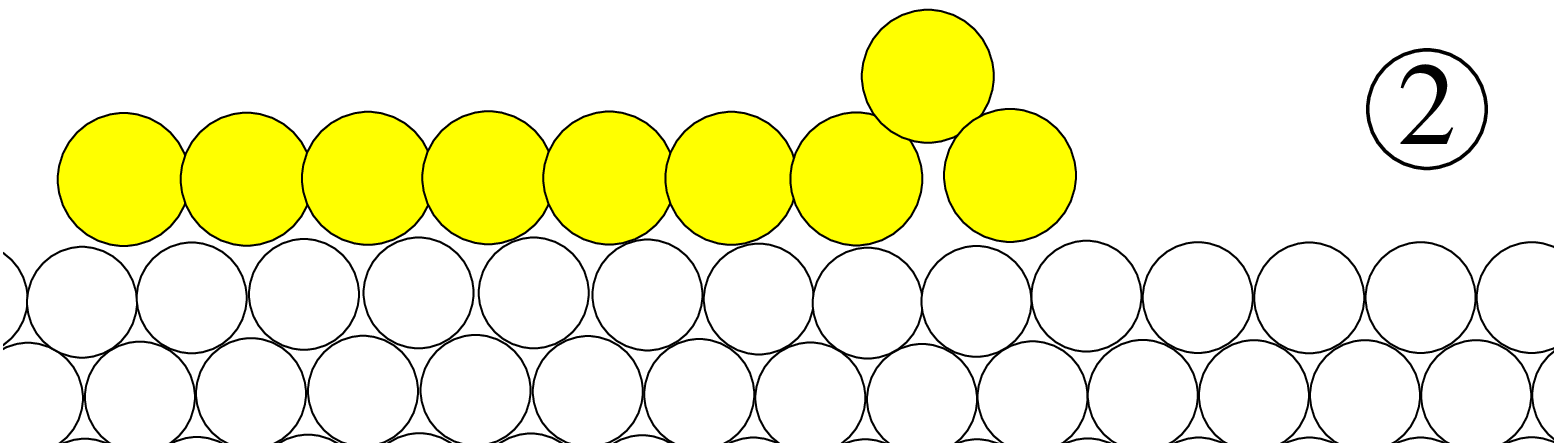}
  \epsfxsize= 0.75\textwidth
  \epsffile{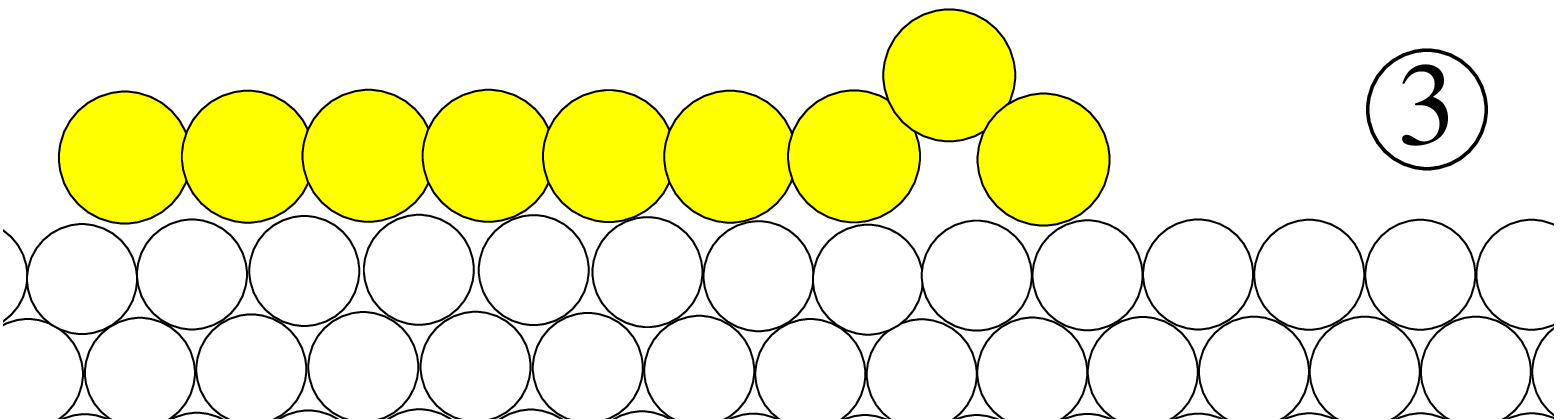}
  \epsfxsize= 0.75\textwidth
  \epsffile{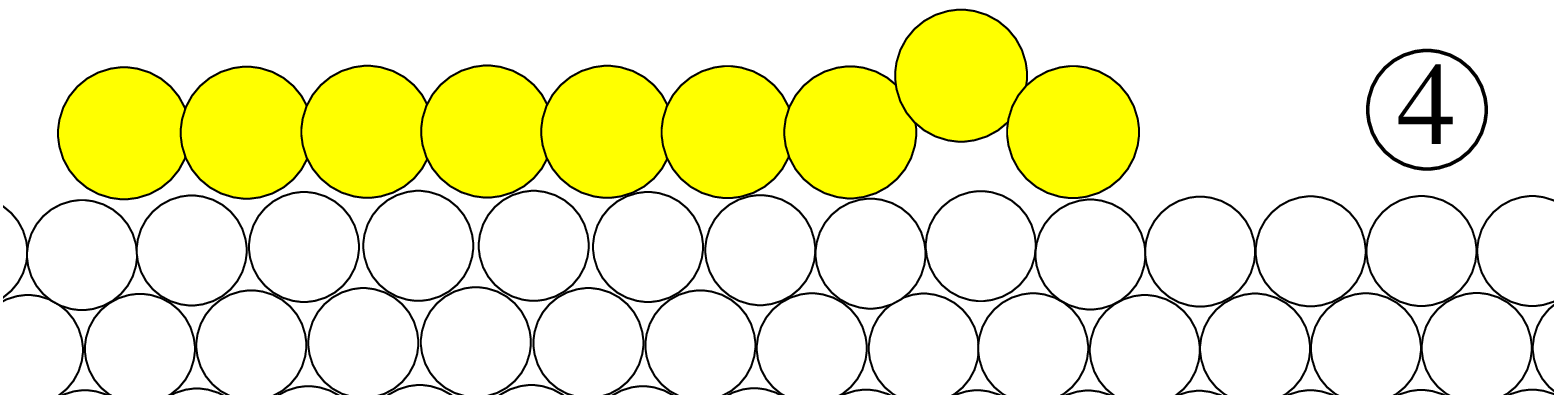}
\end{minipage}
\hfill
\begin{minipage}{0.49 \textwidth}
 \epsfxsize= 0.75\textwidth
  \epsffile{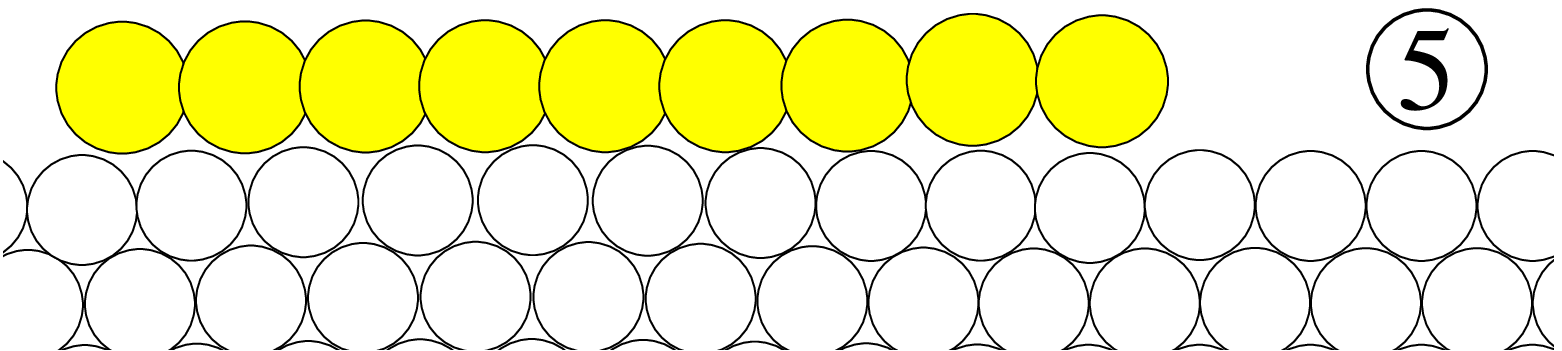}
  \epsfxsize= 0.75\textwidth
  \epsffile{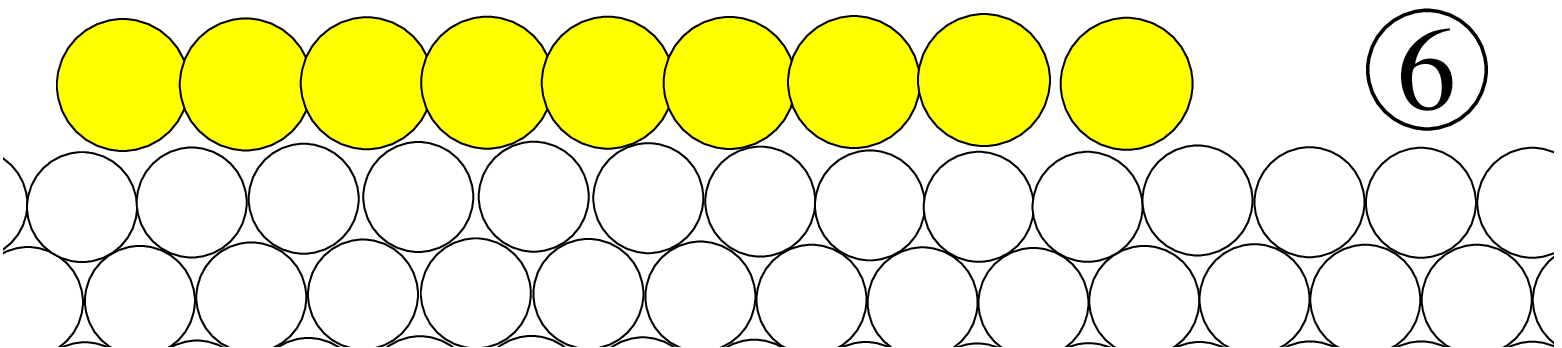}
  \epsfxsize= 0.75\textwidth
  \epsffile{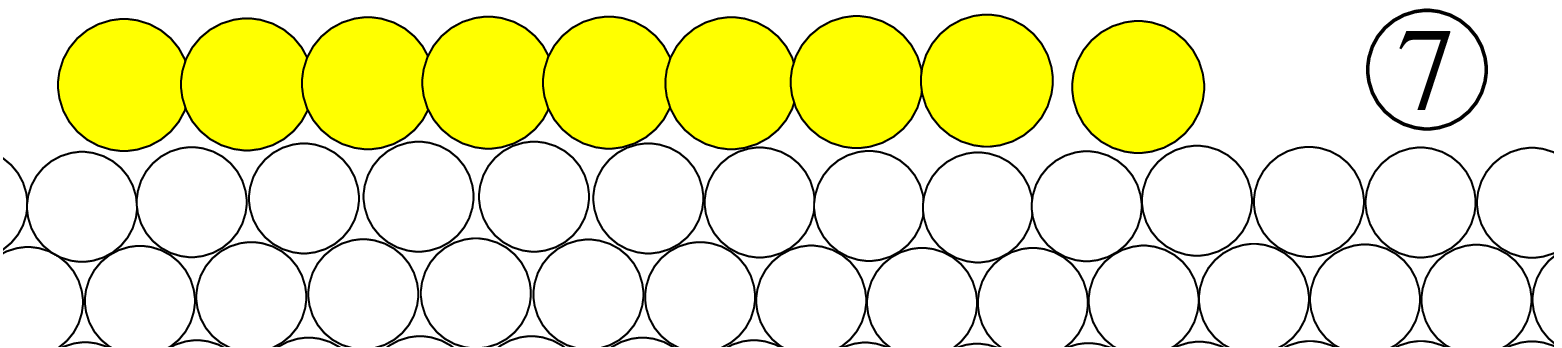}
 \end{minipage}}
\caption{The particle positions during the exchange process corresponding to the PES given in figure \ref{LEN12_12_8}.
Further explanations are given in the text.}
\label{LEN12_12_8_ISLANDS}
\end{figure}  

\begin{enumerate}
\item[\textcircled1] Figure \ref{LEN12_12_8_ISLANDS} \textcircled1 gives the positions of the particles before the exchange 
diffusion process starts. This state corresponds to a local minimum of the PES. 
Note that the edge particle on the right hand site of the island is nearly
on top of an underlying substrate particle.
\item[\textcircled2] The edge particle is drawn over the top of the underlying substrate particle 
indicated by a local maximum of $E_{tot}$.
\item[\textcircled3] Now the edge particle has reached a bridge site between two particles
of the substrate. Due to the high positive misfit also the particle on top of the island runs across
a favorable binding position. This leeds to a further local minimum in the PES and corresponds to the formation
of a dislocation at the edge of the monolayer island.
\item[\textcircled4] In the following the edge particle has to overcome another top site resulting in a local maximum of 
$E_{tot}$.
\item[\textcircled5] Finally the exchange diffusion is finished resulting in an $l=9$ monolayer island.
\item[\textcircled6] Due to the high misfit the bond of the edge particle is weak and a further jump
away from the island's edge involves an only small barrier.
\item[\textcircled7] Because of the spatial proximity of the island the second next binding place away from the island
results in a rather deep local minimum of the total energy of the system.
\end{enumerate}

We conclude from this investigation on the exchange mechanism that in the large misfit and large island regime due 
to the unfavorable binding position of the edge particle exchange diffusion becomes more likely. But one has 
to consider that at least in case of positive misfits the system has to pass through a further binding state, 
which is related to the introduction of a misfit dislocation on the island edge. 
\subsection{Influence of the potential}
We analyze now the influence of the used potential $U_{ij}$ on the calculated barriers.
Since the exchange move involves a lot of stretching and compressing of the participating
particles (see also fig. \ref{LEN12_12_8_ISLANDS}), the exchange barrier should be especially susceptible
to the characteristics of the used interaction $U_{ij}$ around its equilibrium distance. 

This assumption is confirmed by figure \ref{EX_IG_7}(a). Here the exchange barrier $E_{a,ex}$ for a island
of size $l=7$ is shown as a function of the misfit. As one would expect  exchange diffusion involves the highest 
barriers close--by $\varepsilon=0$ for all used potentials.
In comparison to  figure \ref{EX_IG_7}(b) it becomes clear that the steeper the potential around the 
equilibrium distance, the higher becomes the barrier for exchange diffusion. 
\begin{figure}[hbt]
\centerline{
\begin{minipage}{0.49 \textwidth}
  \epsfxsize= 0.98\textwidth
  \epsffile{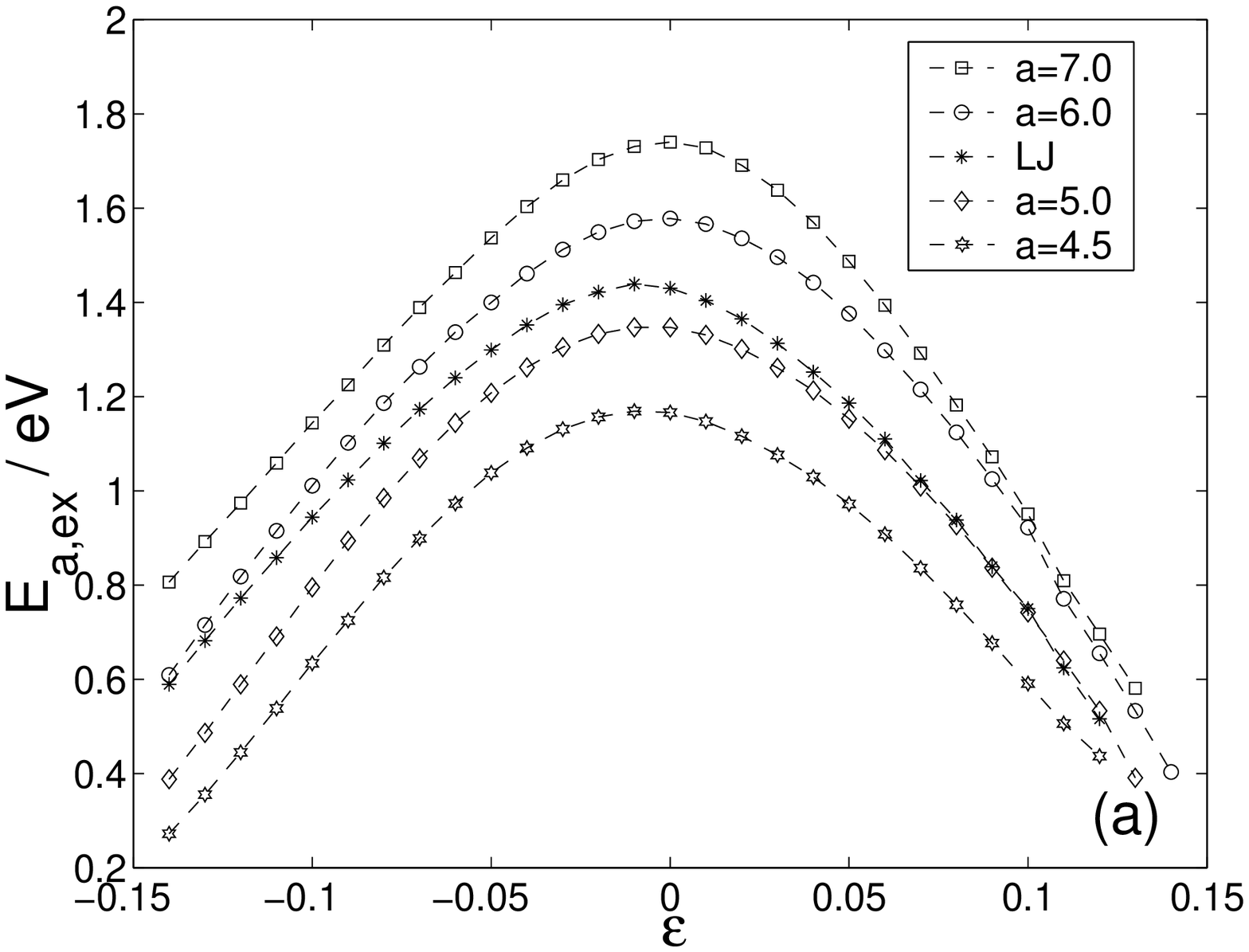}
\end{minipage}
\hfill
\begin{minipage}{0.49 \textwidth}
 \epsfxsize= 0.96\textwidth
  \epsffile{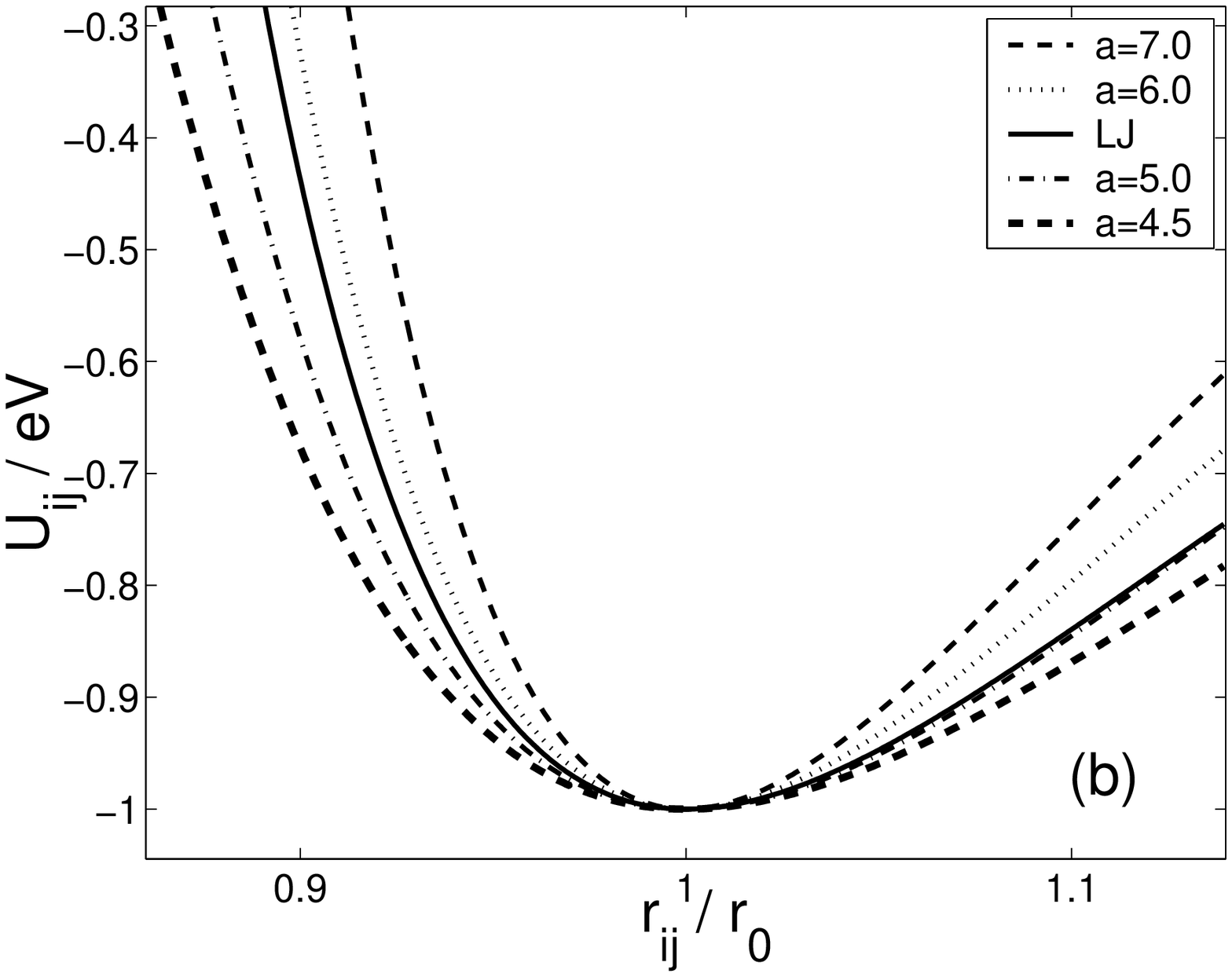}
\end{minipage}}
\caption{(a) $E_{a,ex}$ as a function of the misfit for several Morse and the Lennard--Jones $12,6$ potential. 
(b) The characteristics of the used pair--potential $U_{ij}$ around their equilibrium distance. For a better
comparability the minimum of the Lennard--Jones potential is shifted to $r_{ij}=1.0$.}
\label{EX_IG_7}
\end{figure}  
Therefore the Morse potential with $a=7.0$ gives the highest and the Morse $a=4.5$ potential the lowest exchange barriers
at the same misfit. Of particular interest is the case of the Lennard--Jones interaction: as figure 
\ref{EX_IG_7}(b) displays the Lennard--Jones potential overlaps to a high degree
with the $a=5.0$ Morse potential for particle distances greater than the equilibrium distance.  
This results in the collapse of the exchange barriers of both potentials
for large positive values of $\varepsilon$. For $r_{ij}<r_0$ it behaves more like a Morse $a=6.0$ potential 
which is reflected by the shift to the Morse $a=6.0$ results for highly negative values of $\varepsilon$.

It is also seen from figure \ref{EX_IG_7}(a) that for a given absolute value of the misfit $|\varepsilon|$ 
the exchange diffusion is more favorable for positive than for negative $\varepsilon$.
This is due to the fact that the used potentials are steeper in compression than in tension.
\begin{figure}[h]
\centerline{
\begin{minipage}{0.49 \textwidth}
  \epsfxsize= 0.95\textwidth
  \epsffile{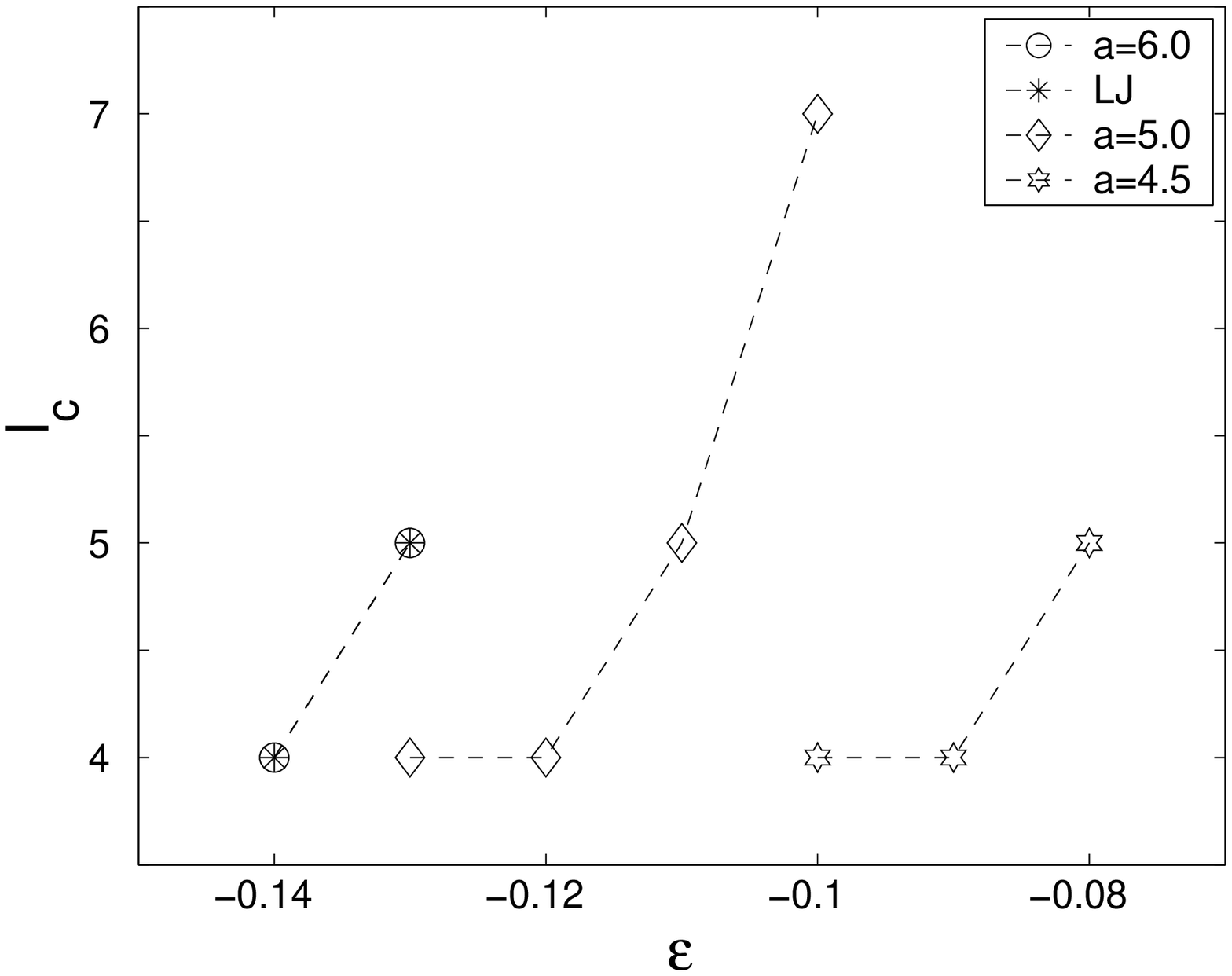}
\end{minipage}
\hfill
\begin{minipage}{0.49 \textwidth}
 \epsfxsize= 0.95\textwidth
  \epsffile{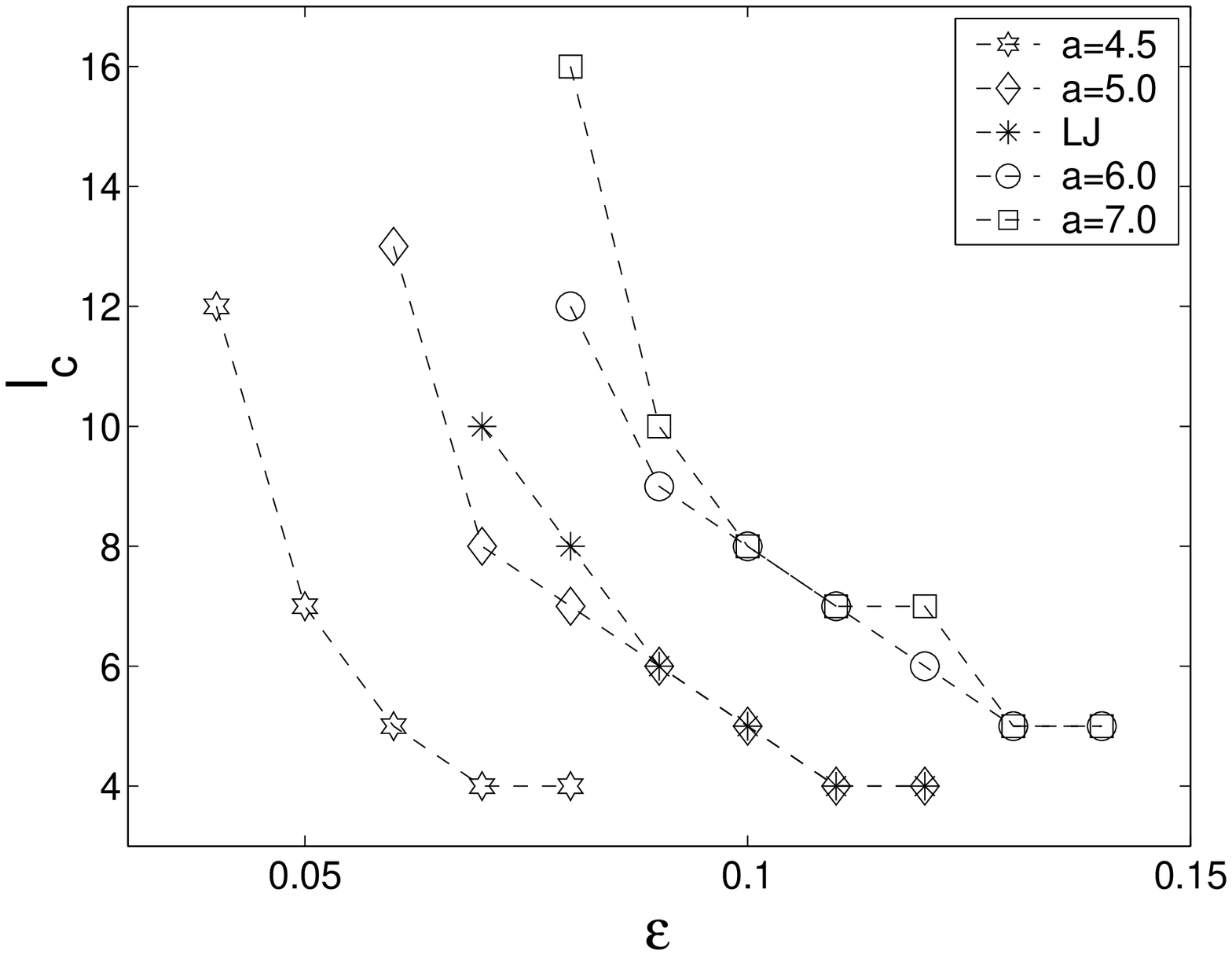}
\end{minipage}}
\caption{$l_c$ for (a) negative and (b) positive values of the misfit $\varepsilon$. Only values of $l_c \geq 4$ are
taken into account.}
\label{LC}
\end{figure}  

Finally we have a look at the mentioned critical island size $l_c$ and critical misfit $\varepsilon_c$
from which on exchange is favored over hopping diffusion.
Figure \ref{LC} shows the critical island size as a function of the misfit for the used pair--potentials.
For both the positive and the negative branch of the misfit exchange diffusion becomes favorable at 
a critical misfit $|\varepsilon_c|$ which is the greater the steeper the potential is. 
At a given potential $|\varepsilon_c|$ is greater for negative misfits than for positive misfits. 
This reflects the fact that the barriers
for exchange are higher in this region of the misfit.

For all considered potentials the critical island size starts to drop from $|\varepsilon_c|$  with increasing  
misfit $|\varepsilon|$. In the high misfit region exchange becomes the determining diffusion mechanism
already for small island sizes. 

In conclusion we find that at a given misfit a steeper potential leads to higher barriers for the exchange diffusion.
This corresponds well to the fact that a steeper potentials delays the introduction of misfit dislocations 
\cite{Vey:Diplom} due to a higher activation barrier \cite{Trushin:2003:EAM}. 

\section{Calculations for the three--dimensional case}
In this section we present calculations for the $3d$ case. Due to the high computational demand and some
conceptual problems regarding the application of the Molecular Static method to a $3d$ surface 
- which we will discuss later - the following results give a {\it qualitative} analysis of the different 
downward diffusion modes. We will also show that the $2d$ calculations indeed are of some relevance to 
the more realistic $3d$ case.

Due to the isotropy of the used pair--potentials, particles interacting via equation (\ref{LJ_2}) or equation  
(\ref{MORSE_2}) naturally 
arrange into a fcc lattice structure. 
As interaction potential we choose the $a=5.0$ Morse potential. Considering our $2d$ calculations this rather 
soft potential should display a strong dependency of the diffusion barriers on the misfit and the island size.
\begin{figure}[hbt]
\centerline{
\begin{minipage}{0.49 \textwidth}
  \epsfxsize= 0.90\textwidth
  \epsffile{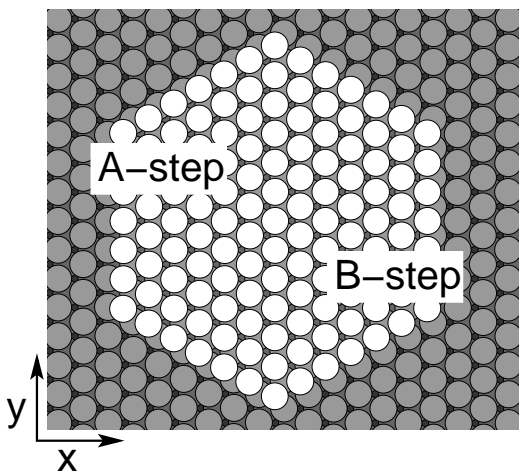}
\end{minipage}
\hfill
\begin{minipage}{0.49 \textwidth}
\caption{Schematic representation of a hexagonal shaped island placed on a fcc(111) surface (top view). Higher particles 
appear in lighter grey.}
\label{HEX_ISLAND}
\end{minipage}
}
\end{figure}

In our calculations we address the fcc(111) surface.
For symmetry reasons we place a hexagonally shaped adsorbate island on a substrate consisting of $h=11$ layers 
(see fig. \ref{HEX_ISLAND}). 
One should notice that there are two different kinds of close--packed island edges on a fcc(111) surface. 
Commonly the $\{100\}$ microfacets of an island are labeled as A edges and the $\{111\}$ microfacets are labeled 
as B edges. Due to the different local arrangement of the particles both kind of edges have to be treated 
separately for the calculation of diffusion barriers on descending steps.
A particle is placed  on top of the island near the edge in question. In order to determine the minimum energy 
path for a descending move the test particle (for hopping: the particle on the top, for exchange: a particle from 
the island edge) is drawn in small steps parallel to the $x$--direction (see fig. \ref{HEX_ISLAND}). At each step the 
test particle is relaxed in the $y$-- and $z$--direction. All other particles are relaxed without restrictions.
A problematic point with $3d$ Molecular Static calculations is, that one can not be sure to really 
find the minimum energy path \cite{Barkema:Mund}. This is due to the huge numbers of possible particle 
positions compared to the $2d$ case. However, the method yields at least a qualitative analysis of the 
different diffusion mechanisms as a function of misfit and island size and allows a comparison of A or B steps.

\subsection{The homoepitaxial case}
We first analyze the barriers for descending diffusion moves from A and B steps in the homoepitaxial case 
($\varepsilon=0$) for an island like shown in figure \ref{HEX_ISLAND}.
Figure \ref{M0_ALLES} shows the energy path for the four possible diffusion types: hopping and exchange 
diffusion on a A and B step, respectively. 
\begin{figure}[hbt]
\centerline{
\begin{minipage}{0.55 \textwidth}
  \epsfxsize= 0.90\textwidth
  \epsffile{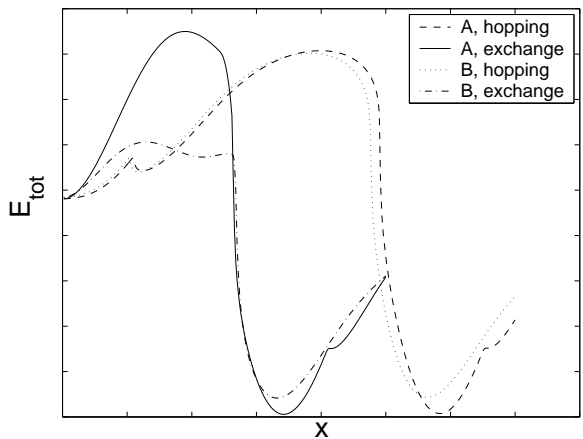}
\end{minipage}
\hfill
\begin{minipage}{0.4 \textwidth}
\caption{Total energy of the system (substrate, island and test particle)
 for hopping and exchange diffusion from A and B steps in the homoepitaxial case ($\varepsilon=0$).}
\label{M0_ALLES}
\end{minipage}
}
\end{figure}

\subsubsection{Hopping diffusion}
As figure \ref{M0_ALLES} displays the energy paths for hopping diffusion down A and B steps look quite similar.
\begin{figure}[h]
\centerline{
\begin{minipage}{0.33 \textwidth}
  \epsfxsize= 0.90\textwidth
  \epsffile{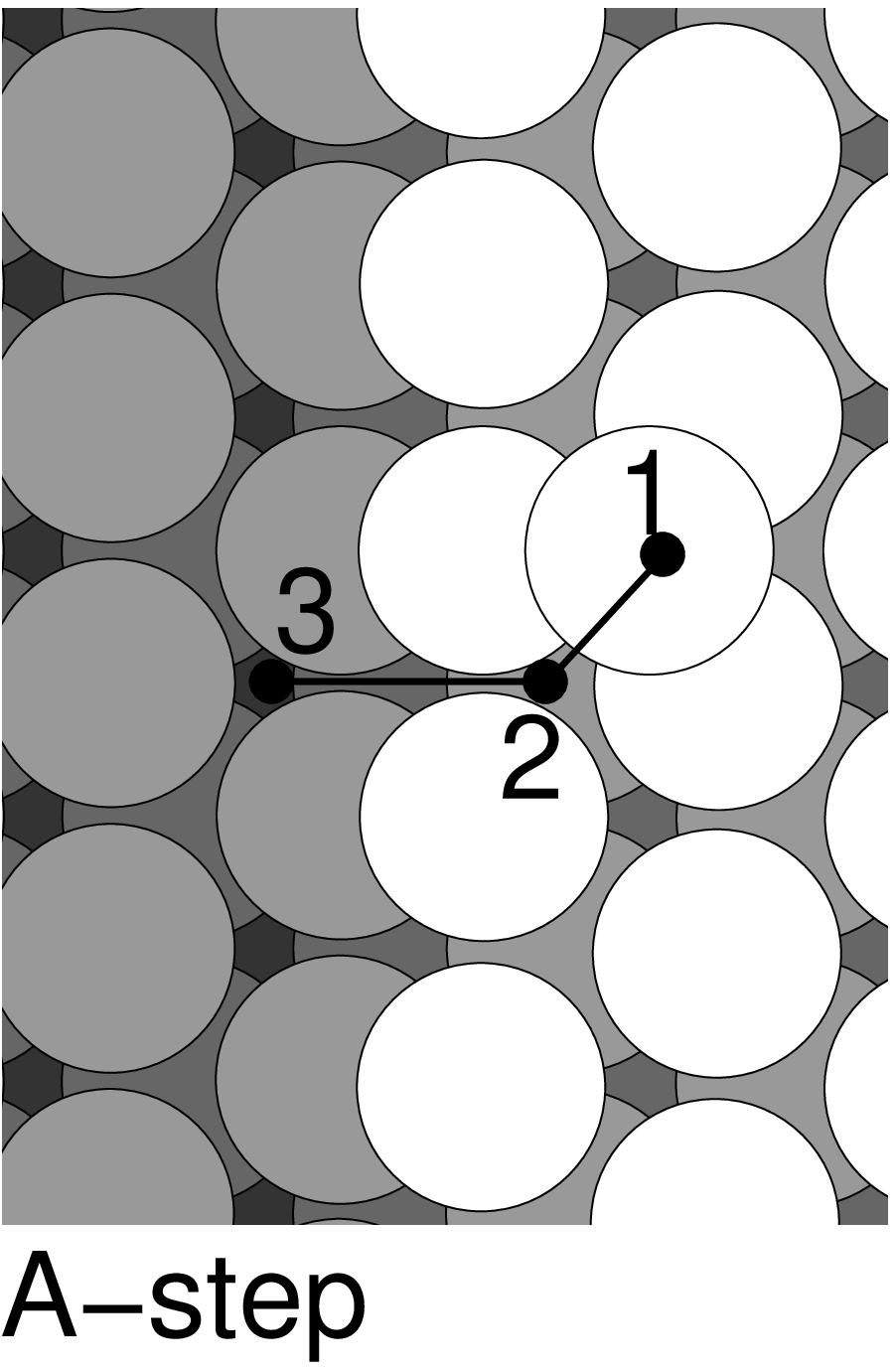}
\end{minipage}
\hfill
\begin{minipage}{0.33 \textwidth}
  \epsfxsize= 0.90\textwidth
  \epsffile{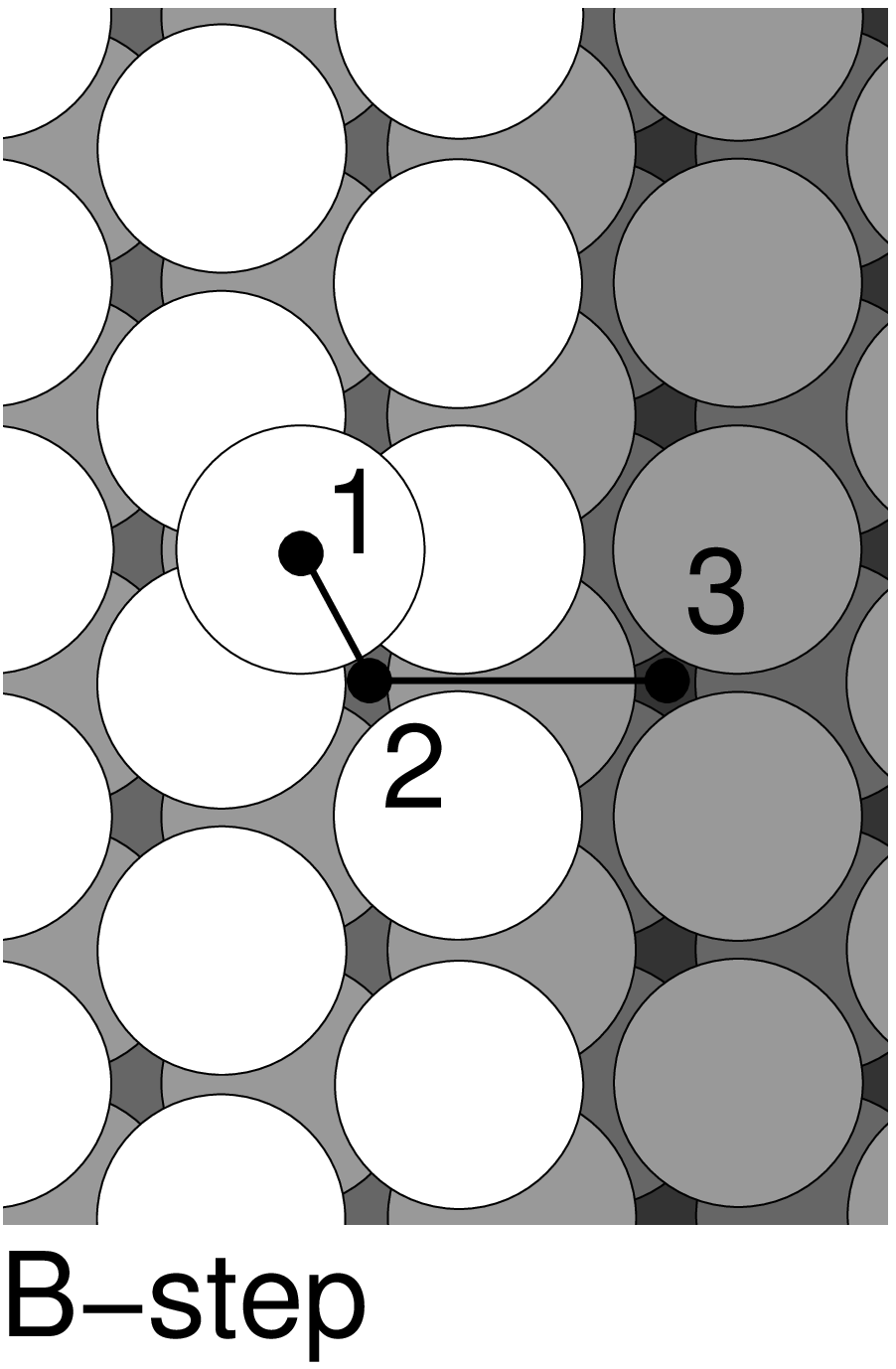}
\end{minipage}
\hfill
\begin{minipage}{0.3 \textwidth}
\caption{Left panel: Way of a particle down an A step. Right panel: Way of a particle down a B step.
The numbers indicate positions of local energy minima. Note that the particle 
starts from an fcc--type binding site on the A step and from a hcp--type binding site on the B step.
}
\label{_HOP}
\end{minipage}
}
\end{figure}
Figure \ref{_HOP} (left panel) shows the way of a descending particle down an A step. The particle starts from 
a fcc--type binding site (minimum $1$). 
Since the particle passes through an additional hcp--type binding site on its way down the step a 
second minimum appears in the energy path (cf. fig. \ref{M0_ALLES}). At the end of the hopping move the 
particle is attached to the edge on a fcc--type binding site (minimum 3).

The path of a particle downwards a B edge is similar (see fig. \ref{_HOP}, right panel). 
The transition starts on a hcp--type binding site, which 
is energetically slightly disadvantageous in comparison to a fcc--type site
(minimum 1). A second minimum arises in the 
energy path, again due to a fcc--type binding site (minimum 2). As figure \ref{M0_ALLES} shows particles 
attached to B steps are clearly weakly bound compared to particles attached to A steps. A similar result was found 
for different pair--potentials in the fcc(111) geometry before \cite{Trushin:1997:EBS}. 
In conclusion - as a result from the similarity of the diffusion paths - 
the hopping barriers downward A and B steps are quite the same.

\subsubsection{Exchange diffusion}
The situation changes completely in the case of exchange diffusion.
Figure \ref{_EX} shows the paths for exchange diffusion on an A step (left panel) and a B step (right panel).
\begin{figure}[h]
\centerline{
\begin{minipage}{0.33 \textwidth}
  \epsfxsize= 0.90\textwidth
  \epsffile{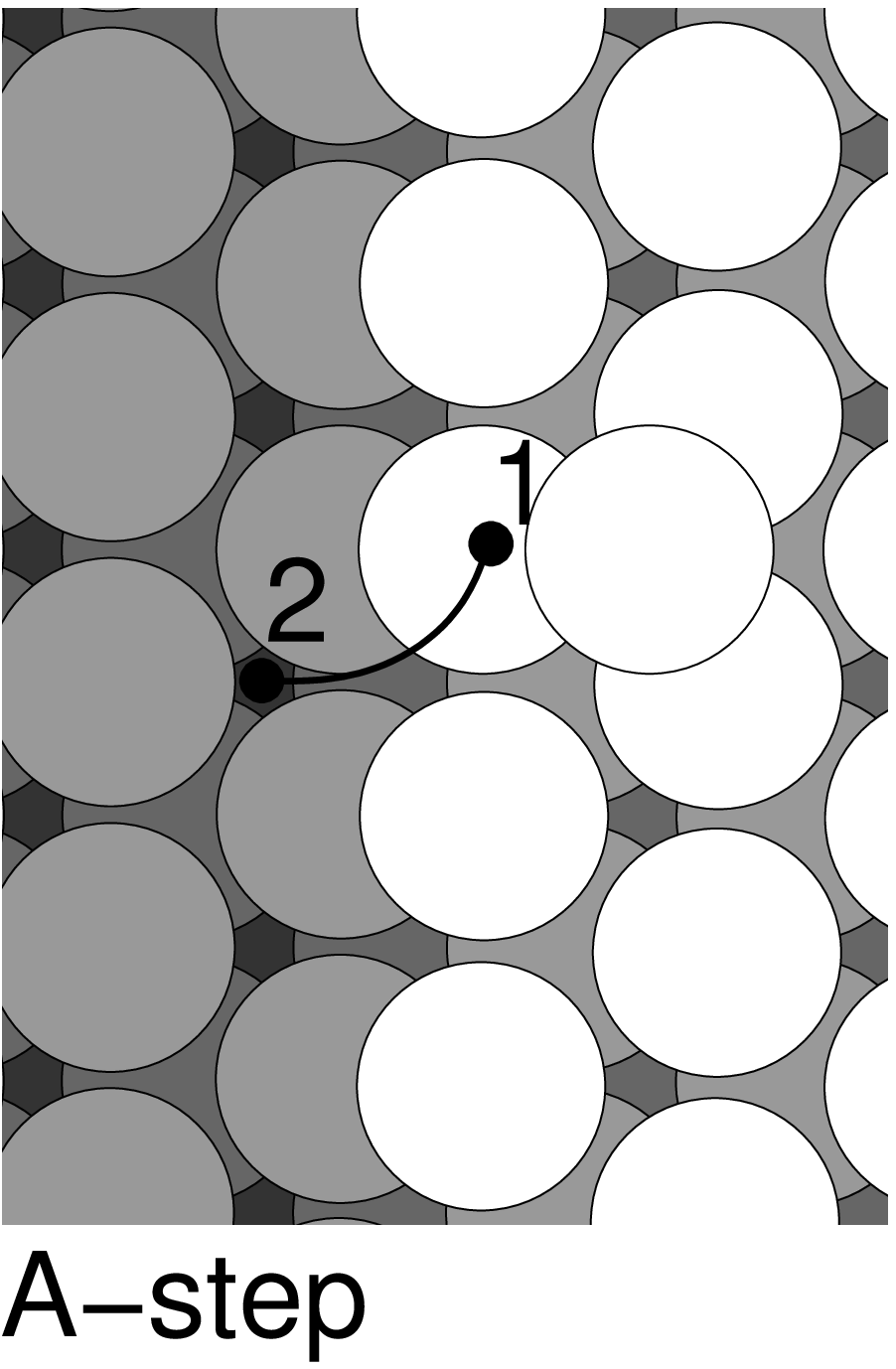}
\end{minipage}
\hfill
\begin{minipage}{0.33 \textwidth}
  \epsfxsize= 0.90\textwidth
  \epsffile{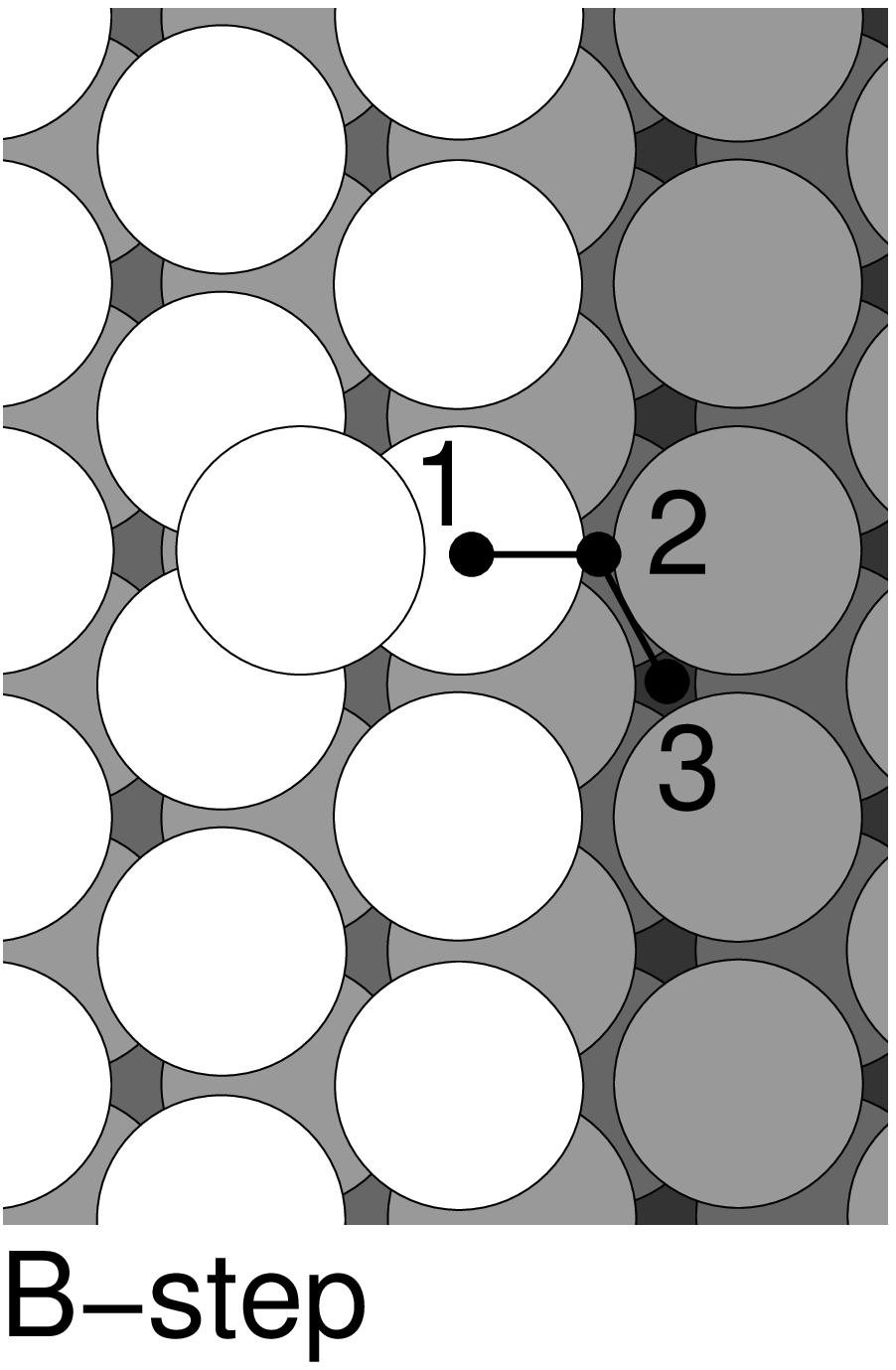}
\end{minipage}
\hfill
\begin{minipage}{0.3 \textwidth}
\caption{Left panel: Exchange diffusion path on an A step. Right panel: Way of a particle down a B step.
The numbers indicate positions of local energy minima.
}
\label{_EX}
\end{minipage}
}
\end{figure} 
For the exchange move on an A step the top of an underlying particle  
lies in front of the test particle (see fig. \ref{_EX}, left panel). 
This leads to the strongly enlarged barrier in the energy path (fig. \ref{M0_ALLES}) for this 
diffusion move. 

On the B step the test particle faces a bridge site between two underlying substrate particles 
(see fig. \ref{_EX}, right panel). 
This leads to a rather small barrier for exchange diffusion on B steps, which is proved for various 
material types 
(see e.g.\cite{Trushin:1997:EBS,Maca:1999:EBD,Maca:2000:EBD,Liu:1993:DBS,Villarba:1994:DMR,Feibelman:1998:ISD,Feibelman:1999:SDA}).
\subsection{The heteroepitaxial case}
From the above made considerations it is clear that the arrangement of the particles in an A step is similar to 
the $2d$ situation: since the test particle has to overcome the top site of an underlying particle 
exchange diffusion is disadvantageous in the homoepitaxial case. Analogous to our findings for the $2d$ case the influence 
of the misfit ($\varepsilon\not=0$) is expected to be especially pronounced on exchange diffusion moves on A steps.
Indeed, the energy path for $\varepsilon=7\%$ (see fig. \ref{P7_ALLES}, left panel) shows a strongly decreased
exchange diffusion barrier for A steps.
\begin{figure}[hbt]
\centerline{
\begin{minipage}{0.55 \textwidth}
  \epsfxsize= 0.90\textwidth
  \epsffile{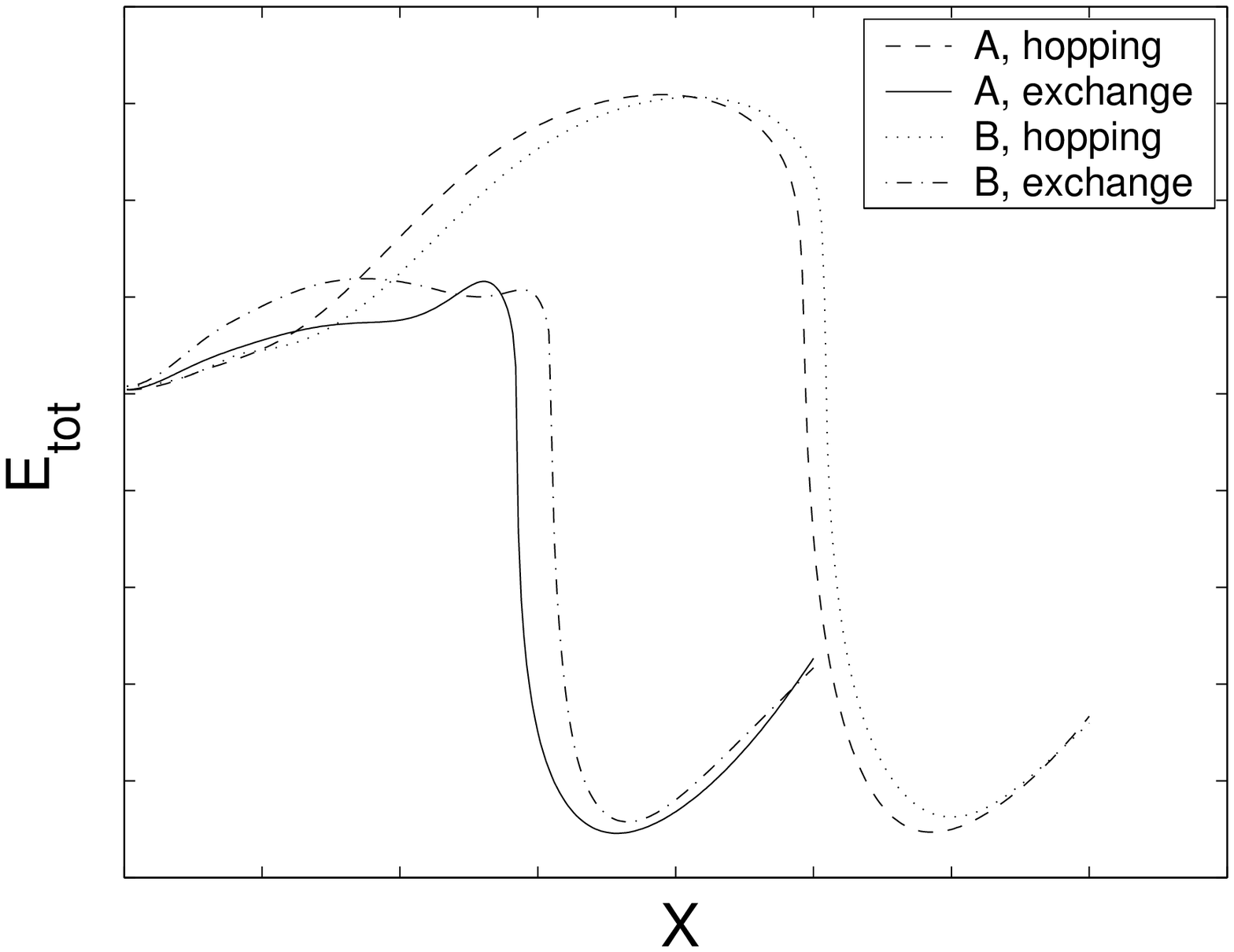}
\end{minipage}
\hfill
\begin{minipage}{0.33 \textwidth}
  \epsfxsize= 0.90\textwidth
  \epsffile{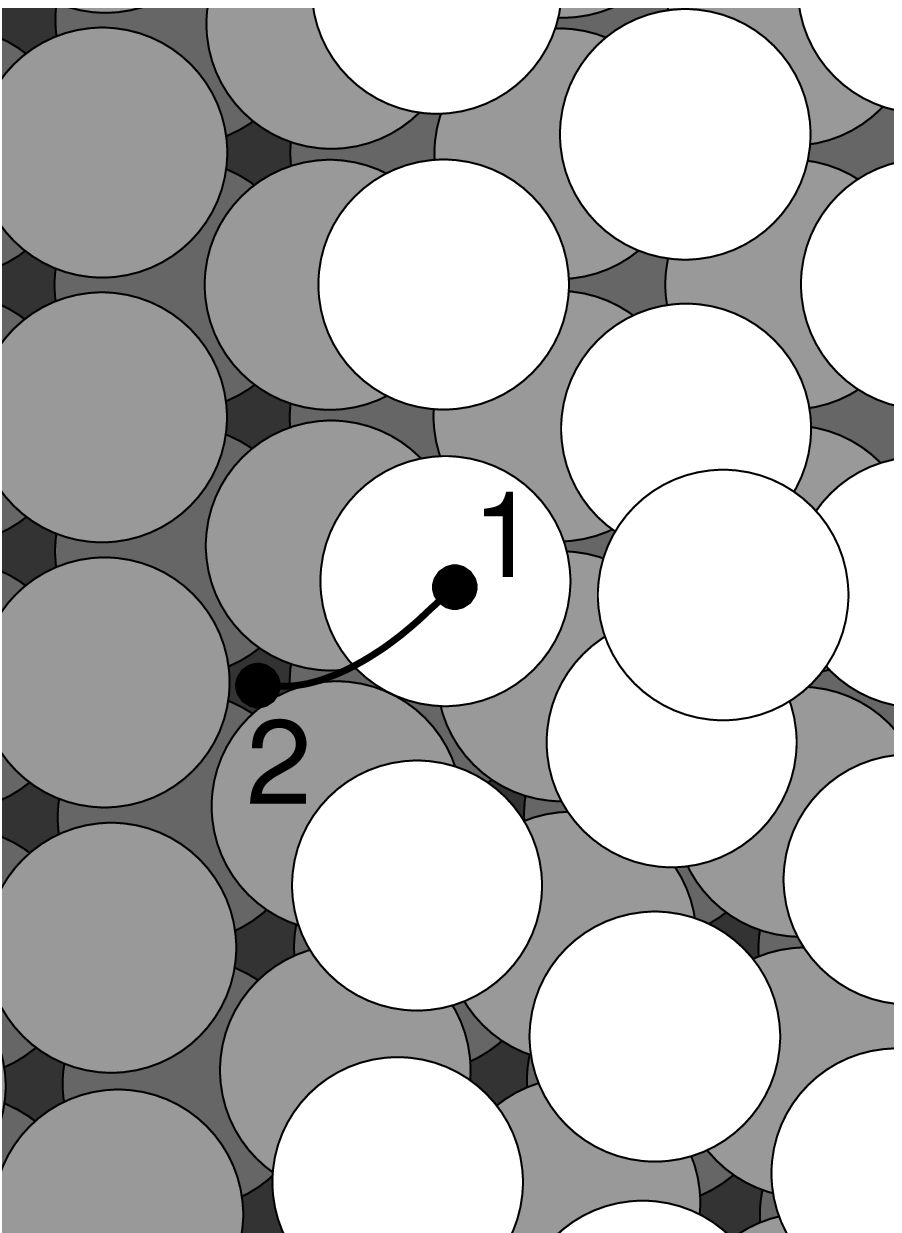}
\end{minipage}
}
\caption{Left panel: Total energy of the system (substrate, island and test particle)
for hopping and exchange diffusion from A and B steps in the heteroepitaxial case ($\varepsilon=7\%$).
Right panel: Way of a particle down an A step for $\varepsilon=7\%$.}
\label{P7_ALLES}
\end{figure}
As figure \ref{P7_ALLES} (right panel) in comparison to the homoepitaxial case (see fig. \ref{_EX}, left panel) 
displays this is due to the arrangement of the particles: the misfit 
causes a shift of the step particles over the top site of the underlying substrate particles. The arrangement of 
the A step particles becomes similar to that of B step particles, resulting in a rather low barrier for 
exchange diffusion.

\section{Conclusions and outlook}
In this chapter we investigated diffusion barriers from island edges. To this end a Molecular Static method 
was applied in two and three dimensions. The calculations have been performed for various Morse and 
the $12,6$ Lennard--Jones potential. Our $2d$ calculation have yielded the following results:
\begin{itemize}
\item The barrier for exchange diffusion moves down an island edge strongly depends on island size and 
misfit. The barrier decreases with increasing misfit and increasing island size, generally.
\item In the large misfit regime the barrier for exchange diffusion becomes smaller than the barrier for 
hopping diffusion.
\item The steeper the used pair--potential is, the larger is the critical misfit at which exchange diffusion 
becomes favorable.
\end{itemize}
For the $3d$ case we performed calculations for the fcc(111) surface using a Morse potential. Here, two results are 
of particular interest:
\begin{itemize}
\item We have shown 
that the behavior of the diffusion barriers on A steps is qualitatively the same as in the $2d$ case: 
an increasing misfit favors the exchange over hopping diffusion on A step. 
\item In the large misfit regime exchange 
barriers on A steps become of approximately the same size as on B steps. 
\end{itemize}
Despite the fact that the Molecular Static method is a frequently used tool even in material specific 
barrier calculations, it implies some problems, especially regarding calculations in three dimensions.
As mentioned above - considering the 
large number of particles one has to take into account - it is difficult to find the minimum energy path with this method. 
Given the interatomic potential the {\it activation--relaxation technique} (ART) 
\cite{Barkema:1996:EBR,Mousseau:1998:TTP,Malek:2000:DLJ} would perhaps be a more appropriate method for the calculation 
of diffusion barriers in three dimensions.

One has also to bear in mind that both barrier {\it and} attempt frequency determine the rate for a diffusion event.
Molecular Dynamics (MD) studies show that the attempt frequency for exchange diffusion is usually smaller than 
for hopping diffusion \cite{Shiang:1994:MDSa,Voter:1984:TST,Shiang:1994:MDSb,Fu:1996:SDD,Iijima:1997:MDS}. 
This favors hopping diffusion mainly at low temperatures.

  \cleardoublepage
\chapter{Off--lattice Kinetic Monte Carlo simulations}
\label{KAP-3} 
As described in chapter \ref{KAP-1} atomic size mismatched systems always try to
relax the strain imposed by the misfit.
This can either be done by a rearrangement of the surface particles into mound--like structures 
or the introduction of misfit dislocations.
Both relaxation mechanisms have in common that the distance between neighboring particles 
can vary over a wide range. For that reason the representation of the system by a rigid lattice is not 
longer practical for the simulation of heteroepitaxial growth.
In order to model these relaxations in simulations it is crucial to allow for continuous
particle positions. 

Given at least an approximation for the interatomic potential Molecular Dynamic (MD) simulations
would be the most realistic way to do so (cf. chapter \ref{KAP-1}). Indeed there are several studies on the 
formation and mobility of misfit dislocations (see e.g. \cite{Dong:1998:SRM,Schiotz:1998:SMS}).
However, this method suffers generally from the restrictions to short physical time scales ($\leq 10^{-6}s$).
In particular MBE relevant time scales  of seconds up to minutes are even with modern computers not yet feasible.
In conclusion, the most promising way to simulate heteroepitaxial growth today is by means of 
off--lattice Kinetic Monte Carlo
(KMC) simulations.

Historically, the first off--lattice KMC methods were the so--called ball and spring simulations, which
are used with some success up to now \cite{Madhukar:1983:FEV,Ghaisas:1986:RSM,Orr:1992:MSI,Barabasi:1997:SAI,Khor:2000:QDS,Lam:2002:CRM}.
In this kind of simulations strain is applied to the system by assuming a harmonic interaction ({\it spring})
between the particles ({\it balls}). However those models fail for example in the simulation of   
misfit dislocations and need somewhat artificial bond--counting rules for the calculation of
binding energies (see chapter \ref{KAP-1} for details).

Here we follow a different direction based on work by Schindler and Wolf \cite{Schindler:1999:TAG}. We apply the method to 
three different problems of heteroepitaxial growth.
We assume that the interaction of the 
particles is given by a pair--potential which is a function of the {\it continuous} 
distance between the particles.
The main idea of the method is to compute the activation barriers from the potential
for each hopping event and to use the obtained rates in a standard rejection--free KMC
simulation. In the following we will discuss the method in detail along with its advantages and
restrictions.

\section{Calculation of the activation energy}
We consider a system consisting of $n$ particles interacting through a pair--potential
$U_{ij}$. The continuous distance between the two particles $i$ and $j$ is given by $r_{ij}$, where
$\vec{r}_{ij}=\left(x_{ij},z_{ij}\right)^T$, $\vec{r}_{ij}=\left(x_{ij},y_{ij},z_{ij}\right)^T$ for the
$2d$, $3d$ case, respectively.
According to the symmetry implied by the potential the particles arrange into a crystal. 
Since all potentials used in this work (cf. appendix \ref{AP-1}) decrease rapidly
towards zero with increasing particle distance a {\it cut--off} distance $r_{cut}$ with $U_{ij}=0$ for
$r_{ij}>r_{cut}$ is assumed. In most cases the cut--off is chosen to be $r_{cut}=3r_{0}$, where
$r_0$ is the equilibrium distance of the particles.

As discussed in chapter \ref{KAP-2} a particle deposited on
the surface of a crystal moves in an energy landscape consisting of local minima (binding states) separated
from each other by saddles (transition states) and maxima. One should notice that according to the definition of 
the potential energy surface (PES) 
given in chapter \ref{KAP-1} in the $2d$ case the transition sites are given by maxima. However, if
one considers the test particle's $z$--coordinate (perpendicular to the surface) as a free parameter the
coordinates $x$ and $z$ along with 
the potential energy of the system reassemble again an energy surface where the transition site is given by a 
first order saddle point (see also fig. \ref{ART_FIG}).
In order to keep a clear notation  in the following we will always refer to saddles 
when transition sites are considered.

Since  a particle hopping from one local minimum with energy $E_b$ 
to another has to overcome such saddles with energy $E_t$, 
the goal is to calculate $E_b$ and $E_t$ and, hence, to
obtain the activation energy $E_a=E_t-E_b$ for each diffusion event in a KMC simulation.

\subsection{Searching for the saddle point}
Due to the influence of the adatom on the crystal one would
actually need to relax the whole system in every step during the search for the transition site
(like it was discussed in the last chapter).
However, it was shown in \cite{Schroeder:1997:DSS,Schindler:1999:TAG} that a {\it frozen} crystal during the
saddle search only shifts the 
activation barriers uniformly to slightly higher values
(about $10\%$ higher). Since calculations for a frozen crystal save a lot of computer time we restrict our method to 
this simplification during the calculation of $E_t$. A further advantage is that during the 
barrier calculations one only has to keep track of the adatom energies, instead of the energy
of the entire system.
  
We now introduce the method used in our simulations for the calculation of the transition energy.
To this end an iterative algorithm for the saddle point search introduced in 
\cite{Barkema:1996:EBR,Mousseau:1998:TTP,Malek:2000:DLJ} is applied to our problem.
This so--called activation--relaxation technique (ART) was originally developed to obtain 
energy minimized structures in glassy materials: a system is allowed to evolve by  
following well--defined paths over saddles between local energy minima.
From this method we implement the saddle search algorithm:
\begin{itemize}
\item First the adatom is slightly displaced from its binding position in the wanted direction 
(i.e. the direction of diffusion). This causes the force $\vec{F}$ acting on the particle to become
nonzero.
\item By iterative application of the redefined force $\vec{G}$ given by
\begin{equation}
\label{G_FORCE}
 \vec{G} =\ \vec{F}-\left(1+\alpha \right)\left(\vec{F}\vec{e}\right)\vec{e}
\end{equation}
with $\alpha>0$ the adatom is moved in small steps in direction of $\vec{G}$ toward the nearby saddle. 
Here $\vec{e}$ is the unit vector
pointing from the last local minium to the current position of the particle.
\item The iteration ends when the saddle point is reached and $\vec{G}=\vec{F}=0$.
\end{itemize}
Since the redefined force $\vec{G}$ is opposite in sign to $\vec{F}$ in the direction parallel to  
$\vec{e}$ and equal to $\vec{F}$ in any direction perpendicular to $\vec{e}$, the particle
is  forced in small steps a valley up--hill in the PES (see fig. \ref{ART_FIG}).
The positive number $\alpha$ controls the increment in each iteration step. As 
figure \ref{ART_FIG} shows,
a too large value of $\alpha$ results in missing the relevant saddle. On the contrary, for small values
of $\alpha$, more iteration steps are needed. In our simulations a good
value of $\alpha$ has to be found to reach the right saddle with as little iteration steps as possible.
Values between $\alpha=0.5$ and $\alpha=2.0$ have yielded the best performance.
\begin{figure}[hbt]
\centerline{
\begin{minipage}{0.49 \textwidth}
  \epsfxsize= 0.95\textwidth
  \epsffile{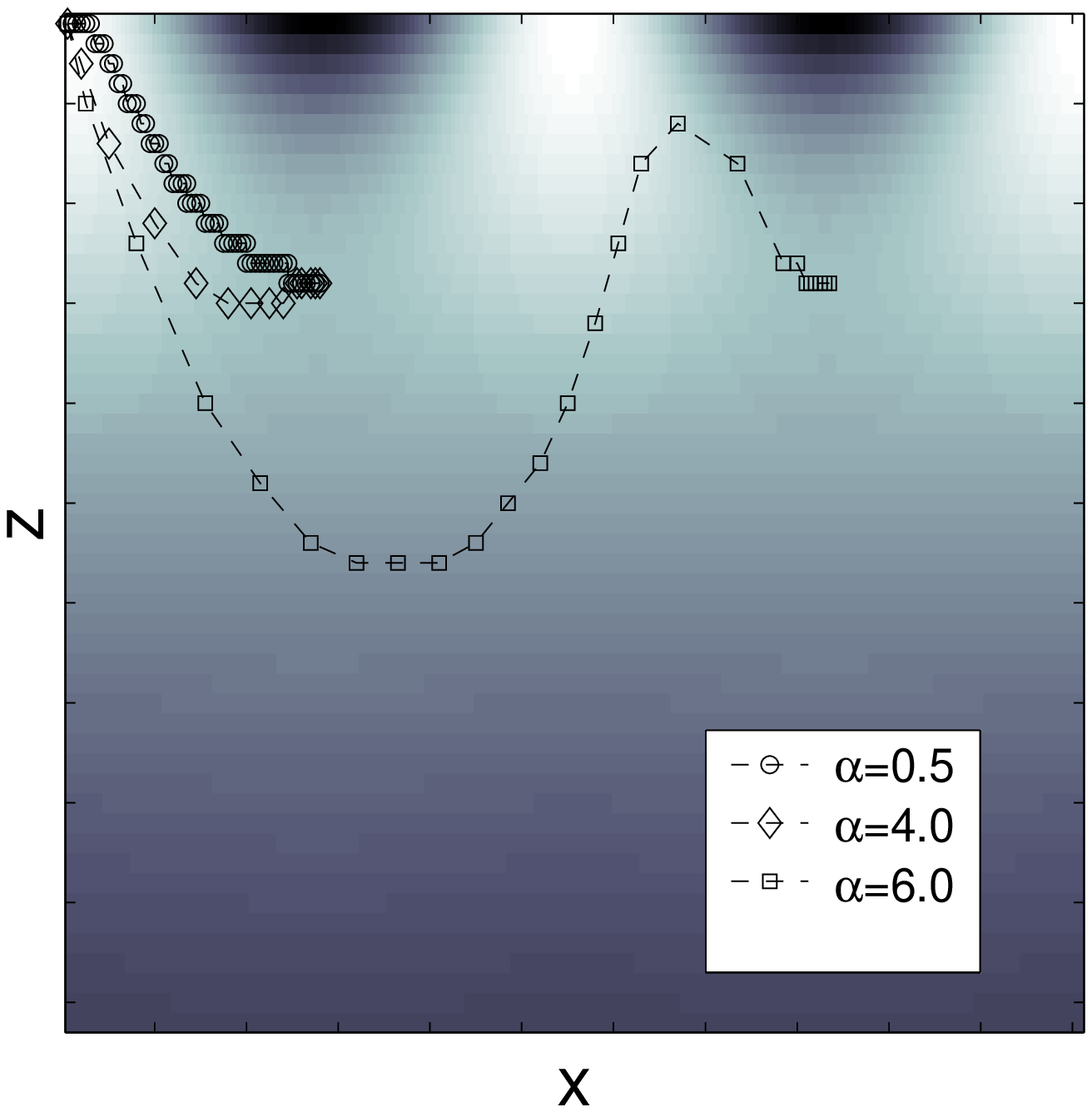}
\end{minipage}
\hfill
\begin{minipage}{0.49 \textwidth}
 \epsfxsize= 0.95\textwidth
  \epsffile{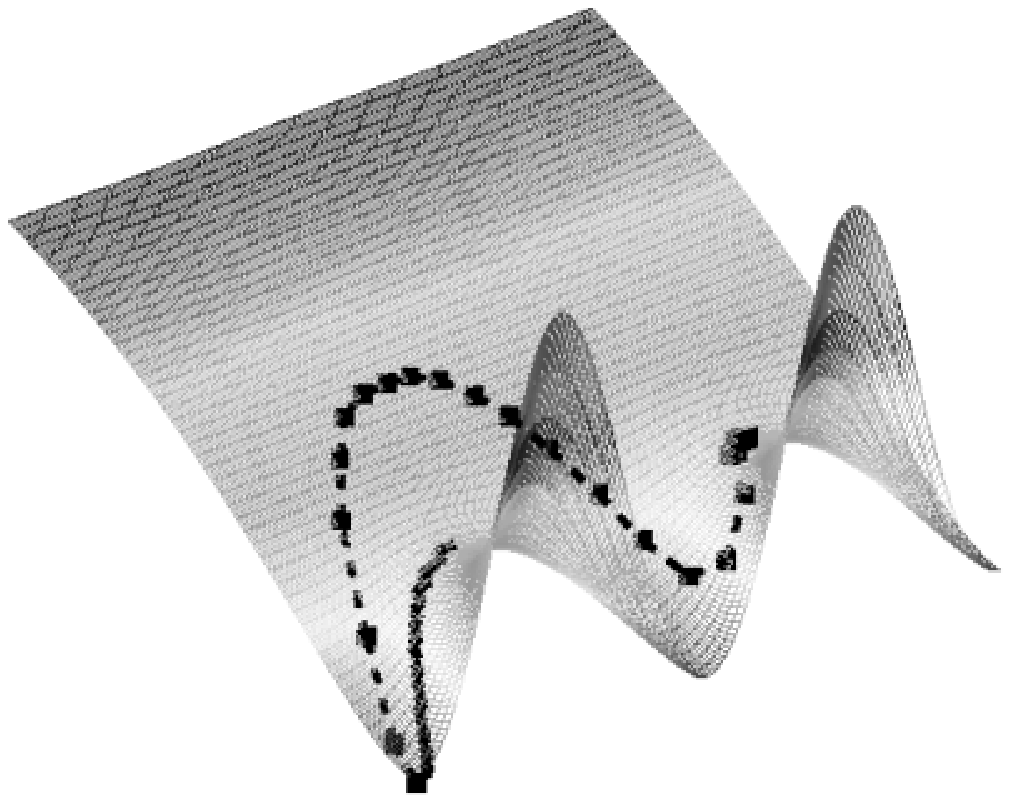}
\end{minipage}}
\caption{Energy surface for the $2d$ case given as contour plot (left panel) and surface plot (right panel).
The trace of the iteration (\ref{G_FORCE}) is shown for different values of $\alpha$. Note that for
a too high value of $\alpha$ (here for example $\alpha=6.0$) the method fails to end up in the relevant saddle.}
\label{ART_FIG}
\end{figure}  

Formally, local maxima also fulfill the stopping criteria $\vec{G}=\vec{F}=0$ 
of the iteration (\ref{G_FORCE}).
But the iteration can only end up in a local maximum if, by accident, a maximum is chosen as starting point.
Otherwise, the components of $\vec{G}$ 
in direction of  $\vec{F}$ drive the adatom downhill away from the local maximum.

In earlier simulations for the $2d$ case a different method has been used for the barrier calculation. As mentioned 
above for the $2d$ case the transition site is given - per definition of the PES - by a local maximum, with 
respect to the $x$--coordinate only.
Therefore $E_t$ can be calculated by minimizing the adatom's potential energy as a function of $z_i$ at fixed 
horizontal coordinates $x_i$ and maximizing this energy with variation of $x_i$.
This is done by using a one--dimensional minimization method 
\cite{Press:1992:NRC}.

Both methods yield the transition site in comparable computing time. 
The iterative method has the advantage to work also for the $3d$ case.
Moreover, the ART--like method proved itself more reliable in detecting the
relevant saddle point.

\subsection{Calculation of the binding energy}
The calculation of the binding energy is computationally less demanding than the saddle search.
In most cases it is sufficient to guess the position of the binding site
by geometrical considerations. For example in the $2d$ case - when the particles 
arrange in a triangular lattice - at moderate misfit the atom in question will occupy the
bridge site between two underlying particles. After the adatom is placed to the guessed
position in a forthcoming relaxation of the system (see next section) the particle ends up
at its binding site with $E_b$.
Only in the high misfit regime, when dislocations have to be considered this can fail. 
In this case, the particle is moved in direction of the binding site and a additional minimum
search is performed. This can be done with the same methods as described for the saddle search.

With the obtained binding and transition energies we are able to calculate the activation barrier  
for hopping diffusion rather fast. However, the described methods do not allow for the calculation of
barriers of concerted moves, like the exchange diffusion  examined in chapter \ref{KAP-2}. For this reason
 we restrict ourselves in the following to situations were these moves can be neglected. However, for 
a more comprehensive description of heteroepitaxial growth concerted moves will have to be 
incorporated in the simulation method in some way.

\section{Deformations of the crystal}
An adatom hopping on the crystal surface from binding site to binding site  surely influences
the surface around its current position. Due to the additional binding partner the nearby 
particles will slightly change their positions. This effect is the more pronounced, the greater the
strain in the system is.

In order to account for these local deformations of the crystal, after each microscopic event 
- i.e. deposition and
diffusion - the system is relaxed. 
This is done by minimizing the total potential energy of the system, given by
\begin{equation}
\label{E_TOTAL}
 E_{tot} = \sum_{i,j}U_{ij}
\end{equation}
, $r_{ij}<r_{cut}$ 
with respect to the particles' coordinates using a standard conjugate gradient method, taken from \cite{Press:1992:NRC}.
As a simplification, most of the time this relaxation is performed in a local way: 
only particles within a sphere of radius $r_{cut}$ around the locus of the last event 
are allowed to change their position during the procedure
(see also fig. \ref{LOCAL_RELAX}).
This restriction saves a lot of computer time and is a valid simplification since the influence of a single 
microscopic event should be locally limited, due to the fast decreasing pair--potentials used in our simulations.
\begin{figure}[hbt]
\centerline{
\begin{minipage}{0.50 \textwidth}
  \epsfxsize= 0.90\textwidth
  \epsffile{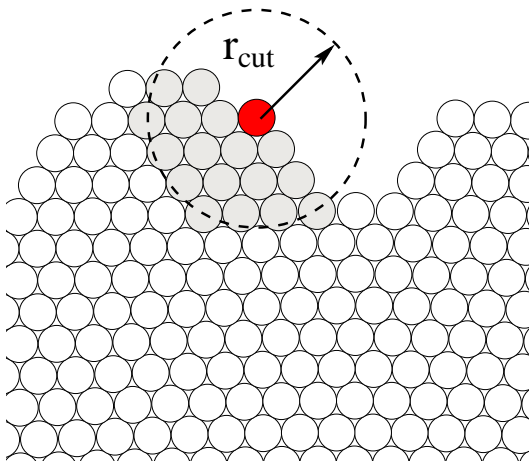}
\end{minipage}
\hfill
\begin{minipage}{0.42 \textwidth}
\caption{Local relaxation in the $2d$ case: a circle of radius $r_{cut}$ is drawn around the active 
(here darkest) particle. The (light gray) particles within this circle are allowed to change their
position during the procedure.}
\label{LOCAL_RELAX}
\end{minipage}
}
\end{figure}

However one may argue that especially at the edge of the considered sphere the local relaxation could add 
artificial strain to the system. For that reason a global relaxation with respect to all particle coordinates 
is performed after a certain number of microscopic events. This number depends on the specific problem
(e.g. value of the misfit, temperature, deposition rates) and is guessed from preliminary simulation runs. 
As a rule of thumb the maximal change of the diffusion barriers 
due to the global relaxation should not exceed more than about $1\%$. 

One should notice that both mentioned types of relaxation do not lead to a substantial rearrangement of the crystal 
but the coordinates and activation energies of the affected particles are changed slightly. The relative positions
of the particles remain unchanged.

\section{Rejection--free Kinetic Monte Carlo simulations}
The binding energy $E_{b,i}$ and transition energy $E_{t,i}$ have now to be calculated for each possible diffusion event
$i$. The activation energy amounts to $E_{a,i}=E_{t,i}-E_{b,i}$ (eq. (\ref{E_A})).
Since we assume that the diffusion process follows an Arrhenius dynamics, see chapter \ref{KAP-1}, the proper
rates are given by $R_i=\nu_0 e^{-\frac{E_{a,i}}{kT}}$.

The obtained diffusion rates along with the rate for deposition of a new particle $R_d$ are now used in
a rejection--free KMC algorithm \cite{Ahr:2002:SPE,Newman:1999:MCM}. 
Because of the fact that 
especially in the low temperature / high barriers regime the probability of non--modifying steps becomes significant  
non rejection--free techniques like the Metropolis algorithm result in a considerable slowing down of the simulation. 
Since in our simulations a diffusion 
barrier increases fast with increasing number of binding partners it pays to take the additional administration effort of
a rejection--free continuous time algorithm: every iteration step of the algorithm a modifying event $k$ is drawn 
and performed with the correct probability
\begin{equation}
\label{P_K}
 p_k=\frac{R_k}{R_d+\sum_iR_i}.
\end{equation} 
Then, the time interval $\tau$ between two modifying steps is drawn randomly 
according to the probability of these steps. The time interval is given according to 
an {\it exponential} distribution by 
\begin{equation}
\label{TAU}
 \tau=-\ln\xi/(R_d+\sum_iR_i),
\end{equation}
where $\xi$ is a uniformly distributed random number $0 < \xi \leq1$.
Following the work of Ahr \cite{Ahr:2002:SPE} the random selection of events is performed 
using a {\it complete binary tree}.

Applied to our off--lattice method the algorithm works like this:
\begin{enumerate}
\item An event $k$ is drawn according to its probability $p_k$ (\ref{P_K}) and performed.
\item The crystal is locally relaxed around the location of the event.
\item The rates for all diffusion events affected by the local relaxation (that is at least the
particles within the relaxation sphere) are newly calculated. The search tree is
updated accordingly.
\item The system time is increased by the time interval $\tau$, equation (\ref{TAU}).
\item If the condition for a global relaxation is met the whole system is relaxed and all diffusion
barriers are newly calculated. Accordingly the search tree is updated.
\item The iteration starts from step $1$ unless a stopping condition - for example when a certain system time is reached 
or a maximum number of particles has been deposited - is fulfilled. 
\end{enumerate}
We should stress here that considering the used computer time the administration of events plays
a minor role in our simulations. The time consuming part of our simulations 
is the calculation of the interaction potential between the particles.

\section{Lattice based method}
 Since calculations of the potential energy always involve a cut--off distance it is possible to predict at 
the beginning of each iteration run which particles will participate in the following energy calculations.
So one could - when a event is drawn - search the whole crystal once for the relevant particles.
In fact this saves a lot of computer time compared to a new online search through the whole crystal  
for every calculation of the interaction potential. But still all particles of the system have to be considered once per
iteration which involves many expansive calculations of distances.

However, in the case of moderate misfit, if the lattice structure is only bent but not broken, one can exploit 
the advantages of a lattice based method.
This makes a double entry book--keeping necessary: on one hand each particle of the crystal is assigned to 
a lattice position in a perfect crystal - in the 3d case indexed by a triple of integer numbers $(i,j,k)$.
On the other hand each particle has still its coordinates given by real numbers $(x,y,z)$. The condition for
the adaptability of the method is that a non--ambiguous mapping between $(i,j,k)$ and $(x,y,z)$ remains possible
throughout the whole simulation. That means in particular that the formation of dislocations is prohibited.

The big advantage of the lattice based method is that the positions of the relevant particles are given by
basic arithmetical operations of integer numbers. Calculations of the particles' real distances are only done
when necessary which saves a lot of computer time during the simulations.
 A nice side effect is that one knows a lot about the structure of the growing crystal. 
This can, for example, be used to detect island edges during the simulations or allows to use pre--calculated
diffusion barriers for recurrent situations.

However, if one wants to simulate heteroepitaxial growth in the large misfit regime - depending on the choice of
the potential - the lattice based method is inoperative. One way here to speed the simulations up is to divide the
system into boxes of side length $r_{cut}$. During the energy calculations for a certain particle only the box which
contains this particle together with the neighboring boxes  has to be  considered. Thus only a small fraction of the 
system has to be searched for relevant particles. It is understood that this requires again a book--keeping over all 
boxes and their particles. This box method is applied in \cite{Vey:Diplom} to the off--lattice simulations and 
accelerates the simulations of dislocation formation significantly.

\section{Conclusions}
We have presented an off--lattice KMC algorithm which allows for the simulation of heteroepitaxial growth
over a wide range of misfits. The following chapters will show that this simulation technique can be applied
to various problems of heteroepitaxial growth and describes them quite successfully.

We have to stress that within this method we are not yet aiming at the simulation of distinct material
systems but heteroepitaxial growth is examined in a qualitative way.
To get a more realistic description of certain materials one would need to use material specific potentials.
On the one hand such potentials for semiconductor (see e.g. \cite{Abraham:1998:MOA,Tersoff:1988:NEA}) 
or metallic (see e.g. \cite{Abraham:1998:MOA}) systems are generally 
numerically much more demanding. Since this potentials are many body potentials and include long--range interactions 
(leading to a greater cut--off distance) they are currently out of scope. On the other hand, most of the material specific
potentials are fitted to properties of the bulk like particle distances and elastic constants and most of them
fail to describe properties of the crystal surface properly.

In the following the aim is to gain general insights in strain--related phenomena rather than to 
obtain material specific results. We focus on phenomena observed in a various number of heteroepitaxial systems 
and therefore should not depend on a particular choice of potential. We restrict ourselves to fundamental questions
like the influence of the misfit or the steepness of the used pair--potential on the described phenomena.
  \cleardoublepage
\chapter{Simulation of misfit dislocations }
\label{KAP-4} 
In this chapter we focus on the simulation of incoherent heteroepitaxial growth where misfit dislocations
appear in the growing film. 
At moderate misfits ($|\varepsilon|\ll 1$) the adsorbate first grows coherently with 
the substrate i.e. the topology is that of a perfect crystal. However, the thicker the adsorbate 
film becomes, the higher becomes the elastic energy stored in the film. At the so--called critical thickness $h_c^d$ the
strain is relieved by the introduction of misfit dislocations (see also chapter \ref{KAP-1}).
In this new incoherent state the crystal topology is perturbed near the substrate/adsorbate interface.
With further deposition of adsorbate material the dislocations are buried and the lattice can grow with 
the adsorbate's lattice constant. In technical applications the formation of dislocations have 
desired as well as undesired effects.

One example is the fabrication of II--VI semiconductor lasers which emit light in the  blue and green region
\cite{Pinardi:1998:CTS}.
Here, for example ZnSe has to be deposited on a GaAs substrate implying a significant lattice mismatch of
about $0.27\%$ at room temperature. But the introduction of misfit dislocations can lead to unstable devices of poor
reliability. In order to avoid the dislocation formation one needs to know the critical thickness under the given growth 
conditions and for the used alloy layers.

On the other hand the impact of buried dislocation structures for example on the fabrication of magnetic
nano--particles has been discussed, recently. One goal here is to grow patterned arrays of Co pillars on a metal
substrate.
To this end a thin Pt film is deposited on a sapphire substrate. The lattice mismatch between sapphire and Pt 
induces the introduction 
of dislocations in the Pt film. Since these dislocations repel each other the equilibrium configuration is given by
a network of rather equally spaced dislocation arrays.  
If the Pt film is thin enough  the variation of the lattice constant due to the
dislocations causes modulations of the activation barriers for Co atoms diffusing on the film's surface. Since - like 
mentioned in chapter \ref{KAP-1} - in metallic systems the diffusion is  decelerated 
in regions of a tensile strained surface, the regular network of dislocations yields patterned nucleation of Co islands 
\cite{Sabiryanov:2003:SDG}. In this way the formation of dislocations opens a pathway for the self--assembly of 
novel magnetic data storage devices.

In conclusion the knowledge about dislocation formation mechanisms, the dependence on the properties of the 
used materials (especially the misfit) and the influence of formed dislocations on further growth of the crystal
play an important role for technical applications.

In this chapter we show that the introduced method (cf. chapter \ref{KAP-3}) is capable of simulating the 
formation of dislocations over a wide range of misfits. 
In the first part of the chapter we focus on the simulation of dislocation formation in the 
high misfit region $|\varepsilon| \geq 3\%$. We examine under which conditions the two mentioned
types of dislocations - climb and glide (see chapter  \ref{KAP-1}) - are formed.
We determine the critical thickness as a function of the misfit and show that our results
may be fitted well by a simple power law.
In the second part of the chapter we address phenomena concerning buried dislocations for a 
region of relatively small misfit ($1.4\% \leq \varepsilon \leq 2.2\%$). 
The simulation results on the evolution of the lattice constant as a function of the adsorbate film 
thickness are compared to experimental results, showing a good qualitative agreement. 
\section{Growth in the high misfit region}
\label{KAP-4_HMR}
As described in chapter \ref{KAP-3}, the simulation of dislocations does not allow for the use of a 
lattice based method, which would reduce the computational effort a lot. 
For that reason we have to restrict our simulations in the following to rather small system sizes in 
$1+1$ dimensions. Each simulation run starts with six atomic layers of substrate with a fixed 
bottom layer. Due to the fixed bottom layer no dislocations 
are introduced in the substrate during growth. The system size $L$ (number of particles in the substrate's upper 
layer) is between $L=100$ and $L=200$. 
Within this range we found no significant dependence of the results on $L$. 
Adsorbate particles are deposited on the crystal's surface with a deposition rate $R_d=1ML/s$.

Because of its numerical feasibility all simulations are carried out for the Lennard--Jones $12,6$ potential
\begin{equation}
\label{LJ}
 {U}_{ij} =\ 4 E_{ij} \left[\left(\frac{{\sigma}_{ij}}{r_{ij}}\right)^{12}
-\left(\frac{{\sigma}_{ij}}{r_{ij}}\right)^{6}\right] 
\end{equation}
(also see appendix \ref{AP-1}).
We choose the same potential depth for all types of particle--interactions and set $E_{ij}=U_0= 1.3125eV$.
In the homoepitaxial case ($\varepsilon=0$) this choice results
in a rather high activation barrier for surface diffusion of about $0.90eV$.
In order to save computer time the interaction potential 
$U_{ij}$ is cut off for distances $r_{ij}>r_{cut}$ with  $r_{cut}=3r_0$, where $r_0$ 
is the equilibrium distance between two nearest neighbors in the lattice. 
The interaction strength at $r_{cut}$ is 
less than $1\%$ of the value at $r_0$ and can therefore be neglected. 
The interaction of two substrate particles is given by $U_{ij}\left(\sigma_{s}\right)$. Two adsorbate 
particles interact via $U_{ij}\left(\sigma_{a}\right)$ whereas we assume that  
a substrate and an adsorbate particle interact via  
$\frac{1}{2}\left(U_{ij}\left(\sigma_{s}\right)+U_{ij}\left(\sigma_{a}\right)\right)$.
Measuring lengths in units of $\sigma_{s}$, 
$\sigma_{a}$ is chosen between $0.85$ and $1.11$, so we can simulate 
heteroepitaxial growth for misfits
\begin{equation}
\label{misfit}
\varepsilon=\frac{\sigma_{a}-\sigma_{s}}{\sigma_{s}}
\end{equation}
between $-15\%$ and $+11\%$. 
For each value of $\varepsilon$ between $5$ and $10$ independent simulation runs are carried out.

Our calculations in chapter \ref{KAP-2} showed that in the case of the Lennard--Jones $12,6$ potential one should take
exchange diffusion for downward movement at step edges for values of the misfit $\varepsilon \leq -13\%$ 
and $\varepsilon \geq 7\%$  into account. However, it is conceptually rather complicated to implement concerted moves
like exchange diffusion events in our off--lattice simulations. 
We therefore neglect exchange diffusion here. This is partly justified by the rather low temperature and high 
potential depths used in our
simulations ($T=0.03 \frac{U_{0}}{k}$), resulting in comparatively small rates for all interlayer diffusion 
processes.

\subsection{Formation of dislocations}
We now investigate the formation mechanism of dislocations. 
In our simulations we observe two different mechanisms of dislocation formation, leading to two different 
types of dislocations: climb and glide dislocations (see also chapter \ref{KAP-1}, fig. \ref{Climb_Glide}).

\subsubsection{Climb dislocations}
\begin{figure}[p]
\centerline{
\begin{minipage}{0.49 \textwidth}
  \epsfxsize= 0.95\textwidth
  \epsffile{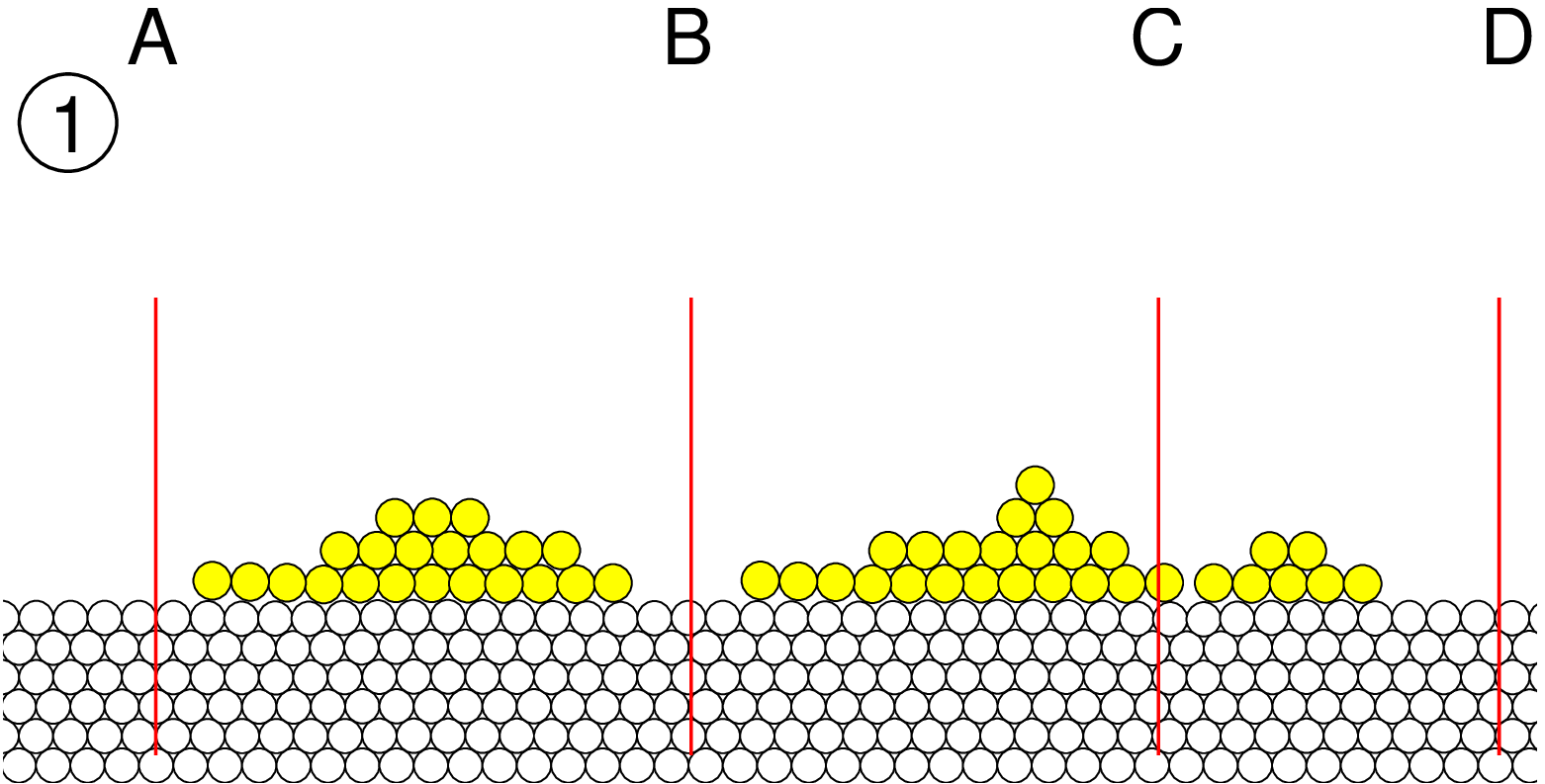}
  \epsfxsize= 0.95\textwidth
  \epsffile{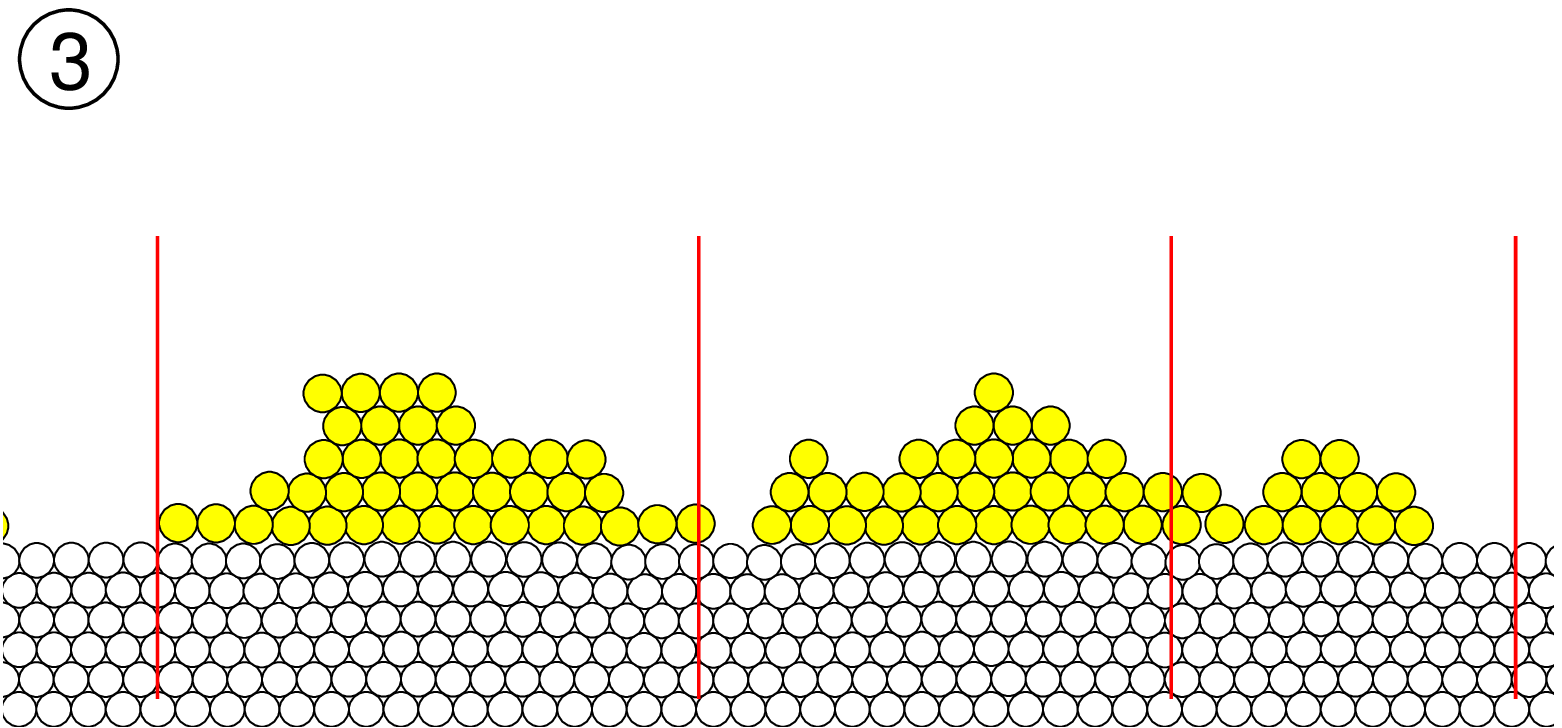}
  \epsfxsize= 0.95\textwidth
  \epsffile{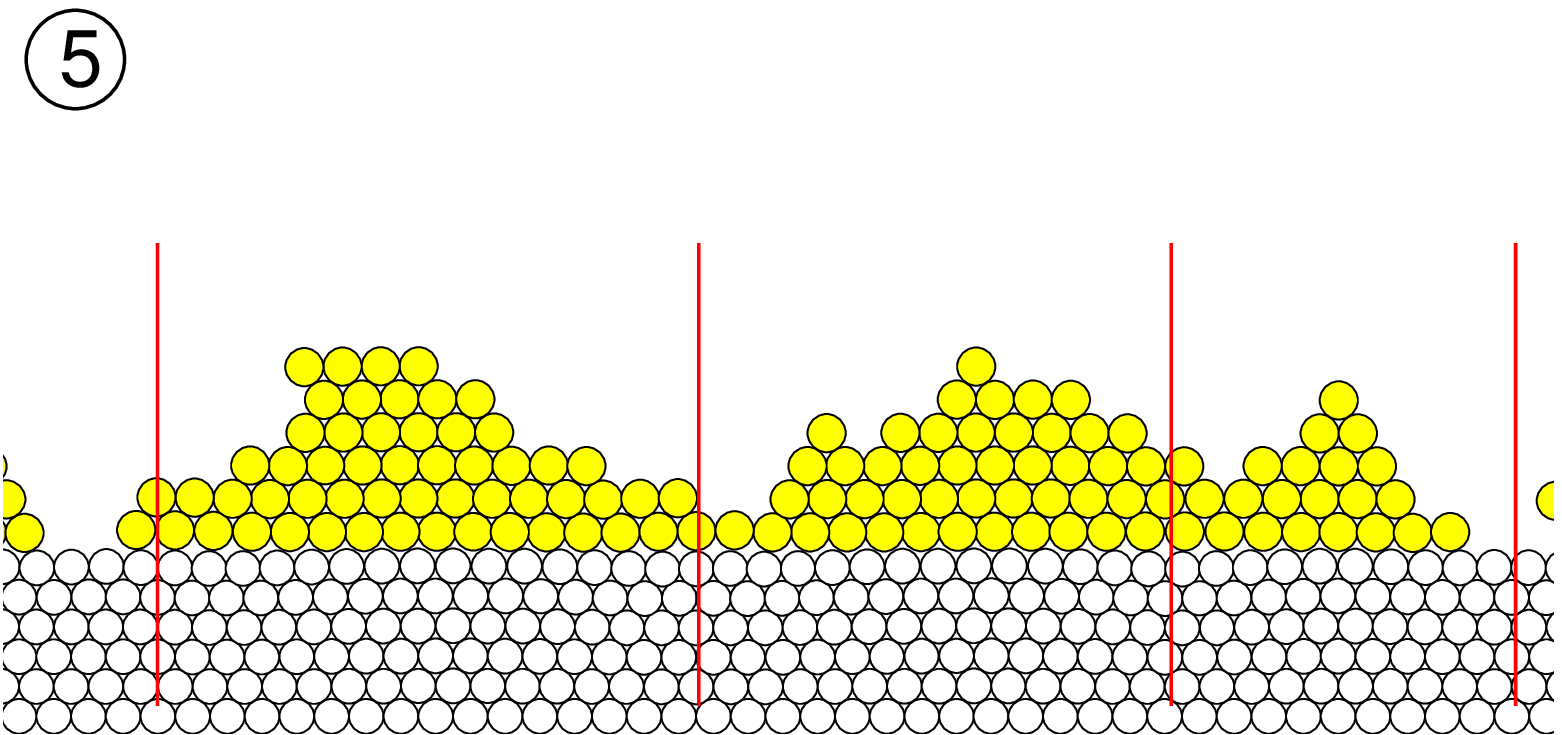}
  \epsfxsize= 0.95\textwidth
  \epsffile{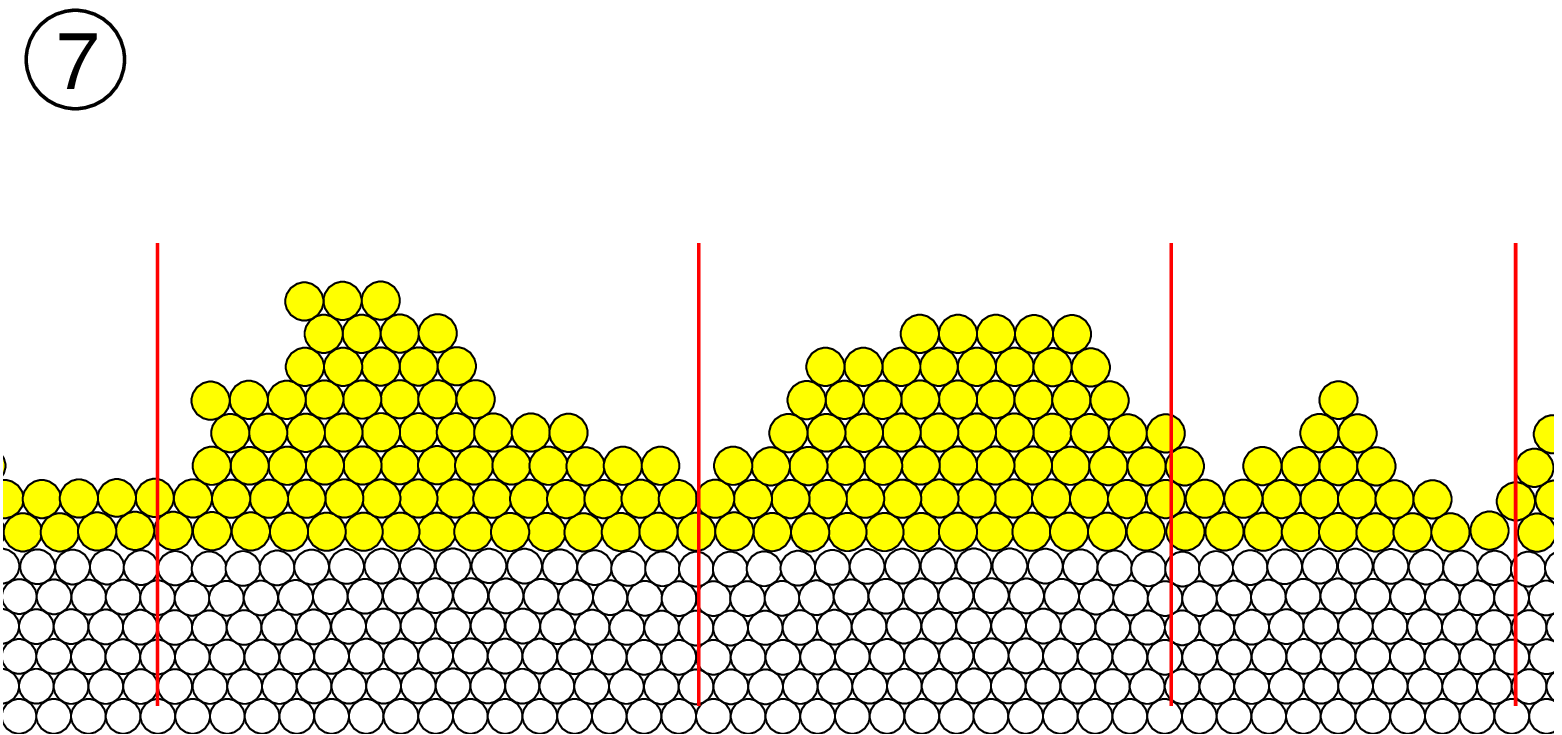}
\end{minipage}
\hfill
\begin{minipage}{0.49 \textwidth}
 \epsfxsize= 0.95\textwidth
  \epsffile{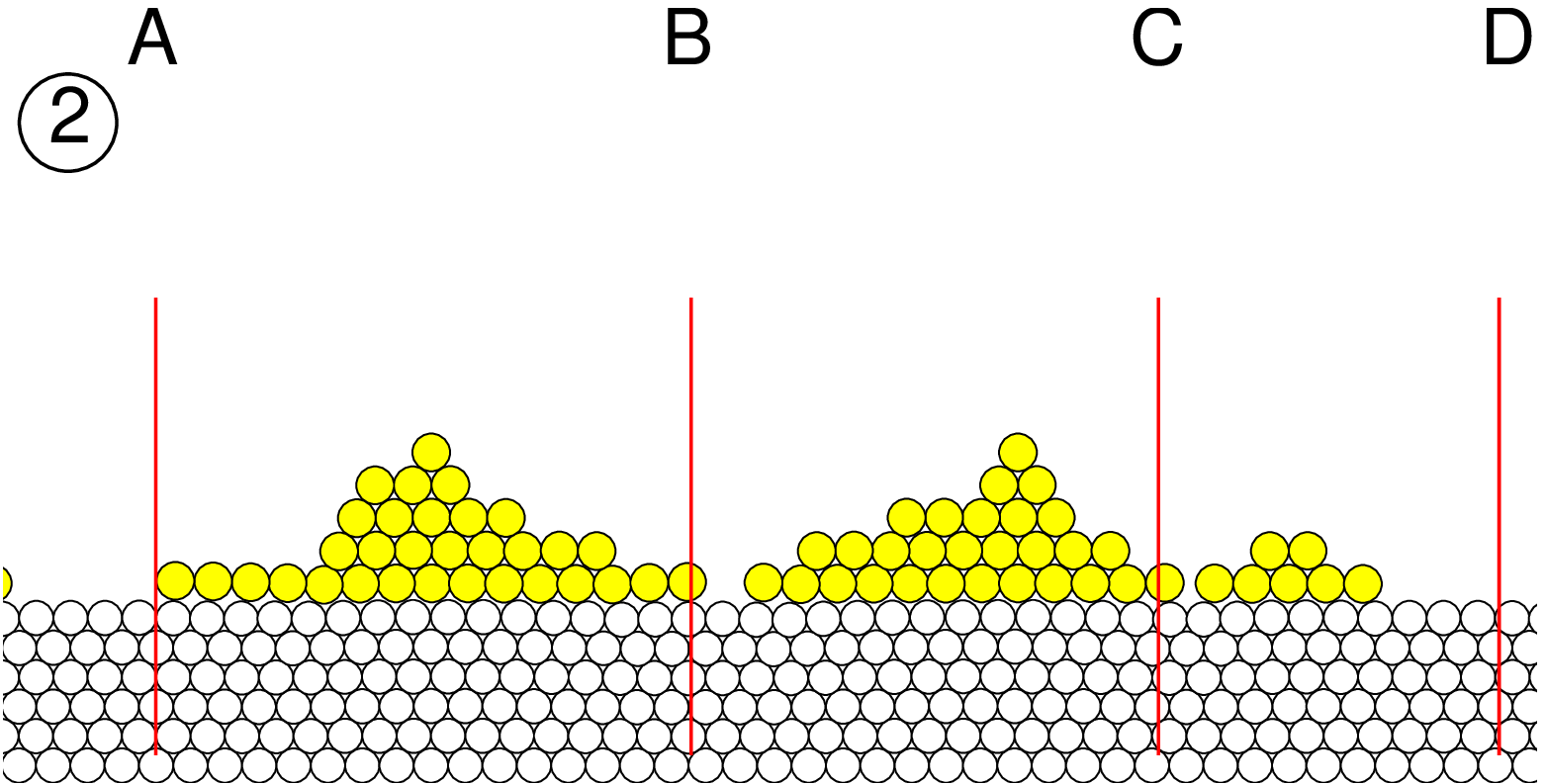}
  \epsfxsize= 0.95\textwidth
  \epsffile{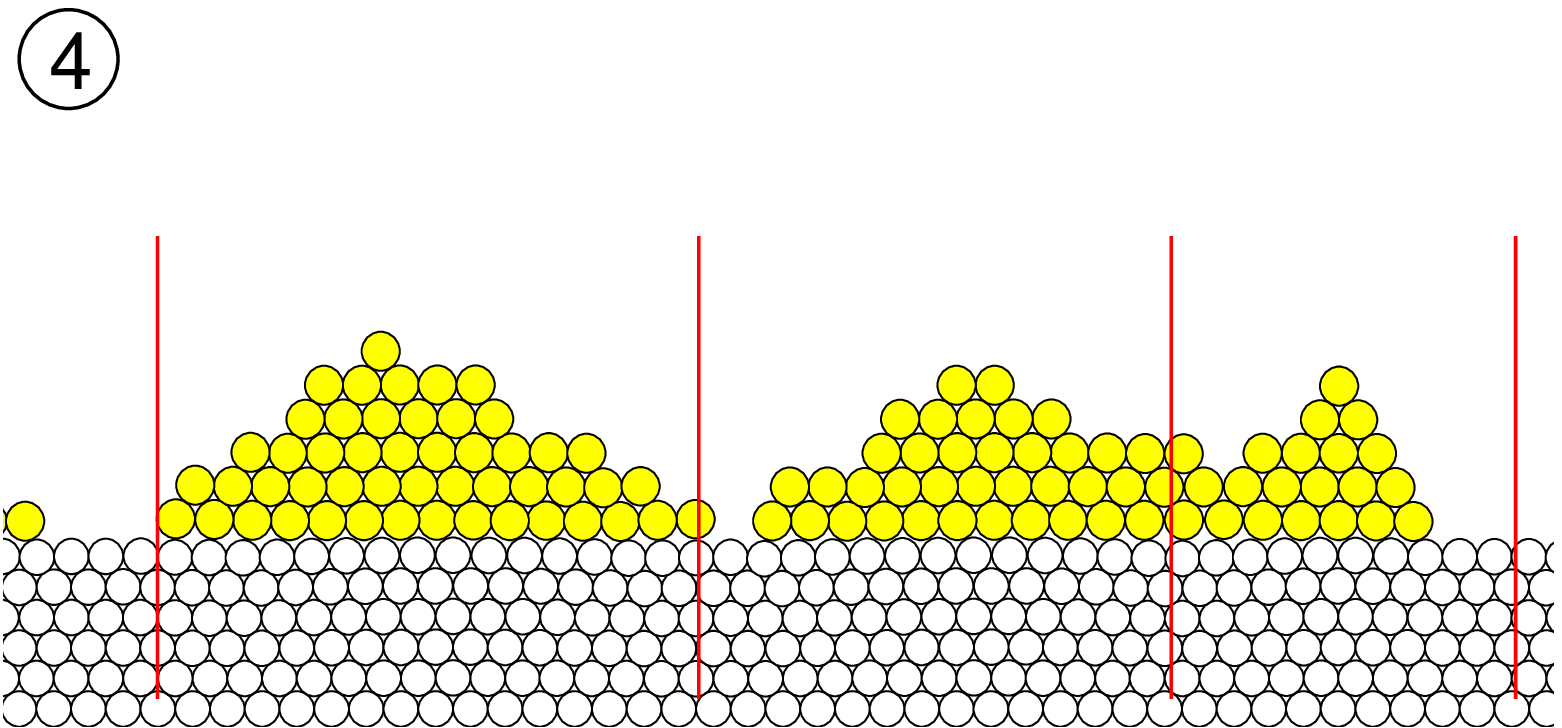}
  \epsfxsize= 0.95\textwidth
  \epsffile{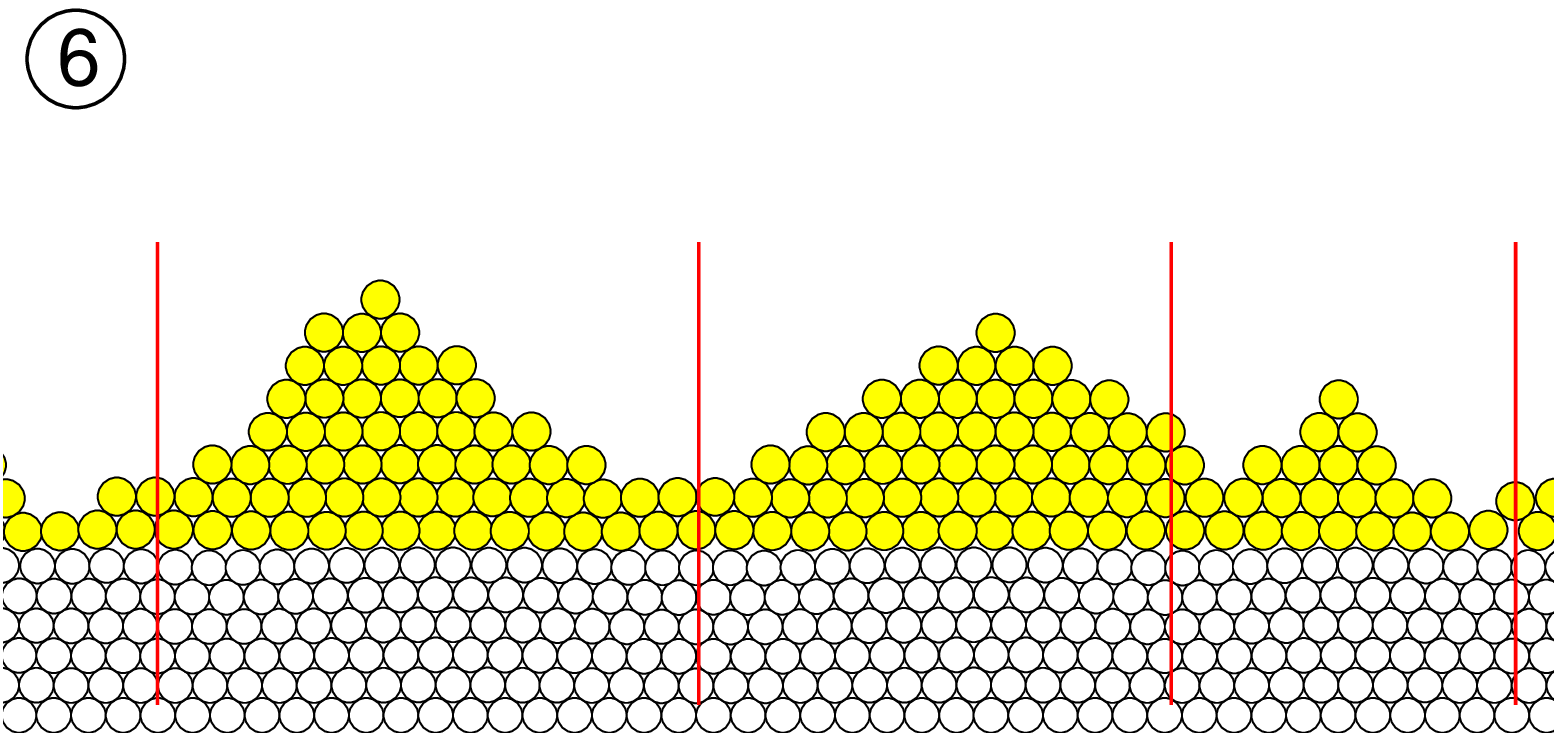}
   \epsfxsize= 0.95\textwidth
  \epsffile{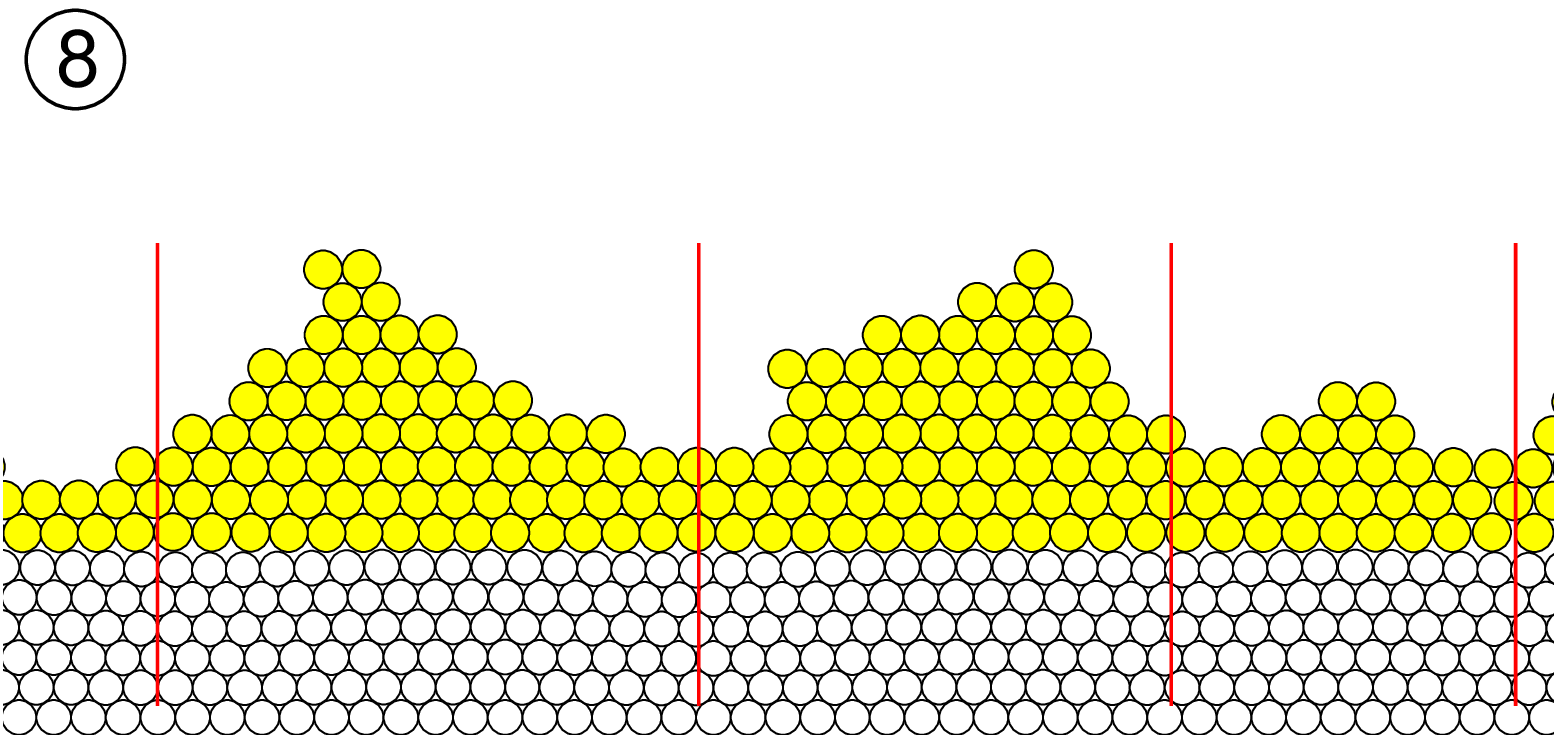}
 \end{minipage}}

\centerline{
   \epsfxsize= 0.49\textwidth
  \epsffile{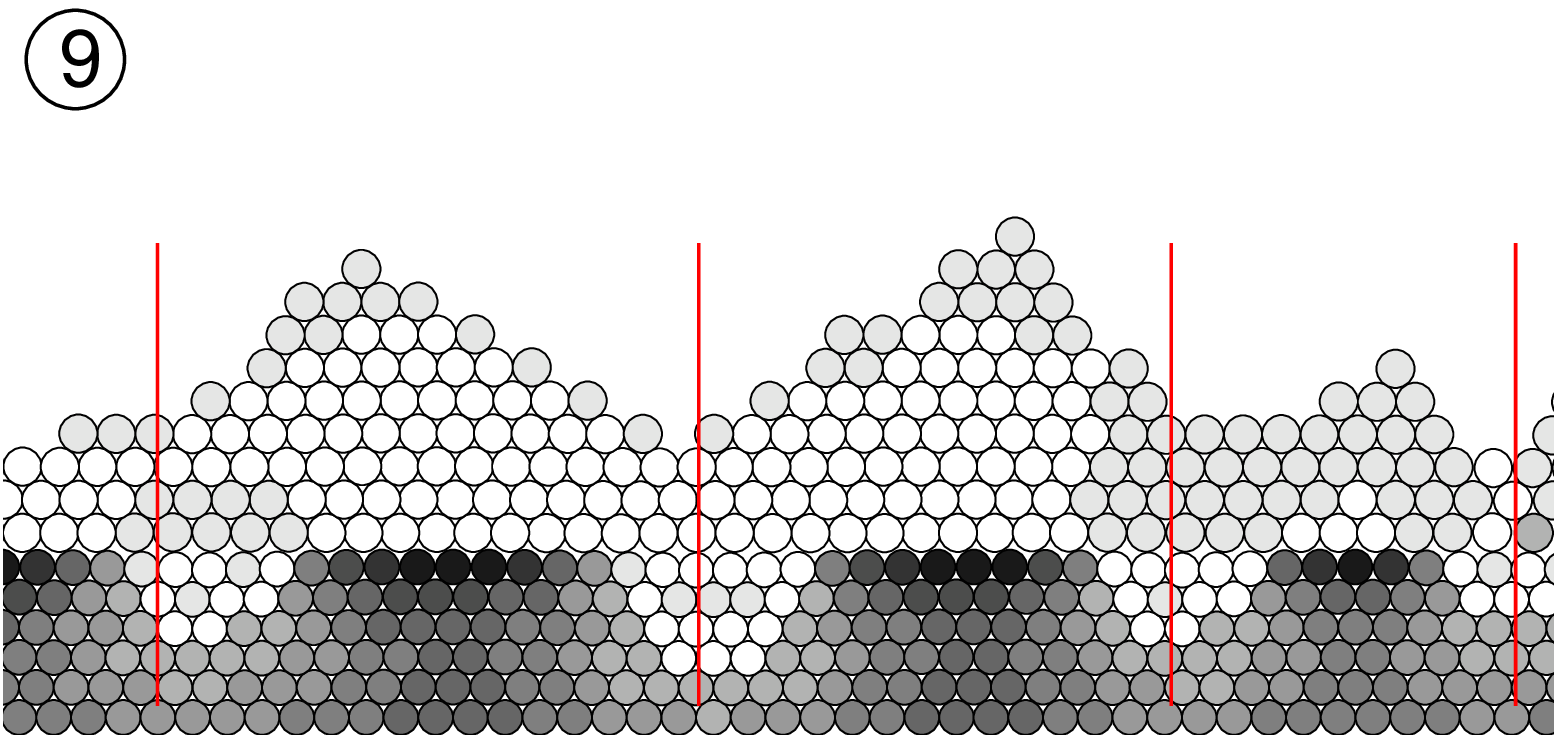}
}
\caption{Formation of climb dislocations: section of a crystal at a coverage with adsorbate particles 
between $1ML$ \textcircled1 and $5ML$ \textcircled9 in steps of half a monolayer. The adsorbate particles 
($\varepsilon=10\%$) appear in
dark grey.  
Final locations of dislocations are indicated by the vertical lines (A to D). 
In figure \textcircled9 the grey level of the particles indicates the particle's average distance to its nearest 
neighbors of the same kind. The lighter its grey level the more is this particle under compression.
}
\label{HIGH_EP_FORM}
\end{figure} 
First we discuss the introduction 
of climb dislocations (see fig. \ref{Climb_Glide}(a)), characterized by a Burgers vector parallel 
to the substrate/adsorbate interface.
To this end, figure \ref{HIGH_EP_FORM} displays the evolution 
of a crystal section for $\varepsilon=10\%$ 
- a rather high positive misfit, where the dislocations are found at the interface of
substrate and adsorbate. The crystal section is shown for different coverage with adsorbate particles 
from $1ML$ (fig. \ref{HIGH_EP_FORM} \textcircled1) to $5ML$ (fig. \ref{HIGH_EP_FORM} \textcircled9) 
in steps of half a monolayer.
The vertical lines (A to D) in the samples  indicate the final positions of the climb dislocations introduced during 
the growth.

In the early stages of growth  (fig. \ref{HIGH_EP_FORM} \textcircled1, \textcircled2) the given section of the
substrate is covered by three well separated adsorbate islands. The formation of islands is here due to the 
rather high Schwoebel barrier (cf. chapter \ref{KAP-2}) in combination with the low growth temperature.
As one can see the islands fit the lattice spacing of the substrate rather well in the island's centers. 
Towards the rims of islands the adsorbate relaxes to its preferred lattice constant. This results in incoherent 
states at the rims indicating already the formation of dislocations.

At a coverage of $2ML$ (fig. \ref{HIGH_EP_FORM} \textcircled3, \textcircled4) at point C two islands start to
merge resulting in a climb dislocation at their contact point. In figure \ref{HIGH_EP_FORM} \textcircled5 the 
same mechanism is observed at point B. Since the given samples are typical for the growth at 
high misfits we conclude that climb dislocations arise preferentially from the merging points of islands.
Samples \textcircled6 to \textcircled9 show that in our model once formed climb dislocation do neither
vanish nor move with further deposition of adsorbate particles.

Figure \ref{HIGH_EP_FORM} \textcircled9 shows the final state of the crystal at $5ML$ coverage. 
Note that the initial three islands can still be identified as the tops of mounds on the crystal's surface. 
Here the grey level for a particle indicates the particle's average distance to its nearest 
neighbors of the {\it same} kind: the lighter its grey level the more is this particle under compression.
In the case of $\varepsilon>0$ adsorbate particles therefore appear darker near dislocations, since they 
are able to approach their preferred (greater) lattice constant there. In regions where the adsorbate
grows coherently with the substrate the compressed adsorbate particles are displayed  brighter.
Note that the substrate is also influenced by the situation of the adsorbate film above: in case of 
positive misfit the substrate is under tension in regions of coherent growth, whereas near dislocations 
the substrate is compressed.

\subsubsection{Glide dislocations}
\begin{figure}[h!]
\begin{minipage}{0.30 \textwidth}
  \epsfxsize= 0.99\textwidth
  \epsffile{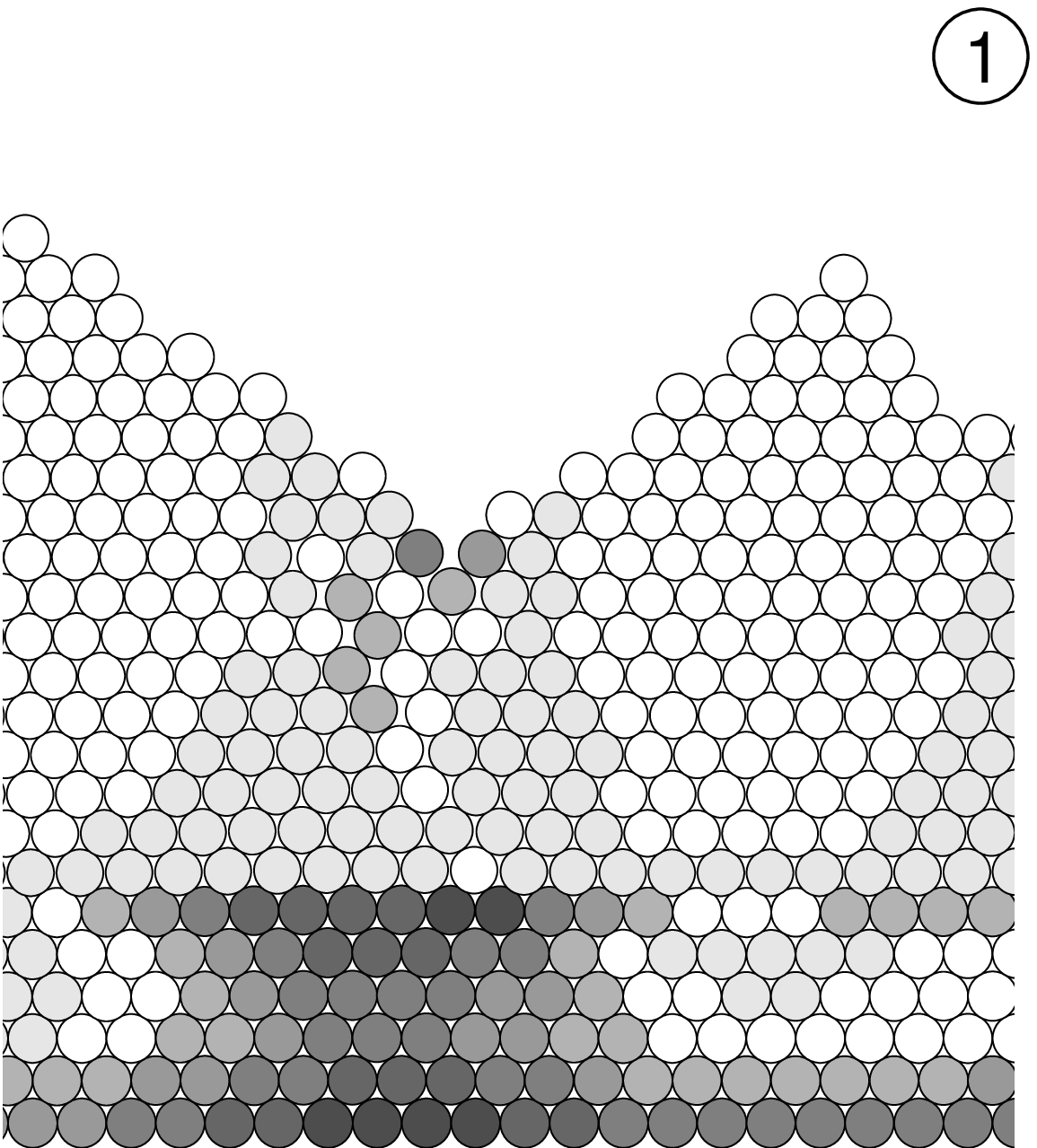}
\end{minipage}
\hfill
\begin{minipage}{0.30 \textwidth}
  \epsfxsize= 0.99\textwidth
 \epsffile{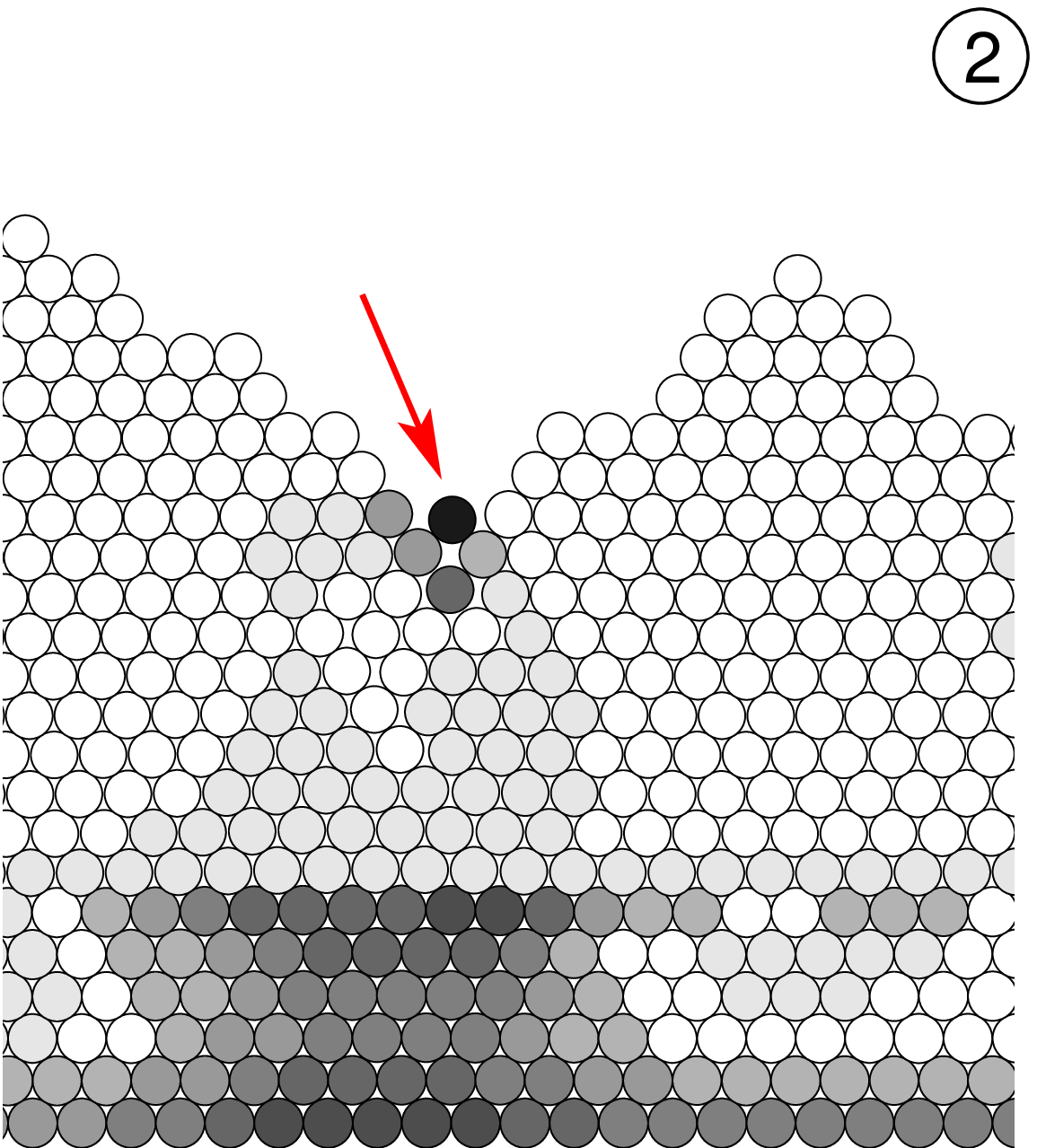}
\end{minipage}
\hfill
\begin{minipage}{0.30 \textwidth}
  \epsfxsize= 0.99\textwidth
 \epsffile{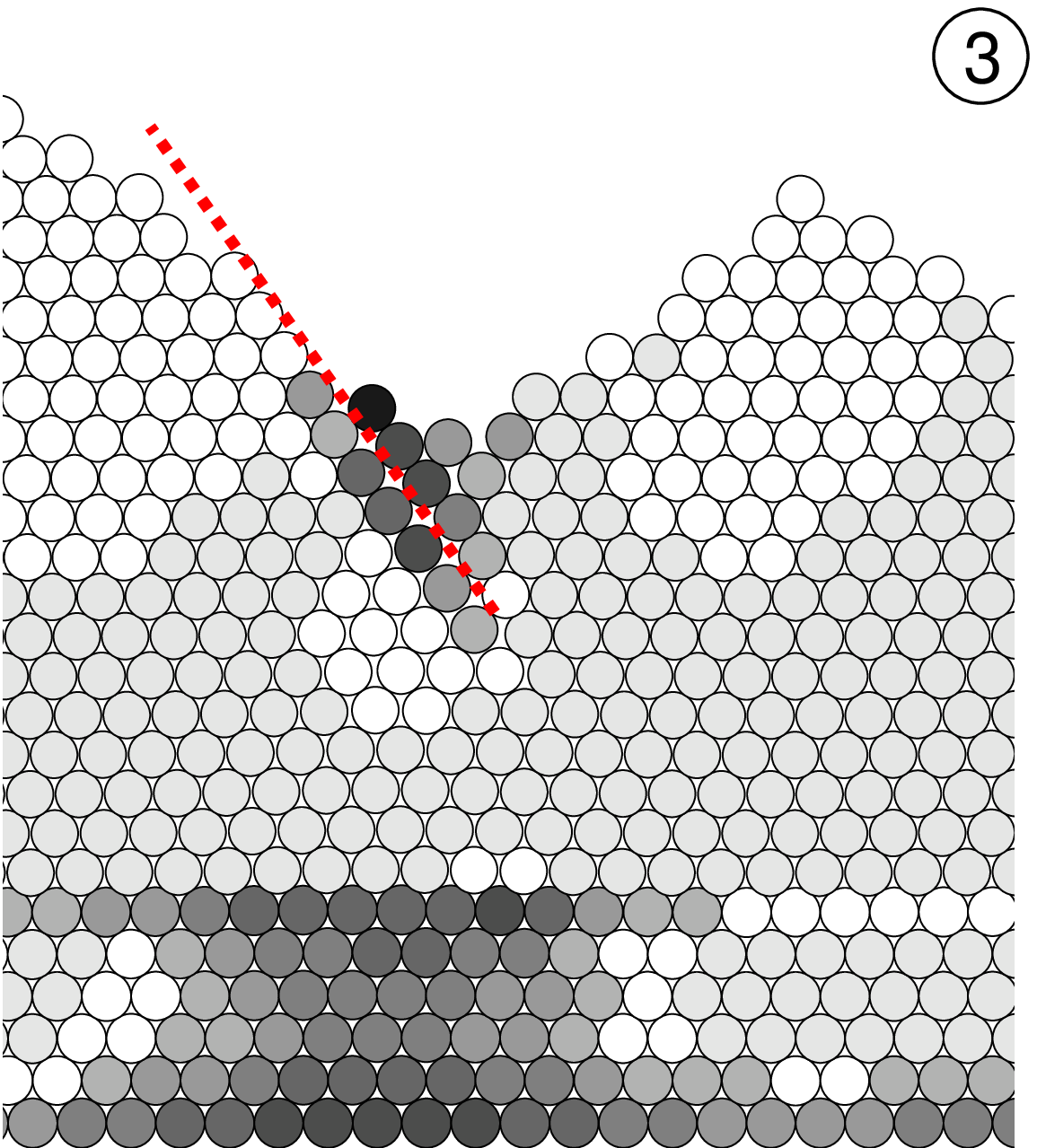}
\end{minipage}
\\
\begin{minipage}{0.30 \textwidth}
  \epsfxsize= 0.99\textwidth
  \epsffile{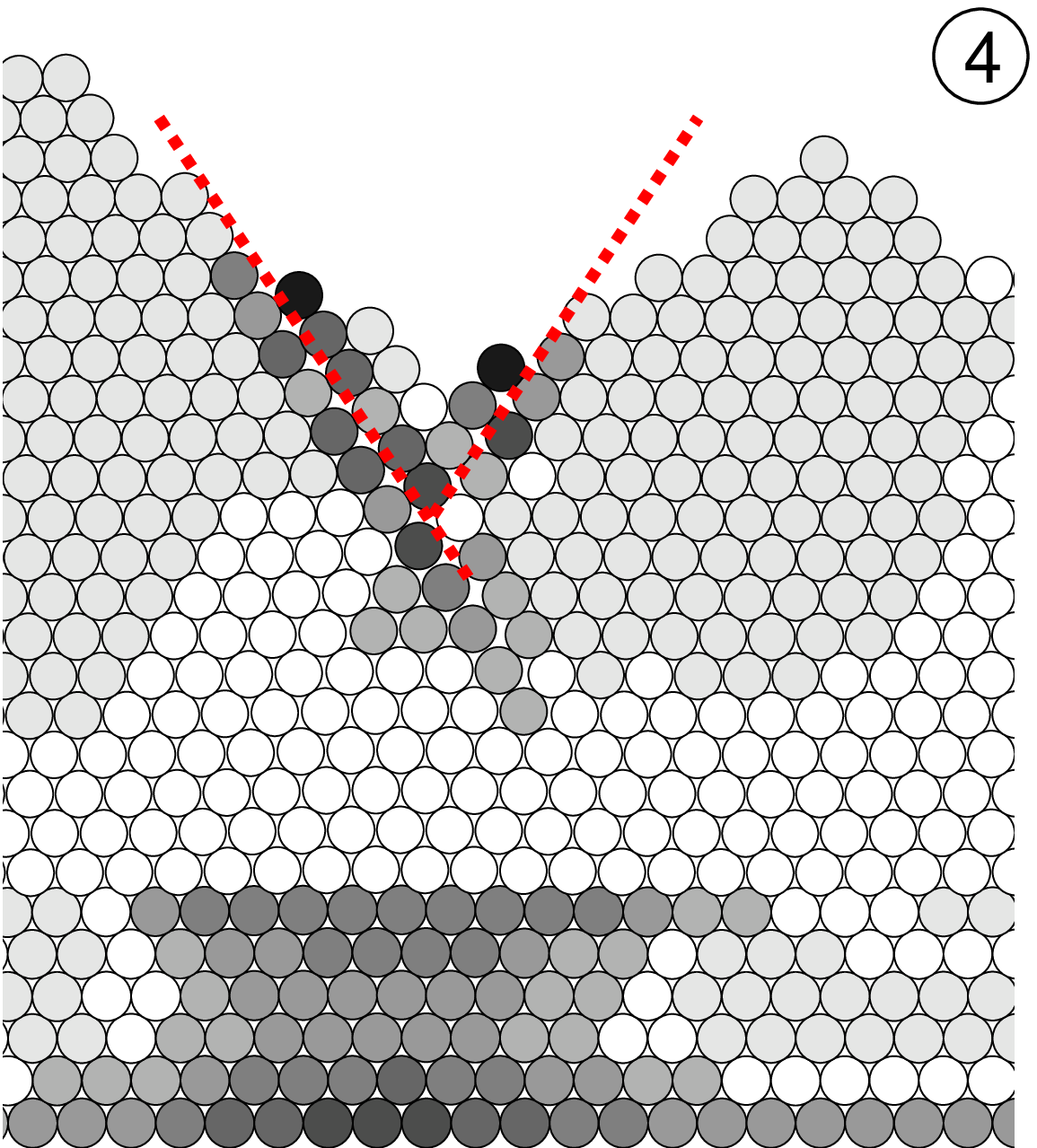}
\end{minipage}
\hfill
\begin{minipage}{0.30 \textwidth}
  \epsfxsize= 0.99\textwidth
 \epsffile{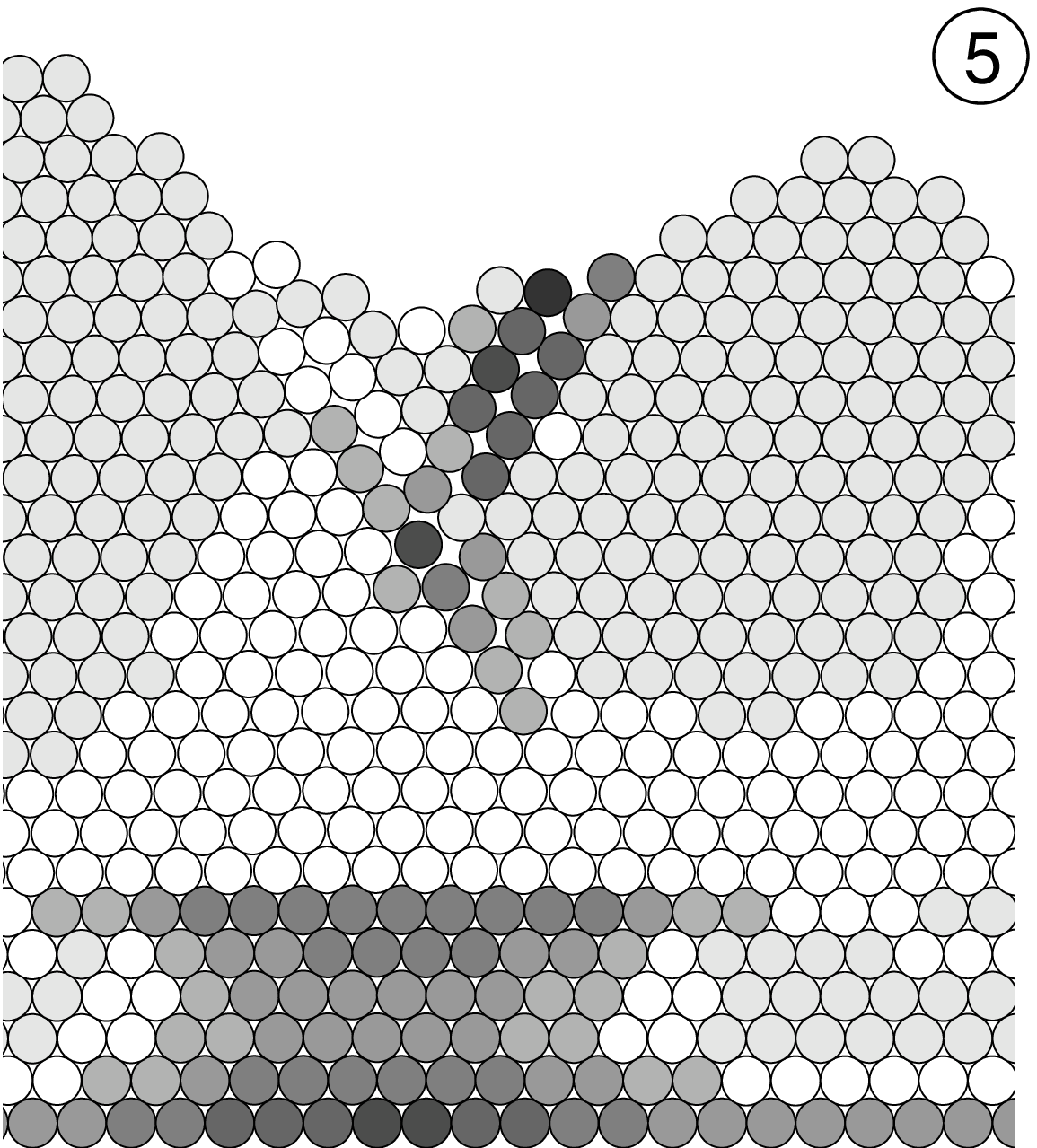}
\end{minipage}
\hfill
\begin{minipage}{0.30 \textwidth}
  \epsfxsize= 0.99\textwidth
 \epsffile{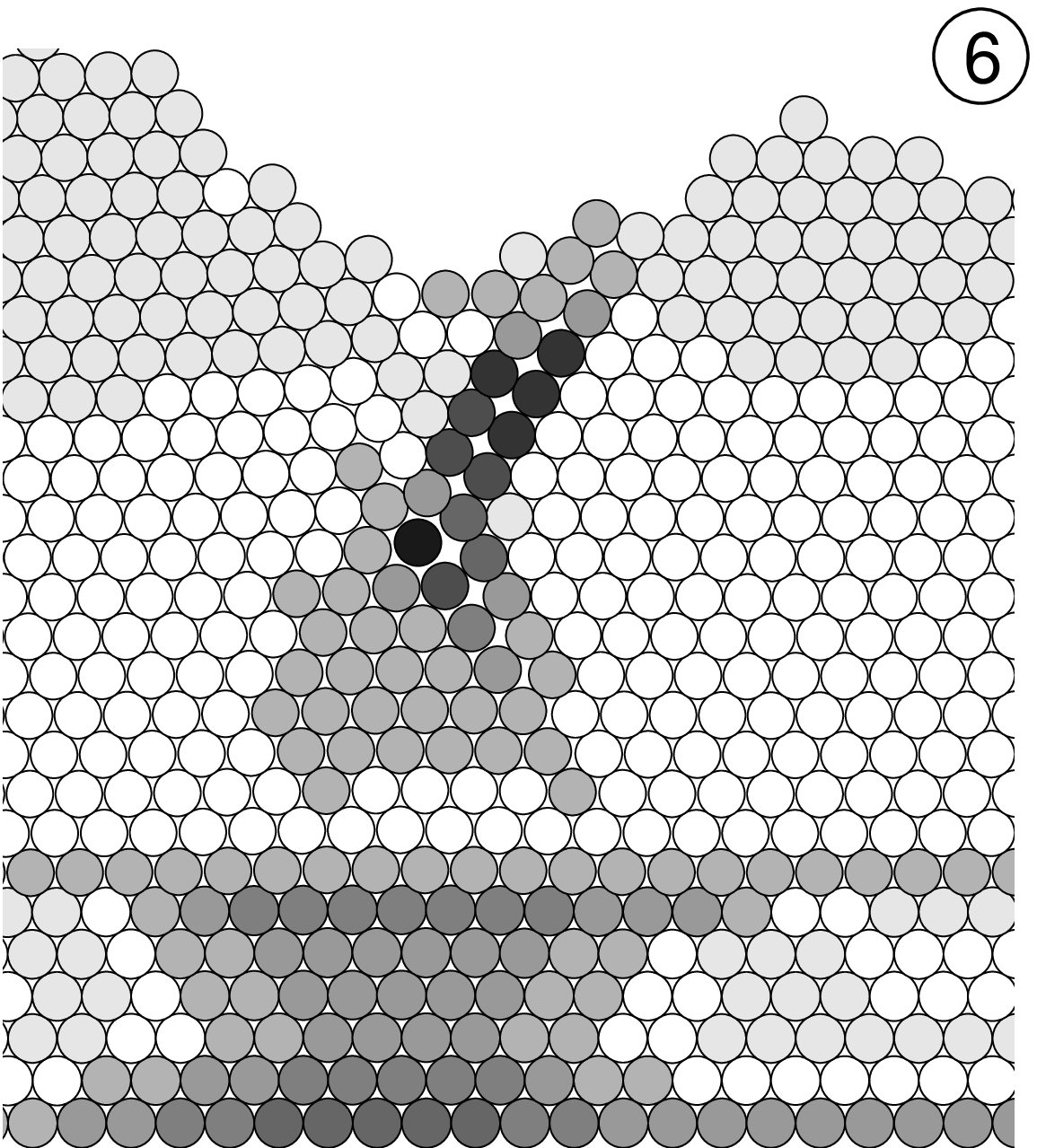}
\end{minipage}

\caption{Formation of glide dislocations for a crystal sections, misfit is $\varepsilon=-5\%$. 
The coverage with adsorbate particles for the samples \textcircled1 to \textcircled6 
is $12ML$, $13ML$, $15ML$, $16.75ML$, $17ML$ and $18ML$, respectively.}
\label{GLIDE_FORM}
\end{figure}
At lower misfits the resulting dislocations can be of a different kind. Here we often find the formation 
of glide dislocations - characterized by a Burgers vector with a component vertical to the substrate/adsorbate
interface which does not contribute to the relaxation (also see fig. \ref{Climb_Glide}(b)). 
Figure \ref{GLIDE_FORM} shows the evolution of a crystal section for a misfit $\varepsilon=-5\%$ 
with a coverage of the substrate from $12ML$ to $18ML$.
For the given value of the misfit dislocation formation does not start at the substrate/adsorbate interface but a 
few monolayers above.

Once again we  observe that the formation of the dislocation starts in the valley between two mounds on 
the crystal's surface. Here the {\it core} of the dislocation is located - indicated by an arrow 
in sample \textcircled2.
From the core the dislocation first grows in two branches shown in \ref{GLIDE_FORM} \textcircled3, \textcircled4.
With further growth \textcircled5 the left branch of the dislocation vanishes with only the right one remaining 
\textcircled6. Note that, although the dislocation is introduced a few monolayers above the substrate/adsorbate
interface, substrate particles are influenced by the morphology of the adsorbate film: the substrate is slightly under
compression in the region where the dislocations arise in the adsorbate film above.

The reason for the misfit dependency of the formation mechanisms is due to the activation 
barrier for the introduction of a dislocation $\Delta E_d$. A recent study by Trushin et. al. 
\cite{Trushin:2003:EAM} showes that in case of the Lennard--Jones system for misfits $\varepsilon \geq 8\%$ 
the climb mechanism is the preferred dislocation formation mechanism. For smaller misfits
both mechanisms - climb and glide - are competitive in energy costs. This findings are in good agreement with our
simulation results, where in the large misfit regime climb dislocations are found to be the 
preferential dislocation type. 

As our simulations indicate dislocations of both types tend to form in valleys between two mounds respectively
islands on the crystal surface. A rough surface therefore seems to facilitate the dislocation formation.
Since particles in mounds can relax their strain up to a certain amount 
(see also chapter \ref{KAP-5}) the binding positions 
in the valleys between them are energetically especially unfavorable.
For example in case of positive misfit particles in a mound's surface can move a little outwards resulting 
in a higher, more favorable distance to their binding partners. But binding sites in the valley where two 
mounds meet are then under compression and serve as seeds for the dislocation nucleation.  
Indeed simulations for smoother surfaces show \cite{Vey:Diplom} that the appearance of dislocations is shifted
to higher thicknesses of the adsorbate film.

\subsection{Number of dislocations}
In our simulations we observe that in each run several dislocations appear quite simultaneously, within the 
range of a few monolayers. 
After the deposition of a few additional monolayers of adsorbate material the number of dislocations stays  
constant. It is clear that, the higher the value $\varepsilon$ of the misfit is, the more dislocations are needed 
in order to relax the strain stored in the crystal. 
In the following we discuss the functional dependence between the number of 
dislocations $n_d$ and $\varepsilon$.

To this end the number of dislocations is counted for each simulation run after the nucleation of dislocations
has stopped. To prove the existence of a dislocation we determine the coordination number $n_{c}$ of each particle 
by calculating the {\it Voronoy} polyhedra \cite{Ashcroft:1976:SSP,Strandburg:1988:TDM}. 
Voronoy polyhedra are a generalization of the 
{\it Wigner--Seitz} cell to a system without a fixed lattice. The number of sides of a Voronoy polyhedron gives the 
coordination number $n_{c}$ ($n_{c}=6$ for a particle in a perfect triangular lattice). 
A {\it Burgers} circuit \cite{Hirth:1968:TD} is drawn  
around regions of the crystal with $n_{c} \not= 6$. 
A non--vanishing Burgers vector then indicates the appearance of a dislocation.
\begin{figure}[hbt]
\centerline{
\begin{minipage}{0.55 \textwidth}
  \epsfxsize= 0.99\textwidth
  \epsffile{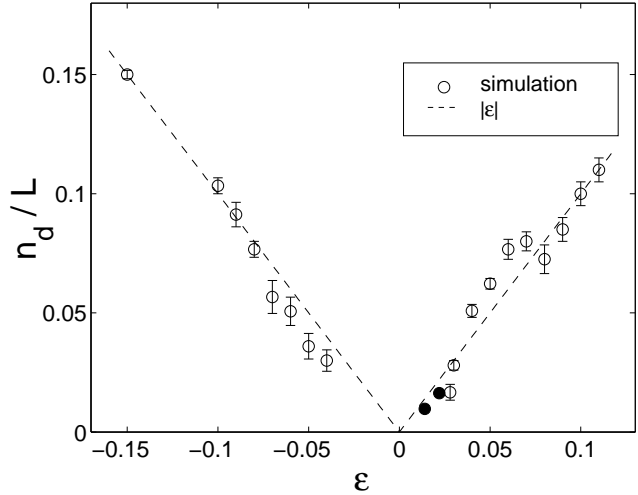}
\end{minipage}
\hfill
\begin{minipage}{0.40 \textwidth}
\caption{Number of dislocations per system size $n_{d}/L$ as a function of the misfit $\varepsilon$. 
  The data points for the low misfit region (filled) are by courtesy of Ch. Vey.
          The error bars represent as the standard error  
          of the simulation results.
          The dashed line gives the theoretical number of {\it perfect} 
          dislocations in a system of size $L$. }
\label{NUMBER}
\end{minipage}
}
\end{figure}

Figure \ref{NUMBER} shows the number of dislocations 
per unit length $n_{D}/L$ counted for each value of $\varepsilon$.
It is calculated after further deposition of six  
monolayers adsorbate material after the first 
appearance of a dislocation in the crystal.
At this thickness of the adsorbate layer the maximum number of dislocations 
should be reached. 
The dashed line gives the theoretical 
number of dislocations in a system of size $L$  under the assumption 
that $n_{D}=L|\varepsilon|$ perfect climb dislocations have to appear
on a rigid substrate in order to fit the adsorbate film onto the substrate. 
{\it Perfect dislocations}  
are those for which the crystal topology far from the substrate/adsorbate interface is the same
as in the coherent state and the Burgers vector is therefore an integer multiple of the lattice
vector.
The formation of glide dislocations causes the 
deviations from the theoretical results for $-7\% \leq \varepsilon \leq -3\%$ 
and $4\% \leq \varepsilon \leq 8\%$. This is due to the fact that glide dislocations are spatially more 
extended than climb dislocations. For this reason in the case $\varepsilon>0$ more and for $\varepsilon<0$ 
less dislocations than $n_{D}=L|\varepsilon|$  have to be built when partial dislocations appear. 
The deviations for small values of the misfit $\varepsilon \leq 3\%$ are mainly due to finite size effects 
and are discussed in detail in \cite{Vey:Diplom}.

\subsection{Critical thickness}
We now discuss
the dependence of the adsorbate thickness  $h_c^d$ where dislocations first appear on the misfit as observed 
in our simulations.
Since the high impact of dislocations on the quality of technical applications  
during the last 50 years several theoretical models have been proposed in order to calculate the critical layer thickness
$h_c^d$ for the formation of dislocations in heteroepitaxial growth. 
\subsubsection{Energy balance model}
The first theoretical treatment was done 
by Frank and van der Merwe \cite{Frank:1949:ODDa,Frank:1949:ODDb} in 1949. They compared the additional interfacial energy 
due to the formation of dislocations at the substrate/adsorbate interface with the elastic energy relieved by this
dislocations. The film thickness where both energies balance each other is then identified with $h_c^d$. This calculation 
implies that only interfacial dislocations may appear. 
Since this approach does only consider the energies of the relaxed and the strained state it neglects actual 
the mechanisms of dislocation formation. 
\subsubsection{Force balance model}
In order to account for these mechanisms in 1974 Matthews and Blakeslee \cite{Matthews:1974:DEM} 
proposed  a force balance model which takes a dislocation
nucleation mechanism of special importance for technical applications into account:
they assumed that preexisting dislocations in the substrate propagate into the adsorbate film (so--called 
threading dislocations). Once the force $F_{\varepsilon}$, acting to elongate the threading dislocation 
in the interface due to the misfits, balances $F_l$, the dislocation line tension resisting the elongation 
of the dislocation, $h_c^d$ is reached \cite{Liu:1999:FMD}. This yields an implicit equation for the 
critical thickness
\begin{equation}
\label{HC_THEO_MB}
 h_c^d=\frac{bC}{\varepsilon}\left(\ln\frac{h_c^d}{b}+1\right),
\end{equation}
where $b$ is the Burger's vector and $C$ is a constant depending on the orientation of the dislocation line and 
the elastic properties of the adsorbate material. 

Both the thermodynamic and the force balance approach yield identical results for the calculation 
of $h_c^d$ \cite{Dong:1998:SRM,Matthews:1975:DAA} as a function of material parameters like the shear modulus or the 
Poisson ratio. However, since for a misfit greater than $4\%$ the critical thickness becomes of the order of the lattice
constant these continuous approaches are no longer valid and the discrete nature of the atomic layers should be
taken into account \cite{Politi:2000:ICG}. Furthermore even in the low misfit regime experimental values 
for $h_c^d$ are usually larger
then the calculated ones \cite{Politi:2000:ICG,Dong:1998:SRM,Cohen-Solal:1994:CTH,Pinardi:1998:CTS}. Possible reasons for this
poor agreement (with discrepancies as high as an order of magnitude) are that the continuous approaches take only
dislocations near the substrate/adsorbate interface into account, 
different growth conditions (e.g. varying flux or temperature) are disregarded
and - most importantly - kinetic barriers for the introduction of dislocations are neglected.
As was shown in \cite{Trushin:2003:EAM} the incoherent, relaxed state is separated from the coherent, strained one 
by a activation barrier $\Delta E_d$ for the introduction of dislocations. This barrier stays finite 
even if the relaxed state is energetically favorable. The experimentally observed 
$h_c^d$ may therefore exceed the calculated values. Experimental results on the temperature dependence of 
the critical thickness also support the idea of strain relaxation as 
an activated process \cite{Tsao:1987:CSS,Zou:1996:TDG}.

\subsubsection{A power law for the critical thickness}
In order to compare our results on the critical thickness with experimental works we 
follow here a different approach
proposed by Cohen--Solal and coworkers \cite{Cohen-Solal:1994:CTH,Bailly:1995:SMD}.
There, an energy balance 
model is proposed for calculating the critical layer thickness in heteroepitaxial growth of semiconductor 
compounds. To this end the classical strain energy, without any change of the substrate 
or dislocation formation, 
and the deformation energy due to a full system of interfacial misfit dislocations were compared.
The energies were calculated using Keating's valence force field approximation \cite{Cohen-Solal:1994:CTH}. 
The method yields a power law dependency of $h_c^d$ on $\varepsilon$: 
 \begin{equation}
\label{HC_THEO}
 h_c^d=a^{*} \varepsilon^{-3/2},
\end{equation}
where $a^*$ is a material specific fit parameter.

Despite the fact that the model neglects both the mechanism of dislocation nucleation and the kinetics of 
strain relief it shows excellent agreement with experimental data.
The authors themselves fitted the power law to measured critical thicknesses 
for IV--IV, III--V and II--VI semiconductor compounds revealing an excellent agreement
with values of $a^{*}$ between $a^{*}=0.12$ and $a^{*}=0.50$.
Pinardi et al. \cite{Pinardi:1998:CTS} demonstrated in an independent study 
that equation (\ref{HC_THEO}) shows much better agreement with
experimental data for II--VI semiconductor compounds (with $a^{*}=0.45$) then the above described force balance method.

\subsubsection{Simulation results and discussion}
Figure \ref{HC_FIG} shows the critical layer thickness $h_c^d$ 
plotted versus the absolute value of the misfit $\varepsilon$. 
\begin{figure}[h]
\centerline{
\begin{minipage}{0.55 \textwidth}
  \epsfxsize= 0.95\textwidth
  \epsffile{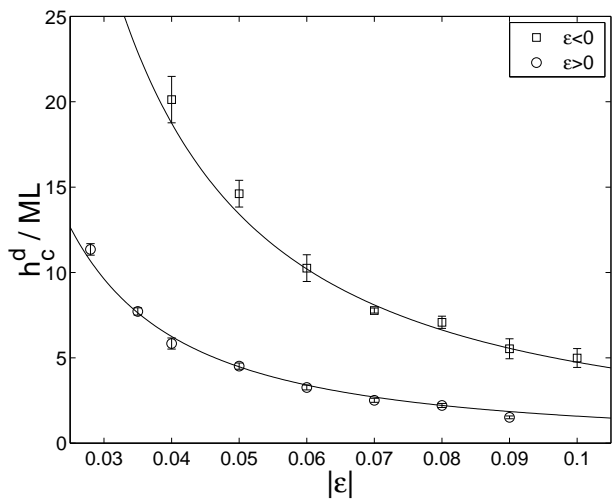}
\end{minipage}
\hfill
\begin{minipage}{0.40 \textwidth}
\caption{Critical thickness $h_c^d$ versus misfit $|\varepsilon|$ for
  $\varepsilon<0$ (upper curve) and $\varepsilon>0$ (lower curve). The error bars are obtained as the standard error  
  of the simulation results. The solid lines are 
  calculated using equation (\ref{HC_THEO}) with $a^{*}=0.15$ for $\varepsilon<0$ and 
  $a^{*}=0.05$ for $\varepsilon>0$. }
\label{HC_FIG}
\end{minipage}
}
\end{figure}
For $-0.03 < \varepsilon < 0.02$ the critical thickness is too large 
to be observed in our simulations. 
The simulation results show a strong dependence of $h_c^d$ on the sign of 
the misfit. This was found before by L. Dong et al. \cite{Dong:1998:SRM}. We believe this dependence
is due to the fact that, e.g., the Lennard--Jones potential is not harmonic. 
The potential is steeper in compression ($\varepsilon>0$) than in tension ($\varepsilon<0$), so that for $\varepsilon>0$
it becomes favorable to form a dislocation for smaller values of $h_c^d$. 
This corresponds well to the fact that in case of the $12,6$ Lennard--Jones potential the activation barrier  
for the introduction of dislocations $\Delta E_d$ is found 
to be generally higher for adsorbate films under tension then for those under compression \cite{Trushin:2003:EAM}.

Our simulation results agree well with the power law equation (\ref{HC_THEO}) (see solid lines in fig. \ref{HC_FIG}).
A nonlinear fit of our results yields 
$a^{*}=0.15$ for $\varepsilon<0$ and $a^{*}=0.05$ for $\varepsilon>0$. 

Thus, our quite simple $1+1$ dimensional model shows qualitatively the same dependence of the critical layer thickness on 
the misfit as semiconductor samples grown in molecular beam epitaxy. The values of $a^{*}$ obtained from our simulations 
show the same order of magnitude as the experimental ones but are expected to depend on the dimension and the applied 
interacting potential \cite{Cohen-Solal:1994:CTH,Bailly:1995:SMD,Vey:Diplom}. 

\section{Simulations for the low misfit region}
In this section we examine the development of the lattice constant during heteroepitaxial growth for the region
of low positive misfits ($\varepsilon<3\%$). 
Of special interest is the dependency of the lattice constant on the film thickness and the misfit.
The examinations are mainly motivated by experimental work for the 
strained layer growth of ZnSe on GaAs.
\subsection{Development of the lattice constant during ZnSe/GaAs heteroepitaxy}
Recently it has become possible \cite{Bader:2003:RTS,Faschinger:2003:IVH} to monitor the vertical lattice constant 
$\bar{a}_{\perp}$ (lattice 
constant in the direction of growth) averaged over the whole adsorbate film during MBE growth.
To this end a novel  X--ray diffraction method was used which allows - unlike conventional methods - for the
real--time in--situ X--ray diffraction (RIX).

Conventional methods for the X--ray diffraction - like the $\omega - 2 \theta$ scan at a symmetrical Bragg
reflection - have two important disadvantages which restrict their application during MBE growth. First an extremely 
exact sample adjustment is required which is barely achievable under growth conditions. More importantly, a 
time--consuming angular scan by rotating both sample and detector has to be performed which takes 
up to 5 minutes per shoot, which simply eliminates real--time monitoring.

\begin{figure}[hbt]
\centerline{
\begin{minipage}{0.49 \textwidth}
  \epsfxsize= 0.99\textwidth
  \epsffile{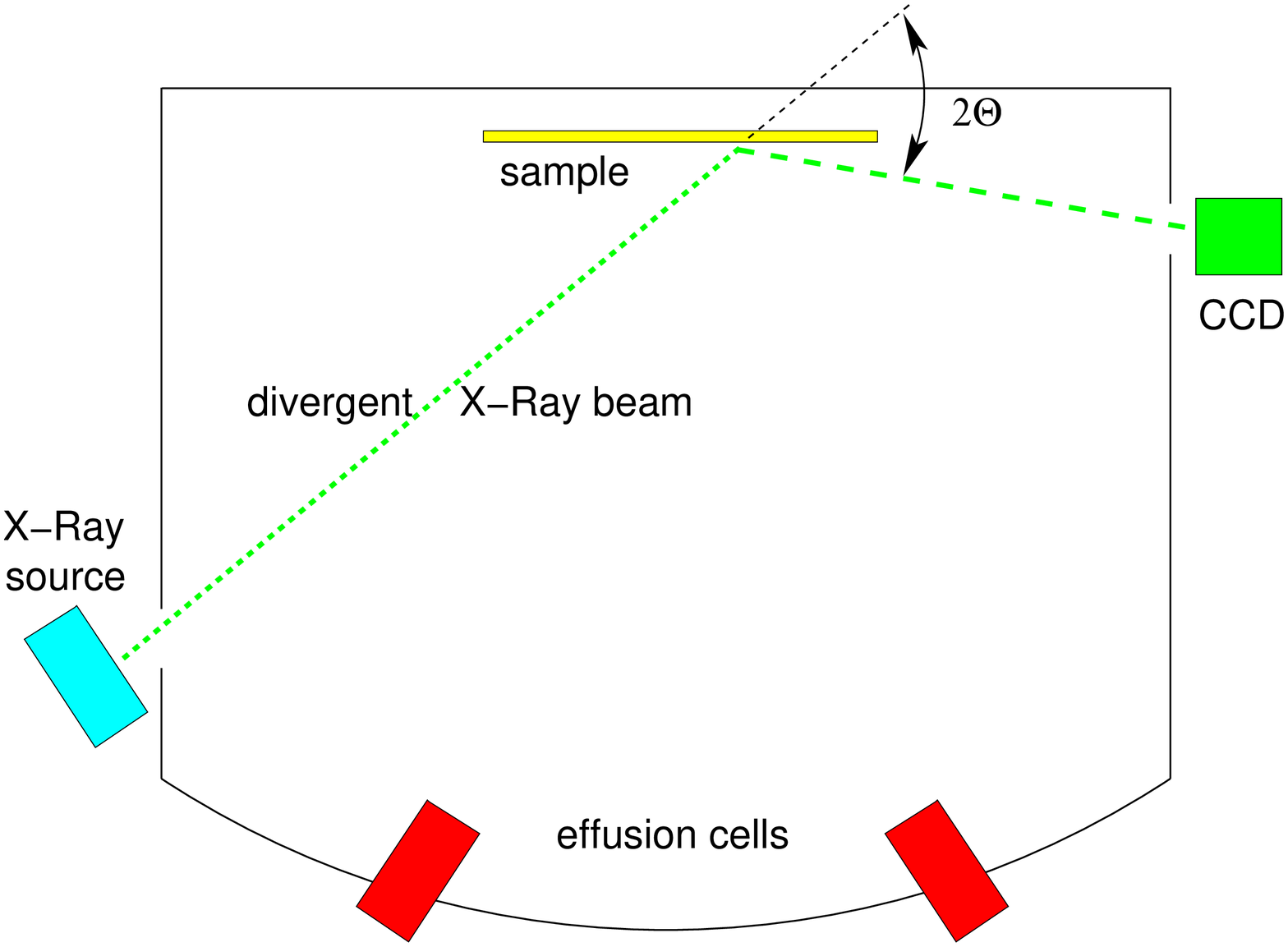}
\end{minipage}
\hfill
\begin{minipage}{0.49 \textwidth}
  \epsfxsize= 0.99\textwidth
  \epsffile{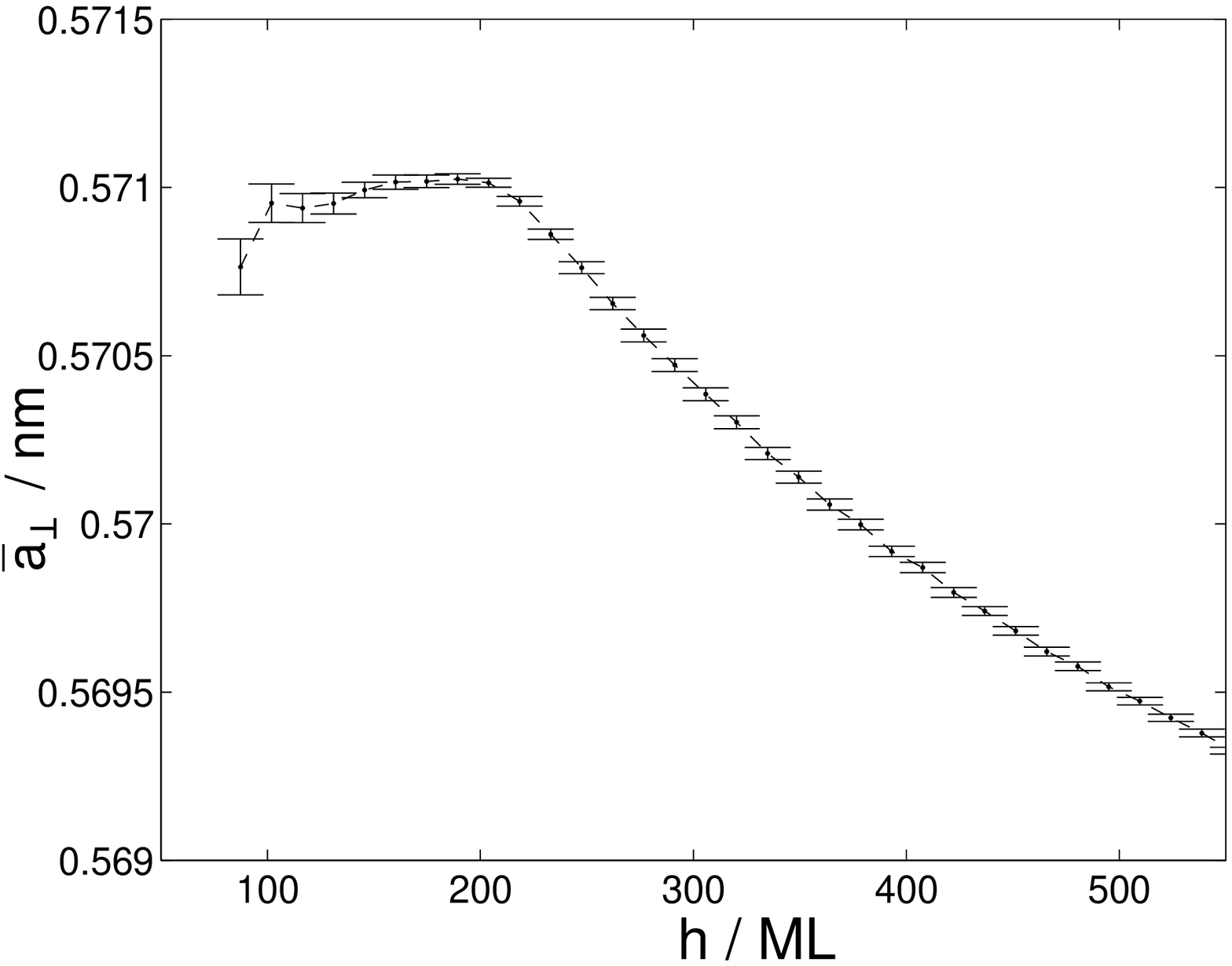}
\end{minipage}
}
\caption{Left panel: Experimental setup for RIX, consisting of a conventional MBE chamber with a standard X--ray source,
providing slightly divergent X--rays and a CCD camera.
Right panel: Development of the ZnSe film averaged vertical lattice constant with increasing layer thickness as 
obtained from RIX. The curve shows the initial coherent growth and the subsequent gradual progress of relaxation. Data is  
provided by courtesy of A. Bader \cite{Bader:2003:RTS}.}
\label{RIX}
\end{figure}
The new RIX method circumvents these problems by using an asymmetric Bragg reflection (here the $113$ reflection, 
resulting in an especially small exit angle) which yields the same
results for a $2 \theta$ scan of the reflection. 
In order to avoid the movement of the sample/detector during the measurement a 
slightly divergent X--ray beam is used which is detected by a multichannel CCD camera (also see fig.\ref{RIX} left panel).
This enables to cover the whole $2 \theta$ range within a few seconds by taking a single image with the CCD camera.
The method yields two peak positions for the adsorbate and the substrate, respectively.

By analyzing the distances between both peaks the vertical adsorbate lattice constant, 
averaged over the whole adsorbate film can be determined. Together with the growth rate, evaluated by 
RHEED (Reflection High Energy Electron Diffraction) 
oscillations, this gives $\bar{a}_{\perp}$ as a function of the film thickness.

Figure \ref{RIX} (right panel) displays the development of $\bar{a}_{\perp}$ as measured for the growth of ZnSe on GaAs 
for a growth temperature of $520K$. It can be seen that in the early stages of growth
$\bar{a}_{\perp}$ first remains approximately constant. This is believed to be due to pseudomorphic (coherent) growth
of substrate and adsorbate in this region. The lateral compression of the larger ZnSe adsorbate
(the misfit is $\varepsilon=0.31\%$ at growth temperature) 
causes a increased vertical lattice constant of 0.57nm which is in agreement with this misfit. 
At a layer thickness $h$ of about $200ML$ ($110nm$) $\bar{a}_{\perp}$ begins to drop.
This indicates the relaxation of the ZnSe film due to the introduction of misfit dislocations: due to the reduced lateral 
compression of the film the vertical lattice constant moves towards the smaller bulk value.

Note that the introduction of dislocations is guessed here by the development of $\bar{a}_{\perp}$ and can not be
observed directly during growth. In order to gain a deeper insight in the relaxation mechanisms leading  
to the observed behavior of $\bar{a}_{\perp}$ we now adapt our simulation method to meet the requirements
of the described problem.

\subsection{Adaptation of the simulation method}
We have to realize an expanded region of coherent growth in our simulations.
As described above this is in principle possible by choosing lower values of $|\varepsilon|$, yielding 
a higher critical thickness.
However, misfits of the order of magnitude like those found for ZnSe/GaAs growth would require 
very large system sizes in order to avoid finite size effects in our simulations.
Since the computation time necessary for these simulations would be beyond our present capacity we choose 
a different route here. 

On the one hand the misfit is chosen moderately low ($1.4\% \leq \varepsilon \leq 2.2\%$) requiring 
a lateral system size of $L=400$, which is well within our computational range.
On the other hand the surface is smoothed by applying a downward funneling 
method \cite{Ahr:2002:SPE,Evans:1991:FMS,Evans:1990:LTE} to the algorithm.
Since, like mentioned, a rough surface promotes the formation of dislocation this helps to produce thicker coherent
films without causing a severe deceleration of the calculations. 

\begin{figure}[hbt]
\centerline{
\begin{minipage}{0.45 \textwidth}
  \epsfxsize= 0.99\textwidth
  \epsffile{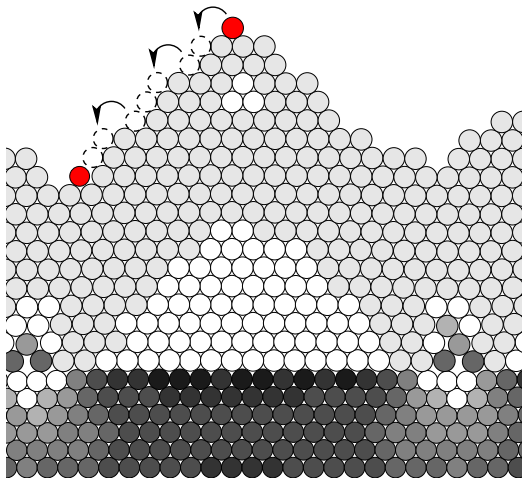}
\end{minipage}
\hfill
\begin{minipage}{0.45 \textwidth}
\caption{Schematic representation of downhill funneling: a newly deposited particle (dark) jumps 
successively to lower neighboring binding sites. In case of two equal low sites at the beginning of the 
process the direction of the first jump is randomly chosen.}
\label{FUN}
\end{minipage}
}
\end{figure}
Figure \ref{FUN} shows a newly arrived particle jumping successively to lower neighboring binding sites 
until the lowest one is reached. The downhill motion is justified by the assumption that a newly deposited 
particle has a considerably higher kinetic energy. This leads to an increased mobility until the excess energy 
is dissipated. At the end of this process the particle gets bound at an energetically favorable site with a high 
coordination number. Results of {\it Molecular Dynamics} 
simulations (cf. chapter \ref{KAP-1}) confirm this considerations \cite{Gilmore:1991:MDS,Yue:1998:MDS}. 
Various growth models showed that an incorporation of the funneling mechanism yields correct results for 
the topology of growing surfaces (see e.g. \cite{Ahr:2002:SPE,Volkmann:Phd}).

With respect to the higher film thicknesses the substrate thickness is increased to
$11$ layers in the following. All other settings of the algorithm remain unchanged.
Ch. Vey \cite{Vey:Diplom} realized the thus modified algorithm and performed the simulations.  

In the following we present results 
for misfits $1.4\% \leq \varepsilon \leq 2.2\%$. 
For each value of $\varepsilon$ the results are averaged over  $10$ independent simulation runs.
First we present the development of $\bar{a}_{\perp}$ as a function of the film thickness and discuss 
the different stages of growth like observed for one value of $\varepsilon$. Subsequently we examine the
influence of the misfit on the averaged vertical lattice constant.

\subsection{Development of the lattice constant with the film thickness}
Figure \ref{E16} shows the development of $\bar{a}_{\perp}$ with increasing film thickness for
a misfit $\varepsilon=1.6\%$. The characteristics of the curve are in qualitative agreement with the
experimental data (see fig. \ref{RIX}, right panel): first the averaged vertical lattice constant remains constant
followed by a subsequent decline of $\bar{a}_{\perp}$.

\begin{figure}[h]
\centerline{
\begin{minipage}{0.55 \textwidth}
  \epsfxsize= 0.99\textwidth
  \epsffile{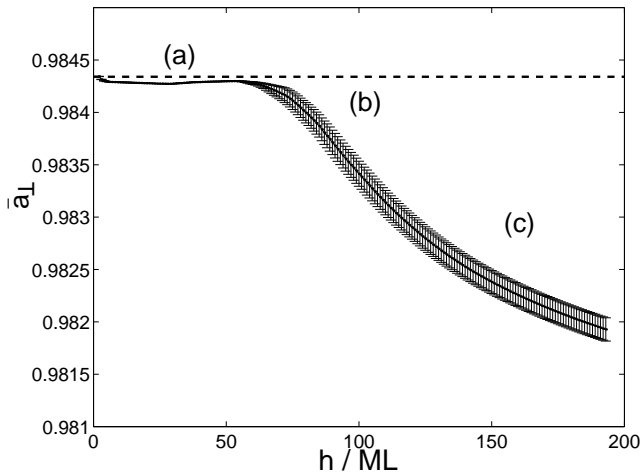}
\end{minipage}
\hfill
\begin{minipage}{0.40 \textwidth}
\caption{Development of the averaged vertical lattice constant with increasing layer thickness for $\varepsilon=1.6\%$, 
showing the initial coherent growth (a), the subsequent gradual progress of relaxation (b) and further growth
of the relaxed adsorbate film (c). The dashed line gives the value of  $\bar{a}_{\perp}^{const}$ according 
to equation (\ref{DELTA}).
The error bars represent the standard error of the simulation results.}
\label{E16}
\end{minipage}
}
\end{figure}
In case of $\varepsilon=1.6\%$ the constant region persists for about $80ML$. 
Figure  \ref{E16_FIGS}(a) shows that within this film thickness the adsorbate indeed grows pseudomorphic 
with the substrate without the introduction of dislocations. Due to the compression of the adsorbate film 
in lateral direction the vertical lattice constant is shifted to  higher values as one would expect for the 
bulk crystal. 
The measured value of $\bar{a}_{\perp}$ can be calculated approximately by 
a simple geometric approach.
\begin{figure}[h]
\centerline{
\begin{minipage}{0.50 \textwidth}
  \epsfxsize= 0.99\textwidth
  \epsffile{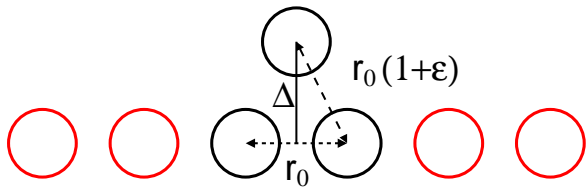}
\end{minipage}
\hfill
\begin{minipage}{0.45 \textwidth}
\caption{Calculation of the distance $\Delta$ between a particle and the underlying compressed layer.}
\label{DREIECK}
\end{minipage}
}
\end{figure}
Consider two particles in a closed layer of the compressed film and a third particle located in the binding position
between them (fig. \ref{DREIECK}). Due to the pseudomorphic growth the lateral 
distance of the two particles within the layer is given by $r_0$ - the equilibrium distance of substrate particles.
The particle on top is free to choose the preferred distance $r_0(1+\varepsilon)$ from its underlying 
nearest neighbors. The distance $\Delta$ between the uppermost particle and the underlying layer results then to
\begin{equation}
\label{DELTA}
\Delta = r_0 \sqrt{\varepsilon^2+2\varepsilon+\frac{3}{4}}.
\end{equation}
Since all layers in the coherent region of growth should have approximately the distance $\Delta$ from each other
this gives a estimated value for $\bar{a}_{\perp}$ as function of the misfit in this region.
Indeed the thus calculated $\bar{a}_{\perp}^{const}$ shows a good agreement with our simulation results (dashed line 
in fig. \ref{E16}).
\begin{figure}[t]
\hfill
\begin{minipage}[b]{0.32 \textwidth}
  \epsfxsize= 0.90\textwidth
  \epsffile{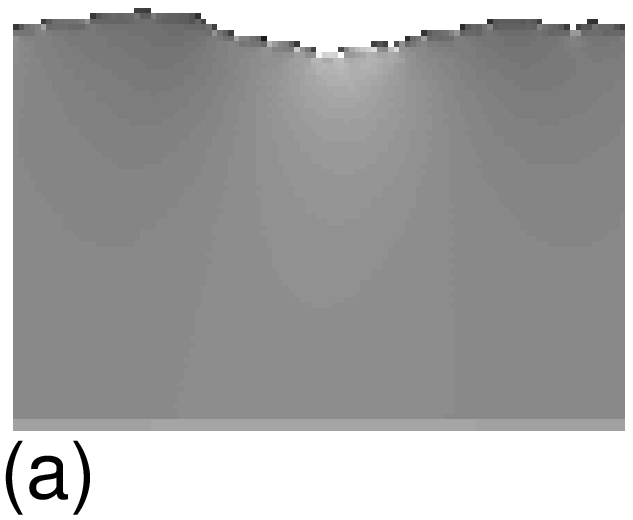}
\end{minipage}
\hfill
\begin{minipage}[b]{0.32 \textwidth}
  \epsfxsize= 0.90\textwidth
  \epsffile{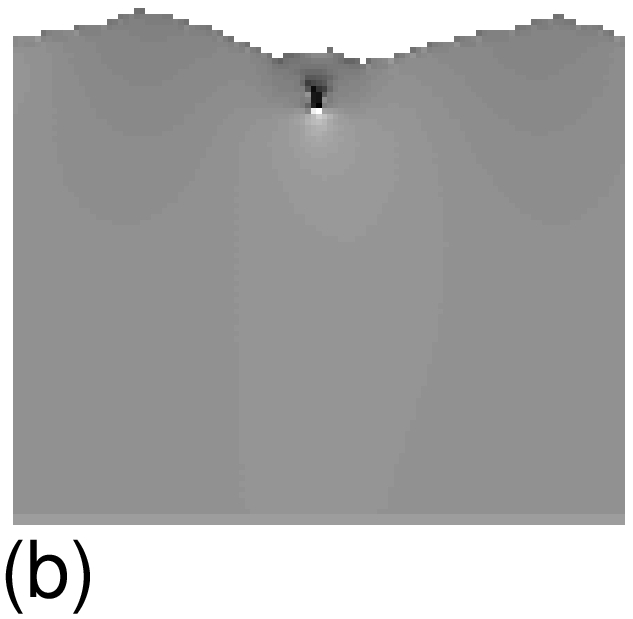}
\end{minipage}
\hfill
\begin{minipage}[b]{0.32 \textwidth}
  \epsfxsize= 0.90\textwidth
  \epsffile{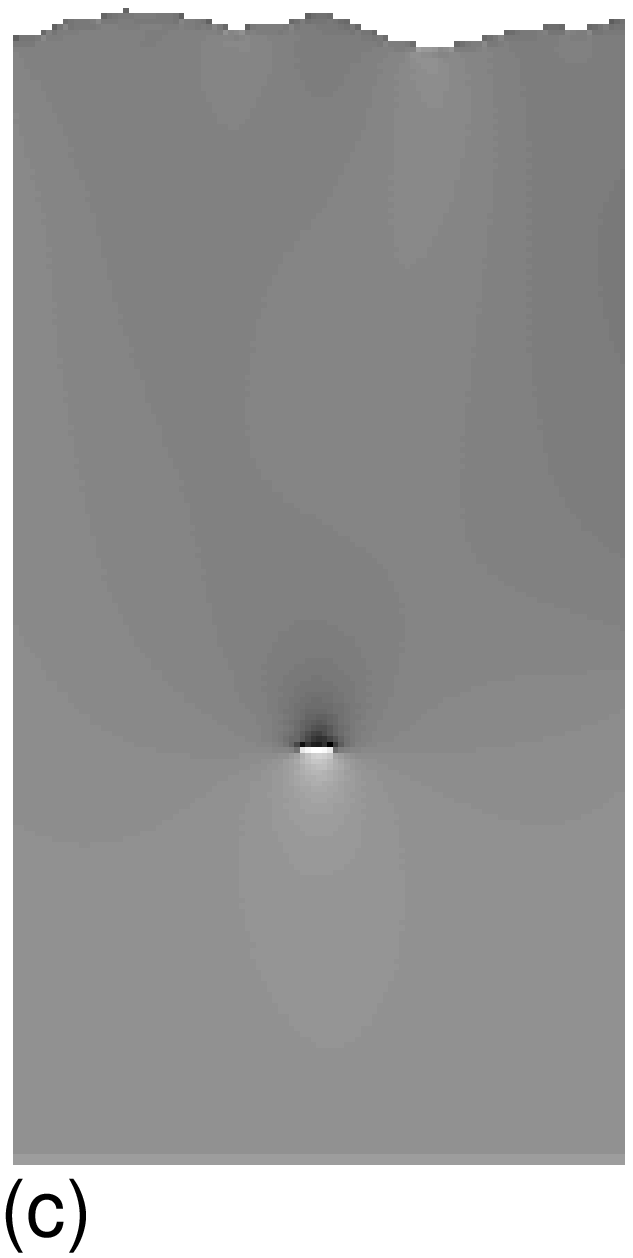}
\end{minipage}
\caption{Section of a simulation run for $\varepsilon=1.6\%$ (the substrate is not shown). 
The three stages of growth: (a) coherent growth ($h=70ML$), (b) formation of dislocations ($h=85ML$) 
and further growth of the relaxed adsorbate film ($h=200ML$).}
\label{E16_FIGS}
\end{figure}

As our simulation results verify (see e.g. fig. \ref{E16_FIGS}(b)) the region of decreasing $\bar{a}_{\perp}$ 
is governed by the introduction of misfit dislocations.
In the coherent film the formation of dislocations leads to a reduction of the compression and therefore 
to higher distances of the particles within the film. According to our simple model this decreases the
distance between subsequent layers and yields thus the reduction of the averaged vertical lattice constant. 

In the third stage of growth after the strain in the adsorbate film is relaxed and the introduction of 
dislocations has finished (see fig. \ref{E16_FIGS}(c)) the distance between two subsequent 
layers $\Delta$ is that of bulk adsorbate.
Since  $\bar{a}_{\perp}$ is averaged over the whole film (compressed and relaxed parts) $\bar{a}_{\perp}$
slowly approaches the adsorbate bulk value. 

In conclusion we are able to reproduce the experimental results on $\bar{a}_{\perp}$ as a function of 
the film thickness qualitatively. Our results confirm the explanation given in \cite{Bader:2003:RTS} 
for the development of $\bar{a}_{\perp}$ with the thickness of the adsorbate film. 
\subsection{Dependence of the lattice constant on the misfit}
\begin{figure}[htb]
\centerline{
\begin{minipage}{0.55 \textwidth}
  \epsfxsize= 0.99\textwidth
  \epsffile{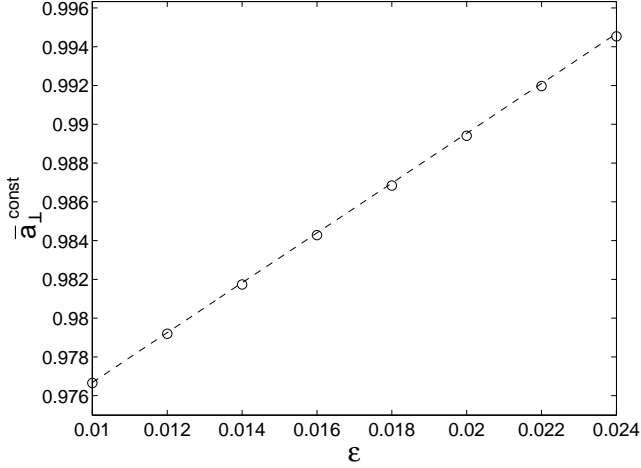}
\end{minipage}
\hfill
\begin{minipage}{0.40 \textwidth}
\caption{Development of the averaged vertical lattice constant as a function of $\varepsilon$.
The dashed line gives the values of  $\bar{a}_{\perp}$ according 
to equation (\ref{DELTA_LIN}).}
\label{GERADE}
\end{minipage}
}
\end{figure}
We now focus on the dependence of $\bar{a}_{\perp}^{const}$ on the misfit between the substrate and the adsorbate layer.
To this end we determine the value of $\bar{a}_{\perp}^{const}$ for 
$1.0 \leq \varepsilon \leq 2.4$.
As figure \ref{GERADE} shows $\bar{a}_{\perp}^{const}$ depends linearly on the misfit. 
Considering equation (\ref{DELTA}) this is
easy to understand. Since we are dealing with rather small misfits here equation (\ref{DELTA}) can be linear approximated
with
\begin{equation}
\label{DELTA_LIN}
\Delta \approx r_0 \left(\sqrt{\frac{3}{4}} + \sqrt{\frac{4}{3}} \varepsilon \right).
\end{equation}
As mentioned above in the coherent film $\Delta$ should be the same for all layers. Equation \ref{DELTA_LIN} 
gives therefore an approximation for the functional dependence of $\bar{a}_{\perp}^{const}$ on $\varepsilon$.
The dashed line in fig. \ref{GERADE} shows the calculated vertical lattice constant which agrees well 
with the simulation data.
Figure \ref{SKALIERT}(a) shows the development of $\bar{a}_{\perp}$ scaled with $\bar{a}_{\perp}^{const}$ for
$1.4 \leq \varepsilon \leq 2.2$. As one would expect the region of decreasing $\bar{a}_{\perp}$ is shifted
to higher values of the film thickness the smaller the misfit becomes.
\begin{figure}[htb]
\centerline{
\begin{minipage}[b]{0.45 \textwidth}
  \epsfxsize= 0.98\textwidth
  \epsffile{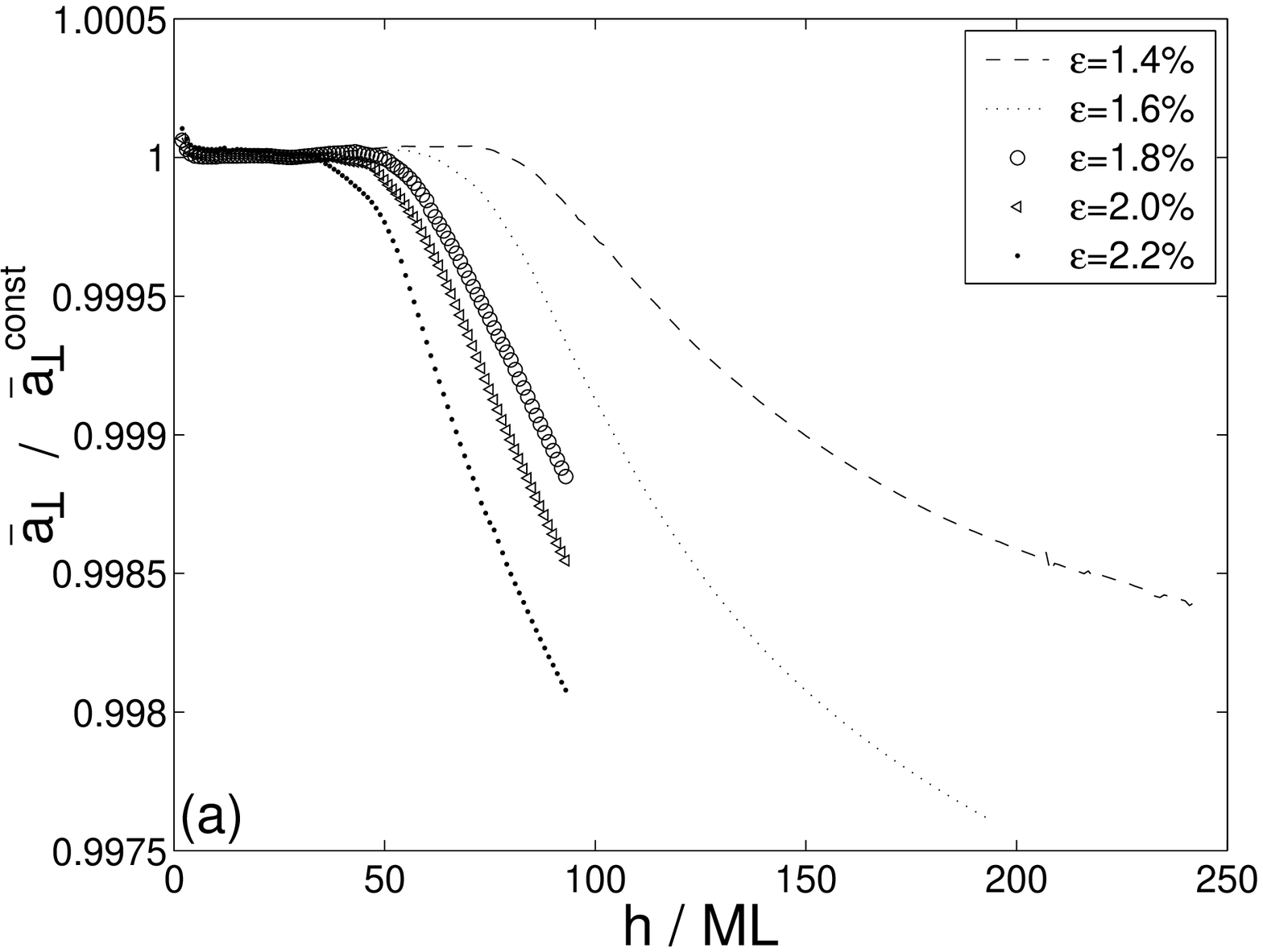}
\end{minipage}
\hfill
\begin{minipage}[b]{0.44 \textwidth}
  \epsfxsize= 0.99\textwidth
  \epsffile{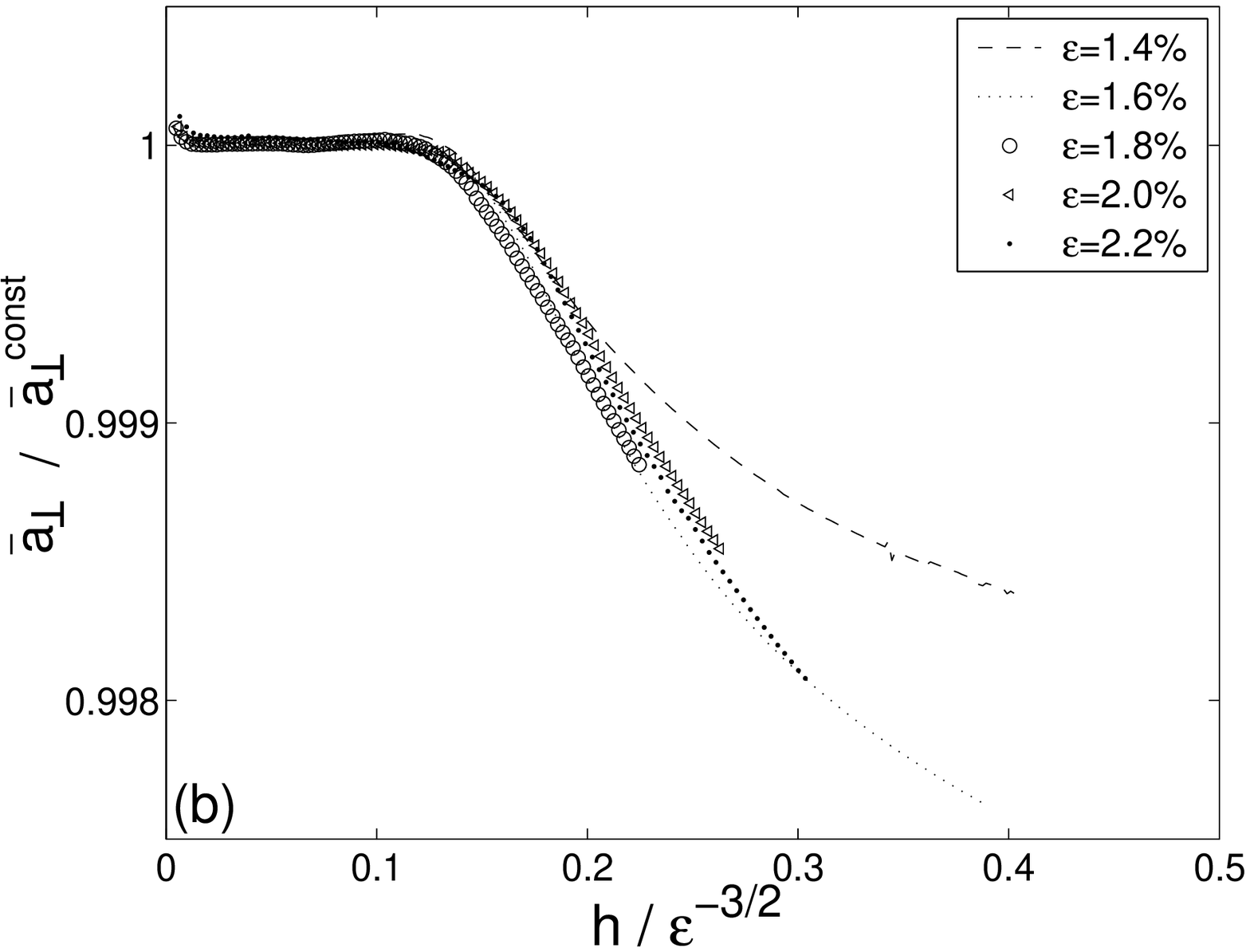}
\end{minipage}
}
\caption{Development of $\bar{a}_{\perp}$/$\bar{a}_{\perp}^{const}$ with increasing layer thickness 
for different values of the misfit. The right panel (b) shows the same curves with the film thickness scaled
by $\varepsilon^{-3/2}$.}
\label{SKALIERT}
\end{figure}
Since this stage of growth is governed by the critical thickness $h_c^d$ for the formation of dislocations it is
obvious to assume that the showed curves scale according to equation (\ref{HC_THEO}) with $\varepsilon^{-3/2}$.
As figure \ref{SKALIERT}(b) shows this guess leads to a rather good collapse of the curves in this region of 
the film thickness. The deviation for $\varepsilon=1.4\%$ is believed to be due to finite size effects (see
 \cite{Vey:Diplom} for details).
\section{Conclusions and outlook}
In conclusion we have shown that within the framework proposed in chapter \ref{KAP-3} 
it becomes possible to examine the appearance of misfit dislocations as one possible
relaxation mechanism in heteroepitaxial growth by the means of KMC simulations.

It was found that - like experimentally observed and theoretically predicted - also in our simulation
dislocations appear at a certain thickness of the adsorbate film $h_c^d$. We detected that $h_c^d$ is 
linked to the misfit by a simple power law, which also was measured for various semiconductor compounds.
Despite the simplicity of the model ($1+1$ dimensions, simple pair--potential as particle interaction) it 
was possible to gain deeper insight into dislocation formation mechanisms. Furthermore we were able to give reasons 
for the experimentally observed development of the vertical lattice constant as a function of the film thickness.

But we also encounter technical problems which limit the application of the algorithm on a realistic 
treatment of e.g. the dislocation evolution. Like mentioned above our model is not capable of treating concerted 
moves. Therefore it does not allow for the calculation of the activation barriers of dislocation movement, which is often 
observed in real materials. Whether it is conceptually possible to solve this problem is not yet clear.

However, there are  numerous further problems concerning dislocations in heteroepitaxial growth which can be addressed 
within the algorithm. 
One important question is certainly the influence of {\it buried} dislocations on the topology of the crystal surface far 
from the disturbance. Preliminary results show that over--grown dislocations influence the evolution of the 
adsorbate over a wide range of the film thickness. Another interesting problem is the influence of the 
steepness of the used potential on the dislocation formation. Especially the dependence of $h_c^d$ on the used
potential should be clarified. Since real MBE growth mostly takes place on stepped substrates this should be 
also examined in simulation which is quite easy to realize in our algorithm.
One important mechanism for the the misfit dislocation formation in real materials is the propagation 
of preexisting substrate dislocations into the adsorbate film. The influence of those threading dislocations 
on the critical thickness is also a problem which could be addressed by the method. 

Last but not least the algorithm should be extended to $2+1$ dimensions. Here one will encounter the 
problem of finding the relevant transition states for the diffusion jumps. It is also not clear if it is 
possible to reach the relevant system size and time scales for this computationally demanding problem.
However, only simulations in $2+1$ dimensions will open the pathway to more realistic lattice structures, realistic  
potentials and thus more material specific predictions.

  \cleardoublepage
\chapter{Simulation of Stranski--Krastanov--like growth}
\label{KAP-5} 
As mentioned in chapter \ref{KAP-1} dislocations (cf. chapter \ref{KAP-4}) clearly dominate 
the strain relaxation in sufficiently thick films and for large misfits.
In material combinations with relatively small misfits an alternative effect 
can govern the initial film growth: if the system manages to overcome the barrier 
$E_{2d-3d}$ instead of growing layer by layer the adsorbate aggregates in $3d$ islands.
The term island is commonly used to indicate that these structures are 
separated - unlike the emergence of mounds due to the Ehrlich--Schwoebel (ES) or
the Asaro--Tiller--Grinfeld (see e.g. \cite{Politi:2000:ICG}) instability.

In chapter \ref{KAP-1} we already discussed the two possible growth modes, known
for the separation of adsorbate material into islands:
the Volmer--Weber growth mode  with islands forming on uncovered substrate and 
the Stranski--Krastanov (SK) growth mode where $3d$ islands are found upon a pseudomorphic 
(dislocation free) wetting--layer of adsorbate material. This SK growth is observed for 
various strained heteroepitaxial systems, always for a quite large, positive misfit
($2\% \leq \varepsilon \leq 7\%$).

In order to avoid conflicts with other definitions and interpretations of the SK growth 
mode in the literature we will resort to the following specifications:
\begin{itemize}
\item The adsorbate film growth in a layer--by--layer way pseudomorphic up to a kinetically controlled
wetting--layer thickness $h_c^*$.
\item $3d$ islands appear suddenly and quite simultaneously, marking the so--called $2d-3d$ transition.
\item The further growth of the $3d$ islands is fed both by capture 
of newly deposited adsorbate particles and the decomposition of the {\it supercritical} wetting--layer.
\item Growth results in well separated $3d$ islands of similar shape and size, on top of the 
stable wetting--layer with reduced stationary thickness $h_c$ of a few monolayers height ($h_c \leq 7ML$).
\end{itemize}
Besides this basic processes a variety of phenomena play important roles for SK growth in real systems.
For example the interdiffusion of adsorbate and substrate material or the segregation of compound adsorbate 
are believed to be of high relevance in many cases \cite{Heyn:2001:CCSa,Cullis:2002:SKT}.
However, SK--like growth is observed in a variety of material systems which may or may not display this 
specific features. For instance, intermixing or segregation should be irrelevant in the somewhat exotic case 
of large organic molecules like PTCDA deposited on a metal substrate, e.g. Ag(111). Nevertheless, this system 
displays SK growth in excellent accordance with the above specifications \cite{Chkoda:2003:TDM}.
Most prominent examples for SK systems are IV--IV (see e.g. \cite{Sutter:2000:NTD,Kastner:1999:KSL,Roland:1993:GGF,Osipov:2002:SDN,Raiteri:2002:SSE,Schittenhelm:1998:SAG,Thanh:1999:FSA}), 
III--V (see e.g. \cite{Penev:2001:ESS,Johansson:2002:KSAa,Johansson:1998:MSD,Johansson:2002:KSAb,Carlsson:1994:STD,Seifert:1997:SGN,Snyder:1991:ESS,Leonard:1994:CLT,Snyder:1992:KCC,Heyn:2001:CCSb,Heyn:2001:FSE,Heyn:2000:FDI,Geiger:1997:OGM,Ruvimov:1995:SCG}) 
and II--VI (see e.g. \cite{Schikora:2000:ISK,Alchalabi:2003:SAS,Schikora:2000:IFK}) semiconductor compounds, see e.g. for InAs/GaAs self--assembled quantum dots
figure \ref{EXINAS}.
\begin{figure}[h]
\begin{minipage}{0.45 \textwidth}
  \epsfxsize= 0.99\textwidth
  \epsffile{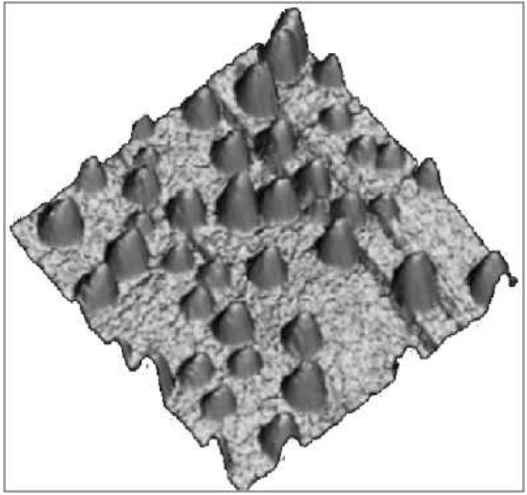}
\end{minipage}
\hfill
\begin{minipage}{0.45 \textwidth}
\caption{$200\times200$ nm scanning tunneling microscopy image of InAs/GaAs 
self--assembled quantum dots \cite{InAs_web}.}
\label{EXINAS}
\end{minipage}
\end{figure} 
Despite the extensive investigation of SK growth, a complete detailed theoretical picture is still lacking, 
apparently. 
This concerns in particular the nature of the $2d-3d$ transition. One problem clearly lies in the richness of 
the phenomenon. On the other hand, the mere diversity of SK--like systems gives rise to the hope that this growth 
scenario might be governed by a few basic universal mechanisms. Accordingly, it should be possible to capture and 
identify these essential features in relatively simple prototype systems.

This hope motivates the investigation of simplifying models without aiming at the reproduction of material specific 
details. Some of the key questions in this context are: under which conditions does a wetting--layer emerge and 
persist? How does its thickness before and after the $2d-3d$ transition depend on the growth conditions? Which 
microscopic processes trigger and control the sudden formation of $3d$ islands? How do the island size and their 
spatial arrangement depend on the parameters of the system?

\section{Simulation model}
The method introduced in chapter {\ref{KAP-2}} is now adapted for the the simulation of SK--like growth in 
$1+1$ dimensions. The pairwise interactions between particles of the system are given according to 
a $12,6$ Lennard--Jones potential (see also appendix \ref{AP-1} for details)
\begin{equation}
\label{LJ_5}
 {U}_{ij} =\ 4 E_{ij} \left[\left(\frac{{\sigma}_{ij}}{r_{ij}}\right)^{12}
-\left(\frac{{\sigma}_{ij}}{r_{ij}}\right)^{6}\right]. 
\end{equation}
The choice of the parameters ${E_{ij},\sigma_{ij}}$ in equation (\ref{LJ_5}) characterizes the different material properties
in our model: interactions between substrate (adsorbate) particles are specified by the sets ${U_s,\sigma_s}$ 
and ${U_a,\sigma_a}$, respectively.  In principle, an independent pair of quantities ${U_{as},\sigma_{as}}$ 
could specify the substrate--adsorbate potential.
For the sake of 
simplicity, aiming at a low number of free parameters,  we use
\begin{equation}
\label{UAS}
 U_{as}= \sqrt{U_{s}U_{a}}
\end{equation}
and ${\sigma}_{as}= \left({\sigma}_s+{\sigma}_a\right)/2$ for the interaction between the two species of particles.

As the equilibrium distance $r_0$ between two Lennard--Jones particles in the bulk is proportional 
to  $\sigma_{ij}$ the lattice misfit $\varepsilon$ becomes
$\varepsilon=\left({\sigma}_a-{\sigma}_s \right)/{\sigma}_s$. 
Since SK growth in real systems is almost exclusively observed for positive misfits \cite{Korutcheva:2000:CSK}, 
we consider only cases with ${\sigma}_a>{\sigma}_s$ here.
If not otherwise specified, we have set the misfit to $\varepsilon=4\%$ - 
a typical value for SK systems like, e.g., in Ge/Si heteroepitaxial growth. 
As demonstrated in chapter \ref{KAP-2} we 
have to take only hopping diffusion into account for such a small misfit. Exchange 
diffusion events can be neglected.
In order to save computer time the potential $U_{ij}$ is cut off
at a distance $r_{ij}>3r_{0}$ where the interaction strength is 
less than $1\%$ of the value at the equilibrium distance.

In our model growth takes place on six atomic layers of substrate with a fixed
bottom layer and periodic boundary conditions in horizontal direction.
The system size $L$ (number of particles per substrate layer) is set to $L=800$
in the following. Earlier simulations of smaller system sizes ($L=400,600$) 
revealed no significant $L$--dependence of the results presented here. 

As we have demonstrated in chapter \ref{KAP-4}, strain relaxation
through dislocations 
is not expected for the chosen misfit within the first few monolayers.
Since we deposit in the following only a few monolayers of adsorbate per simulation run, 
we are able to apply the latticed based method (chapter \ref{KAP-3}).

An important modification concerns interlayer diffusion. Lennard--Jones systems in $1+1$ dimensions display a strong 
Ehrlich--Schwoebel  barrier which hinders such moves at $1d$--island edges (see chapter \ref{KAP-2}). This barrier 
is by far less pronounced in $2+1$ dimensional systems, because interlayer moves follow a path through 
an energy saddle point rather than the pronounced maximum at the island edge. In our investigations of 
the SK scenario we remove the Schwoebel barrier for all interlayer diffusion events by hand. One motivation 
is the above mentioned over--estimation in $1+1$ dimensional systems. More importantly, we wish to investigate 
strain induced island formation without interference of the ES instability. Note that the latter causes  
the formation of mounds even in homoepitaxy \cite{Pimpinelli:1998:PCG,Politi:2000:ICG}. 

\section{Stable wetting--layer thickness}
We will now first investigate the stable wetting--layer thickness $h_c$, which is adjusted after the
process of $2d-3d$ transition is finished.
For semiconductor systems wetting--layer thicknesses from $h_c \approx 1ML$ for the InAs/GaAs system 
\cite{Geiger:1997:OGM} up to $h_c \approx 4ML$ (e.g. Ge/Si \cite{Roland:1993:GGF}) 
or even $h_c \approx 6ML$ for GaInAs/GaAs compounds \cite{Ruvimov:1995:SCG} have been measured. 
Sometimes much thicker wetting--layers ($h_c>50ML$) are reported - however, with respect to the above given 
definition we do not refer to such cases as SK growth.

A kinetic reason for the formation of a stable wetting--layer is the fact that the diffusion of
adsorbate particles is found to be slower on the substrate then on subsequent adsorbate layers. For example in case
of Ge growing on Si \cite{Roland:1993:GGF} the activation energy for the diffusion of Ge on the substrate is about $0.64eV$,
whereas the barrier for Ge diffusing on the first adsorbate layer is about $0.40eV$. A similar behavior is reported 
for the InAs/GaAs heteroepitaxy. Here the activation energy for the desorption of In located on the top of the 
wetting--layer is determined as $3.2eV$ and direct desorption from the substrate corresponds to a barrier of 
$3.4eV$ \cite{Geiger:1997:OGM}. Again the adsorbate particles are better bound to the substrate than to a 
pseudomorphic strained layer of the own species.
In conclusion - at least in some semiconductor systems - 
the wetting--layer is stabilized by a slower diffusion on the substrate and a faster diffusion 
on top of the wetting--layer, as deposited particles will reach and fill gaps on the substrate more easily.

These findings are also confirmed by our simulation results.
For $U_{as}<U_a$ the adsorbate layer thickness decreases from $h_c^*$ to $h_c=0$ with adsorbate islands located directly
on the substrate, eventually. For  $U_{as}>U_a$  a stable 
wetting--layer thickness ${h_c}>0$ persists. 
If we set - at a simulation temperature $T=500K$ - $U_a=0.74eV$, for instance, the choice $U_{as}=2.7eV$ 
(i.e. $U_{s}=10eV$, according to eq. (\ref{UAS}))
leads to a layer thickness 
${h_c} \approx 2ML$, whereas for $U_{as}=0.86eV$  (i.e. $U_{s}=1.0eV$) a single monolayer of adsorbate forms 
the stationary wetting--layer (also see fig. \ref{WL1u2}).
\begin{figure}
\begin{minipage}{0.60 \textwidth}
\begin{minipage}{0.99 \textwidth}
  \epsfxsize= 0.99\textwidth
  \epsffile{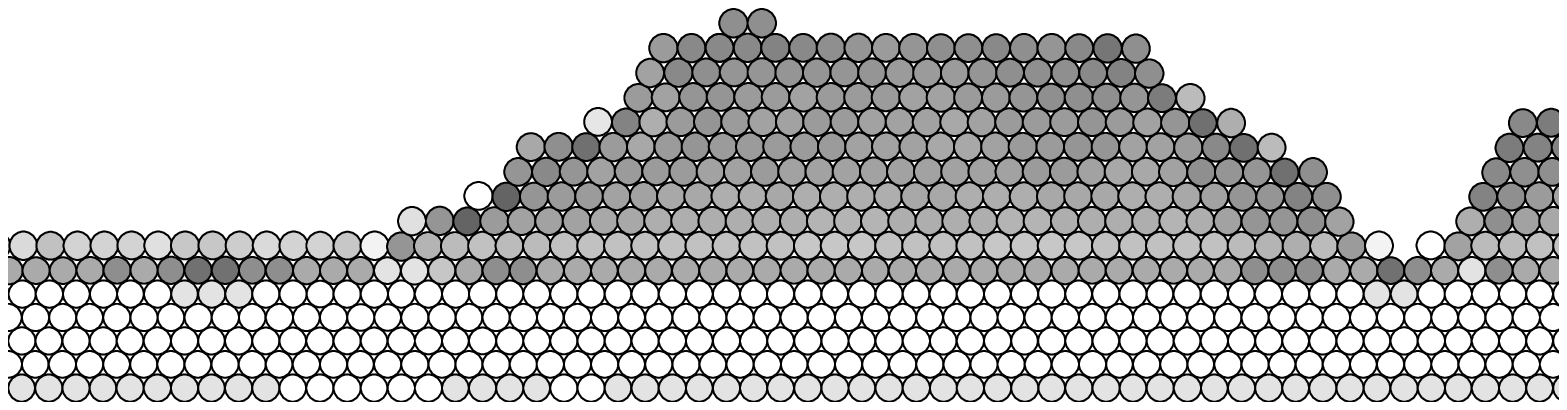}
\end{minipage}

\begin{minipage}{0.99 \textwidth}
  \epsfxsize= 0.99\textwidth
  \epsffile{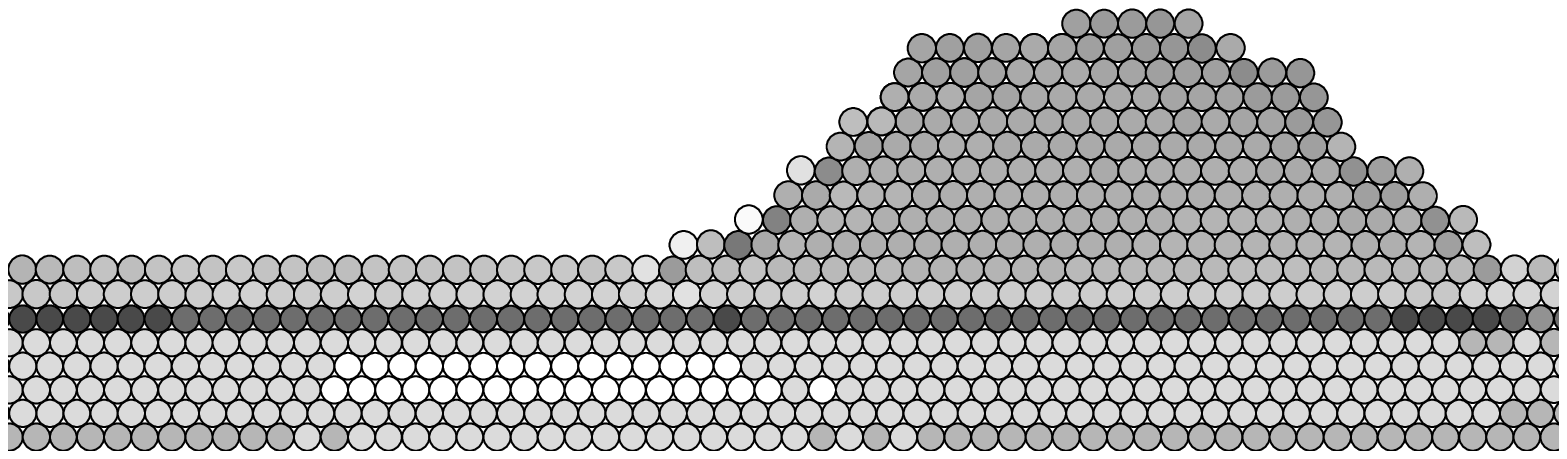}
\end{minipage}
\end{minipage}
\hfill
\begin{minipage}{0.35 \textwidth}
\caption{Snapshots form simulation results for $U_{as}=0.86eV$ (upper panel) with $h_c\approx 1ML$ 
and $U_{as}=2.72eV$ (lower panel) with $h_c\approx 2ML$, both at T=500K. The six bottom layers correspond to
the substrate. The darker the grey level of a particle, the bigger the average distance to its nearest 
neighbors of the same particle type. }
\label{WL1u2}
\end{minipage}
\end{figure}
\begin{figure}[h]
\begin{minipage}{0.50 \textwidth}
\epsfxsize= 0.99\textwidth
\epsffile{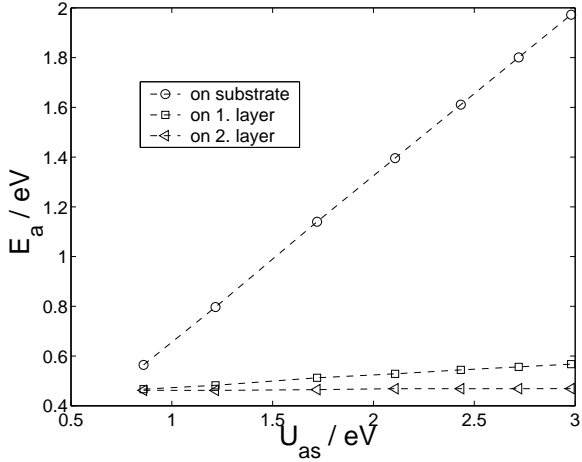}
\end{minipage}
\hfill
\begin{minipage}{0.45 \textwidth}
\caption{Diffusion barrier for adsorbate diffusion on subsequent layers on the surface as a function of the 
adsorbate--substrate interaction strength $U_{as}$.}
\label{WLDIFF}
\end{minipage}
\end{figure} 

Figure \ref{WLDIFF} shows the corresponding activation energies for the diffusion of an adsorbate particle 
on the substrate and the two subsequent adsorbate layers as a function of $U_{as}$.
As one can see in the case $U_{as}=0.86eV$ the barrier for diffusion on the first wetting--layer
becomes $E_a\approx0.47eV$, whereas the barrier for diffusion on the substrate is given by $E_a\approx0.57eV$.
For $U_{as}=2.7eV$ similar values are found for the diffusion on the second and first adsorbate layer,
respectively. We find that an increase of the activation barrier for diffusion on two subsequent layers 
by roughly $0.1 eV$ results in the formation of one additional stable 
adsorbate layer.
In principle this behavior could extend to more then just two adsorbate layers. However, due to the short--range nature 
of the Lennard--Jones potential the influence of the substrate (given by eq. (\ref{UAS})) 
essentially vanishes on wetting--layers of 
three or more layers thickness. In the following we will use $U_{as}=0.86eV$ (i.e. $U_s=1.0eV$) and restrict ourselves to
$h_c\approx1ML$ as a prototype SK scenario. However,  the results presented in this work 
should not depend on the precise value of $h_c$, qualitatively. 

\section{Diffusion process}
Before analyzing the SK--like growth scenario, we compare the barriers for hopping diffusion in various settings on 
the surface. 
One should note here that on the one hand investigations of systems like Ge/Si(001) reveal a very complicated scenario 
due to anisotropies and the influence of surface reconstructions \cite{Roland:1993:GGF,Cherepanov:2002:IMS,Penev:2001:ESS} 
which can not be modeled within our
$1+1$ dimensional system. On the other hand for most semiconductor systems (e.g. Ge/Ge(111) 
\cite{Cherepanov:2002:ISD} or In/GaAs(001) \cite{Penev:2001:ESS}) 
the barrier for hopping diffusion is higher on the surface of a compressed crystal, 
whereas diffusion is faster on the relaxed crystal. 
\subsection{Influence of the island height on diffusion barriers}
As mentioned in chapter \ref{KAP-1} pair--potential systems 
dot not reproduce this feature in general, since 
mechanical compression of the crystal lowers the barrier for surface diffusion
(see also fig. \ref{FWL}).
However, in the mismatched two species system, it is more important to compare diffusion on the substrate, the 
wetting--layer and the surface of partially relaxed islands.
As shown above, the slow diffusion on the substrate in comparison to the fast diffusion on top of the wetting--layer 
stabilizes the wetting--layer. 
\begin{figure}[hbt]
\begin{minipage}{0.50 \textwidth}
  \epsfxsize= 0.99\textwidth
  \epsffile{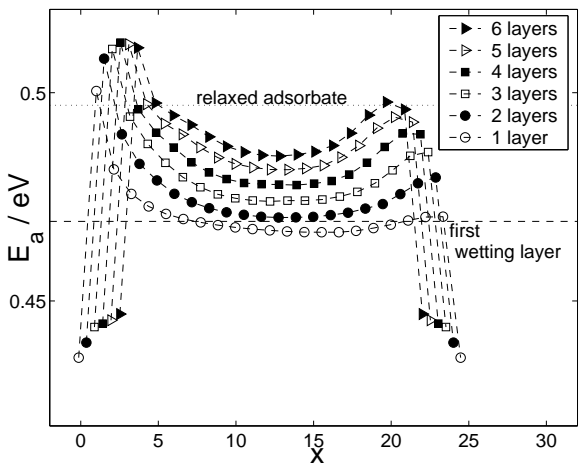}
\end{minipage}
\hfill
\begin{minipage}{0.45 \textwidth}
  \caption{Diffusion barriers as obtained in our model for a single adatom on a flat symmetric multilayer island 
with 24 base particles and one to six layers height. The symbols represent the activation energies for hops from the 
particle position $x$ to the left neighbor site. The island is placed on top of a single wetting--layer. The leftmost 
barriers correspond to downward jumps at the island edge with suppression of the Schwoebel effect. Horizontal lines 
mark the barriers for adatom diffusion on the wetting--layer and on fully relaxed adsorbate material.}
\label{ISCAN}
\end{minipage}
\end{figure}

For the SK--like scenario the diffusion on islands of finite extension is particularly relevant. Figure \ref{ISCAN} shows 
the barriers for diffusion hops on islands of various heights located upon a wetting monolayer. Note two different 
features:
\begin{itemize}
\item Diffusion on top of islands is, in general, slower than on the wetting--layer and the difference increases 
with the island height. In our model, this is an effect of the partial relaxation or over--relaxation in 
the top layer of the island. 
\item Depending on the lateral island size and its height, there is a more or less pronounced diffusion 
bias towards the island center, reflecting the spatially inhomogeneous relaxation (cf. chapter \ref{KAP-1}). Note that
this effect has to be distinguished from the diffusion bias imposed by the Schwoebel barrier, which would be 
present even in homoepitaxy and with particle positions restricted to a perfect undisturbed lattice.
\end{itemize}
Clearly both features favor the formation of islands upon islands and hence play an important role in our simulation of 
SK--like growth. They concern adatoms which are deposited directly onto the islands as well as particles that hop 
upward at 
edges, potentially. As we will argue in the following section, upward diffusion moves play the more important role for the 
$2d-3d$ transition.
\subsection{Influence of the misfit on diffusion barriers}
\begin{figure}[hbt]
\begin{minipage}{0.50 \textwidth}
  \epsfxsize= 0.99\textwidth
  \epsffile{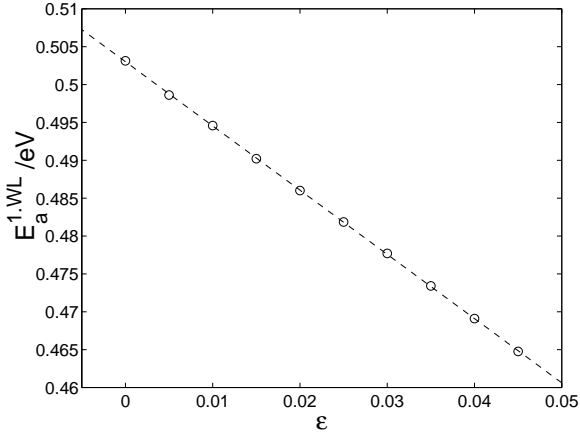}
\end{minipage}
\hfill
\begin{minipage}{0.45 \textwidth}
  \caption{Diffusion barriers $E_a^{1.WL}$ for an adatom on the first wetting--layer 
decreases linearly with increasing misfit $\varepsilon$.}
\label{FWL}
\end{minipage}
\end{figure}

Besides the island height also the misfit is believed to have a strong influence on the diffusion behavior  of 
an adatom diffusing on an island.  
For that reason we measure the diffusion barrier for a single adatom on a flat symmetric island, 
located on one layer of adsorbate,  
for misfits $0\% \leq \varepsilon \leq 4.5\%$.  Note, that with increasing misfit diffusion on the first 
wetting--layer becomes faster, since 
the adatom {\it feels} a more compressed adsorbate layer (cf. chapter \ref{KAP-1}).
Figure \ref{FWL} shows the diffusion barrier $E_a^{1.WL}$ for a particle on the first wetting--layer 
as a function of the misfit. As mentioned above the  
activation energy decreases linearly with increasing misfit (also see \cite{Schroeder:1997:DSS}).
\begin{figure}[hbt]
\begin{minipage}{0.50 \textwidth}
  \epsfxsize= 0.99\textwidth
  \epsffile{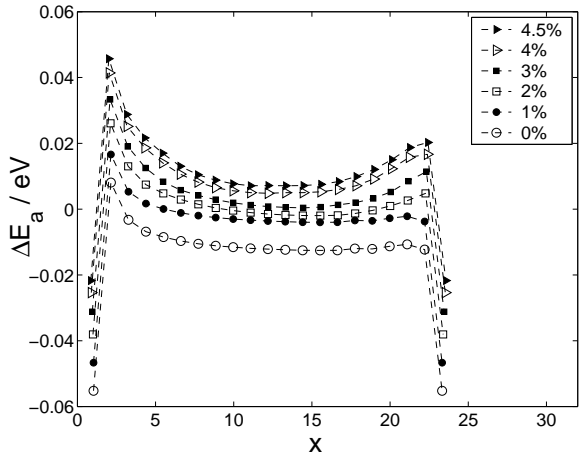}
\end{minipage}
\hfill
\begin{minipage}{0.45 \textwidth}
  \caption{Diffusion barriers minus the barrier for diffusion on the wetting--layer 
as obtained in our model. Calculations are shown for a single adatom on a flat symmetric multilayer island 
with 24 base particles and three layers height. The misfit is chosen between $0\%$ and $4.5\%$. }
\label{EPSILON}
\end{minipage}
\end{figure}

In order to facilitate the comparison of the results for different misfits we subtract $E_a^{1.WL}(\varepsilon)$ 
from the 
activation barriers for diffusion jumps on the island resulting in ${\Delta}E_a$. Again we choose $24$ particles as 
the base size of the exemplary island. Figure \ref{EPSILON} shows results for a three layer high island.
Two important features can be deduced:
\begin{itemize}
\item For misfits $\varepsilon>0$ the diffusion on top of the islands is in general slower than on the 
wetting--layer (${\Delta}E_a >0$). 
The higher the misfit the slower becomes diffusion on the island compared to the diffusion
on the the wetting--layer.
\item The diffusion  bias towards the island center is more pronounced for higher misfits.
\end{itemize}
In conclusion an increasing misfit should enhance island formation. As we will see a higher misfit 
indeed implies a larger island density and smaller mean island base sizes. 
\section{Stranski--Krastanov--like growth scenario}
In our investigation of SK--like growth we follow now a scenario which is frequently studied in experiments
\cite{Johansson:2002:KSAb,Leonard:1994:CLT}:
in each simulation run a total number of $4ML$ adsorbate material is deposited at rates $0.5ML/s \leq R_d \leq 9.0ML/s$.
After the deposition is complete, a relaxation period with $R_d=0$ of about $10^7$ steps follows, corresponding to a 
physical time on the order of $0.3s$. Results have been obtained on average over at least $15$ independent simulation runs 
for each data point.

In our simulation we observe the complete scenario of SK--like growth as described above.
Figures \ref{SKMOVIE1} and \ref{SKMOVIE7} show different stages of growth at $T=500K$ for $R_d=1ML/s$ and 
$R_d=7ML/s$, respectively. Both figures show a section of a $L=800$ system for (top down) $1.5ML$, $2.3ML$, 
$3.0ML$, $4.0ML$ coverage and after the mentioned relaxation time of $10^7$ simulation steps.
\begin{figure}
\begin{minipage}{0.90 \textwidth}
  \epsfxsize= 1.0\textwidth
  \epsffile{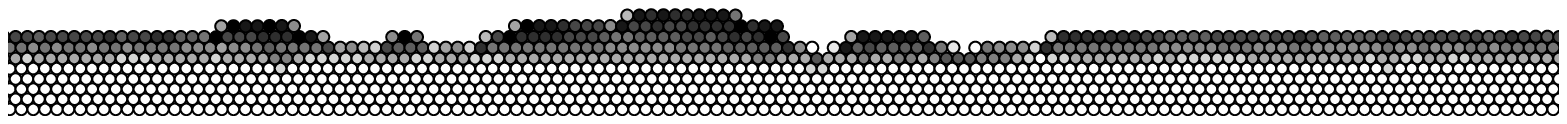}
\end{minipage}

\begin{minipage}{0.90 \textwidth}
  \epsfxsize= 1.0\textwidth
  \epsffile{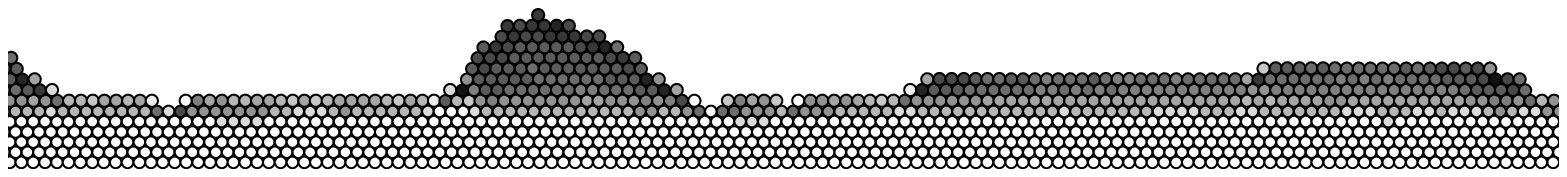}
\end{minipage}

\begin{minipage}{0.90 \textwidth}
  \epsfxsize= 1.0\textwidth
  \epsffile{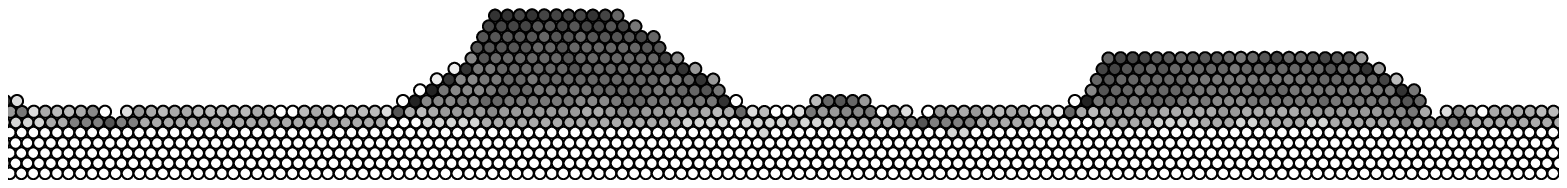}
\end{minipage}

\begin{minipage}{0.90 \textwidth}
  \epsfxsize= 1.0\textwidth
  \epsffile{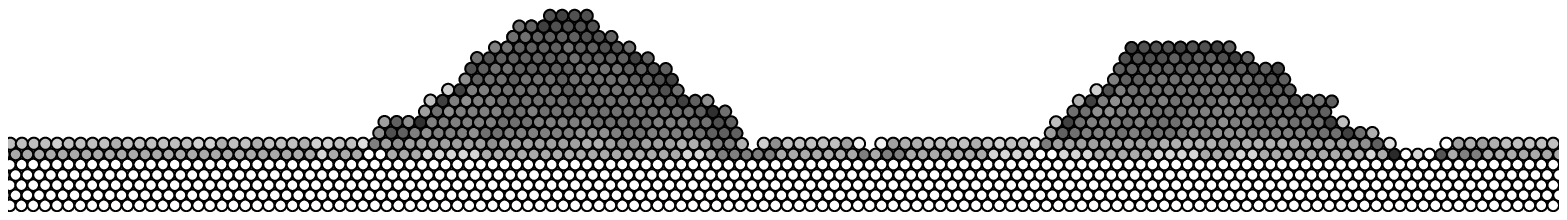}
\end{minipage}
\begin{minipage}{0.99 \textwidth}
  \epsfxsize= 1.0\textwidth
  \epsffile{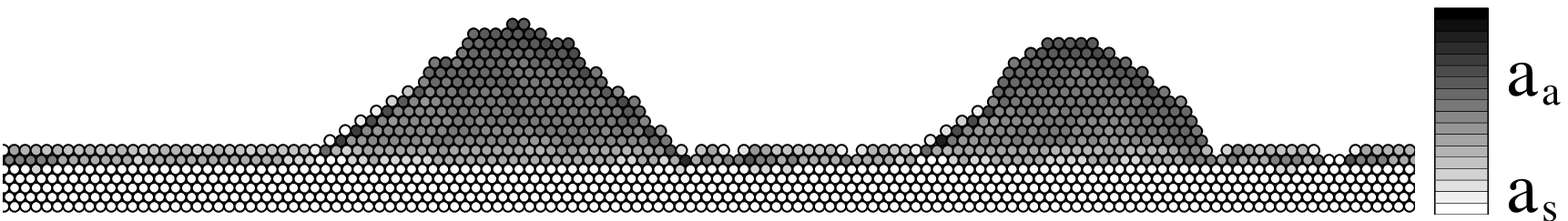}
\end{minipage}
\caption{Sections of a $L=800$ system at $T=500K$ and $R_d=1.0ML/s$ for (top down) $1.5ML$, $2.3ML$, 
$3.0ML$, $4.0ML$ coverage and after $10^7$ additional simulation steps. 
The darker the grey level of a particle, the bigger the average distance to its nearest 
neighbors of the same kind. The color bar gives a legend for the adsorbate particle distances in the downmost picture. }
\label{SKMOVIE1}
\end{figure}
\begin{figure}
\begin{minipage}{0.90 \textwidth}
  \epsfxsize= 0.99\textwidth
  \epsffile{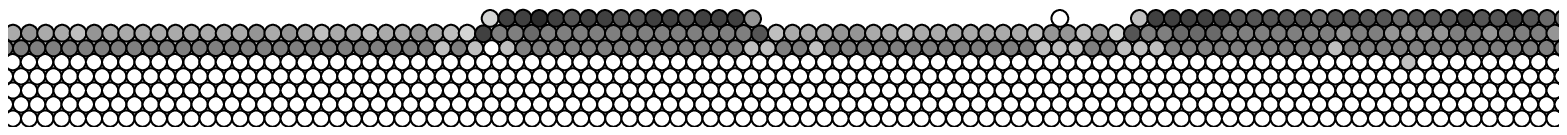}
\end{minipage}
\begin{minipage}{0.90 \textwidth}
  \epsfxsize= 0.99\textwidth
  \epsffile{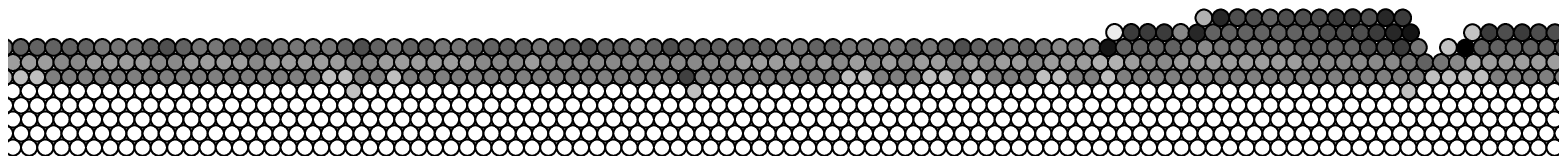}
\end{minipage}
\begin{minipage}{0.90 \textwidth}
  \epsfxsize= 0.99\textwidth
  \epsffile{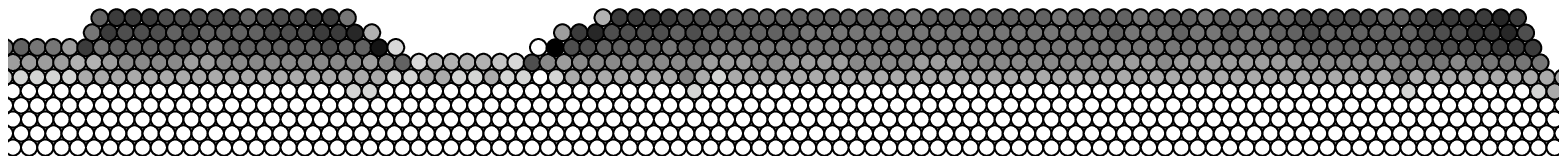}
\end{minipage}
\begin{minipage}{0.90 \textwidth}
  \epsfxsize= 0.99\textwidth
  \epsffile{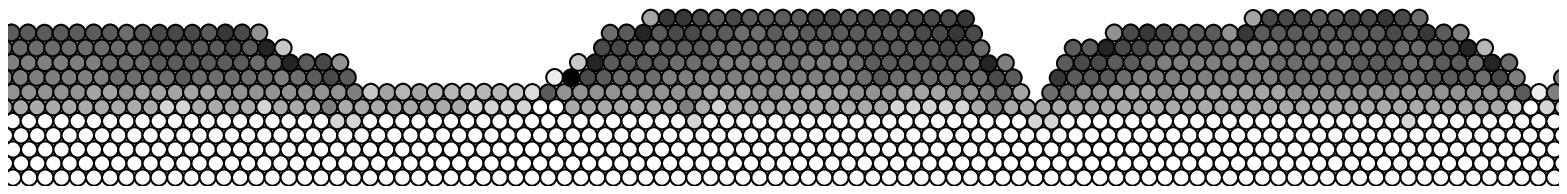}
\end{minipage}
\begin{minipage}{0.99 \textwidth}
  \epsfxsize= 0.99\textwidth
  \epsffile{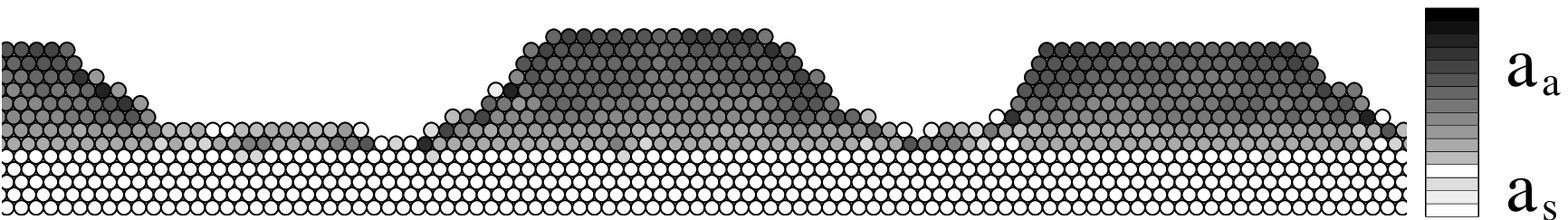}
\end{minipage}
\caption{Same as figure \ref{SKMOVIE1} but for $R_d=7.0ML/s$. Note that the onset of island formation is 
here delayed, resulting in $h_c^*\approx 2.3ML$.}
\label{SKMOVIE7}
\end{figure}
Ultimately, the formation of islands is driven by the relaxation of strain. Material within the $3d$ 
island and at its surface can adapt a lattice constant close to that of bulk adsorbate. On the contrary, particles 
in the wetting--layer are forced to adapt to smaller distances. Note that also the uppermost substrate layer is 
affected slightly by the adsorbate. 

During deposition monolayer islands located on the wetting--layer undergo 
a rapid transition to bilayer islands at a kinetically controlled critical wetting--layer thickness $h_c^*$ 
(first reported in \cite{Snyder:1992:KCC}). 
We identify this transition as the $2d-3d$ transition.
The comparison 
of figures \ref{SKMOVIE1} and \ref{SKMOVIE7} already shows that $h_c^*$ increases with increasing deposition flux. 
For $R_d=1ML/s$ (see fig. \ref{SKMOVIE1}) multilayer islands already start to form for a coverage of $1.5ML$, whereas 
for $R_d=7ML/s$ (see fig. \ref{SKMOVIE7}) the onset of island formation is delayed to about $2.3ML$ coverage.
In particular from figure \ref{SKMOVIE7} it is obvious that islands grow not only by capturing 
newly deposited particles, but also by the decomposition of the adsorbate layer. After the $2d-3d$ transition the 
thickness of the wetting--layer decreases to a stationary value $h_c$. 
Note from figure \ref{SKMOVIE7} that within our simulations it is possible for a {\it very} large island to split into 
well separated smaller ones. 
\subsection{Kinetically controlled critical wetting--layer thickness}
For a systematic determination of $h_c^*$ we follow \cite{Leonard:1994:CLT} and fit the density $\rho$ of $3d$ islands as 
\begin{equation}
\label{IslandDensity}
\rho = \rho_0\left(\theta-h_c^*\right)^\alpha,
\end{equation}
where $\theta$ is the coverage with adsorbate material. The best fit for the data is obtained for 
$\rho_0=7\times 10^{-3}$ and $\alpha\approx 1.5$. 
For example, in the case of InAs grown on GaAs one finds $\alpha\approx 1.76$ \cite{Leonard:1994:CLT}. 
Figure \ref{DENSITY} shows the fit for $T=500K$ and different values 
of the deposition rate.
\begin{figure}[hbt]
\begin{minipage}{0.50 \textwidth}
  \epsfxsize= 0.99\textwidth
  \epsffile{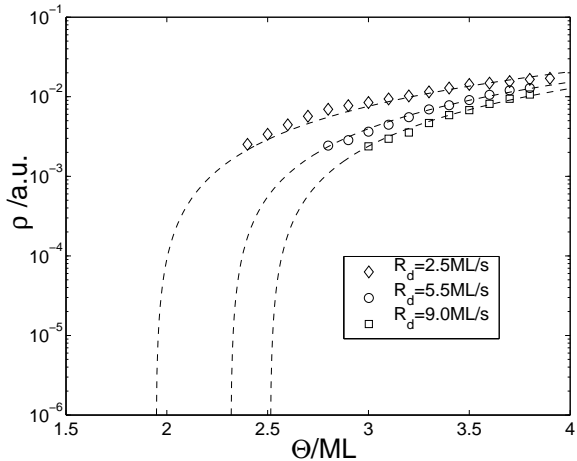}
\end{minipage}
\hfill
\begin{minipage}{0.45 \textwidth}
  \caption{$3d$ island density as a function of the coverage with adsorbate material $\theta$. The dashed lines 
give the results of the fit according to equation (\ref{IslandDensity}) for $\alpha=1.5$ and $\rho_0=7\times 10^{-3}$. 
}
\label{DENSITY}
\end{minipage}
\end{figure}

Figure \ref{HCSTERN} shows $h_c^*$ as a function of 
the deposition flux $F$ for two temperatures.
Increasing values of $h_c^*$ are observed with increasing flux and decreasing temperature. 
Qualitatively the same dependence of the kinetically controlled critical thickness
on $F$ and $T$ was reported in \cite{Johansson:2002:KSAb} for InP/GaAs heteroepitaxial growth. 

This behavior of $h_c^*$ can be understood within our model:
we consider fairly low diffusion barriers at high temperatures 
and suppress the ES effect explicitly. 
As a consequence, layer--by--layer growth is favored and second 
layer nucleation  will play a minor role in the formation 
of mounds.  The limiting effect which sets the characteristic time for
the $2d-3d$ transition is the upward diffusion of particles  
at the edge of existing islands. 
Clearly, the rate for this process will 
decrease drastically with decreasing temperature, because a high activation 
barrier $E_{up}$ has to be overcome.  
This rate is to be compared with the time scale set by the incoming
flux. A high flux $F$ will fill the layer before particles can perform 
an upward hop, and hence it will delay the $2d-3d$ transition.  

These considerations, together with the results of~\cite{Johansson:2002:KSAb},  suggest 
a functional dependence of $h_c^*$ on $F$ and $T$ of the form 
\begin{equation}
\label{hc_eq}
{h_c^*} = {h_0} \left(\frac{F}{R_{up}}\right)^{\gamma}
\end{equation}
where $R_{up}$ is the diffusion rate computed from $E_{up}$ 
of upward hops close to the transition according to an Arrhenius law.
Equation (\ref{hc_eq}) is not expected to  hold for very low fluxes or high temperatures 
where $h_c^*$ nearly coincides with the stationary value $h_c$ \cite{Johansson:2002:KSAb}.

\begin{figure}[htb]
\begin{minipage}{0.50 \textwidth}
  \epsfxsize= 0.99\textwidth
  \epsffile{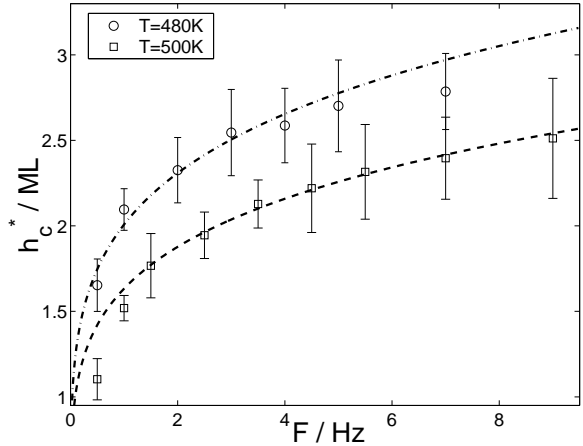}
\end{minipage}
\hfill
\begin{minipage}{0.45 \textwidth}
\caption{
  Kinetically controlled critical thickness $h_c^*$ vs. the deposition flux $F$ for two temperatures. 
  Both curves correspond to equation (\ref{hc_eq}), with $h_o$ and $\gamma$ obtained from
  the data for $T=500K$, only. The error bars represent as the standard error of the simulation results.}
\label{HCSTERN}
\end{minipage}
\end{figure}

In our simulations we observe at the $2d-3d$ transition a typical value
of  $E_{up} \approx 1eV$ for upward hops.
Using this value we have performed a non--linear fit according to
equation (\ref{hc_eq}) with the data for $T=500K$.
It yields very good agreement for $h_o \approx 3 ML, \gamma \approx 0.2$,
cf. figure \ref{HCSTERN} (dashed lower curve).
Setting the temperature to $T=480K$ in equation (\ref{hc_eq}) 
with otherwise unchanged parameters yields the dashed--dotted upper curve in figure \ref{HCSTERN}.
The good agreement with the simulation data strongly supports our assumptions.
Note that $\gamma$ is expected to depend strongly on material systems.
Clearly our data does not allow for a precise determination of $\gamma$, 
however a positive value of $\gamma$ captures the essential $F$  and $T$ dependencies of
$h_c^*$ like observed in \cite{Johansson:2002:KSAb}.    
\subsection{Island properties}
\begin{figure}[p]
\begin{minipage}{0.45 \textwidth}
  \epsfxsize= 0.99\textwidth
  \epsffile{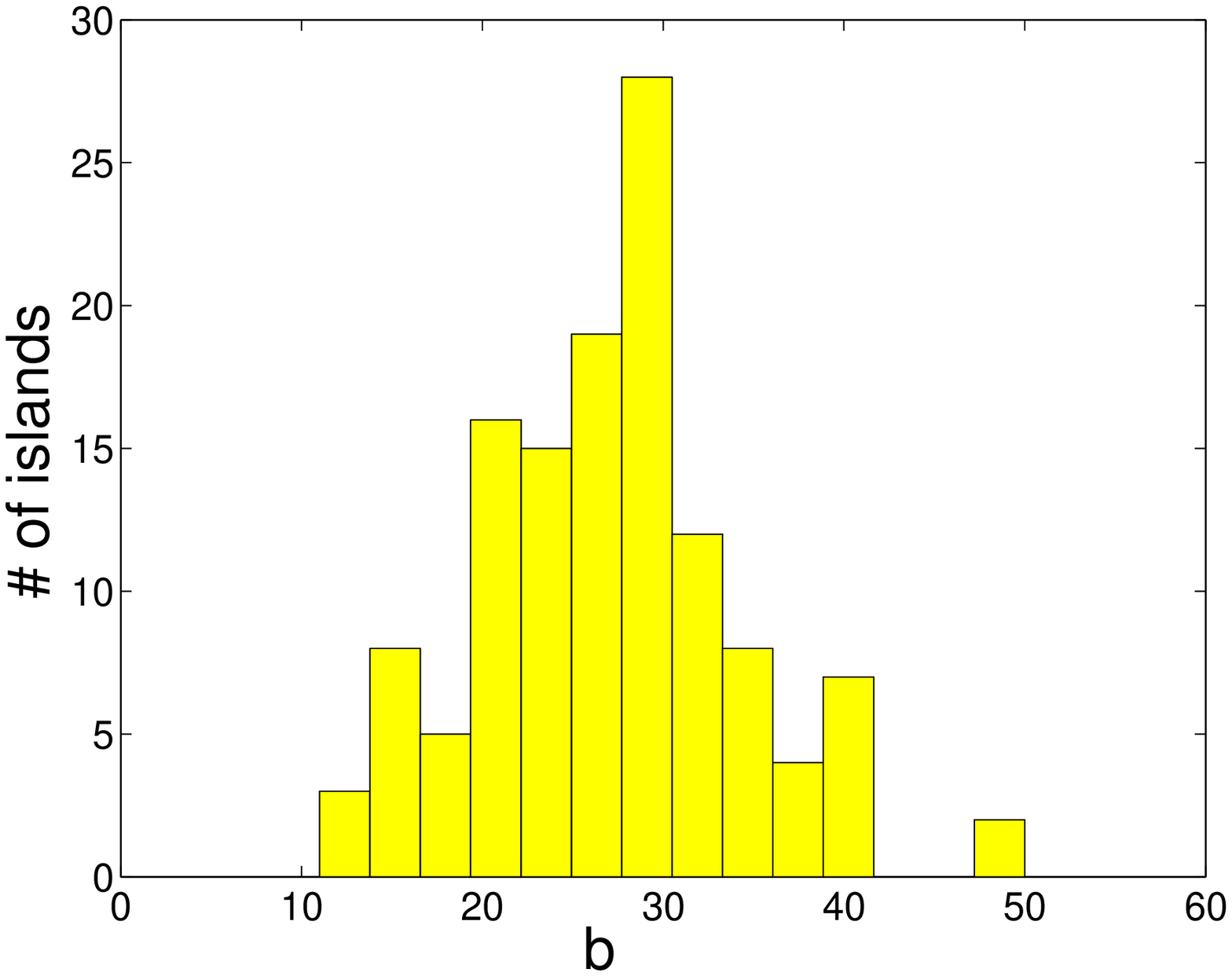}
\end{minipage}
\hfill
\begin{minipage}{0.45 \textwidth}
  \epsfxsize= 0.99\textwidth
  \epsffile{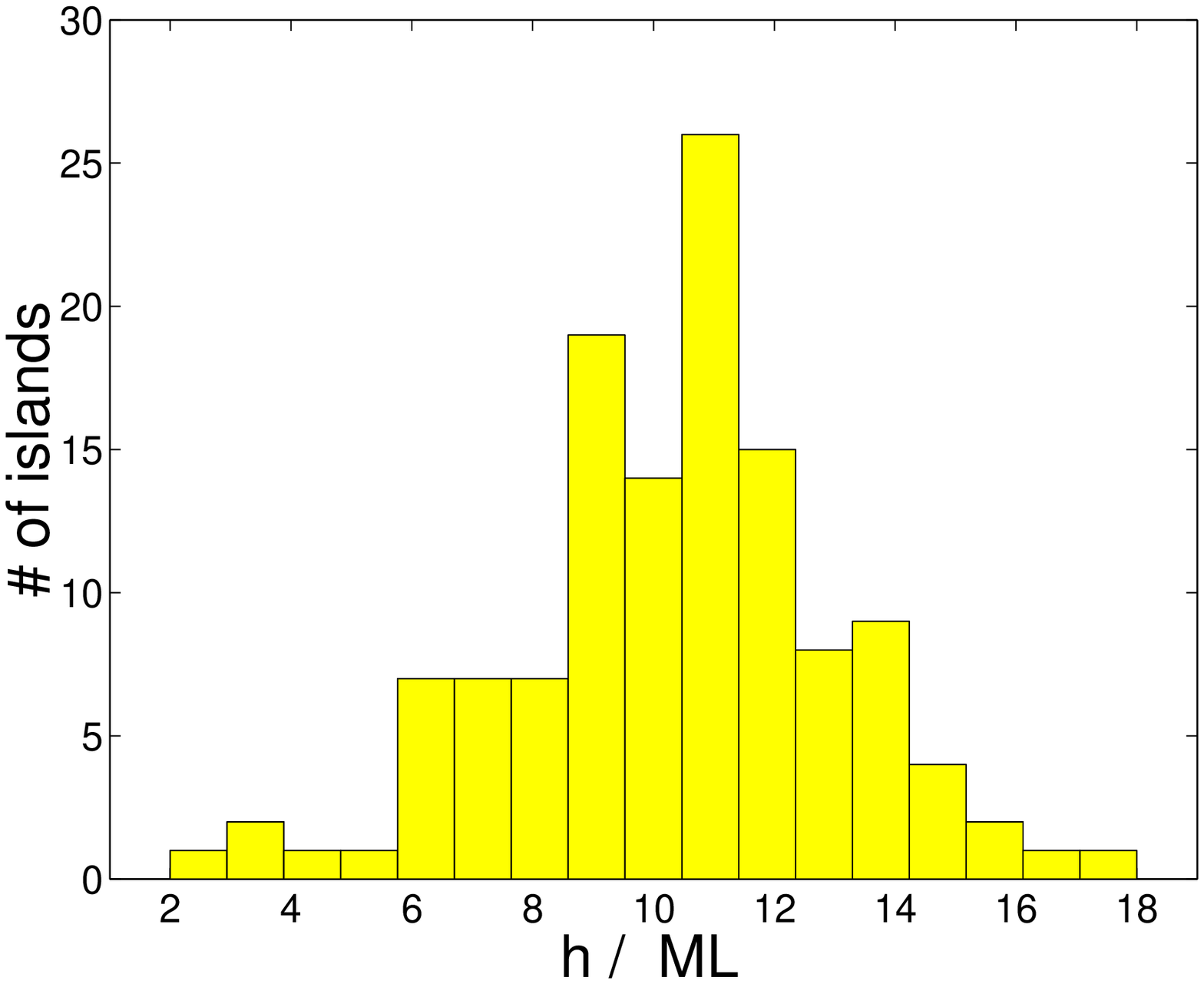}
\end{minipage}
\begin{minipage}{0.45 \textwidth}
  \epsfxsize= 0.99\textwidth
  \epsffile{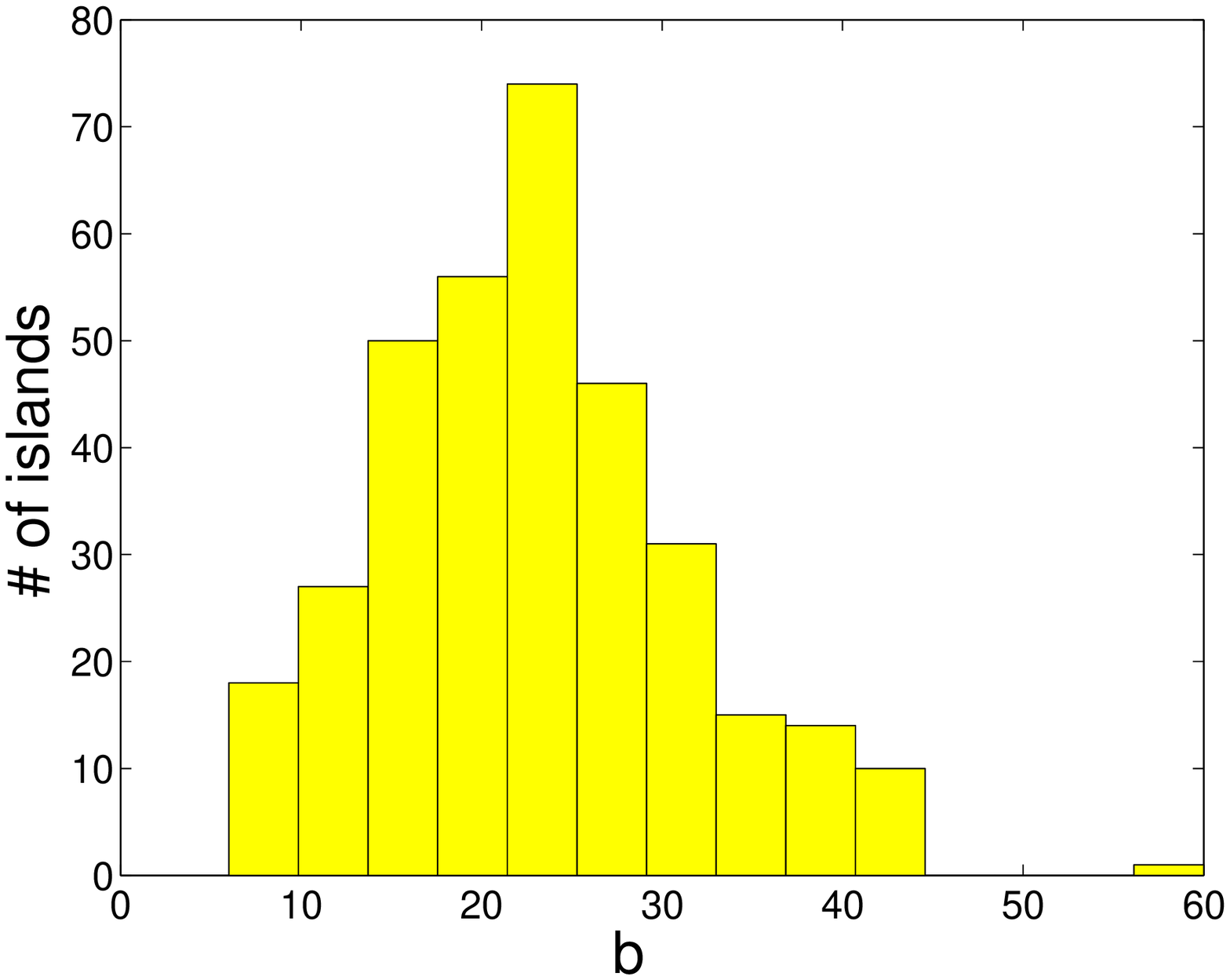}
\end{minipage}
\hfill
\begin{minipage}{0.45 \textwidth}
  \epsfxsize= 0.99\textwidth
  \epsffile{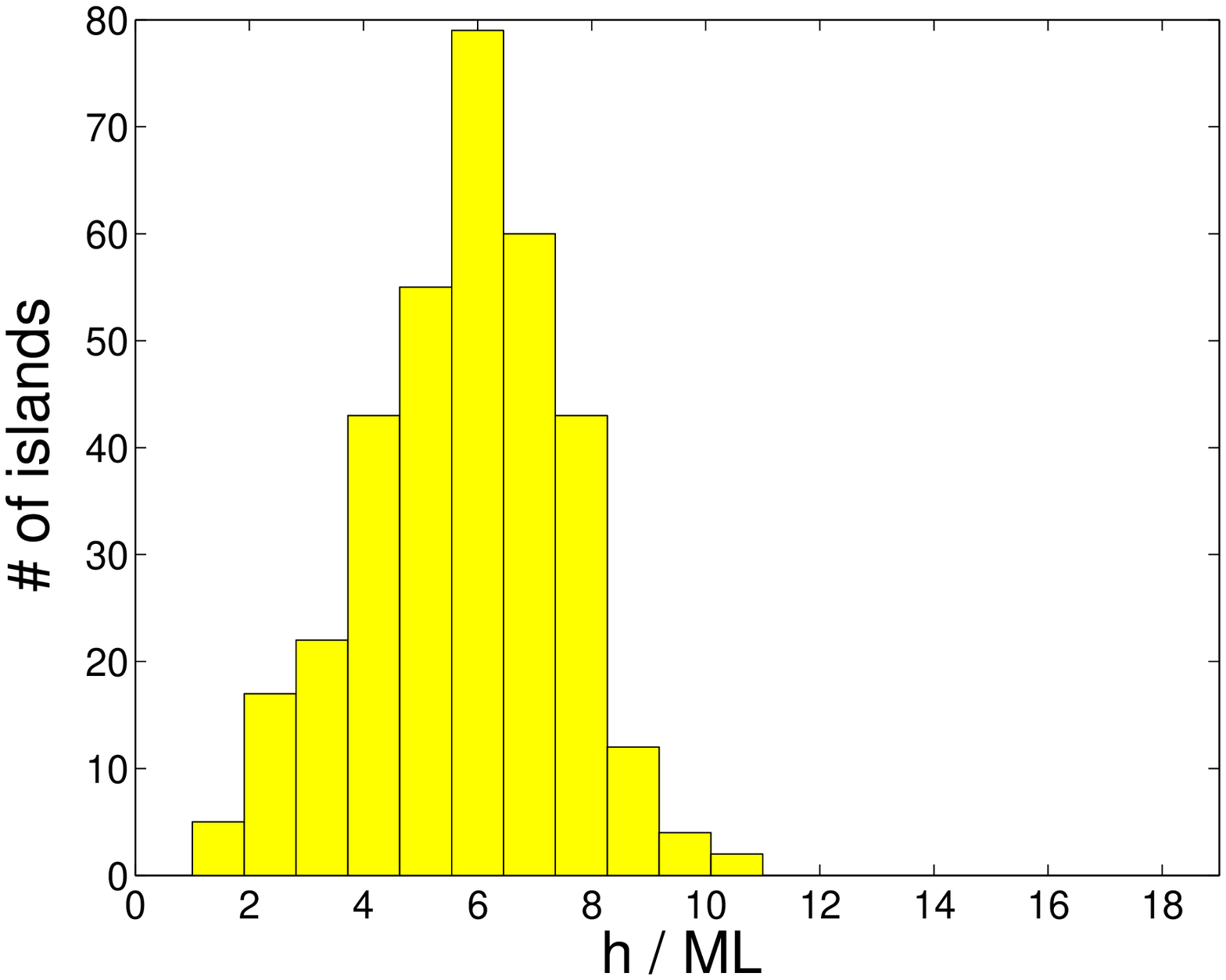}
\end{minipage}
\caption{Base size $b$ and height $h$ distribution of islands for $R_d=1.0ML/s$ (upper row) and 
$R_d=7.0ML/s$ (lower row), at $T=500K$.}
\label{HISTO}
\end{figure}
\begin{figure}[p]
\begin{minipage}{0.48 \textwidth}
  \epsfxsize= 0.99\textwidth
  \epsffile{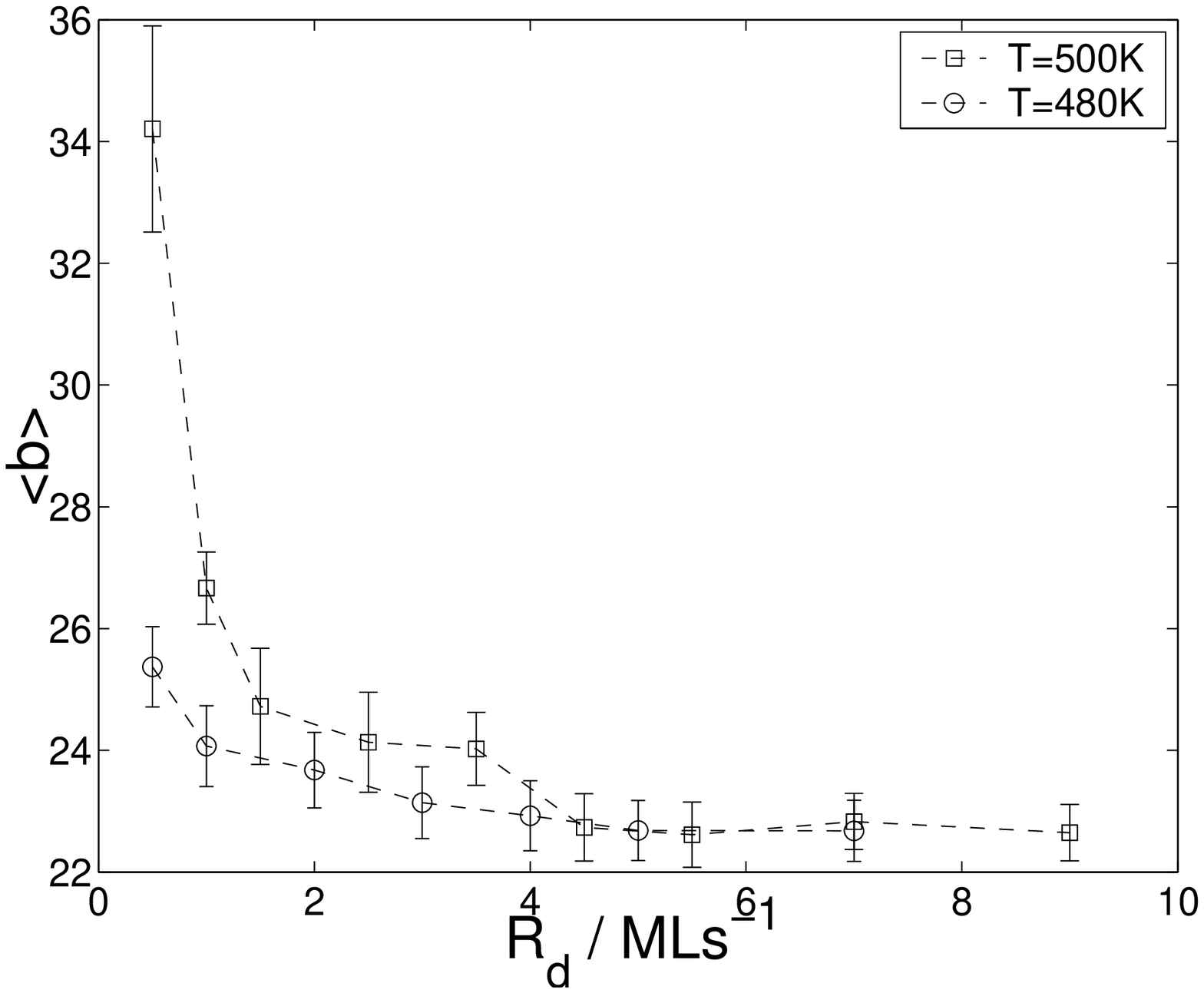}
\end{minipage}
\hfill
\begin{minipage}{0.49 \textwidth}
  \epsfxsize= 0.99\textwidth
  \epsffile{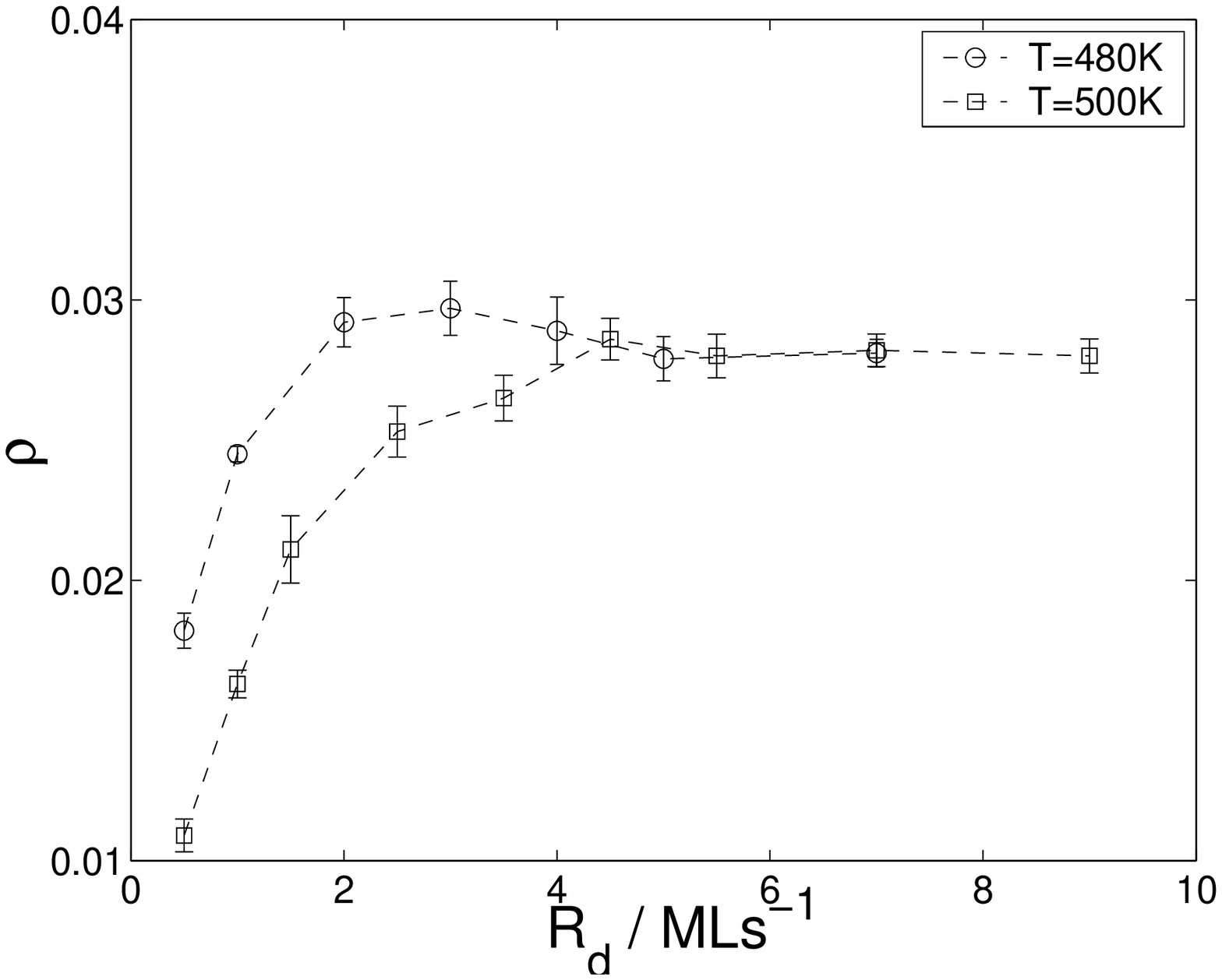}
\end{minipage}
\caption{Left panel: Average base size $<b>$ of multilayer islands vs. the deposition rate $R_d$ for two temperatures.
Right panel: Density of multilayer islands $\rho$ vs. the deposition rate $R_d$ for two temperatures. 
The error bars represent as the standard error of the simulation results.}
\label{BASE}
\end{figure}
In order to characterize the self--assembled islands that emerged during growth, we measure 
their base $b$ given by the number of particles 
in the island bottom layer and their height $h$ in monolayers.
Figure \ref{HISTO} shows base size and height distribution
for $R_d=1.0ML/s$ and $R_d=7.0ML/s$, both at $500K$.
For $R_d=1.0ML/s$ the mean base size is $27$ particles and the mean height is $10ML$.
For $R_d=7.0ML/s$ the mean base size and mean height are $23$ particles and $6ML$, respectively.

As one can further conclude from figure \ref{BASE} (left panel) the mean
base size first decreases with increasing particle flux but becomes 
constant and independent of the temperature
at higher values of the deposition rate. All results shown here
are obtained at the end of the relaxation period with $R_d =0$.
Whereas the mean values do not change significantly,
fluctuations decrease with time in this phase. 
We also consider 
the density ${\rho}$ of multilayer islands, i.e. their total number per 
particle in a substrate layer.  Figure \ref{BASE} (right panel)
shows $\rho$ as a function of deposition rate and temperature.
For low deposition rates the  
density rises with increasing $R_d$  while it decreases for 
increasing temperature.

At higher values of the deposition rate the island density becomes constant.
A similar behavior of the island density as a function of
$R_d$ and $T$ was reported for InP/GaAs heteroepitaxial growth \cite{Dobbs:1997:MFT,Johansson:2002:KSAb}.
The constant, temperature independent region of the density is
reminiscent of the behavior of the base size for high fluxes.
The saturation behavior further demonstrates the importance of upward hops vs. aggregation 
of deposited particles on islands. The latter process would yield a continuous increase of 
the island density with the flux.

\subsection{Influence of the misfit}
\begin{figure}[h]
\begin{minipage}{0.50 \textwidth}
  \epsfxsize= 0.99\textwidth
  \epsffile{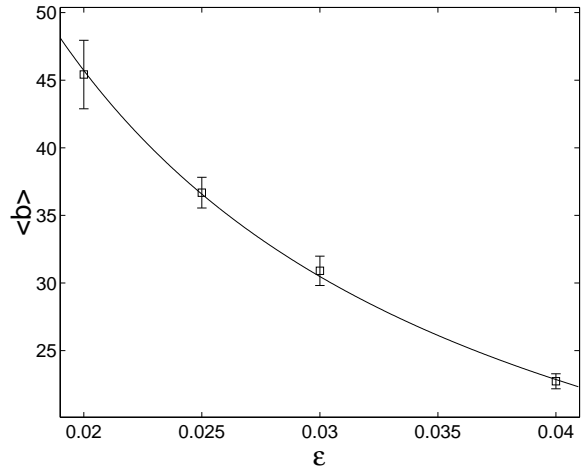}
\end{minipage}
\hfill
\begin{minipage}{0.45 \textwidth}
 \caption{Average base size $<b>$ of multilayer islands vs. the misfit $\varepsilon$ at $R_d=4.5ML/s$ and
$T=500K$.}
\label{MISFITSK}
\end{minipage}
\end{figure}
Since our above calculations indicate that the misfit has an important impact on the island formation,
we finally take a look at the misfit dependence of the island size and density.
From several simulation 
runs for values of the misfit $2\% \leq \varepsilon \leq 4\%$ at $R_d=4.5ML/s$ and $T=500K$ 
we deduce that the base size at high deposition rates is mainly
determined by the misfit and decreases with increasing misfit like $b \approx 0.91{\varepsilon}^{-1}$ 
(see fig. \ref{MISFITSK}). 
Figure \ref{MISFITISLANDS} shows corresponding sections from simulation runs for $R_d=4.5ML/s$ and $T=500K$.
\begin{figure}[htb]
\begin{minipage}{0.60 \textwidth}
\begin{minipage}{0.99 \textwidth}
  \epsfxsize= 0.99\textwidth
  \epsffile{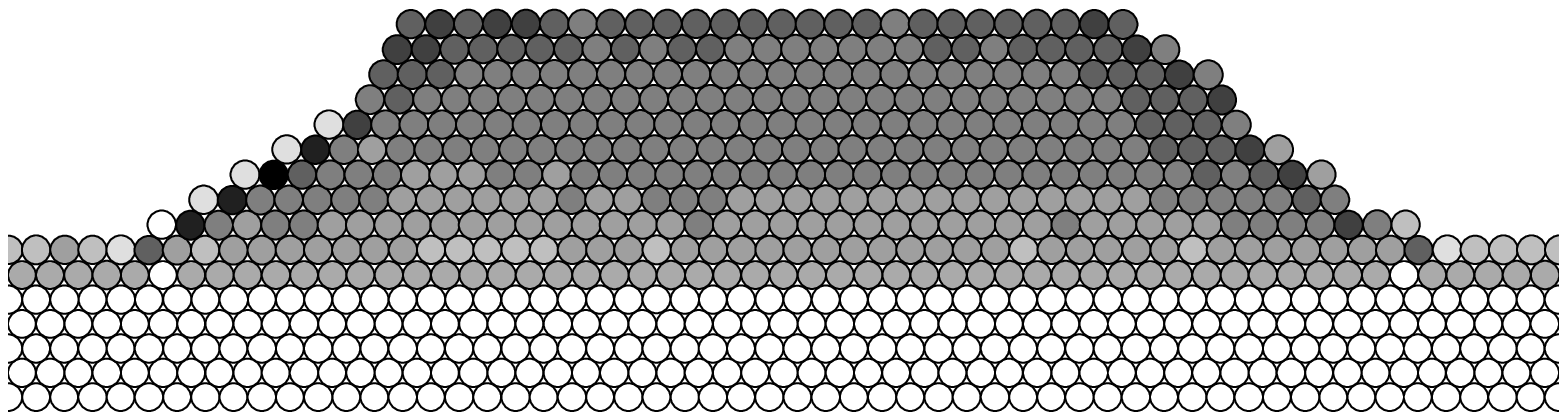}
\end{minipage}
\begin{minipage}{0.99 \textwidth}
  \epsfxsize= 0.99\textwidth
  \epsffile{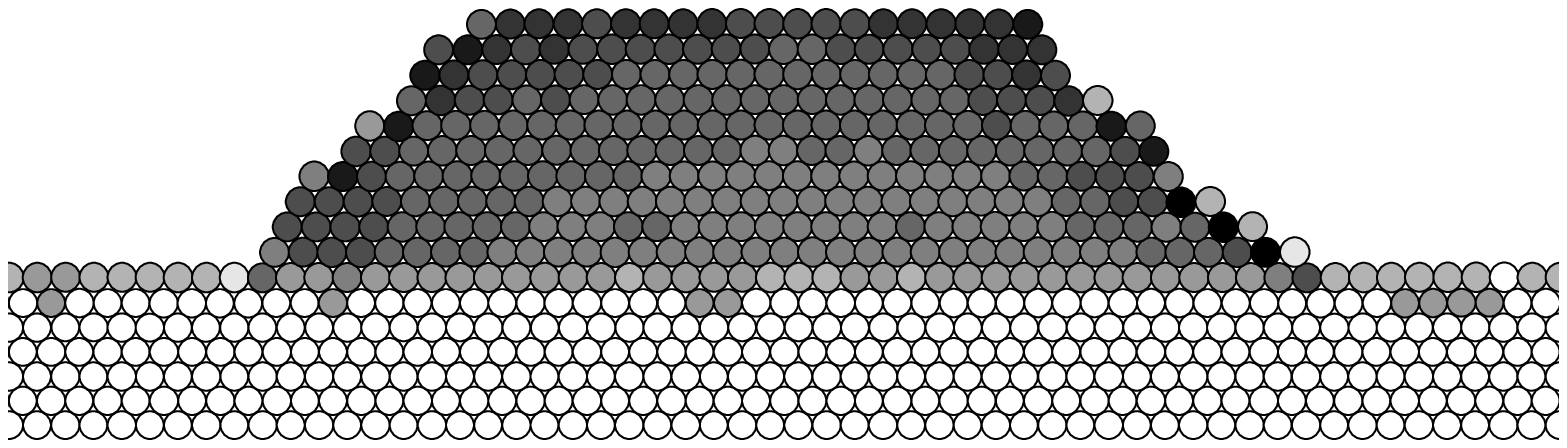}
\end{minipage}
\begin{minipage}{0.99 \textwidth}
  \epsfxsize= 0.99\textwidth
  \epsffile{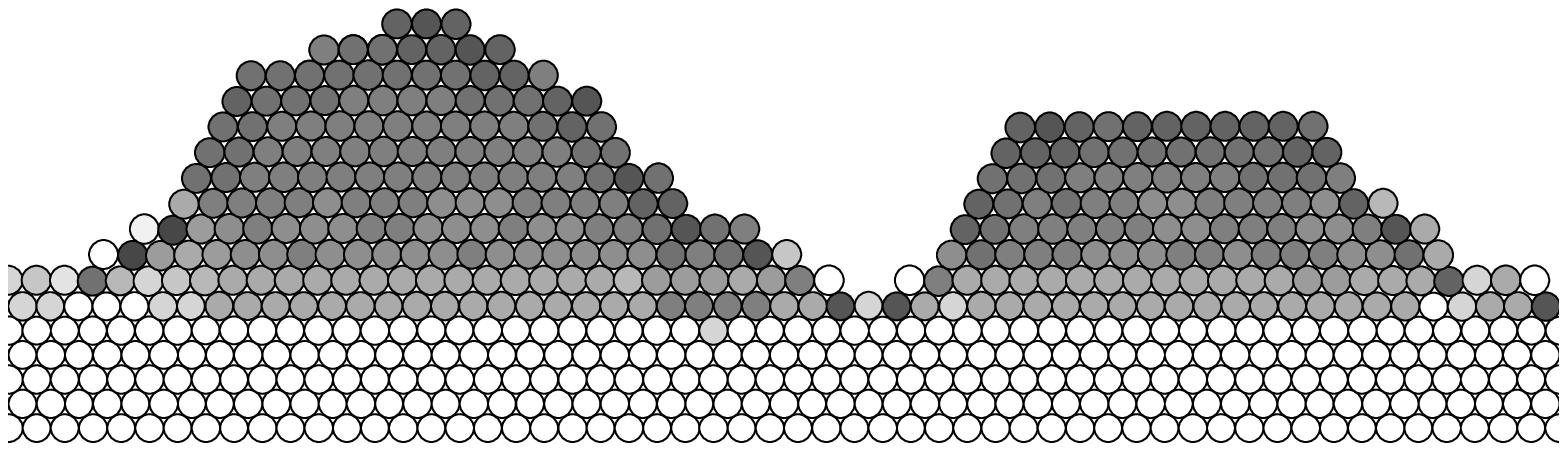}
\end{minipage}
\end{minipage}
\hfill
\begin{minipage}{0.35 \textwidth}
\caption{Typical islands for (top down) $\varepsilon=2\%$, $\varepsilon=3\%$ and $\varepsilon=4\%$ for deposition rate 
$R_d=4.5ML/s$ and temperature $T=500K$. The pictures show sections of the same size from simulation runs with 
$L=800$.}
\label{MISFITISLANDS}
\end{minipage}
\end{figure}
In misfit heteroepitaxy due to the incompatibility of the substrate and the adsorbate lattice, ${\varepsilon}^{-1}$ 
is the relevant length scale (cf. chapter \ref{KAP-4}).
The driving force for the misfit dependence of the island size
is probably the slower diffusion on top of islands and 
the increased bias to the island center for higher values of $\varepsilon$, cf. figure \ref{EPSILON}.
We believe that the observed dependence of the base size on the misfit for high growth rates 
is reminiscent of this fact. High growth rates lead therefore to a narrow size distribution 
around a self-limiting, strain dependent island size. 

Note that in situations close to equilibrium one would expect
an $\varepsilon$--dependence $b\propto \varepsilon^{-2}$, see e.g. \cite{Politi:2000:ICG} 
for theoretical arguments. 
In principle, this regime should be attained within our model
in the limit $R_d \to 0$ where the observed size $b$ and island height increases drastically, 
indeed.  
\section{Conclusions and outlook}
In conclusion we present a model for the simulation of SK--like growth which is capable of reproducing 
various important phenomena observed in experimental studies.
Our work provides a fairly detailed and plausible picture of the SK growth mode.
We have shown that a strong adsorbate--substrate interaction can be the cause for the formation of a stable 
wetting--layer due to the relatively slow diffusion of adatoms on the substrate.

We have demonstrated that the appearance of a kinetically controlled 
critical wet\-ting--layer thickness, like observed in various 
experimental works, can be explained within our model. We argued that high particle fluxes and low temperatures 
stabilize the metastable adsorbate layer.

The formation of islands is traced back to the relaxation of strain in the adsorbate layer: this strain relaxation 
leads to a pronounced bias towards the island center on top of the multilayer island. In addition the diffusion is 
slower on top of the island than on the wetting--layer in our model.
The island size is determined by the particle flux, the temperature and most importantly the misfit between adsorbate
and substrate. For large enough particle fluxes, the island size becomes independent of the temperature and is 
controlled by the misfit only.

Though we were able to gain a first insight into relevant mechanisms of self--assembled island formation various important
questions still remain open. One interesting point is whether 
intermixing between substrate and adsorbate material stabilizes 
the wetting--layer and thus increases the critical wetting--layer thickness. To this end exchange diffusion in the bulk 
has to be considered in the model. Like for all other types of concerted moves it is not yet clear if this is conceptually 
possible within our method.

Ultimately the method should be extended to $2+1$ dimensions, more realistic empirical potentials and realistic
lattice structures. Due to the high diffusion rates and low particle fluxes, which are necessary for the simulation 
of Stranski--Krastanov growth this is currently beyond computational means.
  \cleardoublepage
\chapter{Simulation of multi--component growth}
\label{KAP-6} 
In this chapter we focus on the analysis of multi--component growth by means of Monte Carlo simulations.
To this end we study the submonolayer growth of two types of adsorbate particles on a given substrate.
Our examinations are mainly motivated by experimental studies, where a variety of material systems have been 
found which, though immiscible in the bulk, form stable alloy layers if deposited as a thin film on certain 
substrate materials.

Though this phenomenon is witnessed for various material systems 
(for example CoAg/Ru(0001) \cite{Hwang:1996:CIS,Hwang:1997:STM,Thayer:2001:RST,Thayer:2002:LSS},  
CoAg/Mo(110) \cite{Tober:1998:SAL}, 
FeAg/Mo(110) \cite{Tober:1998:SAL},  
AgCu/Ru(0001) \cite{Stevens:1995:SSA},   
PdAu/Ru(0001) \cite{Sadigh:1999:SRO}) they all have one thing in common:
the types of deposited adsorbate particles show a misfit of opposite sign with the substrate.

Indeed  it was shown \cite{Tersoff:1995:SCA} that in systems dominated by an atomic 
size mismatch surface confined alloying is a possible strain relaxation mechanism 
(also see chapter \ref{KAP-1}). 
This is due to the fact that mixing between the different types of material 
reduces the strain energy in the surface layer.  
Within enhanced models \cite{Daruka:2003:TCF,Krack:2002:DSB} - taking the
competition between chemical binding and elastic energy into account - it was shown by means of 
equilibrium simulations, that this competition can result in a striped structure of the film. 
Depending on the concentration of the adsorbate 
materials and the chemical binding between them, alloying and the formation of misfit dislocations are competing
relaxation processes \cite{Thayer:2001:RST,Daruka:2003:TCF}.

These theoretical findings are also in agreement with various experimental studies, where under certain growth 
conditions both adsorbate types form islands of a regular stripe structure. Furthermore the alloying can lead here
to a change in the morphology of the islands' shape. In case of CoAg/Ru(0001) for example both adsorbate types form 
islands of compact shape when deposited alone on the substrate. But their co--deposition yields islands of pronounced 
ramification riddled with alternating veins of approximately constant width (see fig. \ref{HWANG}).
\begin{figure}[hbt]
\begin{minipage}{0.50 \textwidth}
  \epsfxsize= 0.75\textwidth
  \epsffile{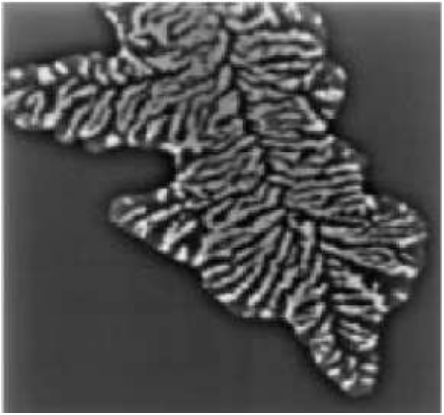}
\end{minipage}
\hfill
\begin{minipage}{0.45 \textwidth}
  \caption{$5000\times5000$ nm scanning tunneling microscopy image of a dendritic CoAg island on Ru(0001)
(by courtesy of R.Q. Hwang \cite{Hwang:1996:CIS}).}
\label{HWANG}
\end{minipage}
\end{figure}

Although the observed striped islands show a remarkable similarity with the equilibrium simulation results
it is not yet clear if strain effects are indeed the driving force for the observed phenomena.
As proposed in \cite{Hwang:1996:CIS} also a difference between the chemical binding energy of equal and unequal adsorbate 
particles would both favor the observed vein structure and the dendritic island shape.

In the following we will present a model which is capable of simulating multi--component growth in $2+1$ dimensions, 
incorporating both strain and binding energy effects. We will first discuss the results of some equilibrium 
simulations and compare them to KMC simulation results afterwards. We will try to distinguish whether the phenomena
observed in KMC simulations are due to elastic or binding interactions.
For that reason we also propose a lattice gas KMC model which is mainly governed by an edge diffusion barrier
and where strain effects are completely neglected.
\section{Simulation model}
Since for conceptional reasons phenomena like the formation of alternating vein structures or the ramified island 
growth can not be mapped to $1+1$ dimensions we have to extend our simulation method to $2+1$ dimensions.
In order to keep the computational effort acceptable we choose the simple cubic (sc) 
symmetry for our simulations, which has two important advantages: 
first one has to take only four possible in--plane transition sites per 
particle into account, in comparison with six transition sites in case of, e.g., the fcc lattice. And more 
importantly, since the sc lattice is not close--packed the number of particles to be considered during the 
relaxation processes is rather small.

But the choice of the sc symmetry implies also an important problem: all of the experimental results presented
in \cite{Hwang:1996:CIS,Hwang:1997:STM,Thayer:2001:RST,Thayer:2002:LSS,Tober:1998:SAL,Stevens:1995:SSA,Sadigh:1999:SRO} are given for metals  grown on a substrate
of fcc/hcp symmetry. However, this should primarily affect the (compact) island shape: in case 
of the discussed materials one finds triangular shaped islands, whereas in the sc symmetry square shaped islands should 
be formed.

In order to stabilize the sc lattice we adapt the method proposed in \cite{Schroeder:1997:DSS} and choose 
\begin{equation}
\label{CUBIC_6}
 {U}_{ij} = \left(0.1 + 8\,\left( \frac{x_{ij}^2}{r_{ij}^2}-\frac{1}{2} \right) 
\left( \frac{y_{ij}^2}{r_{ij}^2}-\frac{1}{2} \right)
\left( \frac{z_{ij}^2}{r_{ij}^2}-\frac{1}{2} \right)\right)V_{ij}.
\end{equation}
as interaction potential between two particles $i$ and $j$ (see appendix \ref{AP-CUBIC} for details).
In the following two kinds of pair--potentials $V_{ij}$ are used during our simulations: 
The Lennard--Jones (LJ) potential discussed in appendix \ref{AP-LJ}   
\begin{equation}
\label{LJ_6}
 {V}_{ij} =\ 4 E_{ij} \left[\left(\frac{{\sigma}_{ij}}{r_{ij}}\right)^{12}
-\left(\frac{{\sigma}_{ij}}{r_{ij}}\right)^{6}\right] 
\end{equation}
and the Morse potential (see appendix \ref{AP-MORSE})
 \begin{equation}
\label{MORSE_6}
 {V}_{ij} =\ E_{ij} e^{a\left({\sigma}_{ij}-r_{ij}\right)} \left(e^{a\left({\sigma}_{ij}-r_{ij}\right)}-2 \right).
\end{equation}
In order to save computer time during energy calculations $U_{ij}$ is cut off for
particle distances greater than $r_{cut}=2r_{0}$, whereas for the calculation of the activation
barriers the cut--off distance is set to $3r_{0}$. Both simplifications are perfectly
justified since the used potentials decline fast towards zero for increasing particle
distance. 
For both types $E_{ij}$ gives the depth of the potential for the interaction of two particles $i$ and $j$ at the
equilibrium distance $r_0$ with $r_0 \propto \sigma_{ij}$. 
In the following the interaction strength between two particles of the substrate is given by 
$E_{ij}=E_S$ and $\sigma_{ij}=\sigma_S=1$. $E_{ij}=E_A$, $\sigma_{ij}=\sigma_A$ and $E_{ij}=E_B$, 
$\sigma_{ij}=\sigma_B$ are chosen for the A--A and B--B interaction, respectively.
Here A and B denotes the two adsorbate species and $S$ labels the substrate material.
For the interaction of adsorbate particles of type $X=A,B$ with the substrate we set $E_{ij}=E_{XS}=\sqrt{E_X E_S}$ and 
$\sigma_{ij}=\sigma_{XS}=1/2(\sigma_{X}+\sigma_{S})$. Likewise, $E_{ij}=E_{AB}=\sqrt{E_A E_B}$ and 
$\sigma_{ij}=\sigma_{AB}=1/2(\sigma_{A}+\sigma_{B})$ holds for the interaction between A and B adsorbate particles.
The misfit $\varepsilon$ ($\varepsilon > 0$) is applied in a symmetric way to the system:
\begin{eqnarray}
\label{MISFIT}
 \sigma_A  =  1-\varepsilon \\
 \sigma_B  =  1+\varepsilon.
\end{eqnarray}
The additional parameter $a$ in equation (\ref{MORSE_6}) 
determines the steepness of the Morse potential around its minimum.
It is used with three different values of the parameter $a$:
$a=5.0$, $5.5$ and $6.0$.

The potential depths $E_{ij}$ are chosen in such a way that they meet two demands:
on the one hand the ratio between $E_S$ and $E_A$, $E_B$ is kept fixed for all potentials:
\begin{equation}
\label{E_S}
E_A=E_B=\frac{1}{6}E_S.
\end{equation}
On the other hand the barrier for diffusion on plain substrate $E_{a,sub}$ in the case of homoepitaxy ($\varepsilon=0$)
should become roughly the same value for all used potentials to facilitate the comparison of the results. 
We choose $E_S$ here in such a way that in the case of homoepitaxy ($\varepsilon=0$) $E_{a,sub} \approx 0.37eV$ -
a typical value for self--diffusion barriers of metals 
(see e.g. \cite{Trushin:1997:EBS,Maca:1999:EBD,Yu:1997:POE,Kellog:1990:SSD}).
The resulting $E_{S}$ for the different potentials are shown in table \ref{TAB_ES}.
\begin{table}
\begin{tabular}{|l||r|r|r|r|} \hline
  potential & LJ & a=5.0 & a=5.5 & a=6.0  \\ \hline 
  $E_S/eV$ & 3.0 & 3.0  & 2.814 & 2.70 \\ \hline
\end{tabular}
\caption{The substrate--substrate interaction strength $E_S$ for the used pair--potentials. The substrate--adsorbate and
adsorbate--adsorbate interactions are then given according to equations (\ref{E_S}) and (\ref{E_AB}), respectively.}
\label{TAB_ES}
\end{table}
\section{Equilibrium simulations}
In order to derive the influence of the potential depth for the A--B interaction we first carry out
canonical equilibrium simulations at full coverage of the substrate, where the number of A and B particles
is kept fixed. The substrate is given as an $100 \times 100$ crystal of $6$ layers height.
Periodic boundary conditions
are applied in the $x$-- and $y$--direction. 
At the beginning of each simulation run the substrate is randomly covered with adsorbate particles
at a certain ratio $n_A/n_B$ between the number of A and B particles. 
Then the system approaches thermal equilibrium at temperature $T$ by means of a 
rejection--free method (see \cite{Ahr:2002:SPE}).
\subsection{Rejection--free method}
Since the simulations are carried out for a range of the misfit where even at full coverage of the substrate 
misfit dislocation are not observed each particle can be allocated at a certain site of the
$100 \times 100$ square lattice.
In order to realize a fixed number of each particle species like required in canonical simulations we follow here 
a method for the rejection--free canonical equilibrium simulation proposed by Ahr \cite{Ahr:2002:SPE,Ahr:2002:FSI}.
In each event an A particle exchanges its binding site with a B particle. We choose a nonlocal 
dynamics where the range of particle jumps is unlimited. This yields considerably faster equilibration compared to
local dynamics. For simplicity, we  permit only exchange jumps 
of particles which are more than $r_{cut}$ away from each other, i.e. exchange of two particles within this 
range is forbidden. Such processes would complicate the calculation of the configuration energies.

If now an A particle located on site $i$ of the square lattice exchanges its binding site with a B particle of site 
$j$ the energy difference between the final and the initial state is given by $\Delta H = \Delta H_j - \Delta H_i$, where
$\Delta H_x=H_x(A)-H_x(B)$ is the energy difference of the system with site $x$ occupied with an A and B particle.
$H_x(A)$ and $H_x(B)$ are calculated in a local way: an A particle is set to the site $x$ and the particles of 
the system within a radius $r_{cut}$ around this site are allowed to relax locally. The local energy 
is registered as $H_x(A)$. Conversely, the same procedure is performed for an B particle 
at site $x$  in order to calculate $H_x(B)$. The rates
\begin{equation}
\label{RF-RATE}
r_{i\to j}=e^{\frac{\Delta H_i - \Delta H_j}{2kT}}
\end{equation}
fulfill the detailed balance condition. Then the probability for an A particle on site $i$ to 
exchange its site with a B particle at site $j$ factorizes, i.e.
\begin{equation}
\label{RF-PROP}
p_{i\to j}=p_i^{A\to B}p_j^{B\to A} \quad\mbox{where e.g. }\quad p_x^{A\to B}=\frac{r_x^{A\to B}}{\sum_xr_x^{A\to B}}.
\end{equation}
Here $p_x^{A\to B}$, $r_x^{A\to B}$ give probability 
and rate for a change from A to B particle at site $x$, respectively.
Analogous $p_x^{B\to A}$, $r_x^{B\to A}$ are given for a change from B to A particle at site $x$.
The rate $r_x^{A\to B}$ is given by $r_x^{A\to B}=\exp\left(\Delta H_x /2kT\right)$ 
if site $x$ is occupied by an A particle and zero otherwise.
Conversely, $r_x^{B\to A}=\exp\left(-\Delta H_x /2kT\right)$ on sites occupied by B particles and zero on 
sites occupied by A particles.

Due to this factorization property one can proceed now in two steps: In the first step, 
using a binary search tree a site $i$ occupied with an
A particle is drawn with the probability $p_x^{A\to B}$. Then a B occupied site $j$ is selected with
probability $p_x^{B\to A}$, using a second search tree. 
If the distance between both sites is greater than $r_{cut}$ the exchange jump is 
performed and the rates of all affected events are updated. 
Otherwise, the event is rejected and the system remains unchanged. Since the number of rejected
events is small for large systems, the loss of speed can be neglected.

In order to avoid artificial strain accumulation due to the local relaxation for the calculation of $\Delta H_x$ 
the system is globally relaxed after a fixed number of simulation steps (here $5000$), all rates are newly calculated
and the search trees are updated accordingly. The system's total energy per particle $E_{tot}$ is registered after 
each global relaxation. All simulation runs are halted after $20$ global relaxation events
(i.e. after $10^5$ simulation steps).
\subsection{Influence of misfit and binding energy}
\begin{figure}[h]
\begin{minipage}{0.24 \textwidth}
  \epsfxsize= 0.99\textwidth
  \epsffile{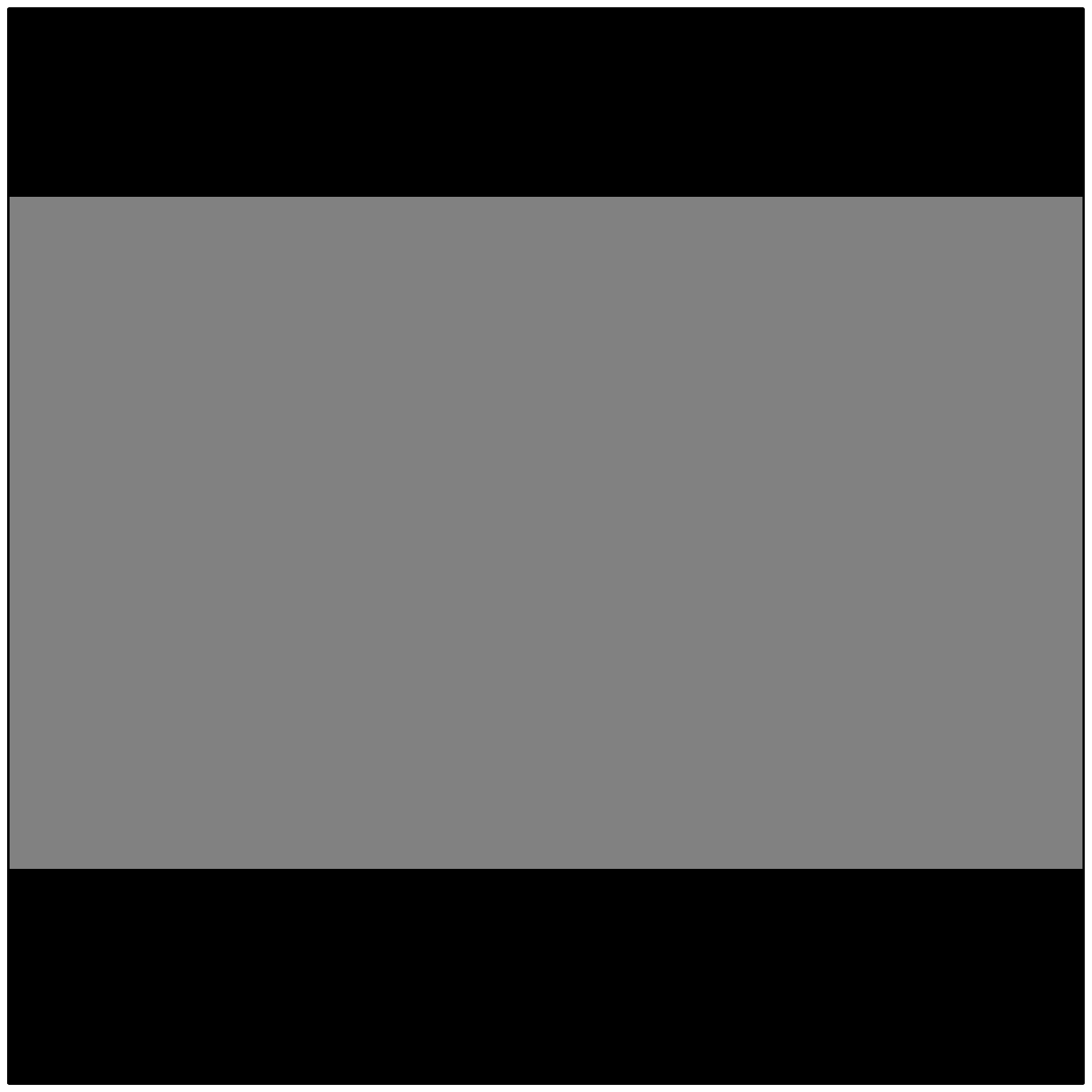}
\end{minipage}
\hfill
\begin{minipage}{0.24 \textwidth}
  \epsfxsize= 0.99\textwidth
  \epsffile{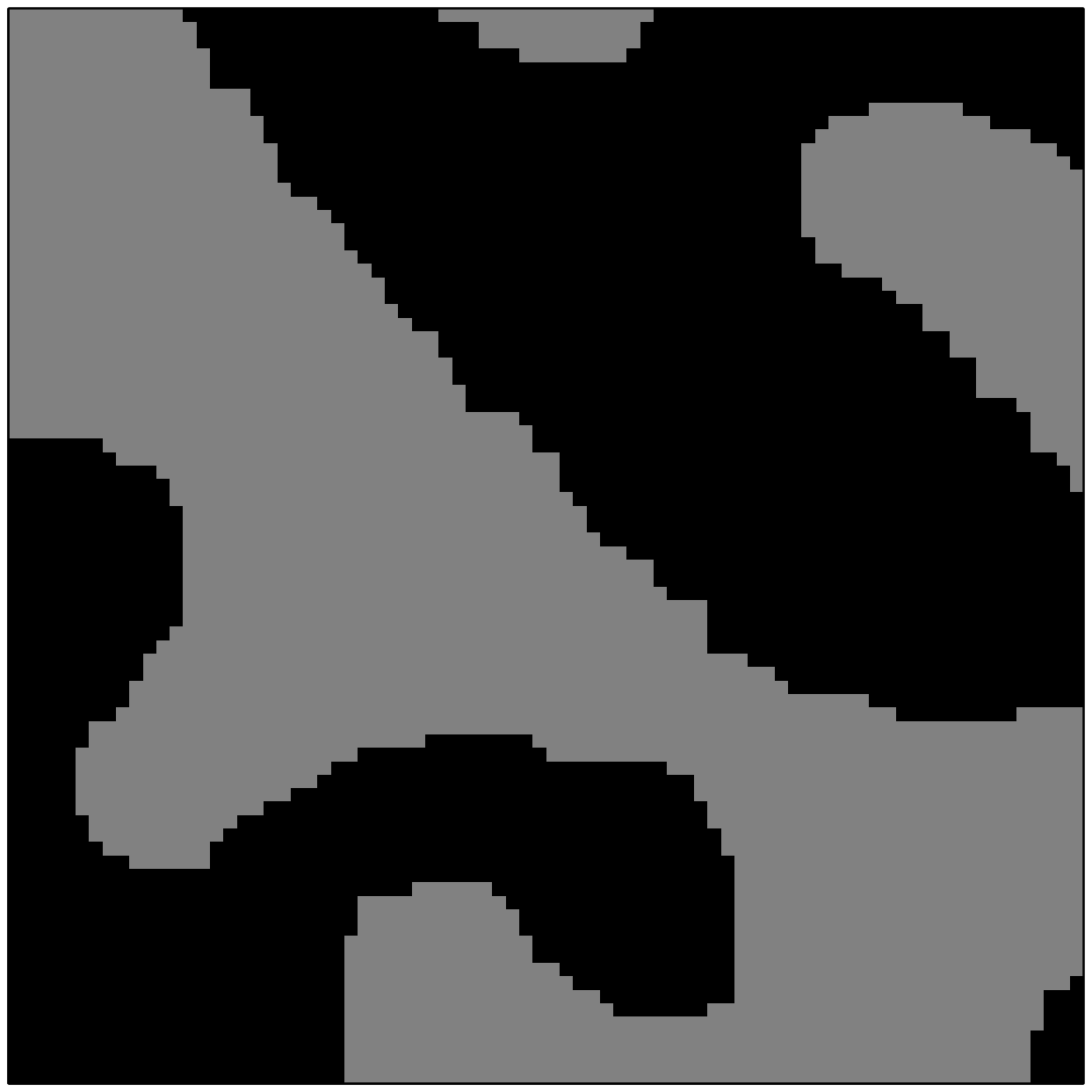}
\end{minipage}
\hfill
\begin{minipage}{0.24 \textwidth}
  \epsfxsize= 0.99\textwidth
  \epsffile{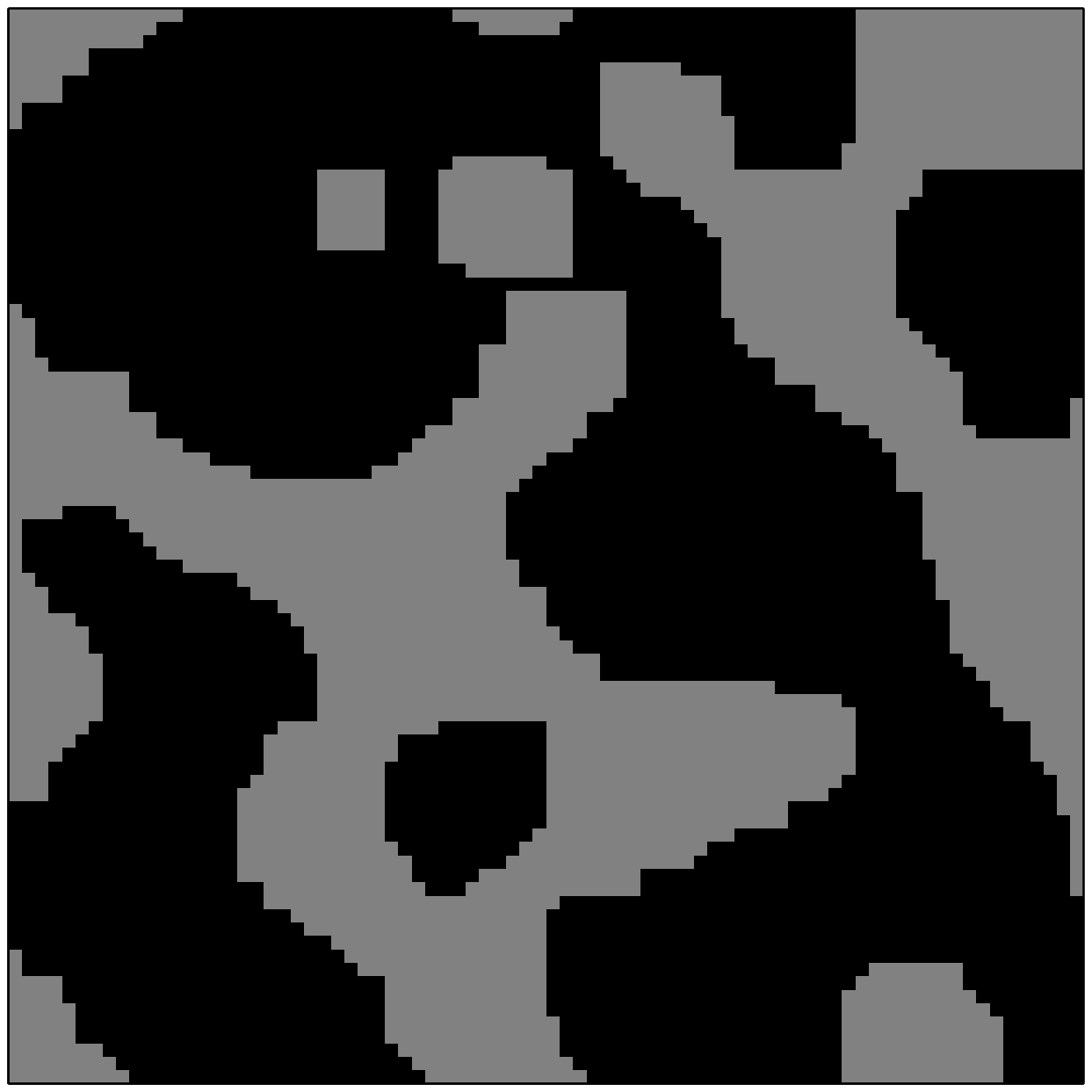}
\end{minipage}
\hfill
\begin{minipage}{0.24 \textwidth}
  \epsfxsize= 0.99\textwidth
  \epsffile{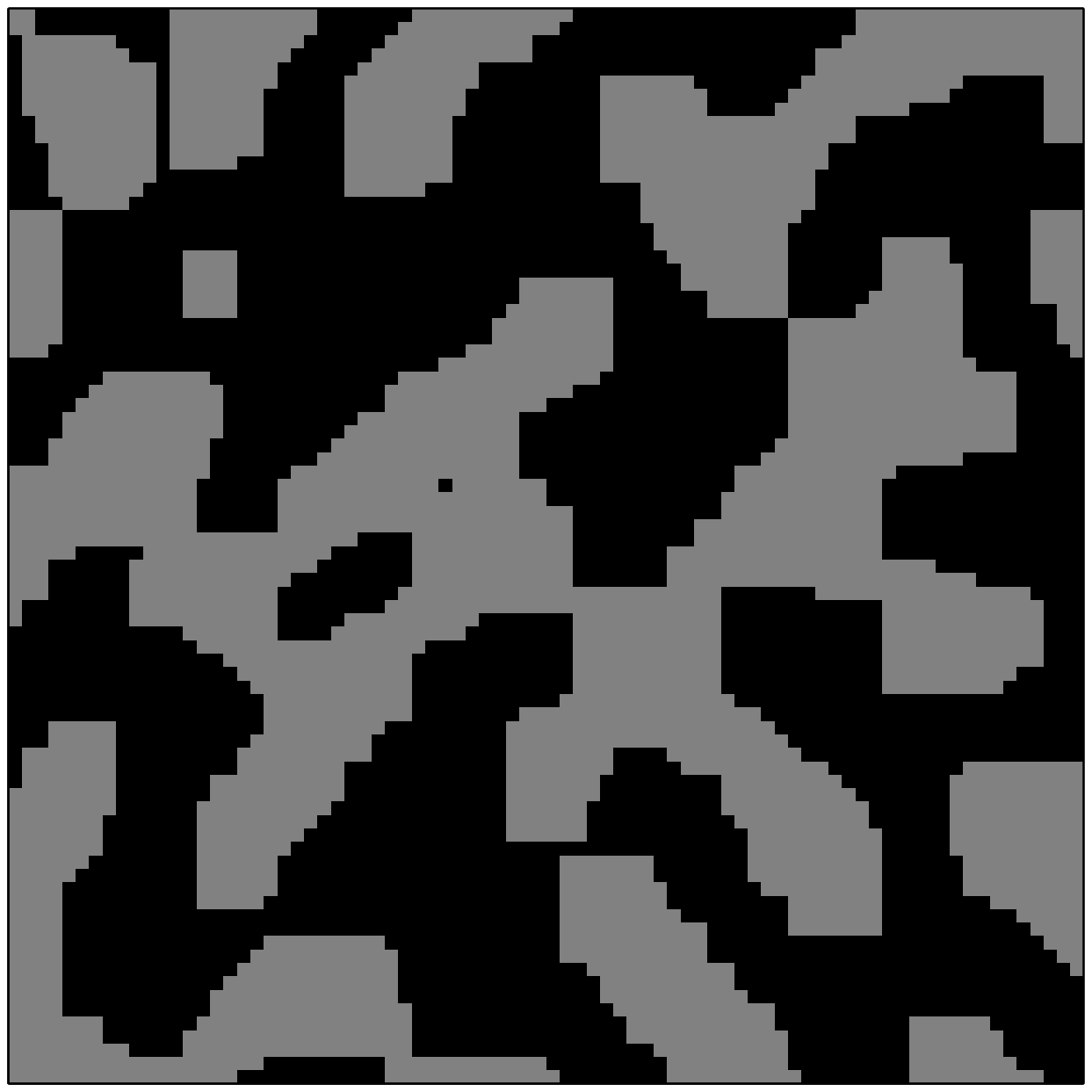}
\end{minipage}
\caption{$80 \times 80$ sections for the Lennard--Jones potential with $E_{AB}=0$ at $T=250K$.
The misfits is (from left to the right) $\varepsilon=0\%$, $\varepsilon=3\%$, $\varepsilon=4.5\%$
and $\varepsilon=5.5\%$. The bigger B particles are shown in light gray.}
\label{EAB0}
\end{figure}
In order to detect the influence of misfit and binding energy we first present some 
results for the cubic Lennard--Jones potential (eq. (\ref{CUBIC_6}), (\ref{LJ_6})).
Figure \ref{EAB0} shows typical sections for simulation runs with $E_{AB}=0$ at $T=250K$.
Since there is no interaction between A and B particles for the homoepitaxial case ($\varepsilon=0\%$)
a fast demixing of the two particle types is observed. A straight line in the $x$-- or $y$--direction
separates both areas minimizing
the energy loss of the system.

From the sample for $\varepsilon=3\%$ it appears, that a straight boundary between the both 
particle areas is less favorable.
This becomes more obvious for higher misfits (e.g.  $\varepsilon=4.5\%$, $\varepsilon=5.5\%$) where several
clusters of both particles types are formed and the boundaries between A and B domains are preferentially 
aligned in the $<11>$ directions.

The reason for the observed $\varepsilon$--dependence is quite clear: the higher the misfit the less
favorable become extended areas of the same particle type. The strain energy in such big clusters 
even exceeds the energy lost due to the enlarged domain wall between regions of A und B particles. 
However, even though the misfit causes separation between 
the particle types at $E_{AB}=0$,  no regularity or ordering of the structures 
(like the alternating veins mentioned before) is observed.

The situation changes completely for the case of  $E_{AB}>0$ (see fig. \ref{GLW_LEN_SNAPS}). 
\begin{figure}[h]
\begin{minipage}{0.24 \textwidth}
  \epsfxsize= 0.99\textwidth
  \epsffile{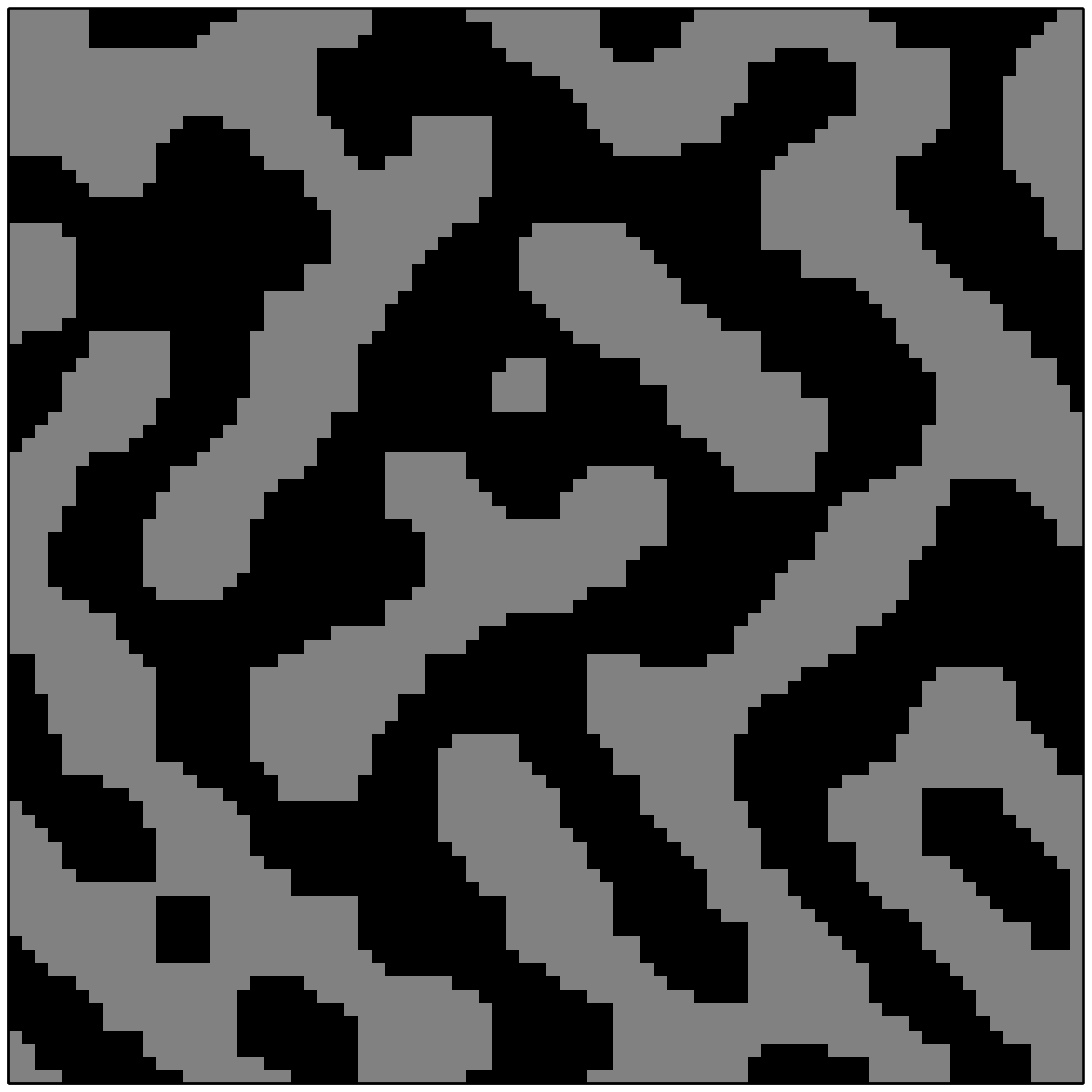} 
\end{minipage}
\hfill
\begin{minipage}{0.24 \textwidth}
  \epsfxsize= 0.99\textwidth
  \epsffile{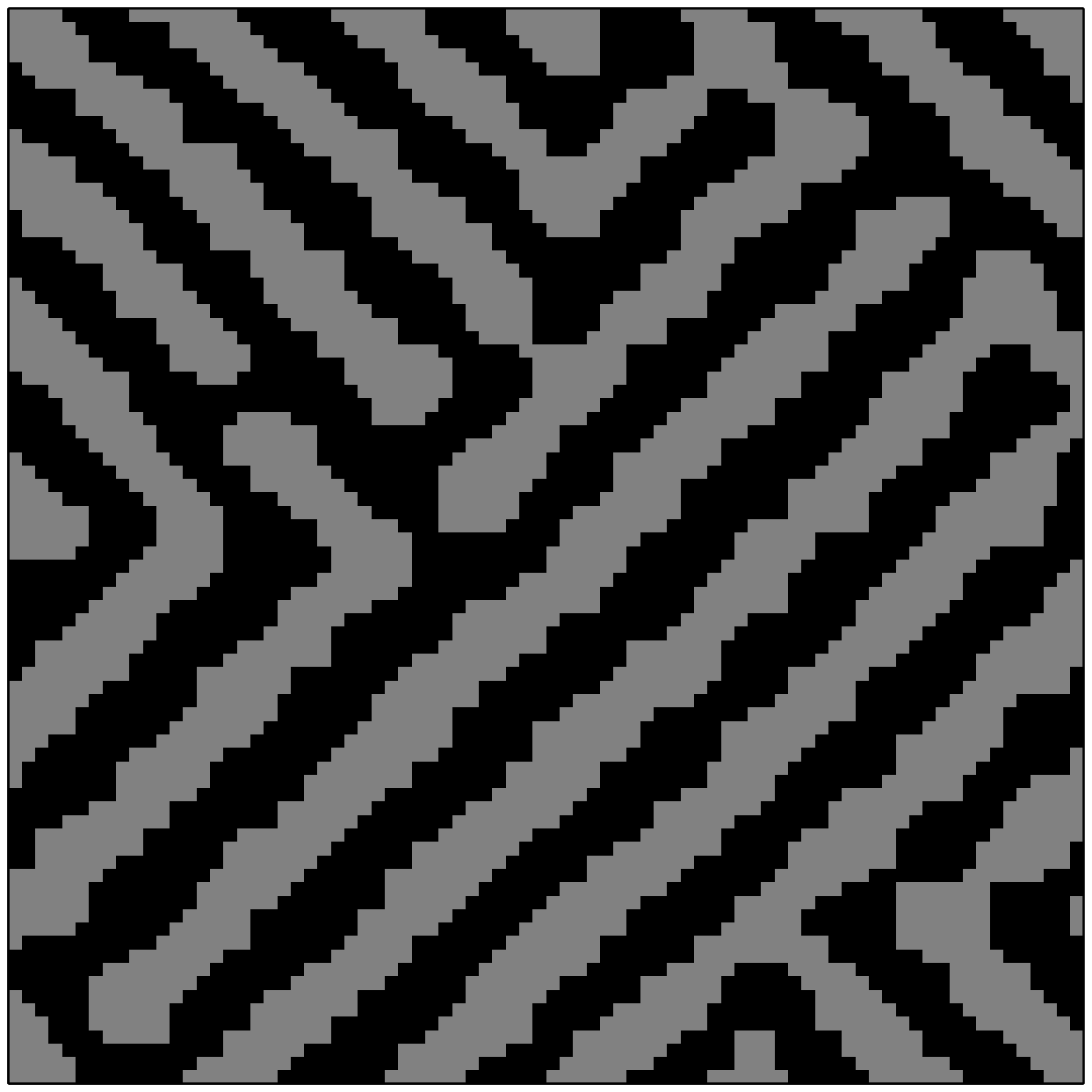}
\end{minipage}
\hfill
\begin{minipage}{0.24 \textwidth}
  \epsfxsize= 0.99\textwidth
  \epsffile{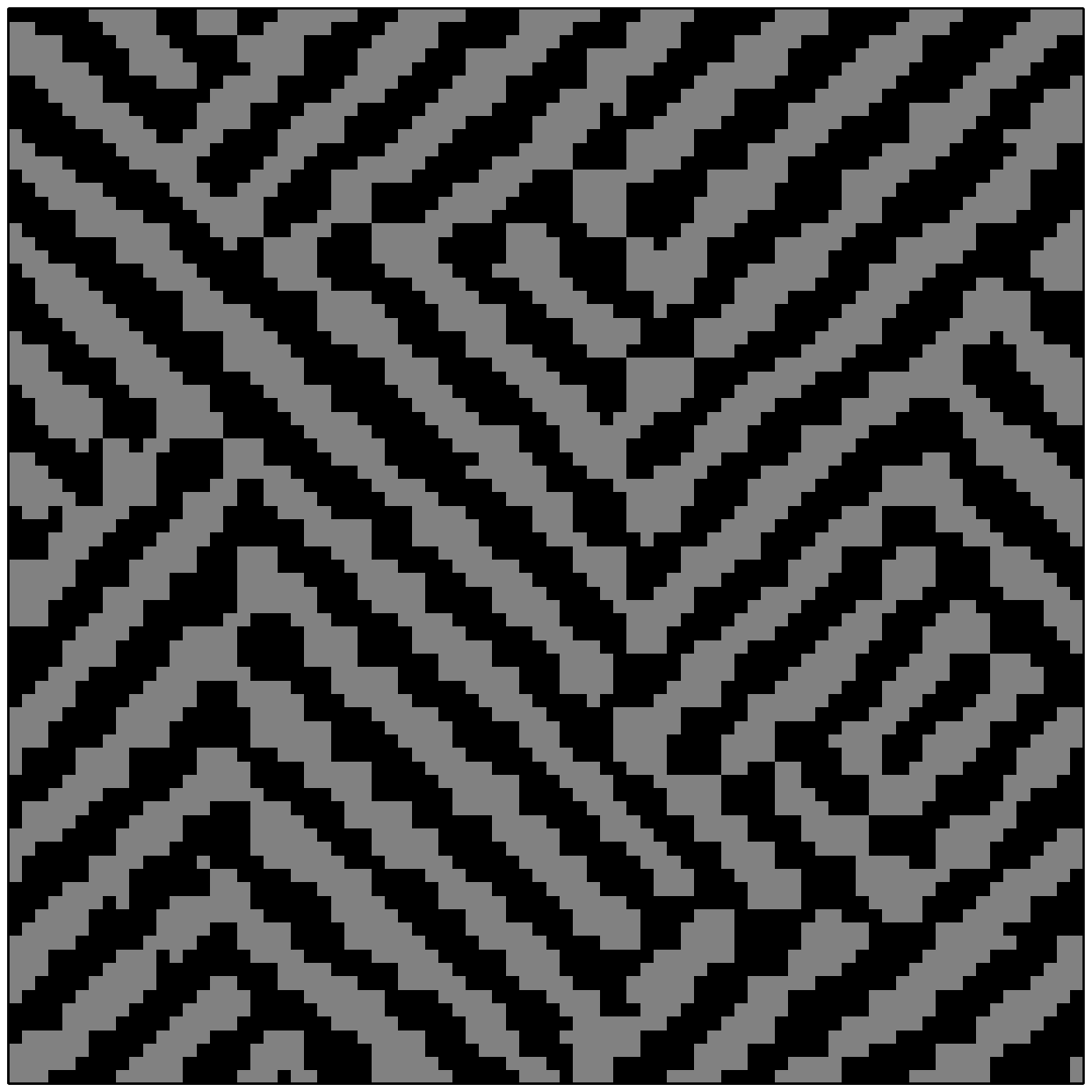}
\end{minipage}
\hfill
\begin{minipage}{0.24 \textwidth}
  \epsfxsize= 0.99\textwidth
  \epsffile{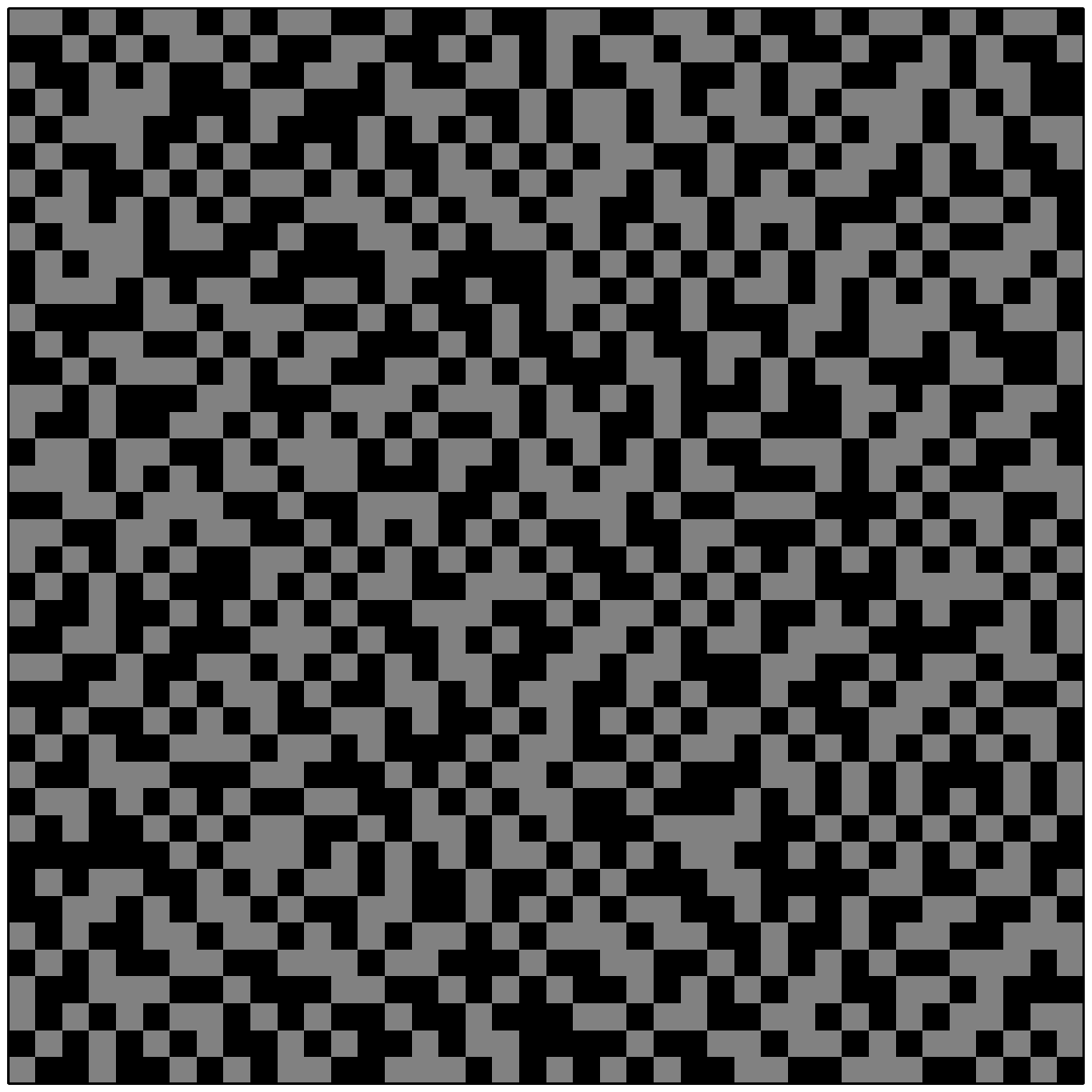}
\end{minipage}

\begin{minipage}{0.24 \textwidth}
  \epsfxsize= 0.99\textwidth
  \epsffile{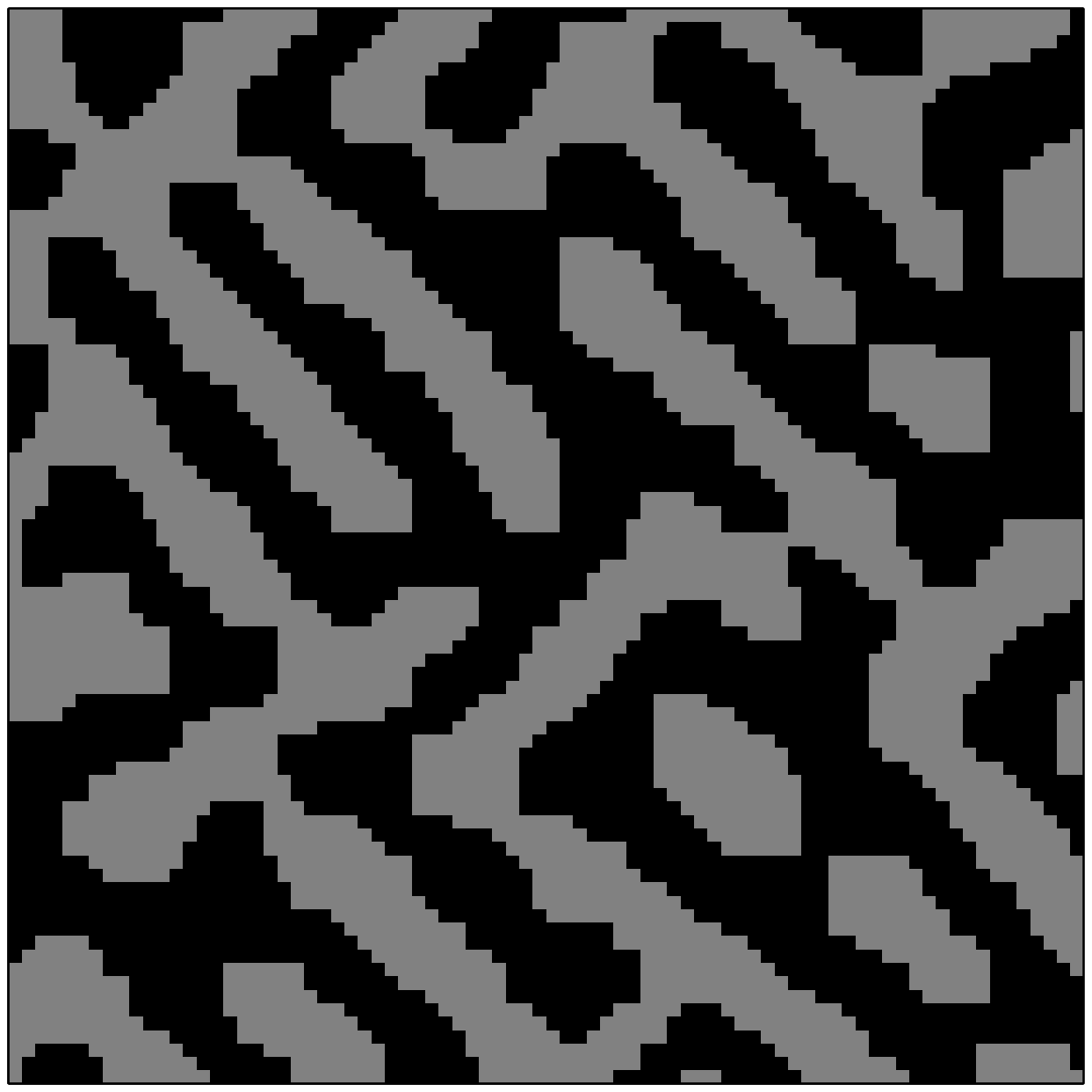}
\end{minipage}
\hfill
\begin{minipage}{0.24 \textwidth}
  \epsfxsize= 0.99\textwidth
  \epsffile{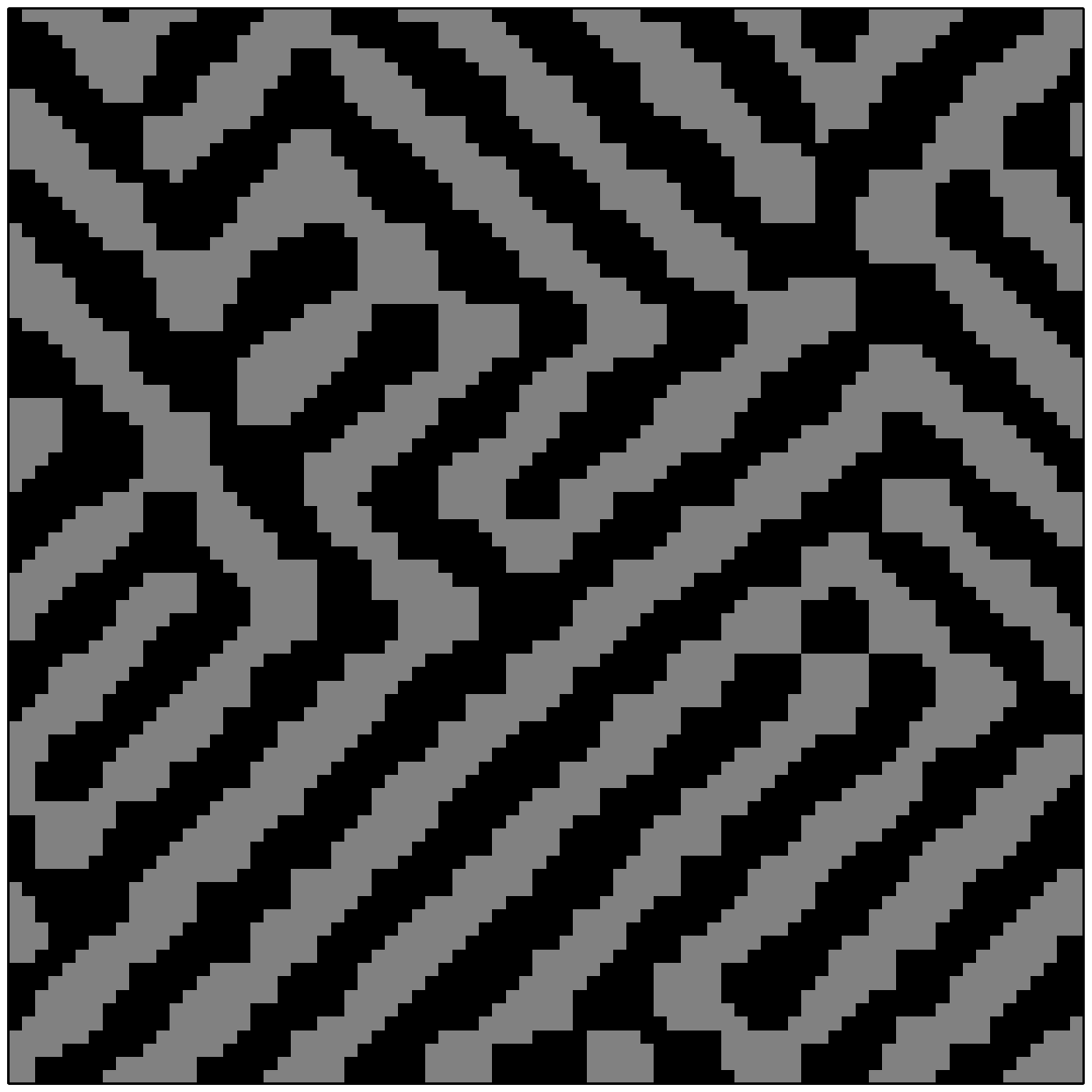}
\end{minipage}
\hfill
\begin{minipage}{0.24 \textwidth}
  \epsfxsize= 0.99\textwidth
  \epsffile{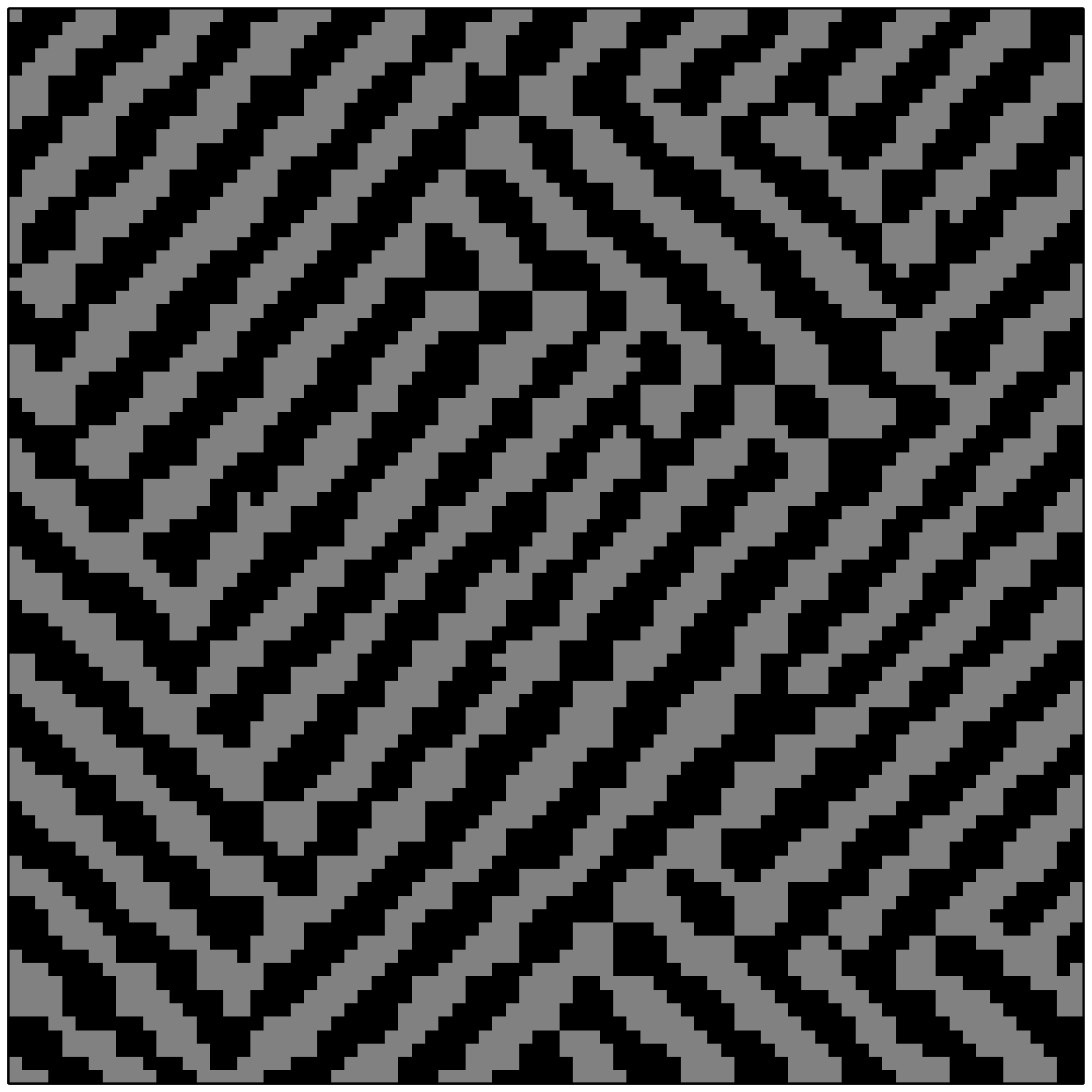}
\end{minipage}
\hfill
\begin{minipage}{0.24 \textwidth}
  \epsfxsize= 0.99\textwidth
  \epsffile{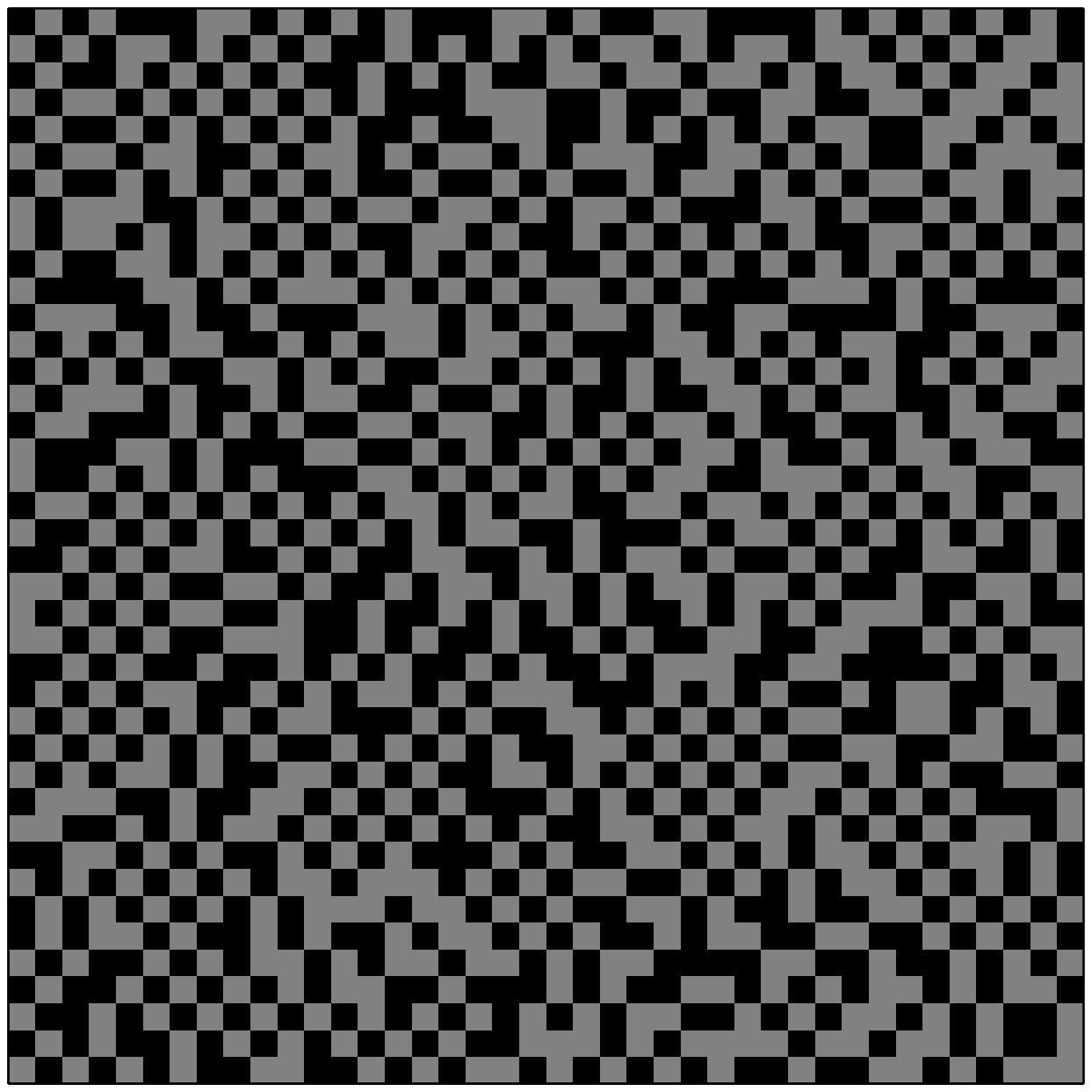} 
\end{minipage}

\begin{minipage}{0.24 \textwidth}
  \epsfxsize= 0.99\textwidth
  \epsffile{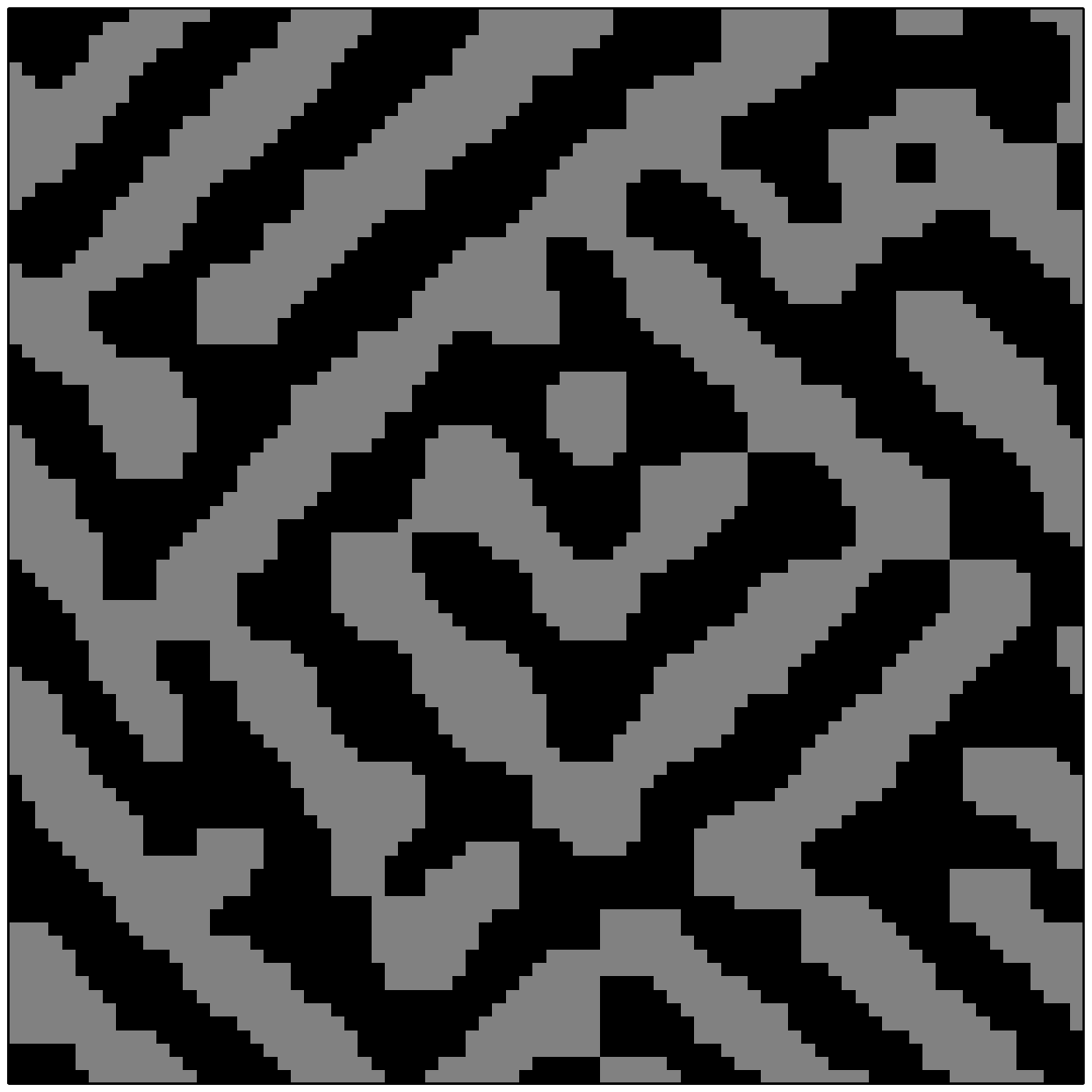}
\end{minipage}
\hfill
\begin{minipage}{0.24 \textwidth}
  \epsfxsize= 0.99\textwidth
  \epsffile{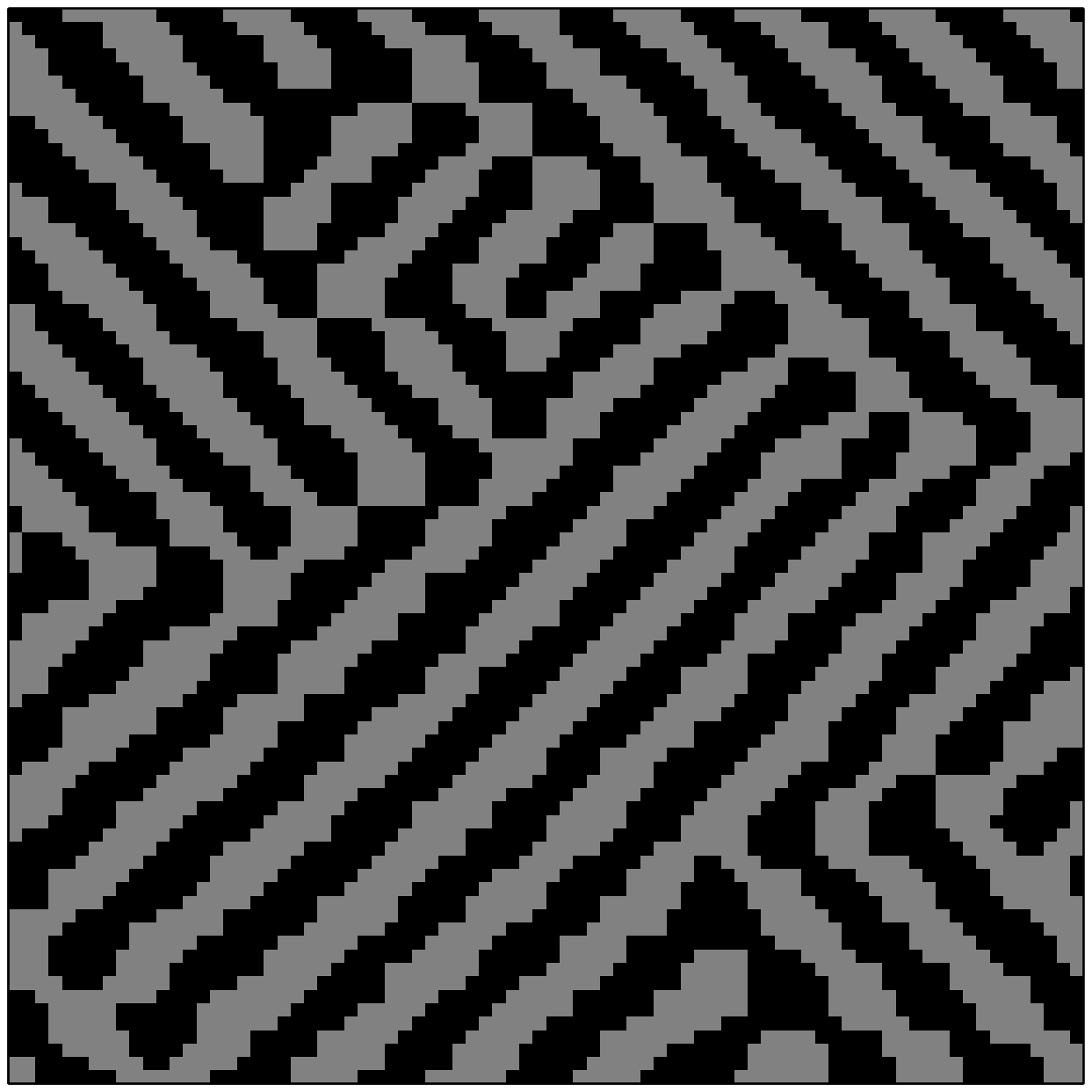}
\end{minipage}
\hfill
\begin{minipage}{0.24 \textwidth}
  \epsfxsize= 0.99\textwidth
  \epsffile{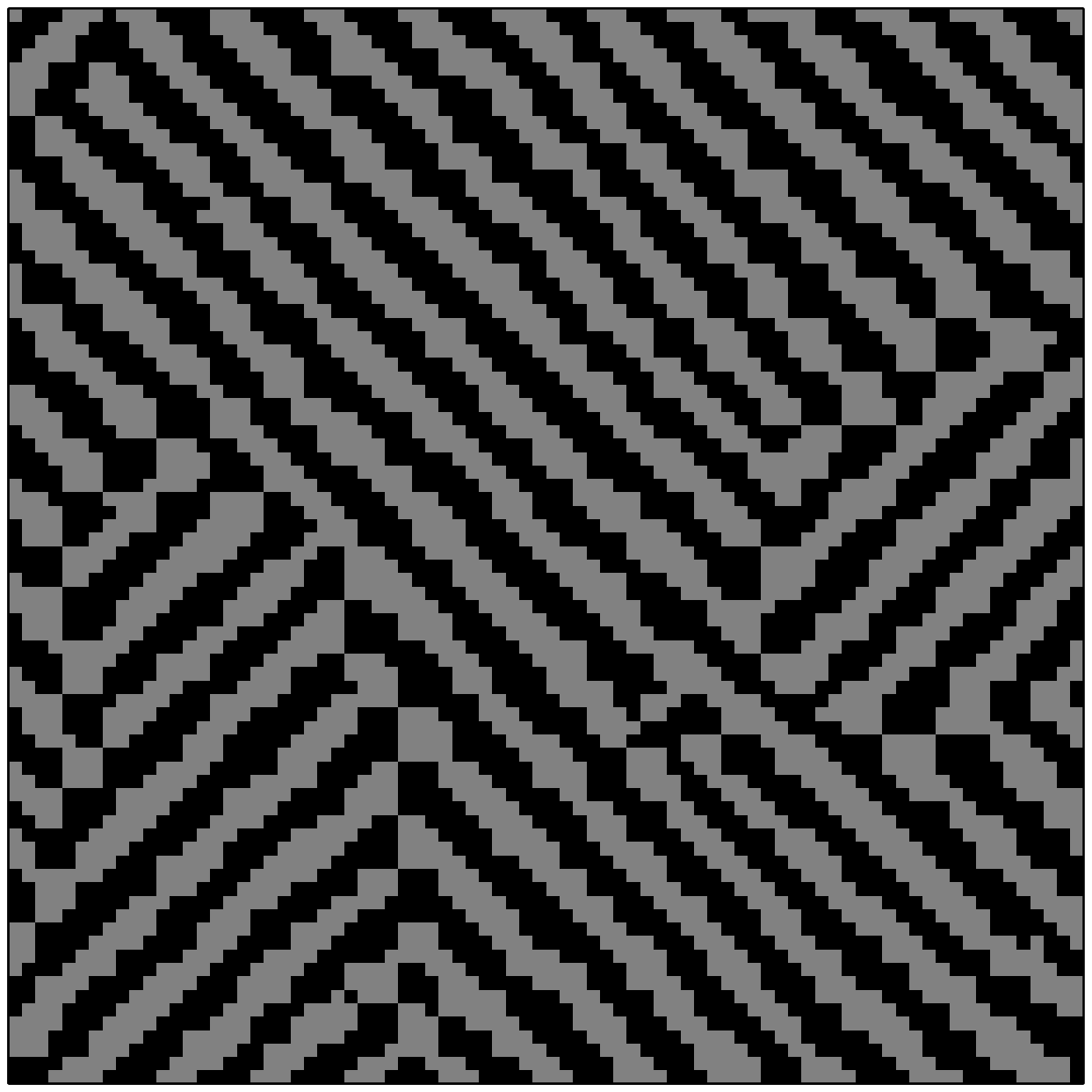}
\end{minipage}
\hfill
\begin{minipage}{0.24 \textwidth}
  \epsfxsize= 0.99\textwidth
  \epsffile{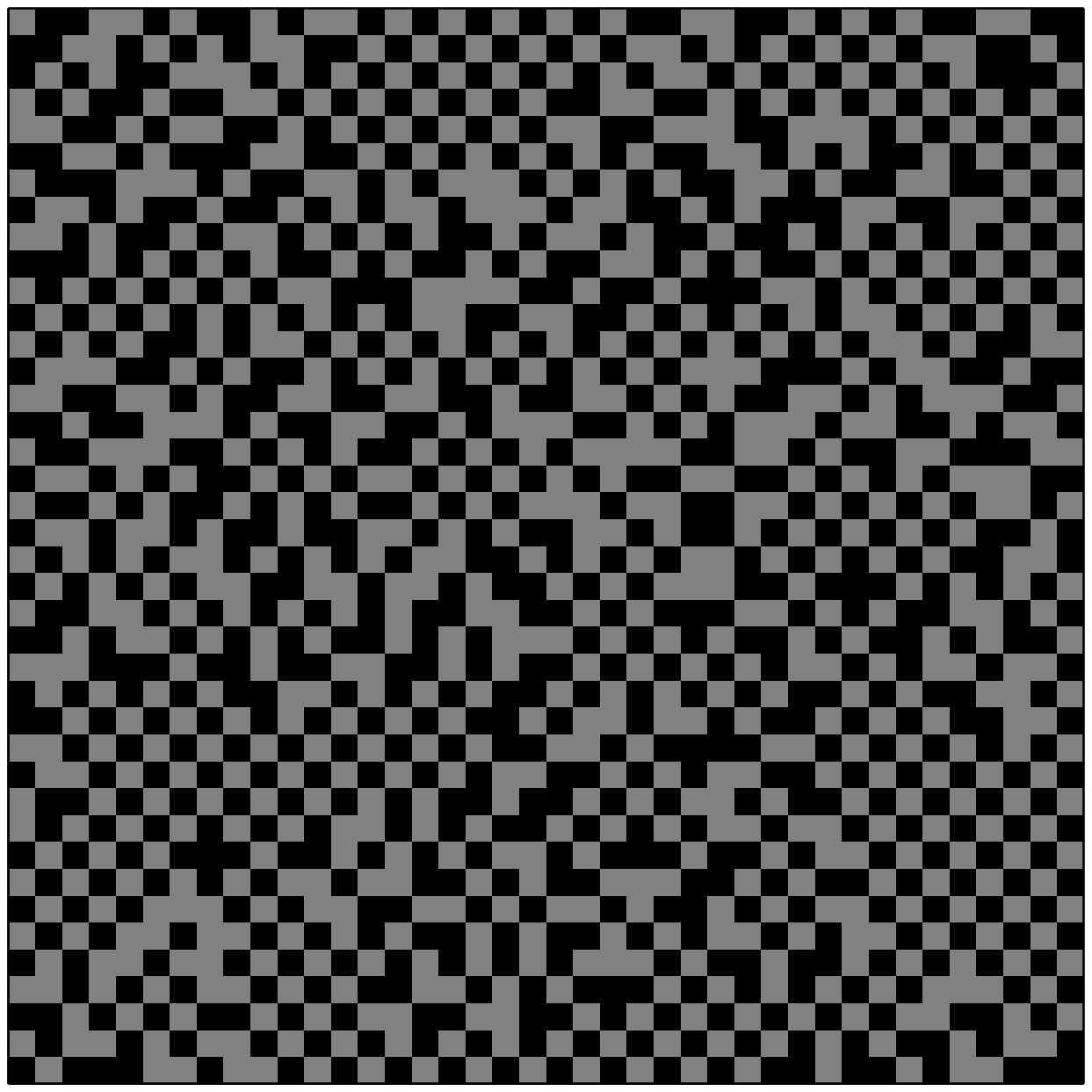}
\end{minipage}
\caption{Snapshots  for the Lennard--Jones potential at $T=250K$ with 
$E_{AB}=0.6E_A$, $E_{AB}=0.8E_A$, $E_{AB}=0.9E_A$, $E_{AB}=1.0E_A$ (from the left to the right) and
$\varepsilon=4.5\%$, $\varepsilon=5.0\%$, $\varepsilon=5.5\%$ (top down). 
The panels for $E_{AB}=1.0E_A$ show $40 \times 40$ sections of the system, the remaining panels show
$80 \times 80$ sections.}
\label{GLW_LEN_SNAPS}
\end{figure}
Now a regular 
arrangement of alternating A  and B stripes, preferentially in the $<11>$ directions is visible.
As already known from other atomistic models with size mismatch \cite{Tersoff:1995:SCA,Krack:2002:DSB}
the competition between binding of the particles and strain energy  is the cause for 
these regular patterns.
Furthermore with, increasing $E_{AB}$ and increasing misfit the stripes become thinner and  
more regular in size and shape.
For the case $E_{AB}=E_A=E_B$ the system approaches a checkered state, i.e. a stripe width of one.

The alignment of the stripes in the $<11>$ direction - already visible for the $E_{AB}=0$ case (see fig.
\ref{EAB0}) - is here due to the cubic symmetry of the used potential: both
particle types try to reach their preferred stripe width $l$ in each lattice direction ($x$ and $y$).
Note, that the used cubic form of the potential (eq. (\ref{CUBIC_6})) has only a weak interaction in the  
$<11>$ direction (also see fig. \ref{CUBIC_FIG} in the appendix \ref{AP-1}).
\begin{figure}[h]
\begin{minipage}{0.47 \textwidth}
  \epsfxsize= 0.99\textwidth
  \epsffile{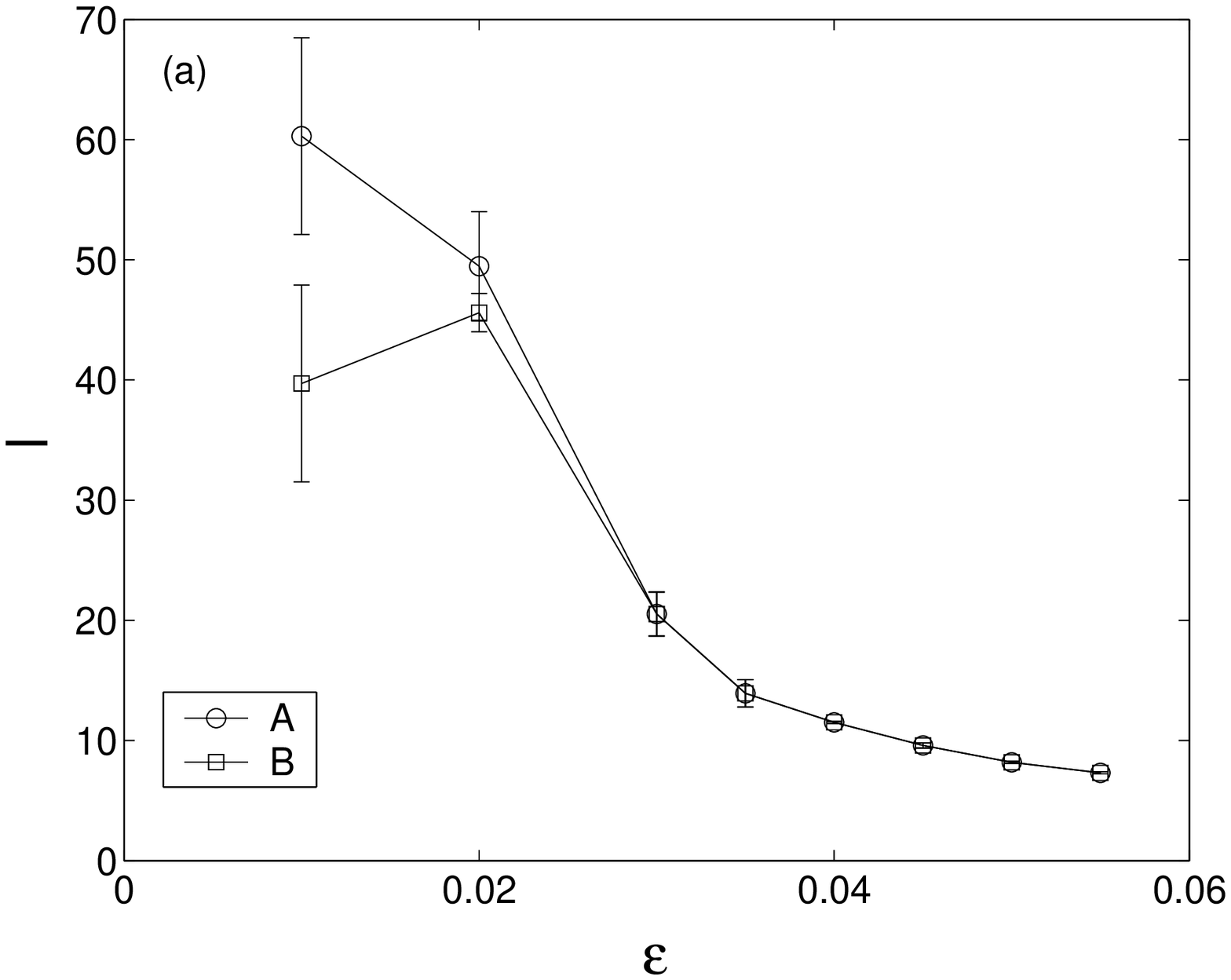}
\end{minipage}
\hfill
\begin{minipage}{0.45 \textwidth}
  \epsfxsize= 0.99\textwidth
  \epsffile{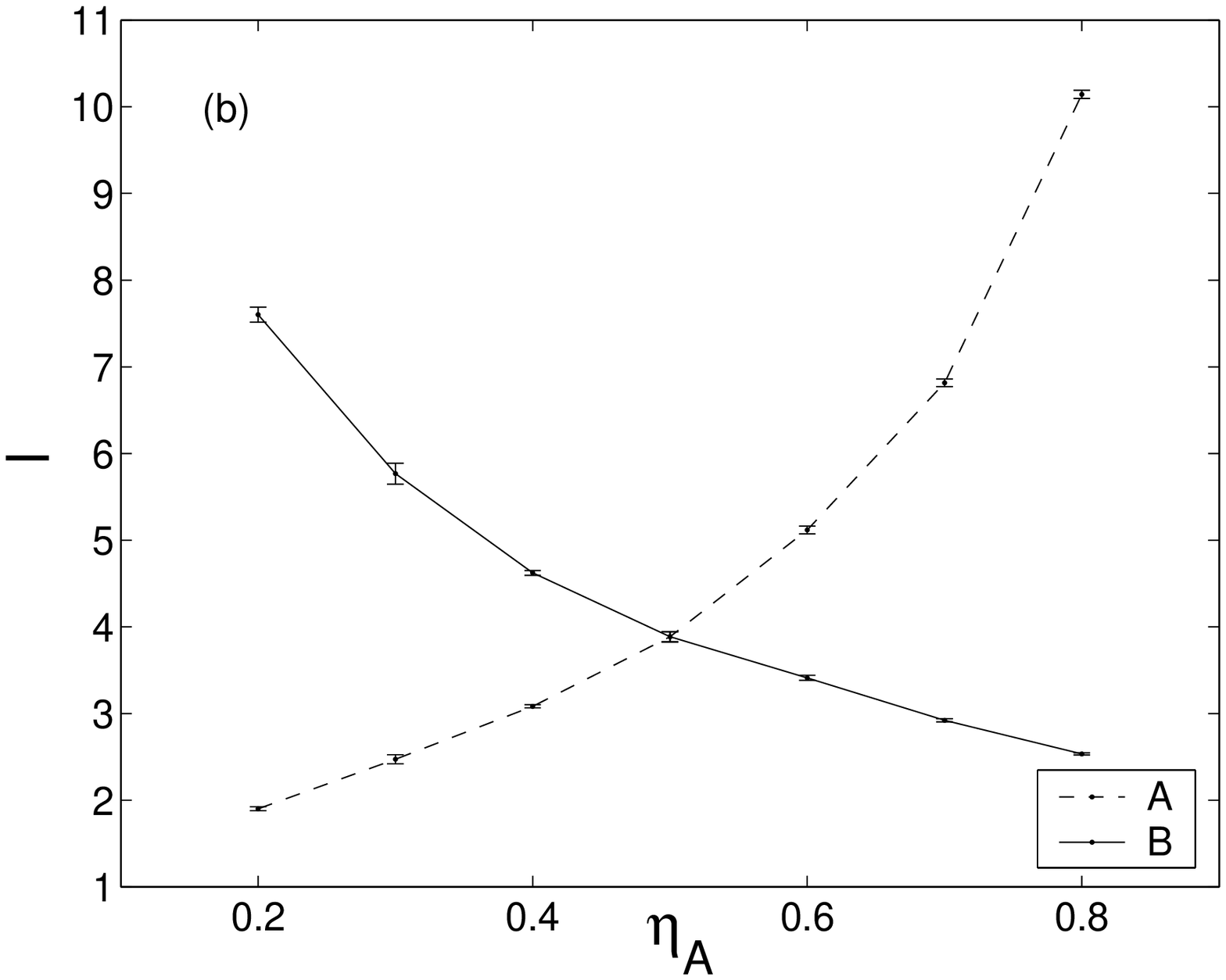}
\end{minipage}
\caption{(a) Stripe width for $E_{AB}=0.6E_{AB}$ and particle concentrations 
$\eta_A=\eta_B=0.5$ as a function of the misfit.
(b) Stripe width for $E_{AB}=0.9E_{AB}$, $\varepsilon=5\%$ as a function of the A particle
 concentration $\eta_A$ ($\eta_B=1-\eta_A$, consequently). Temperature is $T=250K$ for both
pictures. Each value is obtained by averaging over $3$ independent simulation runs. 
The errorbars are given by the standard deviation. }
\label{LEN_W1}
\end{figure}

Figure \ref{LEN_W1}(a) shows the width $l$ of A  and B stripes for $E_{AB}=0.6E_A$ as a function of the 
misfit. Since the concentrations $\eta_A = \eta_B =0.5$ of A, B particles are the same A  and 
B stripes have about the same width, whereas the situation changes completely when $\eta_A \not= \eta_B$.
As figure \ref{LEN_W1}(b) shows for $E_{AB}=0.9E_A$ and $\varepsilon=5\%$ the stripe width increases with 
increasing concentration of the particle type. It is noticeable that the bigger B particles form 
thinner stripes at high B concentration than the smaller A particles at high A concentration.
This is due to the asymmetric pair--potential, which is steeper in compression than in tension and thus 
(compressed) B stripes are a little more restricted in their width than A stripes.
\subsection{Temperature dependence}
We take now a short look at the temperature dependence of the stripe structures. To this end 
simulations for the Lennard--Jones potential are performed for $T=500K$. 
The simulations were halted after the same number of simulation steps as for $T=250K$. 
A comparison of
samples at $T=250K$ and $T=500K$ shows, that in regions of smaller binding energies 
$E_{AB}\leq 0.6E_A$ the higher temperature yields widened stripes. 
\begin{figure}[hbt]
\begin{minipage}{0.24 \textwidth}
  \epsfxsize= 0.99\textwidth
  \epsffile{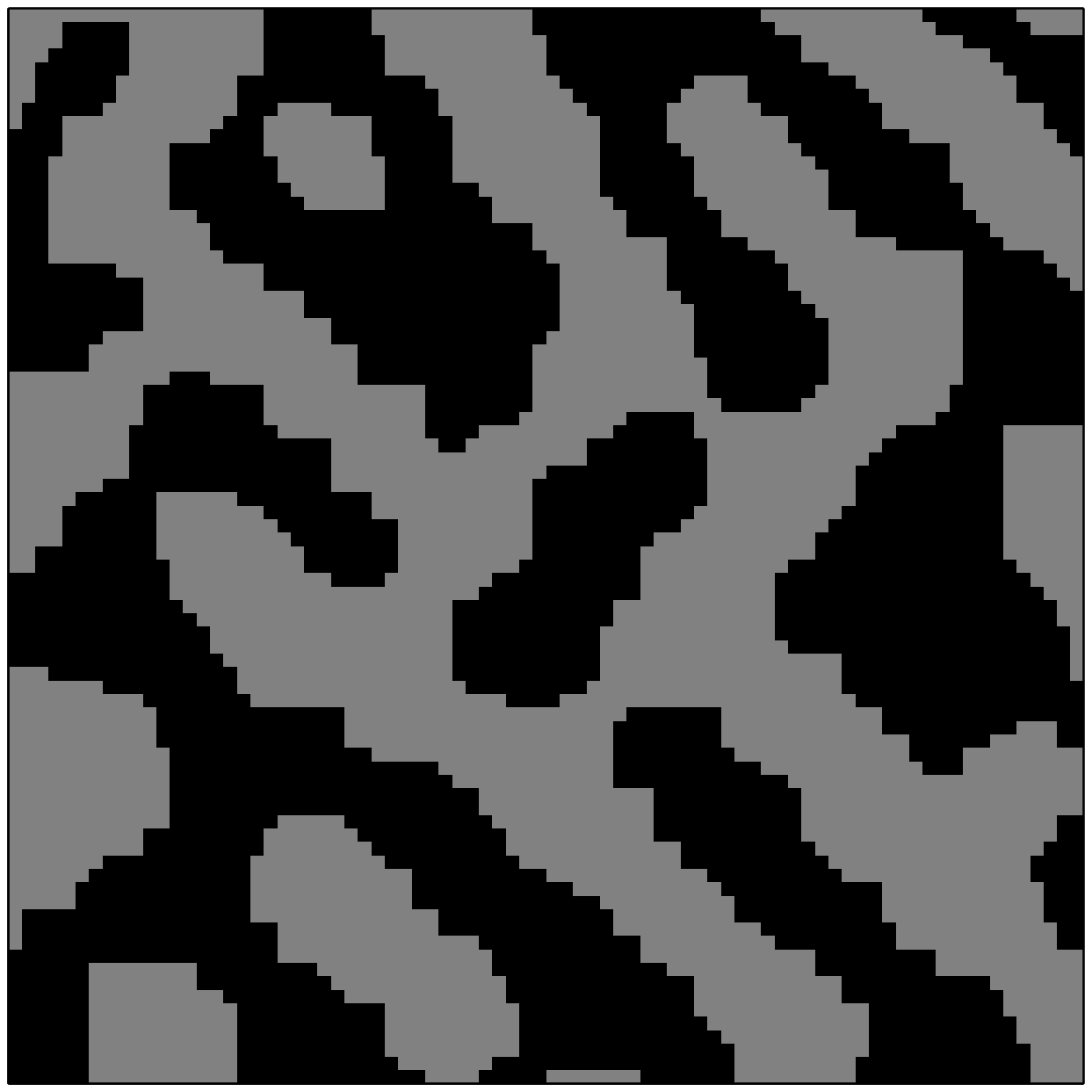}
\end{minipage}
\hfill
\begin{minipage}{0.24 \textwidth}
  \epsfxsize= 0.99\textwidth
  \epsffile{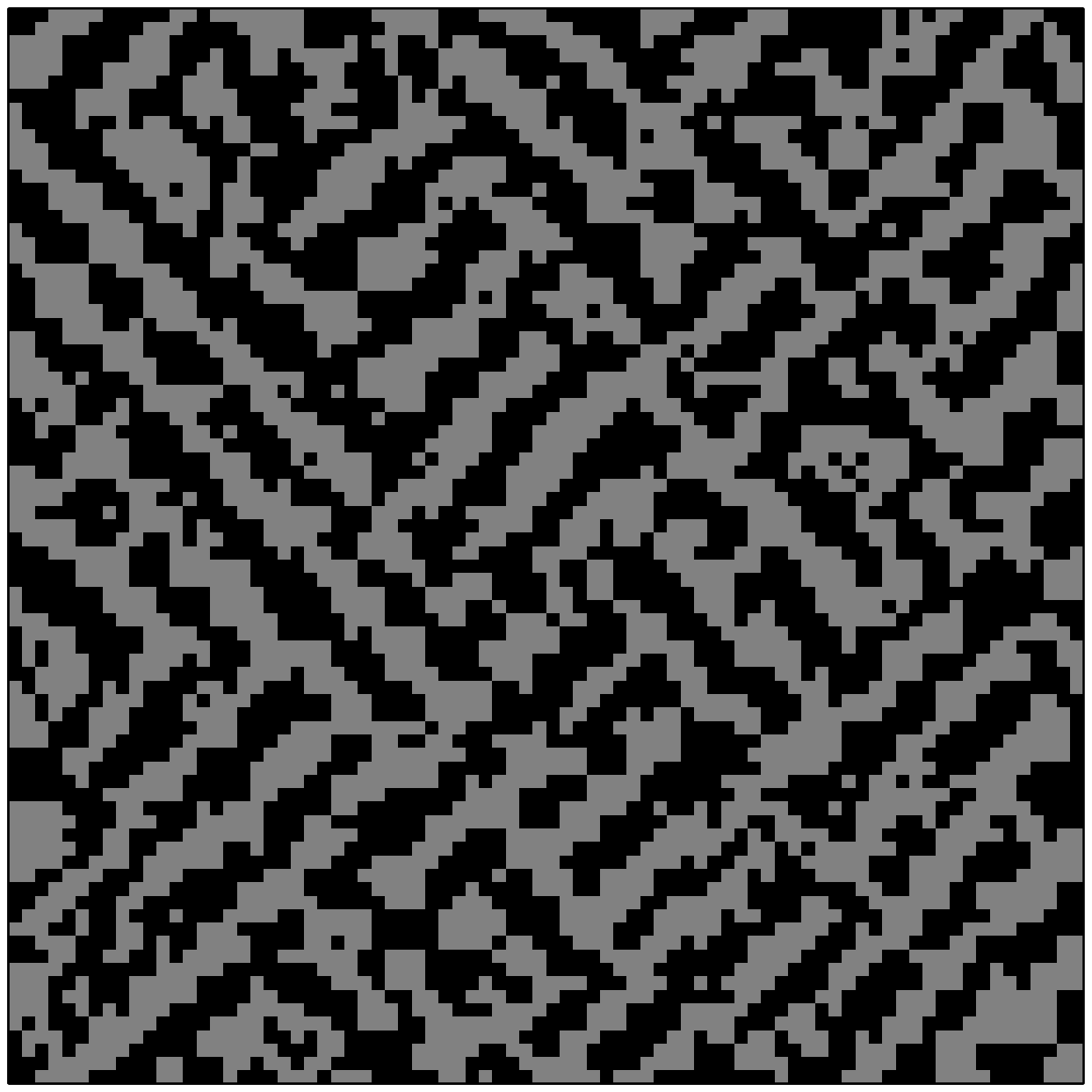}
\end{minipage}
\hfill
\begin{minipage}{0.48 \textwidth}
  \caption{Snapshots ($80 \times 80$ sections) for the Lennard--Jones potential at $T=500K$ at  $\varepsilon=4.5\%$
for $E_{AB}=0.6E_A$ (left panel) and $E_{AB}=0.9E_A$ (right panel).}
\label{GLW_T500_SNAPS}
\end{minipage}
\end{figure}
This is due to the fact that
the system approaches a deeper value of $E_{tot}$ for the higher temperature in shorter times 
(also see fig. \ref{ETOT}), i.e. a smaller number of simulation steps is required as for $T=250K$.
For higher values of $E_{AB}$ the high temperature only results in rather noisy stripes, 
but of about the same width as for $T=250K$.
\begin{figure}[hbt]
\begin{minipage}{0.50 \textwidth}
  \epsfxsize= 0.99\textwidth
  \epsffile{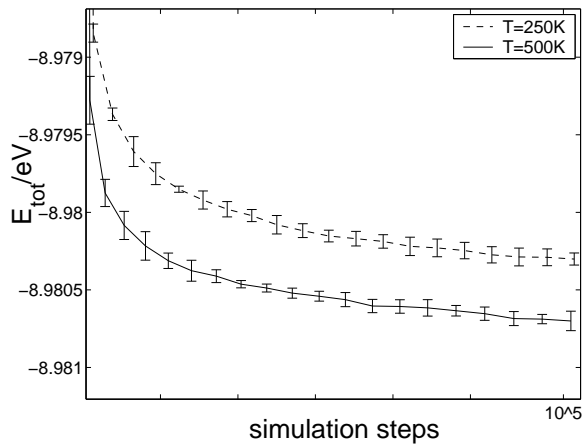}
\end{minipage}
\hfill
\begin{minipage}{0.45 \textwidth}
  \caption{Total energy of the system per particle $E_{tot}$ vs. the number of simulation steps for
$T=250K$ and $T=500K$. $E_{AB}$ is set to $0.6E_A$ and the misfit is $\varepsilon=4.5\%$. Each data point 
is obtained by averaging over $3$ independent simulation runs. The errorbars represent the standard deviation.}
\label{ETOT}
\end{minipage}
\end{figure}
\subsection{Influence of the interaction potential}
With otherwise unchanged parameters the simulations are now performed for the $a=6.0$ Morse potential, which is 
steeper in both - compression and tension - than the Lennard--Jones potential used before
(also see appendices \ref{AP-LJ},\ref{AP-MORSE}). The simulation temperature in set to $T=250K$.
However, as figure \ref{GLW_M6_SNAPS} shows Lennard--Jones and Morse $a=6.0$ potential yield 
quite similar results: again the competition between strain and binding energy causes alternating stripes of
decreasing width with increasing $\varepsilon$. Due to the cubic symmetry the stripes are again
solely aligned in the $<11>$ direction.
\begin{figure}[hbt]
\begin{minipage}{0.24 \textwidth}
  \epsfxsize= 0.99\textwidth
  \epsffile{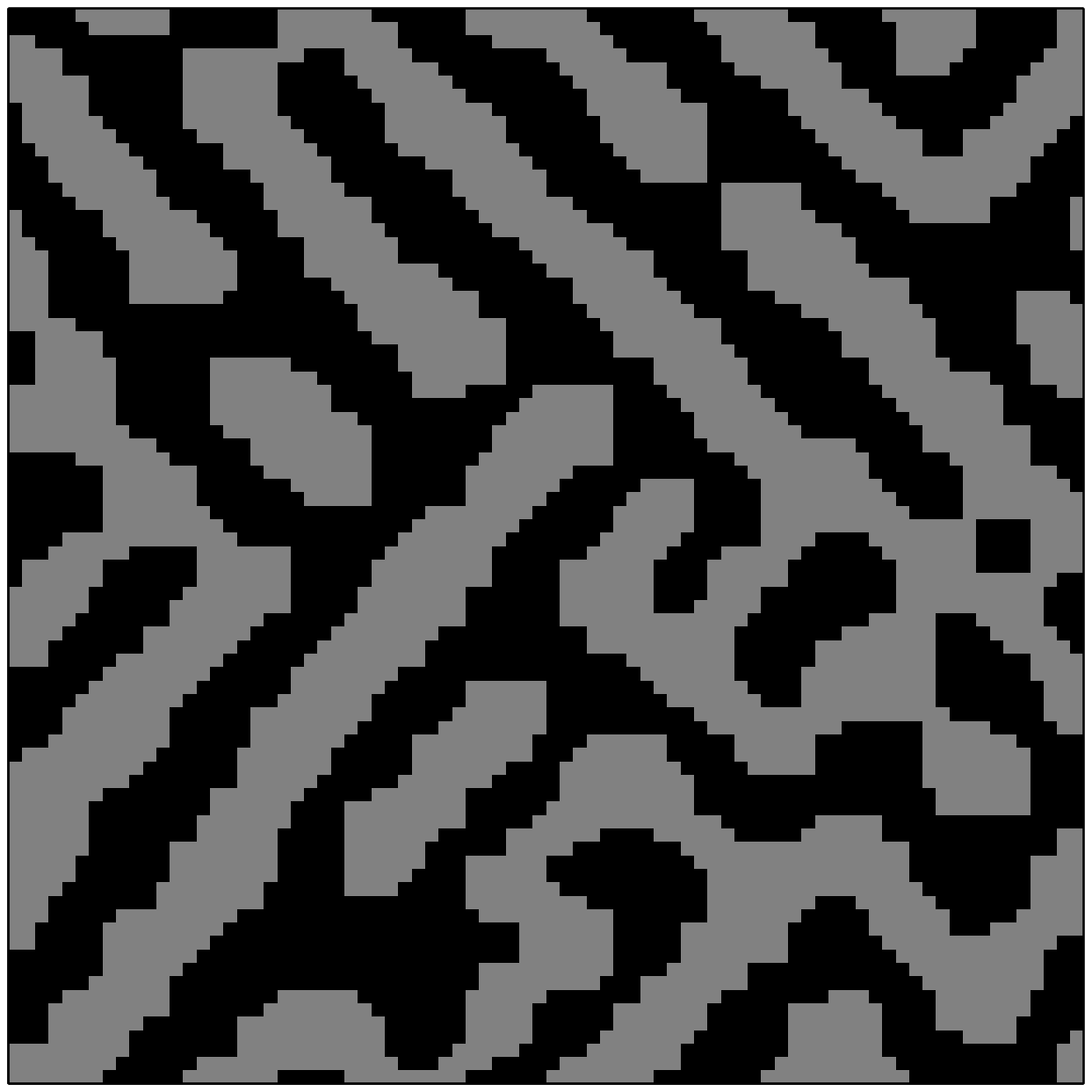}
\end{minipage}
\hfill
\begin{minipage}{0.24 \textwidth}
  \epsfxsize= 0.99\textwidth
  \epsffile{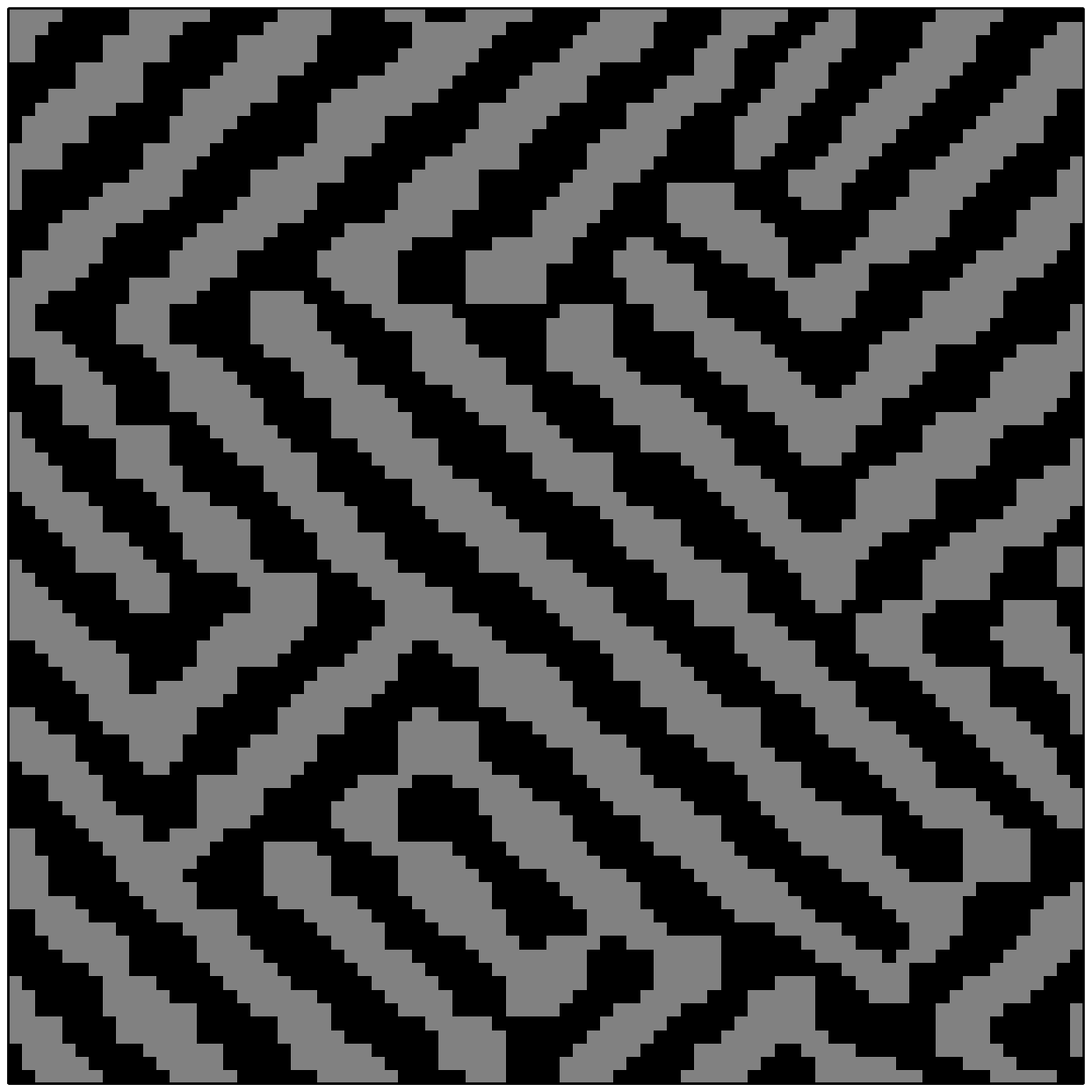}
\end{minipage}
\hfill
\begin{minipage}{0.24 \textwidth}
  \epsfxsize= 0.99\textwidth
  \epsffile{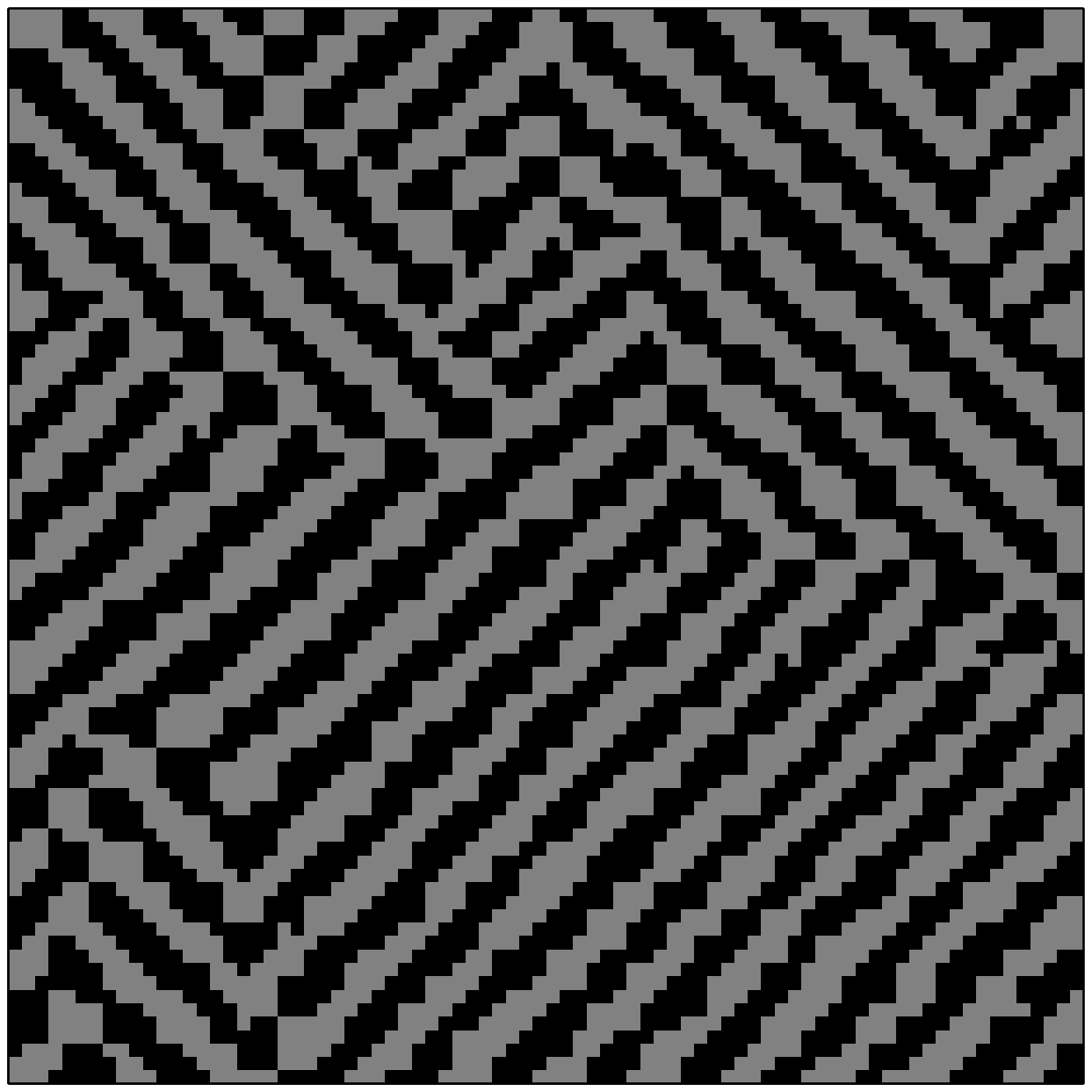} 
\end{minipage}
\hfill
\begin{minipage}{0.24 \textwidth}
  \epsfxsize= 0.99\textwidth
  \epsffile{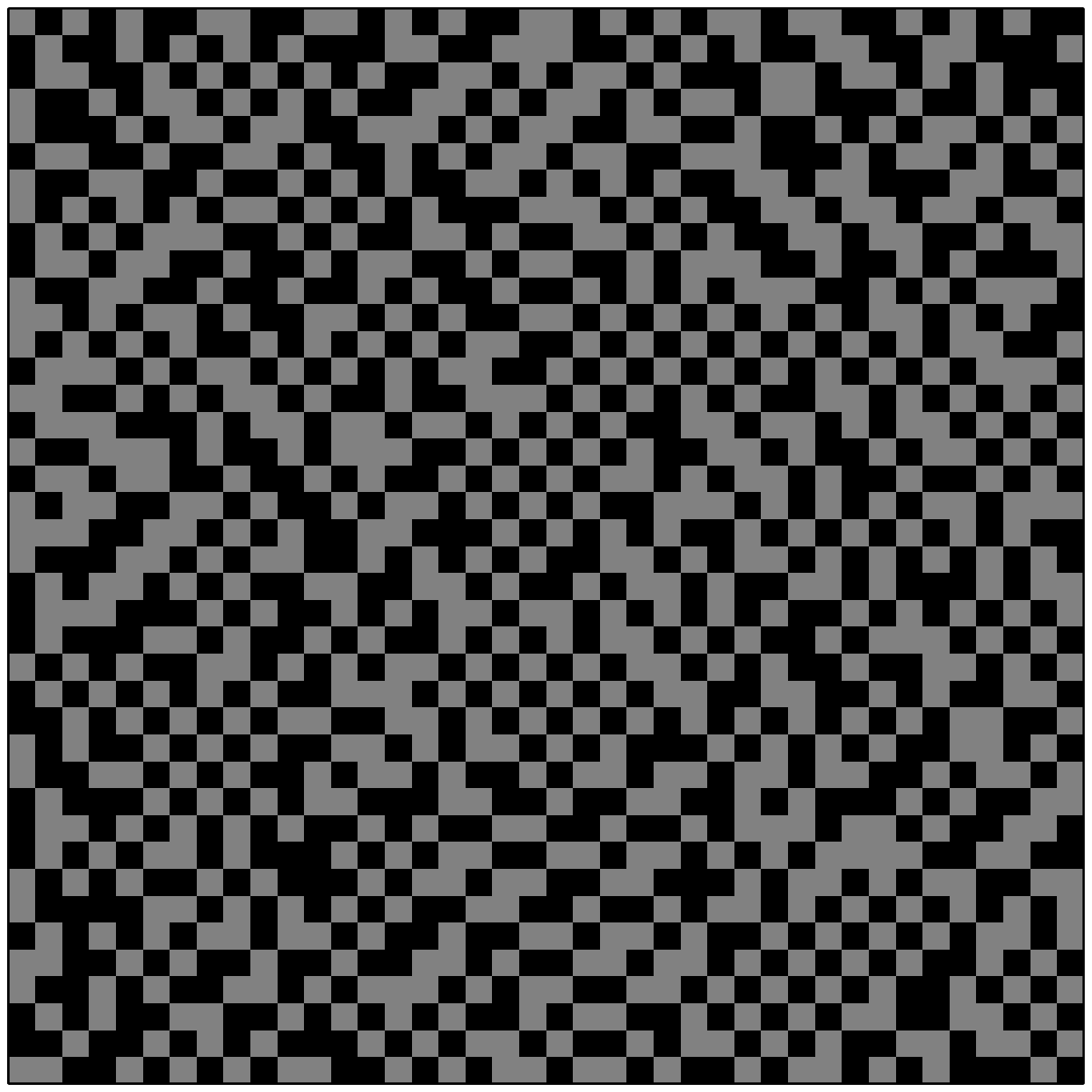} 
\end{minipage}
\caption{Snapshots  for the Morse $a=6.0$ potential at $T=250K$ with 
$E_{AB}=0.6E_A$, $E_{AB}=0.8E_A$, $E_{AB}=0.9E_A$, $E_{AB}=1.0E_A$ (from the left to the right) at 
$\varepsilon=5.5\%$. 
The panel for $E_{AB}=1.0E_A$ shows a $40 \times 40$ sections of the system, the remaining panels show
$80 \times 80$ sections.}
\label{GLW_M6_SNAPS}
\end{figure}

As figure \ref{LEN_M6_VGL} points out the main difference one observes is that for the same misfit and 
$E_{AB}\leq 0.6E_A$ the
stripes for the Morse potential are systematically thicker. Whereas at higher values of $E_{AB}$ the mean 
stripe width is nearly identical for both potentials at a given misfit.
\begin{figure}[hbt]
\centerline{
\begin{minipage}{0.55 \textwidth}
  \epsfxsize= 0.99\textwidth
  \epsffile{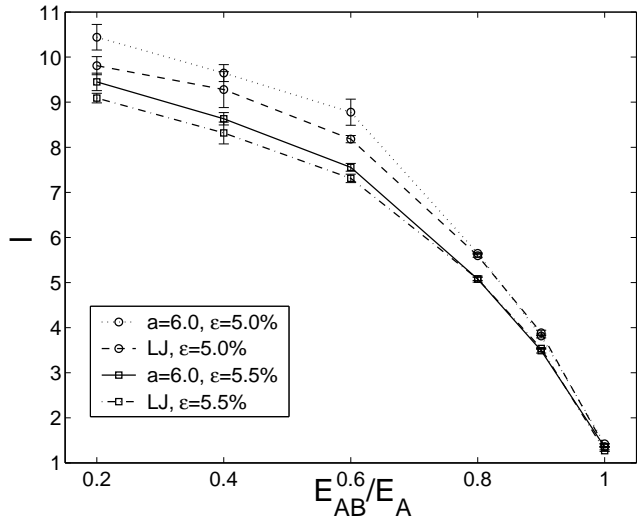}
\end{minipage}
\hfill
\begin{minipage}{0.40 \textwidth}
\caption{Width $l$ of B stripes as a function of $E_{AB}$ for different values of $\varepsilon$.
Each data point is obtained by averaging over $3$ independent simulation runs. 
The errorbars are given by the standard deviation.}
\label{LEN_M6_VGL}
\end{minipage}
}
\end{figure}
However, even at values $E_{AB}\leq 0.6E_{A}$ the deviation are small compared to the influence of the 
particle concentration or the temperature on the stripe width.

In conclusion we learn from the 
equilibrium simulations, that for the heteroepitaxial case ($\varepsilon>0$) 
the state of minimum free energy 
is not given by a complete separation of the different particle types, 
but rather splits the system in several clusters of both species.
A non--vanishing binding energy $E_{AB}>0$ between both particles types together with $\varepsilon>0$ yields  
regular patterns of alternating stripes. 
The width of these stripes is above all controlled by the value of $\varepsilon$ together with 
the binding energy.
The competition between strain and binding energy 
results in structures not unlike the ones observed experimentally. 
However, the question is now if also growth simulations of a 
system, which is governed by these competition, yield a similar 
morphology.
\section{KMC simulations}
In the following we discuss results obtained in off--lattice KMC simulations.
Just like in the equilibrium simulations growth takes place on a $100 \times 100$ substrate 
of $6$ layers height with periodic boundary conditions applied in the $x$-- and $y$--direction.
For all simulation runs the deposition rate for both types of particles is set to 
$0.005ML/s$ resulting in an overall deposition rate of $R_d=0.01ML/s$.
The simulations are halted when half of the substrate is covered with adsorbate particles.
Since we are only interested in the submonolayer regime here, we
disregard second layer nucleation, i.e. particles which are deposited onto other particles will be ignored. This
implies the existence of a rather high Ehrlich--Schwoebel barrier for diffusion across step edges. Note that for the 
same reason jumps of particles onto others are suppressed.
Unless otherwise mentioned the temperature is set to $T=500K$ in the following.  
The same attempt frequency $\nu_0=10^{12}$Hz is considered for all diffusion processes.

\begin{figure}[h]
\centerline{
\begin{minipage}{0.55 \textwidth}
  \epsfxsize= 0.99\textwidth
  \epsffile{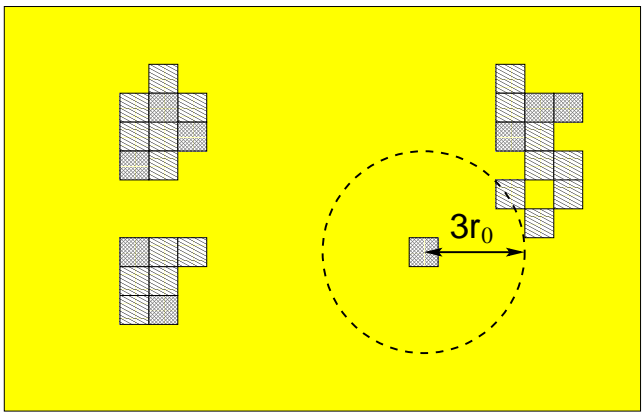}
\end{minipage}
\hfill
\begin{minipage}{0.40 \textwidth}
\caption{Schematic representation of the criterion for barrier calculation. Only if a adsorbate particle 
{\it feels} other particles within a radius of $3r_0$ barriers are calculated. Otherwise a stored value 
for diffusion on plain substrate is selected.}
\label{TRICK}
\end{minipage}
}
\end{figure}
For the interaction between A and B particles we choose 
\begin{equation}
\label{E_AB}
E_{AB}=0.6E_A=0.6E_B=0.10E_S.
\end{equation}
- a value well between the ones investigated during the equilibrium simulations, where rather thick stripes are formed
and the influence of the misfit should be clearly observable.
Furthermore, on the basis of the equilibrium simulation results, we expect to find a marked 
dependence on the choice of the potential for this interaction strength. 
Here again $E_A=E_B$ are given according to equation (\ref{E_S}) 
and $E_{S}$ is chosen from table \ref{TAB_ES} for the different potentials.

Note that this choice of the potential depth yields a higher barrier for edge diffusion 
than for diffusion on plain substrate in our simulations.
However, the barrier for edge diffusion is still smaller than that for detachment from the edge.
So particles attached to an island edge are more likely to diffuse there than to detach.
This is of particular importance since we focus here on phenomena, where edge diffusion is 
supposed to have a strong impact.
Note also that for the cubic lattice (eq. (\ref{CUBIC_6})) diagonal diffusion jumps 
can be neglected since as figure \ref{PES_3d} shows the barrier for a diagonal jump is here represented by a
maximum in the PES.

The realization of the simulation algorithm is as described in chapter \ref{KAP-3}. 
The simulations are carried out for a range of misfits in which even at full coverage 
dislocations are not observed and each particle is allocated at a certain site of the
square lattice. We can therefore take advantage of the lattice based method proposed in chapter \ref{KAP-3}.

Since we simulate submonolayer growth we are able to adopt a further simplification to the method.
Due to the cut--off distance for barrier calculation, at a given misfit the diffusion barrier for an 
adsorbate particle on plain substrate does not change as long as no other adsorbate particle is within 
a radius of $3r_0$ (see also fig. \ref{TRICK}). For this case stored values can be used for the diffusion
barrier of A and B particles on plain substrate, introducing only very little error.  
Additionally to the barrier calculation also the 
local relaxation can be omitted here.  Especially in the early stages of growth this accelerates 
the computations a lot.
Preceding simulation runs with full calculation of all diffusion barriers showed that these fixed barriers 
for diffusion steps on plain substrate have no impact on the results of our simulations.
\begin{figure}[hbt]
\begin{minipage}{0.24 \textwidth}
  \epsfxsize= 0.99\textwidth
  \epsffile{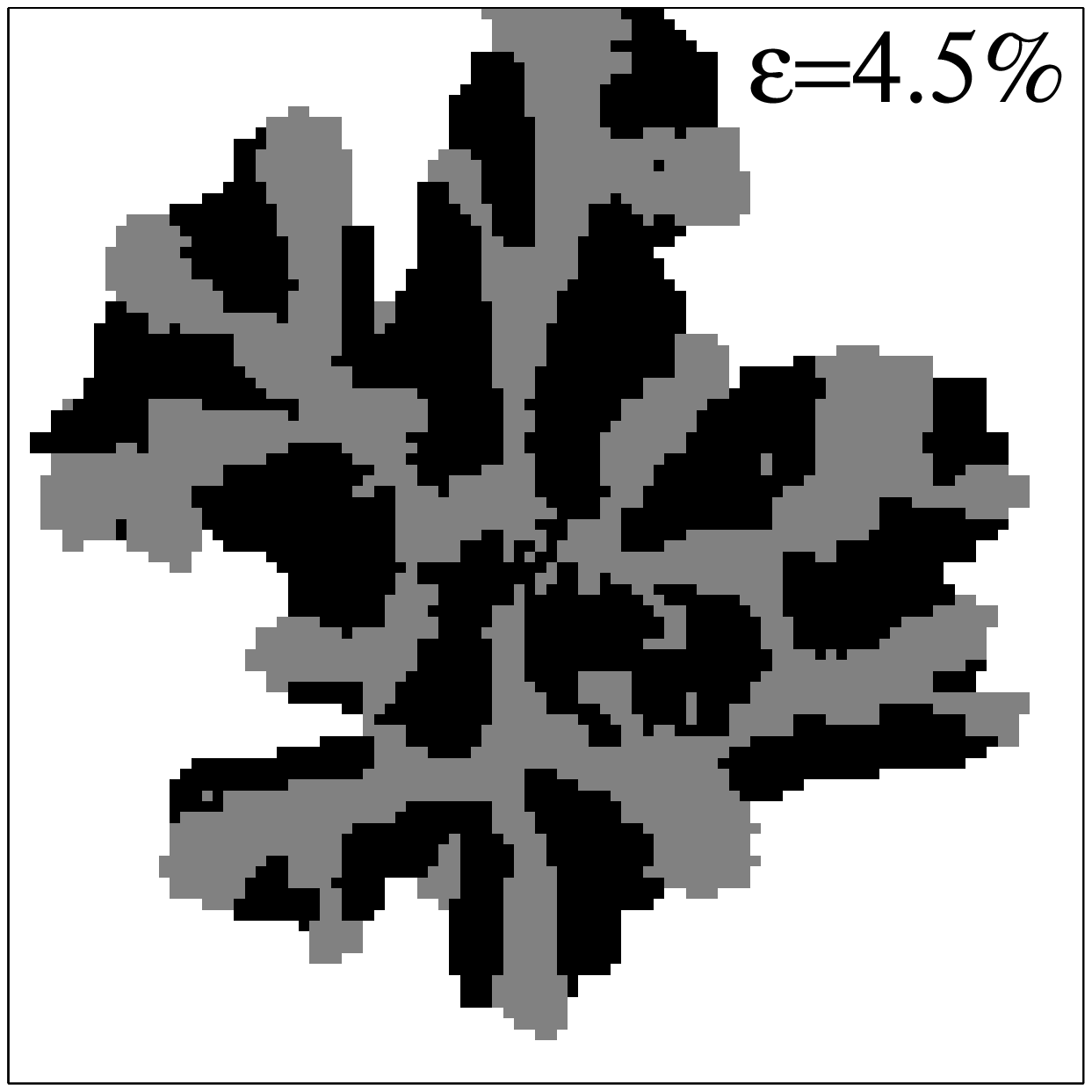} 
\end{minipage}
\hfill
\begin{minipage}{0.24 \textwidth}
  \epsfxsize= 0.99\textwidth
  \epsffile{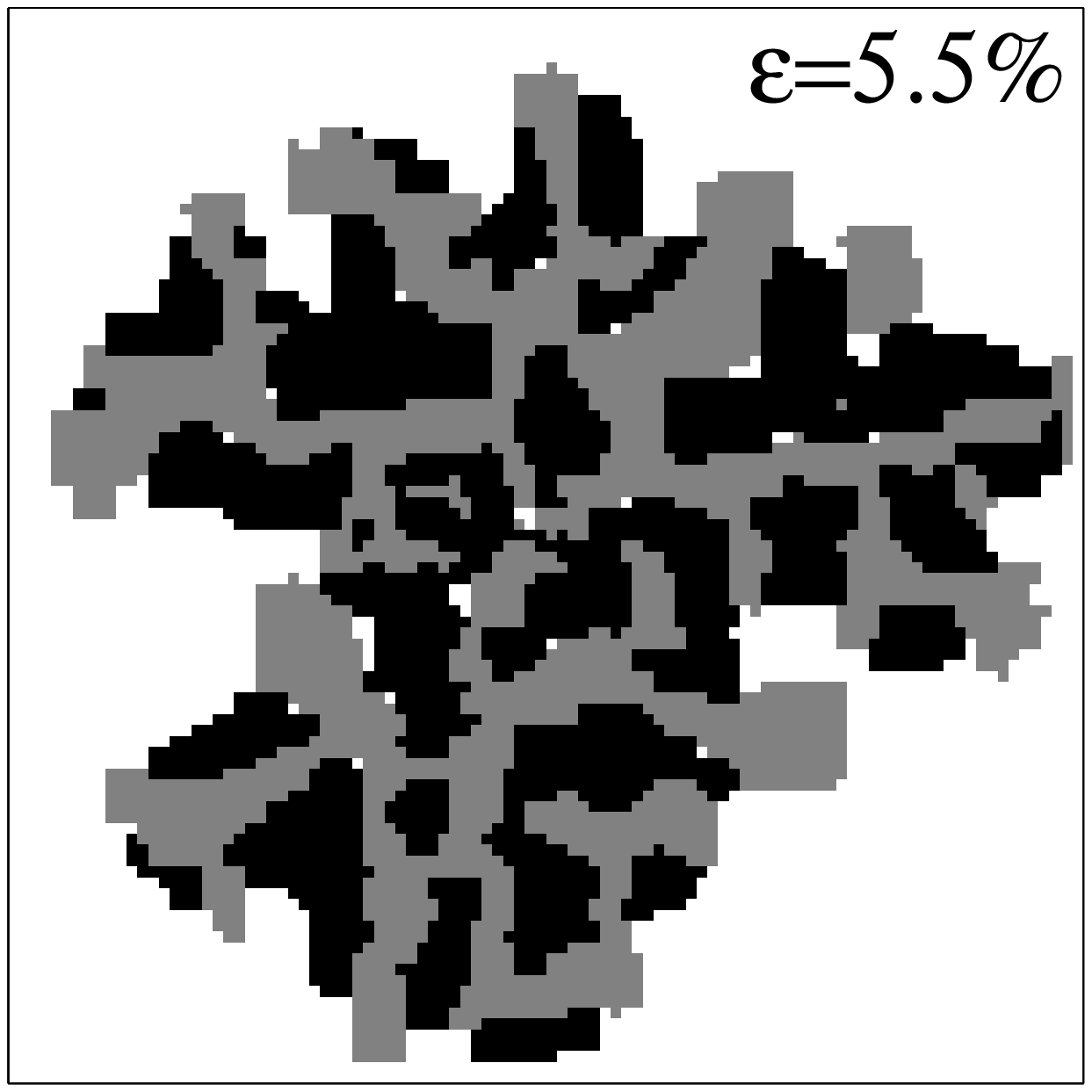}
\end{minipage}
\hfill
\begin{minipage}{0.24 \textwidth}
  \epsfxsize= 0.99\textwidth
  \epsffile{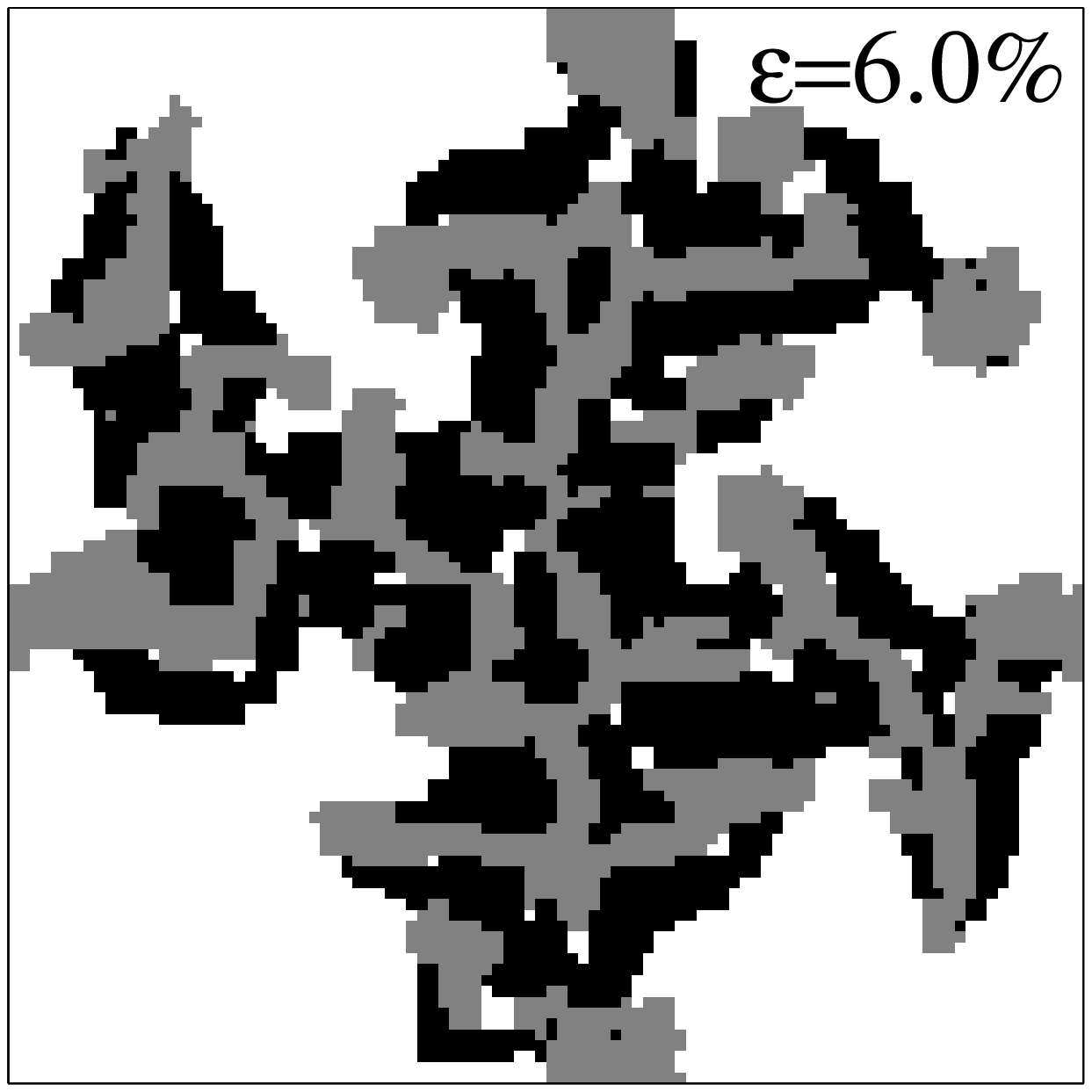}
\end{minipage}
\hfill
\begin{minipage}{0.24 \textwidth}
  \epsfxsize= 0.99\textwidth
  \epsffile{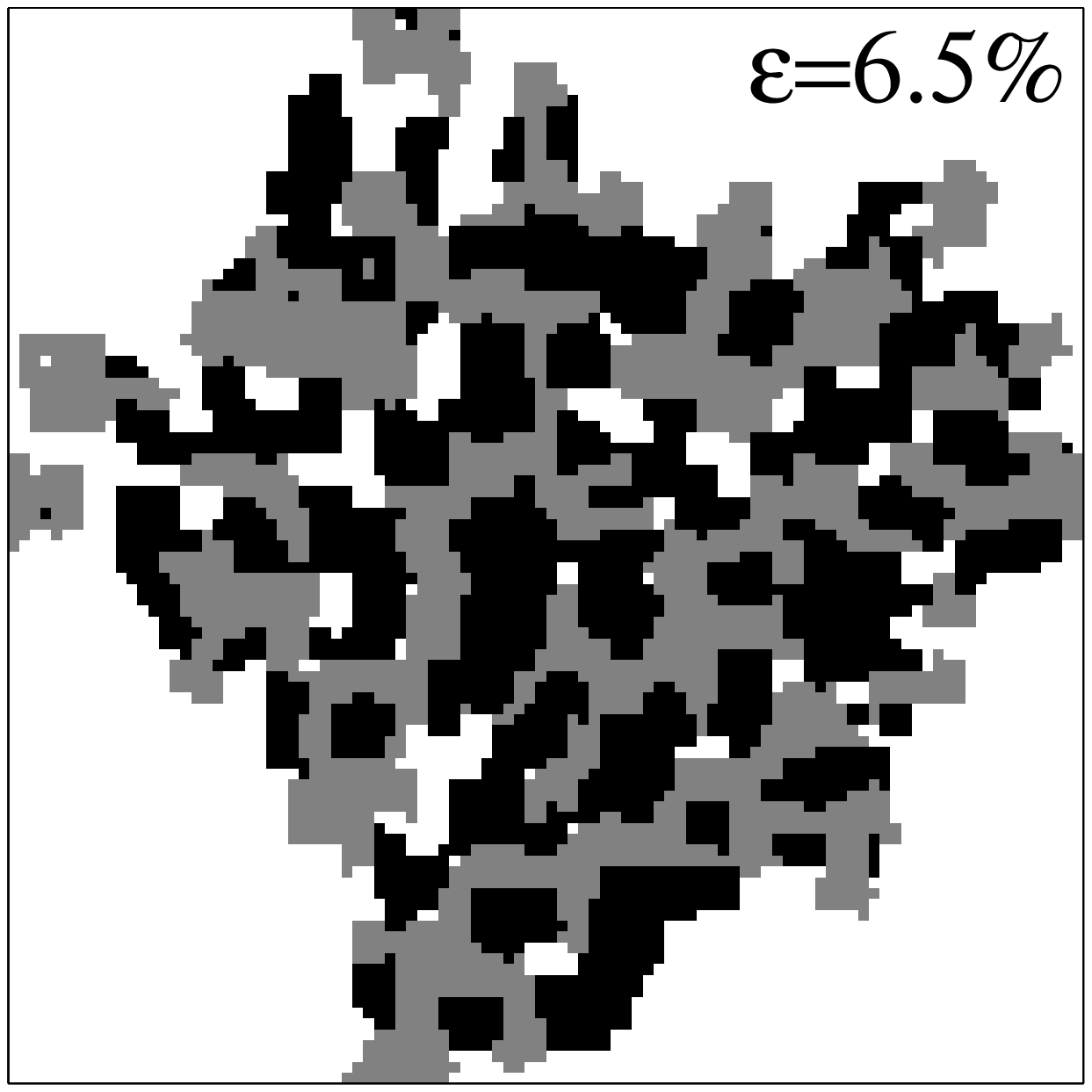}
\end{minipage}

\begin{minipage}{0.24 \textwidth}
  \epsfxsize= 0.99\textwidth
  \epsffile{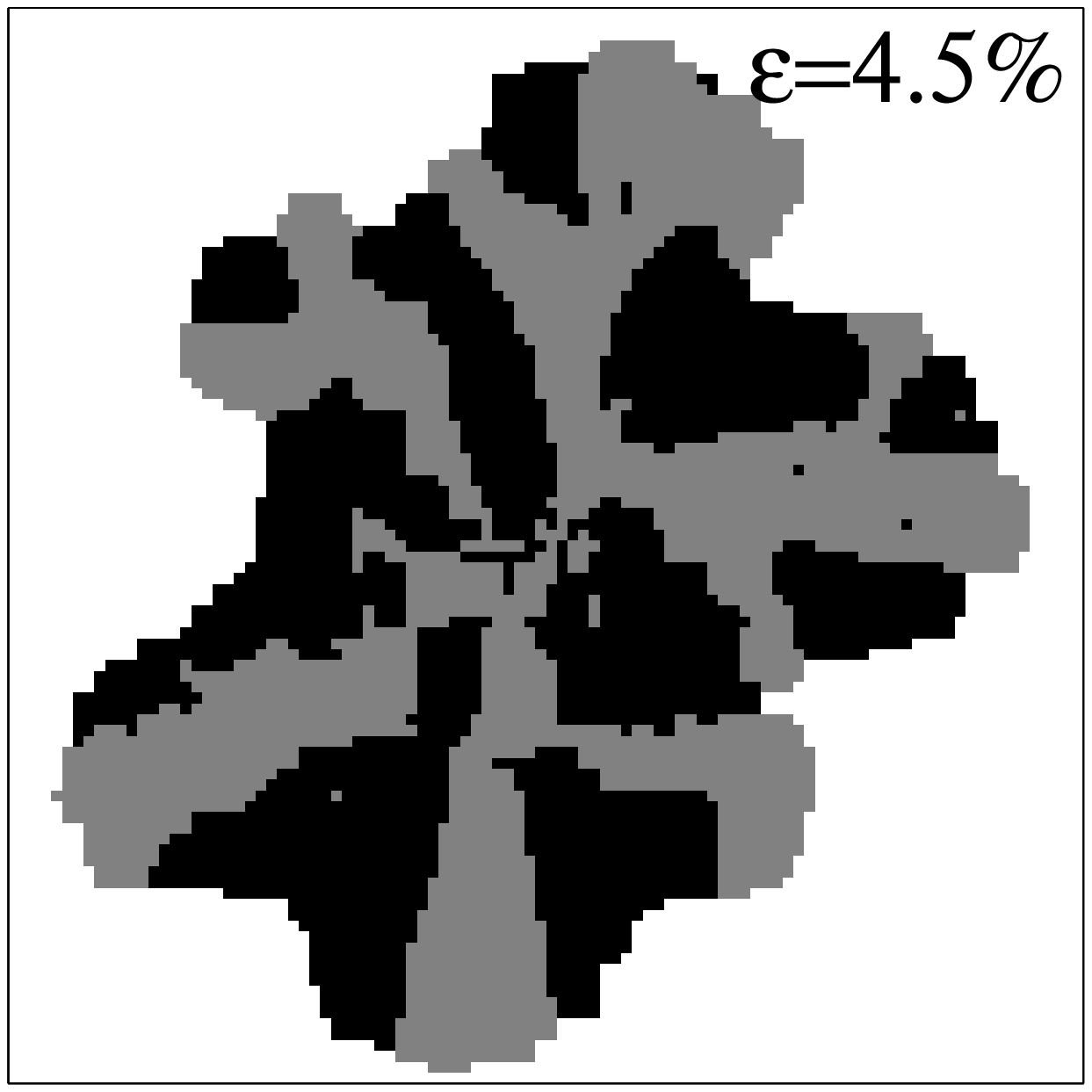}
\end{minipage}
\hfill
\begin{minipage}{0.24 \textwidth}
  \epsfxsize= 0.99\textwidth
  \epsffile{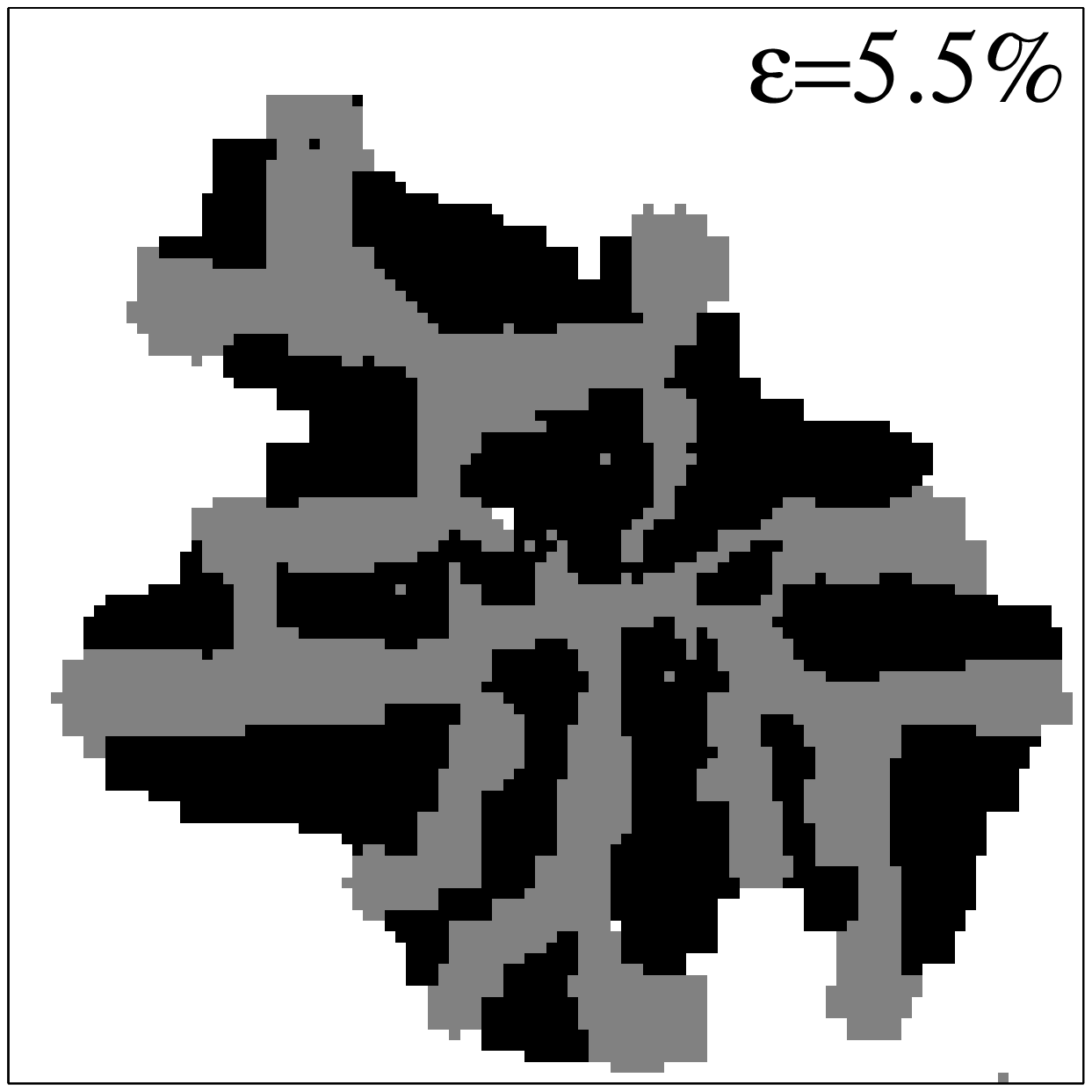}
\end{minipage}
\hfill
\begin{minipage}{0.24 \textwidth}
  \epsfxsize= 0.99\textwidth
  \epsffile{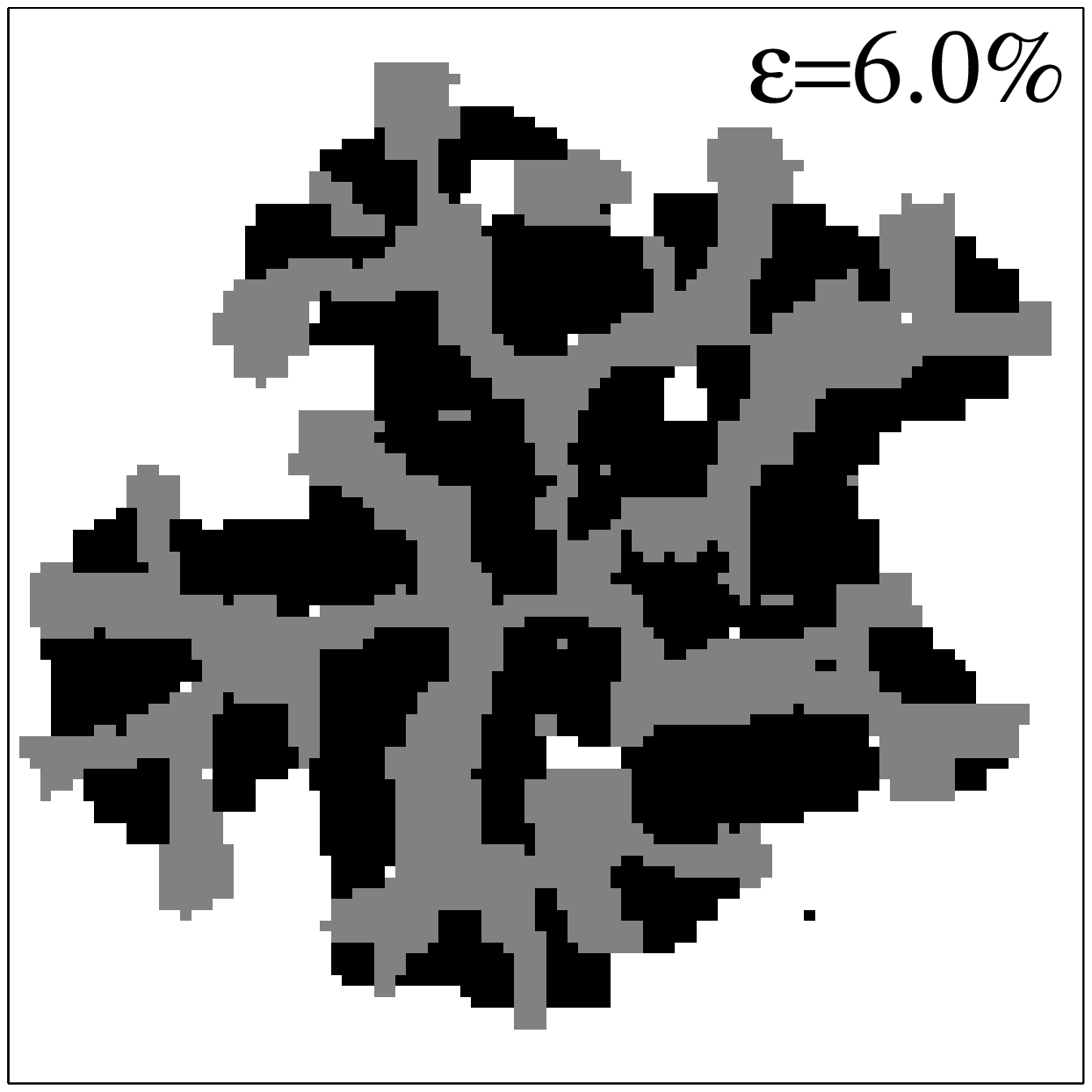}
\end{minipage}
\hfill
\begin{minipage}{0.24 \textwidth}
  \epsfxsize= 0.99\textwidth
  \epsffile{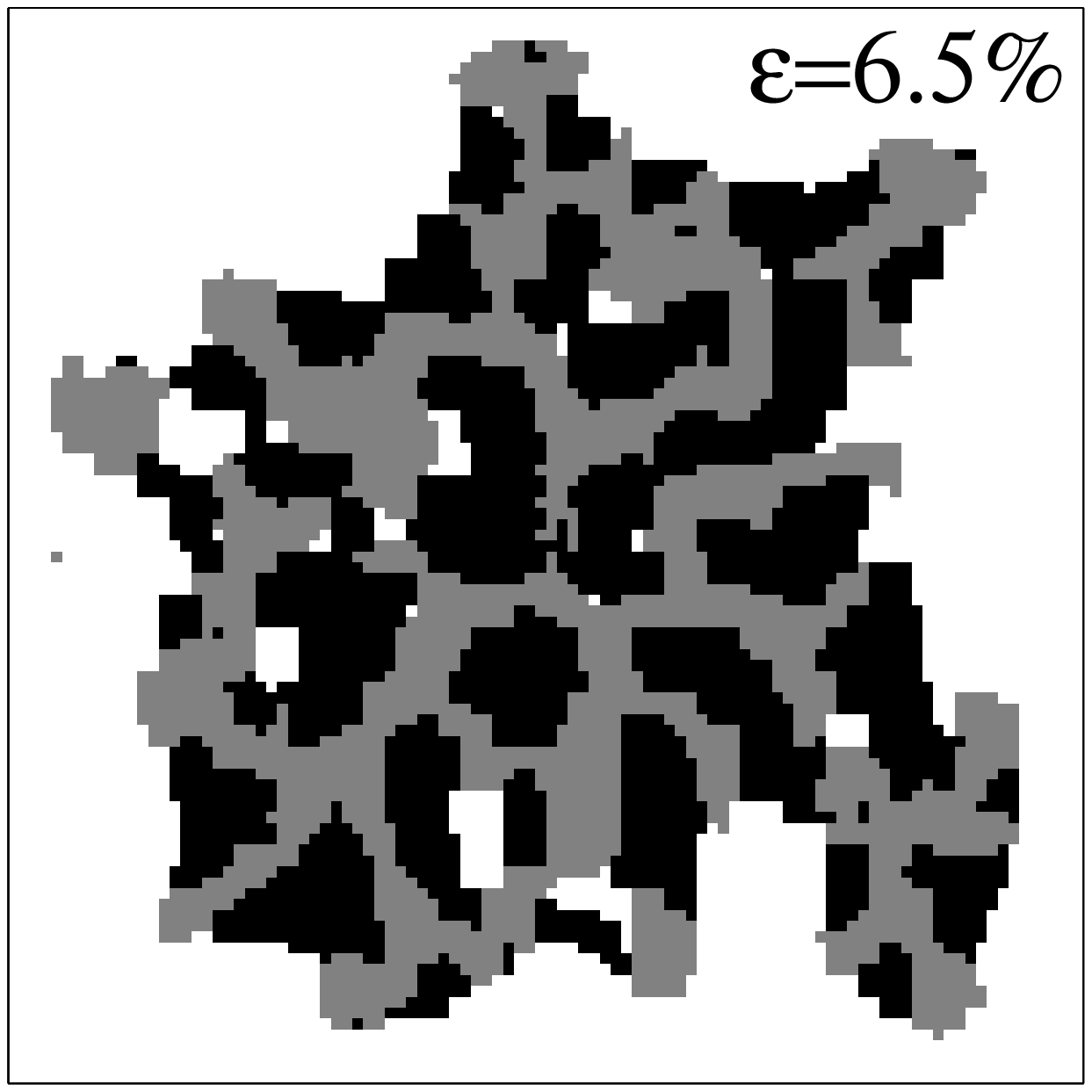}
\end{minipage}

\begin{minipage}{0.24 \textwidth}
  \epsfxsize= 0.99\textwidth
  \epsffile{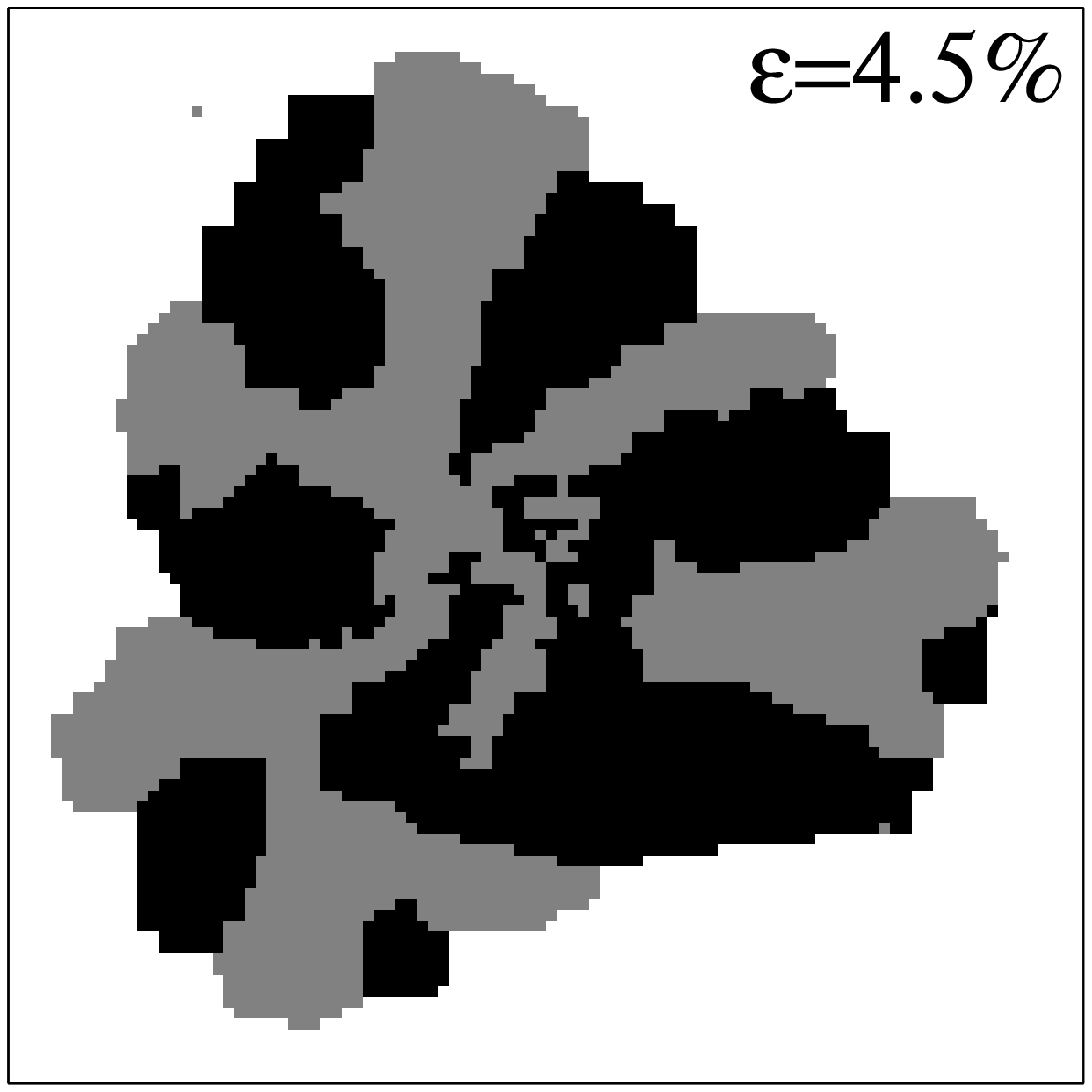}
\end{minipage}
\hfill
\begin{minipage}{0.24 \textwidth}
  \epsfxsize= 0.99\textwidth
  \epsffile{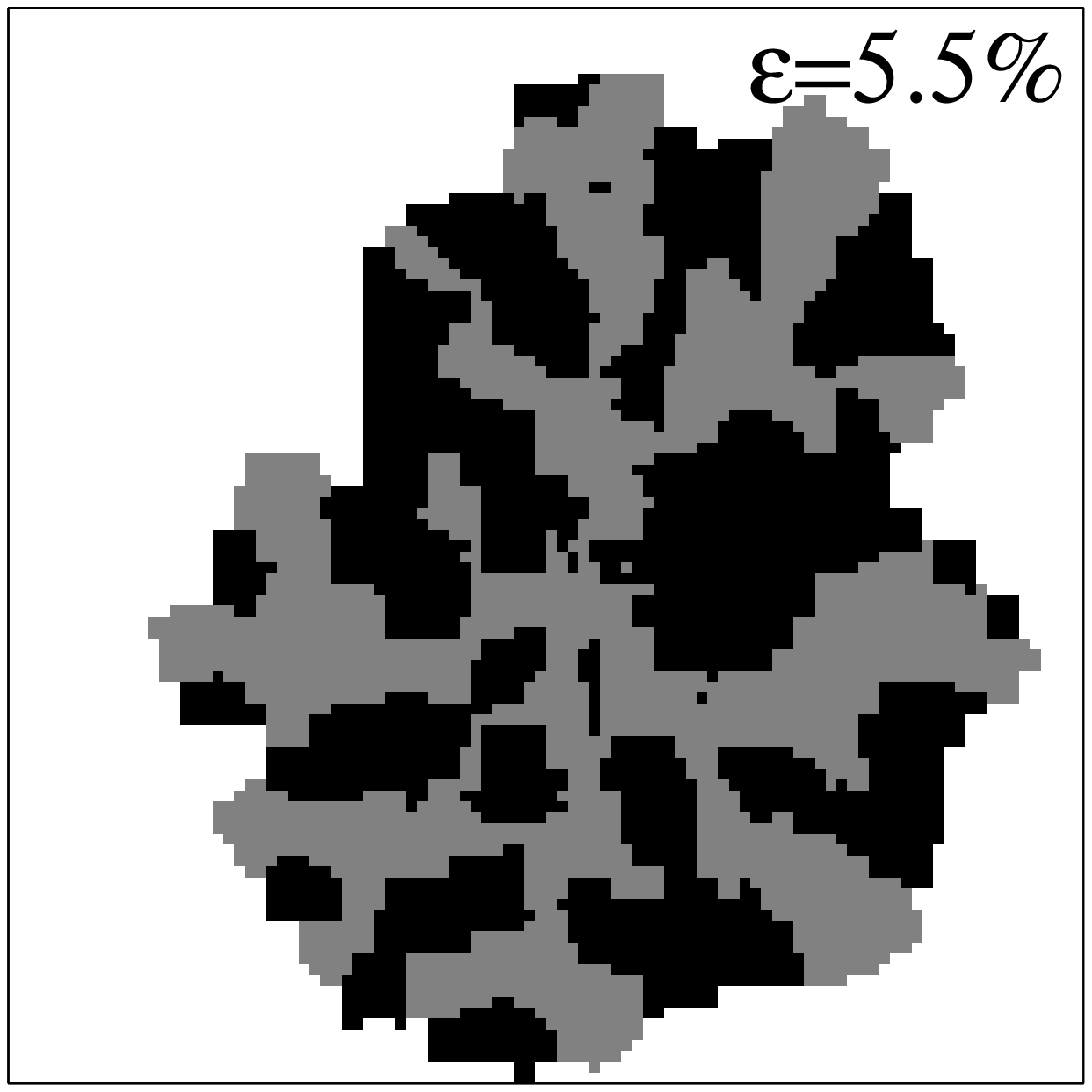}
\end{minipage}
\hfill
\begin{minipage}{0.24 \textwidth}
  \epsfxsize= 0.99\textwidth
  \epsffile{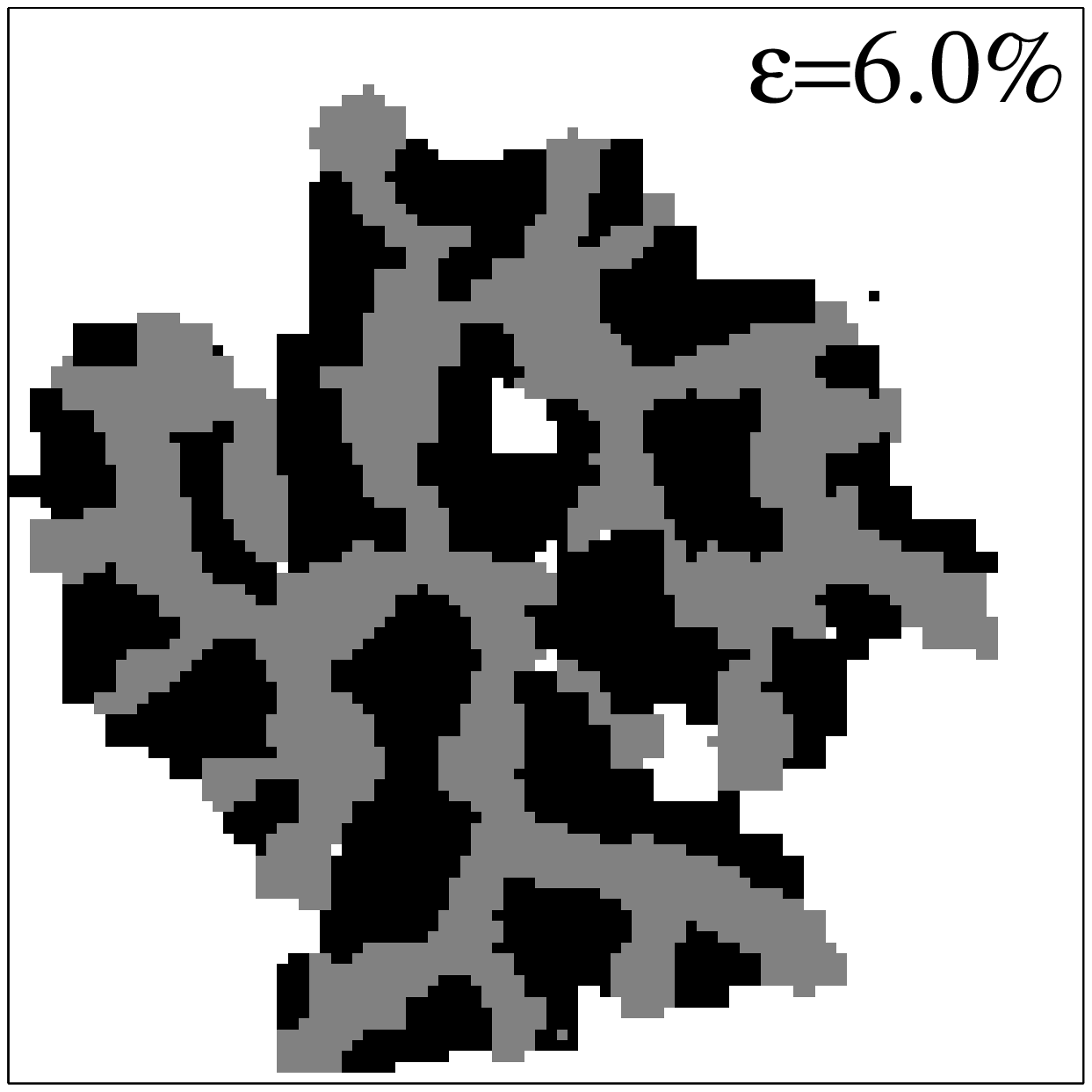}
\end{minipage}
\hfill
\begin{minipage}{0.24 \textwidth}
  \epsfxsize= 0.99\textwidth
  \epsffile{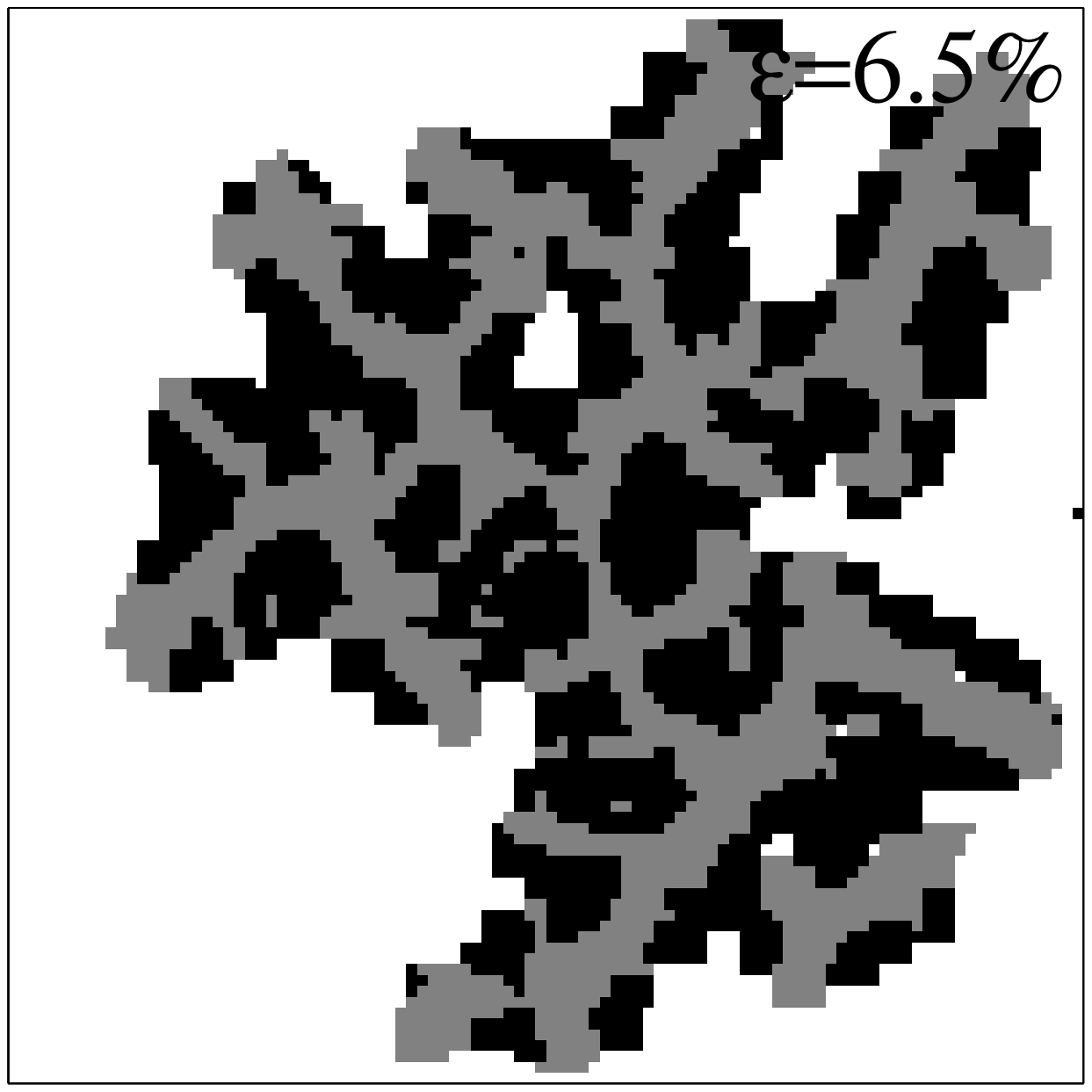}
\end{minipage}
\caption{Snapshots for the Morse $a=5.0$, $a=5.5$ and $6.0$ potential (top down) for different values of $\varepsilon$. The
bigger B particles are shown in light gray.}
\label{OFF_SNAPS}
\end{figure}
\subsection{Influence of misfit and potential}
We present now results on the influence of the value of $\varepsilon$ and the used potential at a temperature 
$T=500K$. Preliminary simulation runs showed that under the same growth conditions both particle types form
rectangular shaped islands if they are deposited alone on the substrate. 
The situation changes completely in the case of co--deposition: 
Figure \ref{OFF_SNAPS} shows snapshots of simulation runs for the Morse $a=5.0$, $a=5.5$ and $a=6.0$
 potential for $4$ different values of $\varepsilon$. 

These structures are exemplary for all simulation results: 
the B particles (shown in light gray) assemble into a few big clusters. With increasing misfit the branches
of these clusters become thinner and of more uniform width. 
The A particles surround these branches without showing a similar shape. 
It is also seen from figure  \ref{OFF_SNAPS} that with increasing misfit the ramification of the structure as a whole 
increases. This is clearly related to the restricted width of the B stripes: a B particle rather attaches to the thin
end of a stripe. 
That means the thinner the stripes the faster is the outwards growth of the light gray branches, leading to an 
increasing ramification of the structure. 

At a given misfit the B branches are the thinner the smaller the value of $a$ in the Morse potential is.
Consequently, at a given misfit the island--ramification is more pronounced for $a=5.0$ than for $a=6.0$.
This is in agreement with the equilibrium simulations where a steeper potential yields thicker stripes.
\begin{figure}
\begin{minipage}{0.45 \textwidth}
  \epsfxsize= 0.99\textwidth
  \epsffile{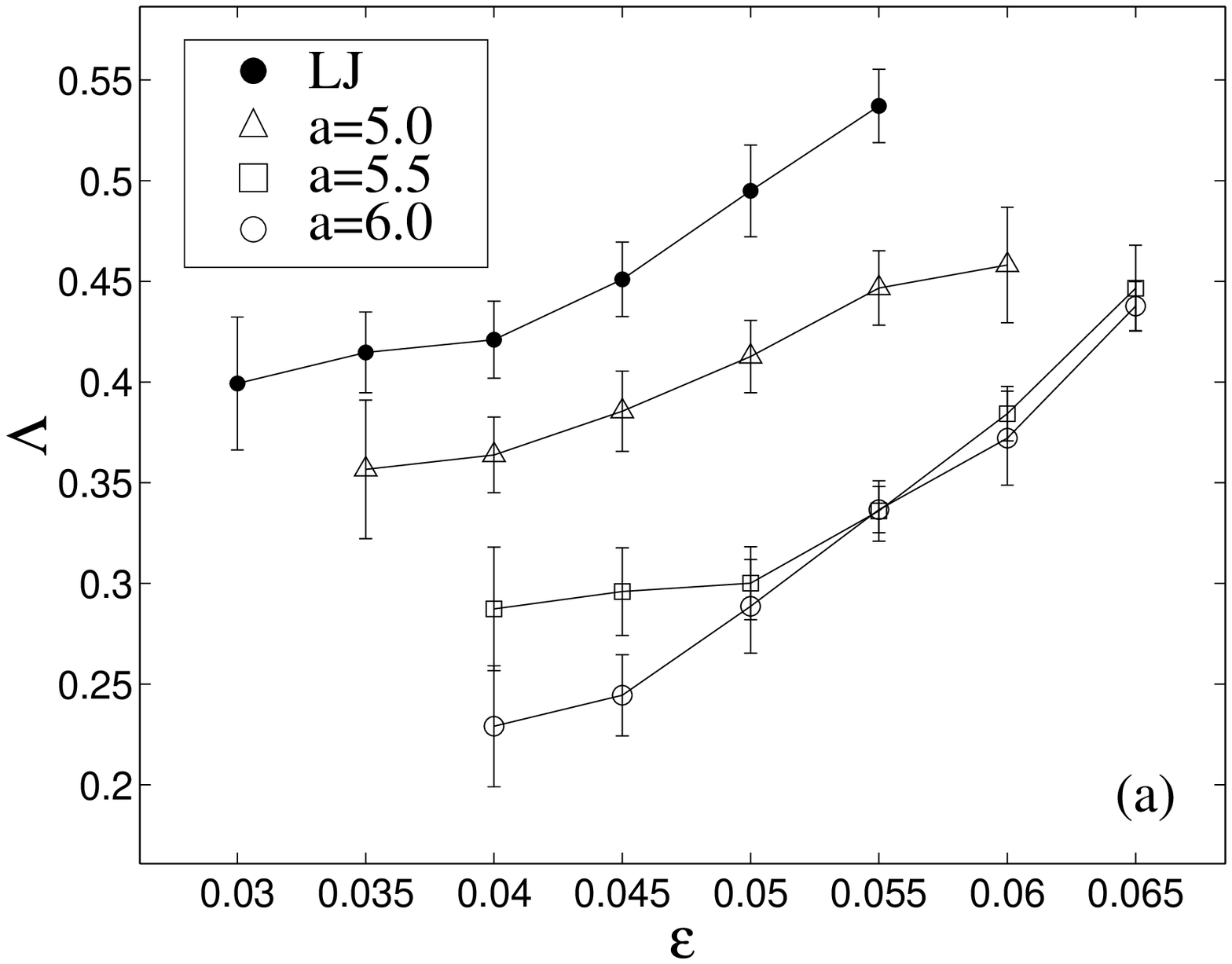}
\end{minipage}
\hfill
\begin{minipage}{0.45 \textwidth}
  \epsfxsize= 0.97\textwidth
  \epsffile{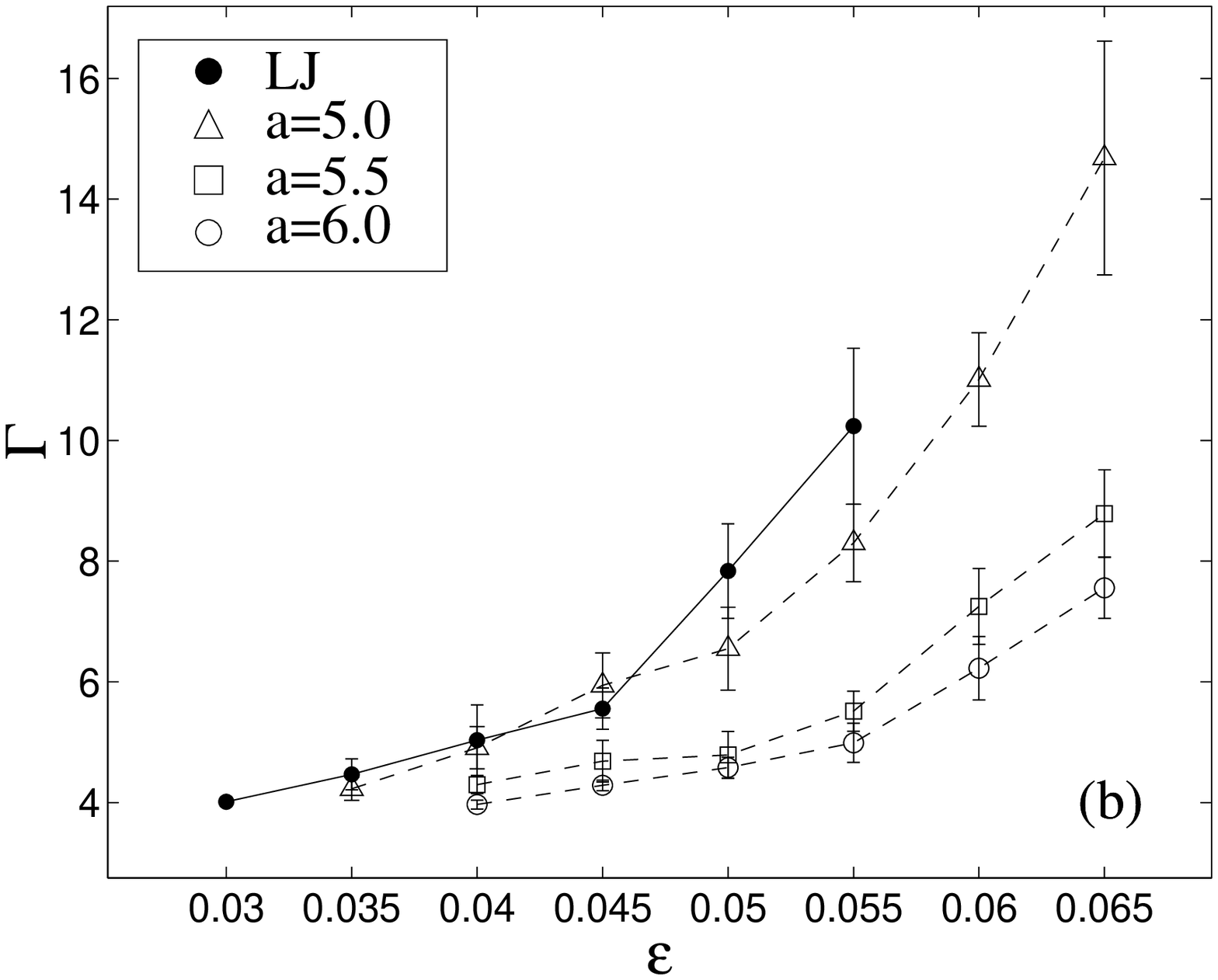}
\end{minipage}
\caption{(a) Ratio $\Lambda$ between perimeter particles and total number of particles in the big B clusters 
for the used potentials.
(b) The number of perimeter particles divided by the square root of deposited particles $\Gamma$ vs. $\varepsilon$.
Each value is obtained by averaging over $10$ independent simulation runs. 
The errorbars are given by the standard deviation.}
\label{QANT}
\end{figure}

In order to quantify the made observations we calculate for each connected cluster of B particles 
the ratio $\Lambda$ between its perimeter length and its volume.
This is done by counting the 
number of perimeter particles together with the total number of particles in the same cluster. 
We take only the {\it backbone} of the structures into account and neglect smaller clusters 
($<700$ particles).
The ratio $\Lambda$ should then give a measure for the
average thickness of this cluster (see fig. \ref{QANT}(a)).
 For example, for a rather thin cluster most of its particles sit at the
edge and therefore $\Lambda$ should be close to 1, whereas with increasing cluster thickness $\Lambda$ should
decrease towards 0.
In addition we measure $\Gamma$ given by the number of particles with less then 
$4$ nearest neighbors divided by the square root of the number of deposited particles (see fig. \ref{QANT}(b)). 
$\Gamma$ provides a measure for the length of the structure's perimeter and therefore the ramification. 
For one perfect quadratic island on the substrate it results to $\Gamma \approx 4$, 
whereas an increasing $\Gamma$ indicates roughening of the island's shape.
The correlation between $\Lambda$ and $\Gamma$ is clearly observable for all used potentials:
$\Lambda$ increases with increasing misfit indicating thinner B clusters. Simultaneously the ramification
increases.  The formation of B branches of well--defined thickness is a common 
phenomenon for the used pair--potentials.
\subsection{Influence of the temperature}
\begin{figure}[hbt]
\begin{minipage}{0.24 \textwidth}
  \epsfxsize= 0.99\textwidth
  \epsffile{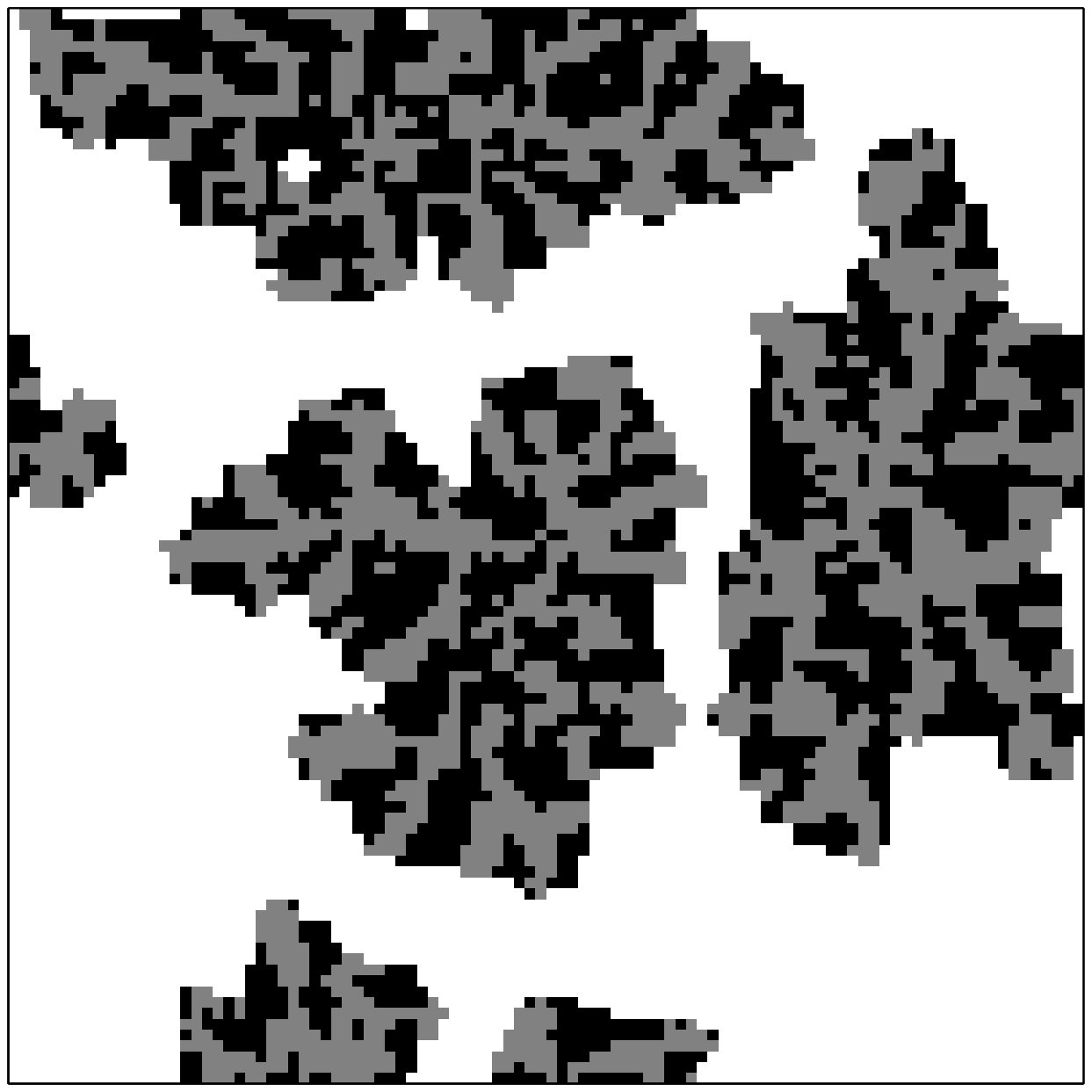}
\end{minipage}
\hfill
\begin{minipage}{0.24 \textwidth}
  \epsfxsize= 0.99\textwidth
  \epsffile{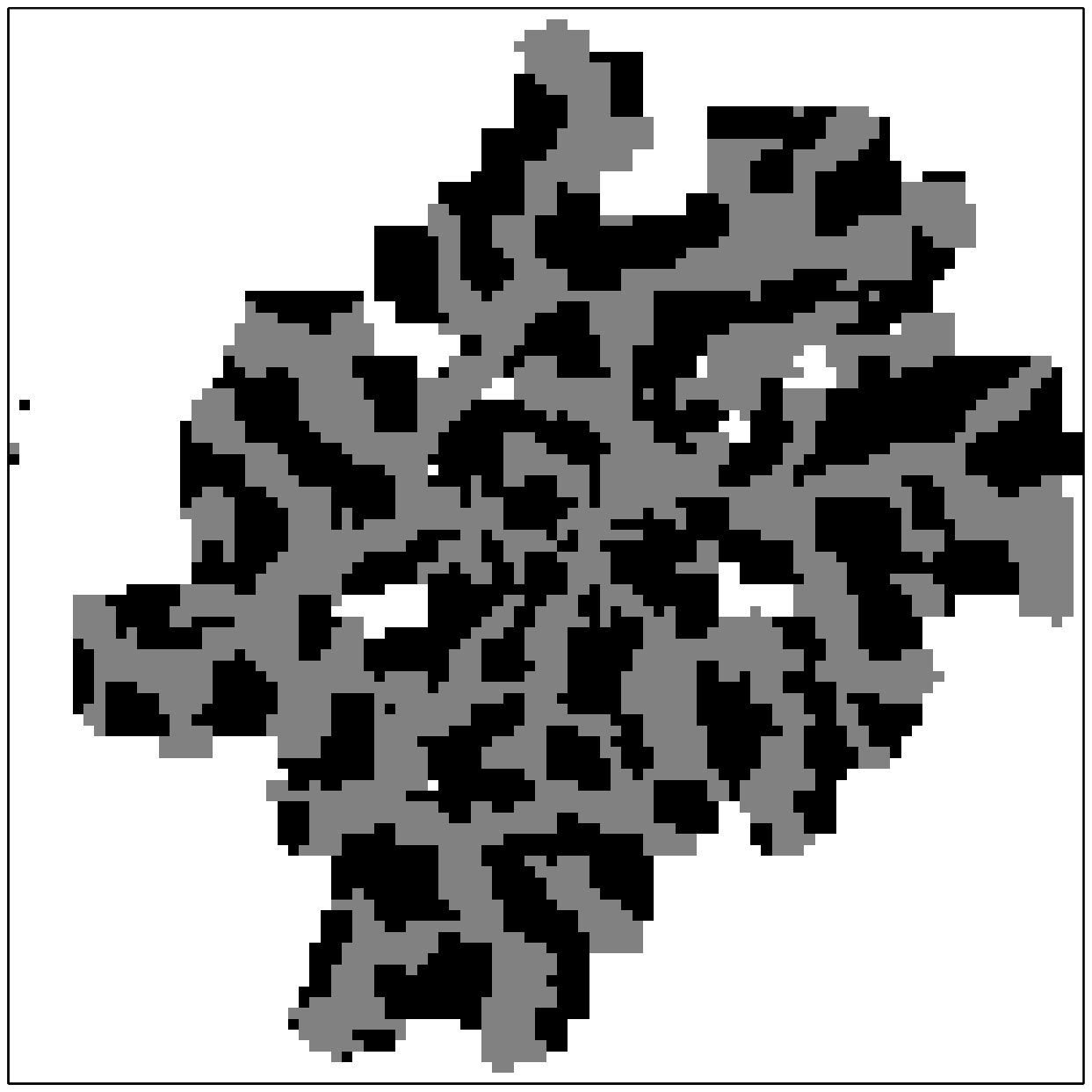}
\end{minipage}
\hfill
\begin{minipage}{0.24 \textwidth}
  \epsfxsize= 0.99\textwidth
  \epsffile{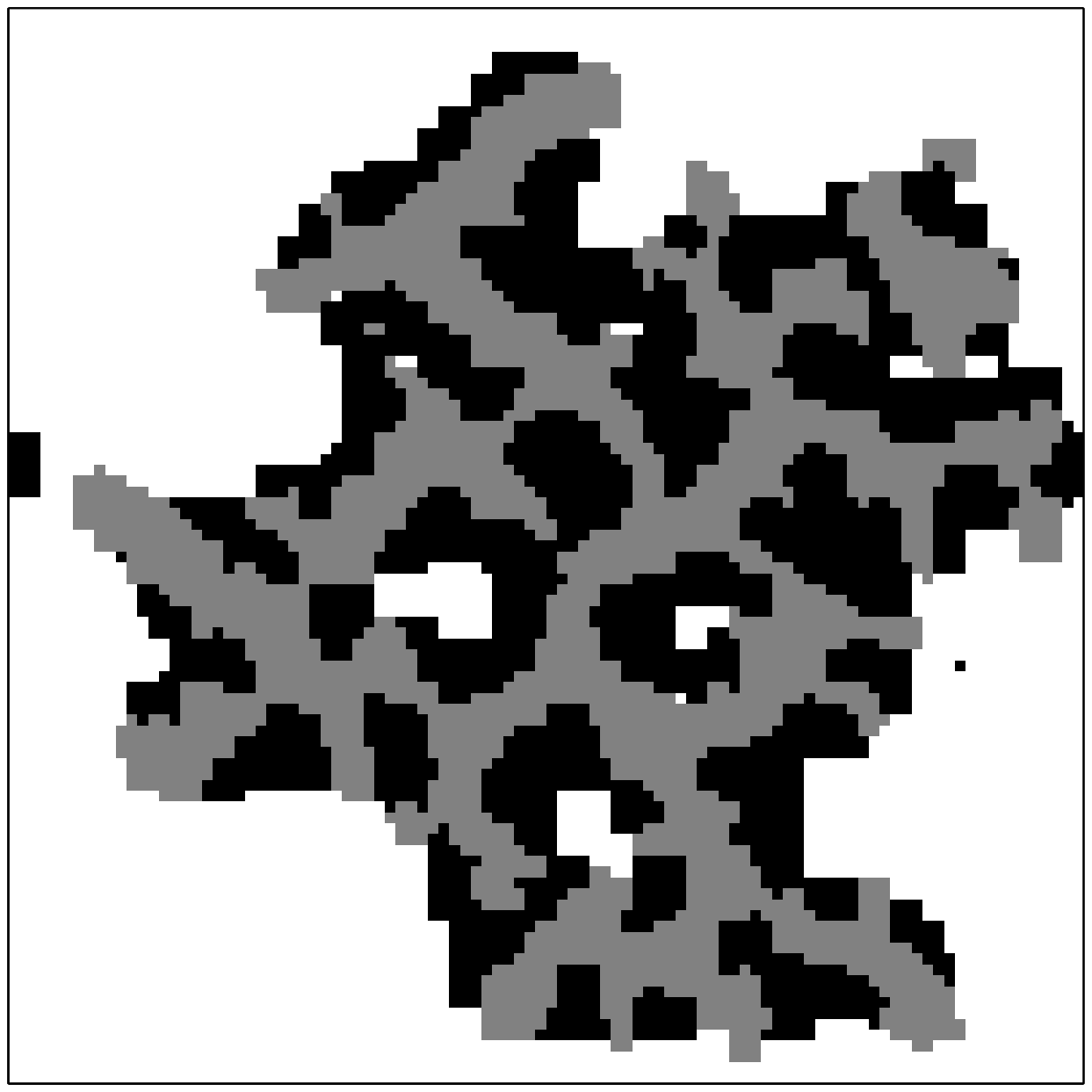}
\end{minipage}
\hfill
\begin{minipage}{0.24 \textwidth}
  \epsfxsize= 0.99\textwidth
  \epsffile{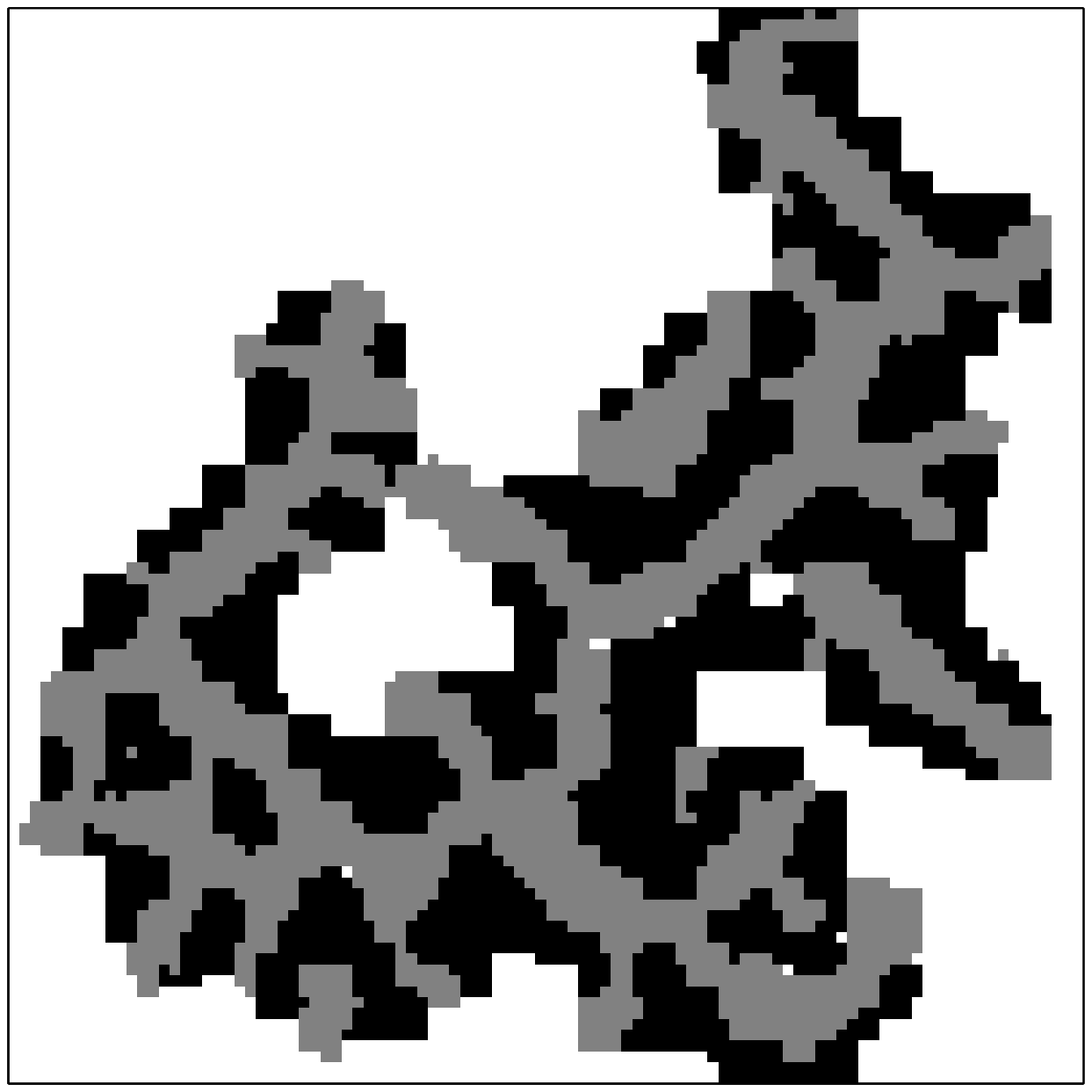}
\end{minipage}
\caption{Snapshots  for the Morse $a=6.0$ potential, $\varepsilon=6.5\%$ at 
$T=400K$, $T=450K$, $T=525K$ and $T=550K$ (from the left to the right).}
\label{M6_T_SNAPS}
\end{figure}
One could now argue that the observed ramification of the islands is due to temperature effects: i.e. 
the used temperature is high enough for the formation of cubic clusters if a single particle type is  
deposited, but enlarged edge diffusion barriers in the co--deposition regime cause dendritic growth.

In order to investigate the temperature dependence of island growth we simulate for the 
Lennard--Jones at $\varepsilon=5.0\%$ and the $a=6.0$ Morse potential 
at $\varepsilon=6.5\%$ growth for different temperatures. 
As shown above for the given parameters strongly ramified islands grow at $T=500K$. Figure \ref{M6_T_SNAPS} shows
some snapshots for $400K\leq T\leq 550K$ for the Morse potential. Similar results are obtained for the Lennard--Jones 
potential.
At $T=400K$ (fig. \ref{M6_T_SNAPS}, left most panel)  two islands emerge due to the reduced diffusion length. Both 
show frayed edges and rather thin and disordered B stripes. With increasing temperature the B stripes become wider and 
more regular in shape, the island edges become smoother. 
Also $\Lambda$ and $\Gamma$ as functions of $T$ (see fig. \ref{LGvsT}(a),(b)) exhibit this behavior. 
It is remarkable here, that the ramification $\Gamma$ does not decreases monotonously 
with increasing temperature (as one would 
expect), 
but passes for both potentials 
through a minimum at $T\approx 475K$. Then $\Gamma$ slowly increases again with increasing $T$. 
This confirms that the observed ramification is not due to the low growth temperature.
\begin{figure}
\begin{minipage}{0.45 \textwidth}
  \epsfxsize= 0.99\textwidth
  \epsffile{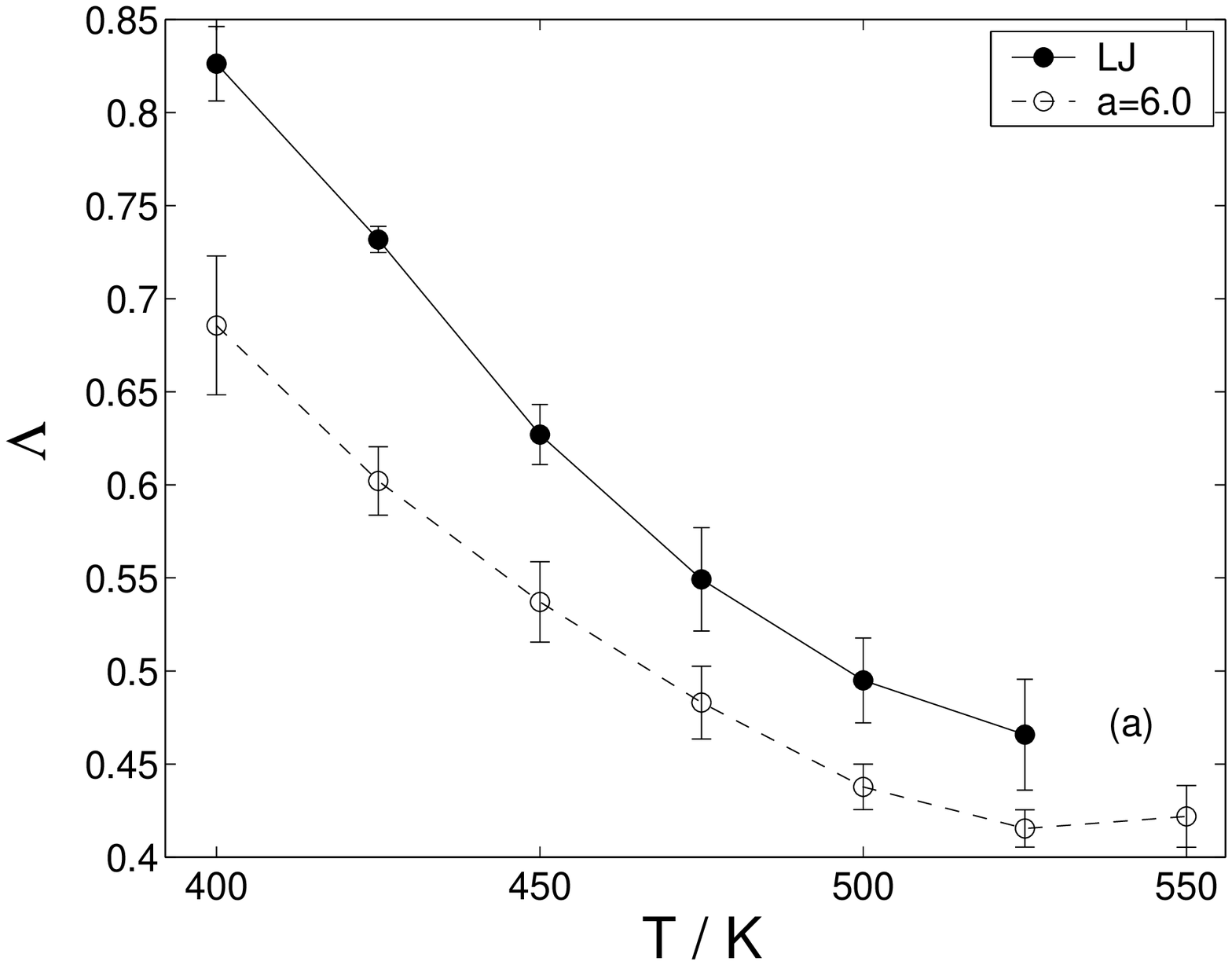}
\end{minipage}
\hfill
\begin{minipage}{0.45 \textwidth}
  \epsfxsize= 0.97\textwidth
  \epsffile{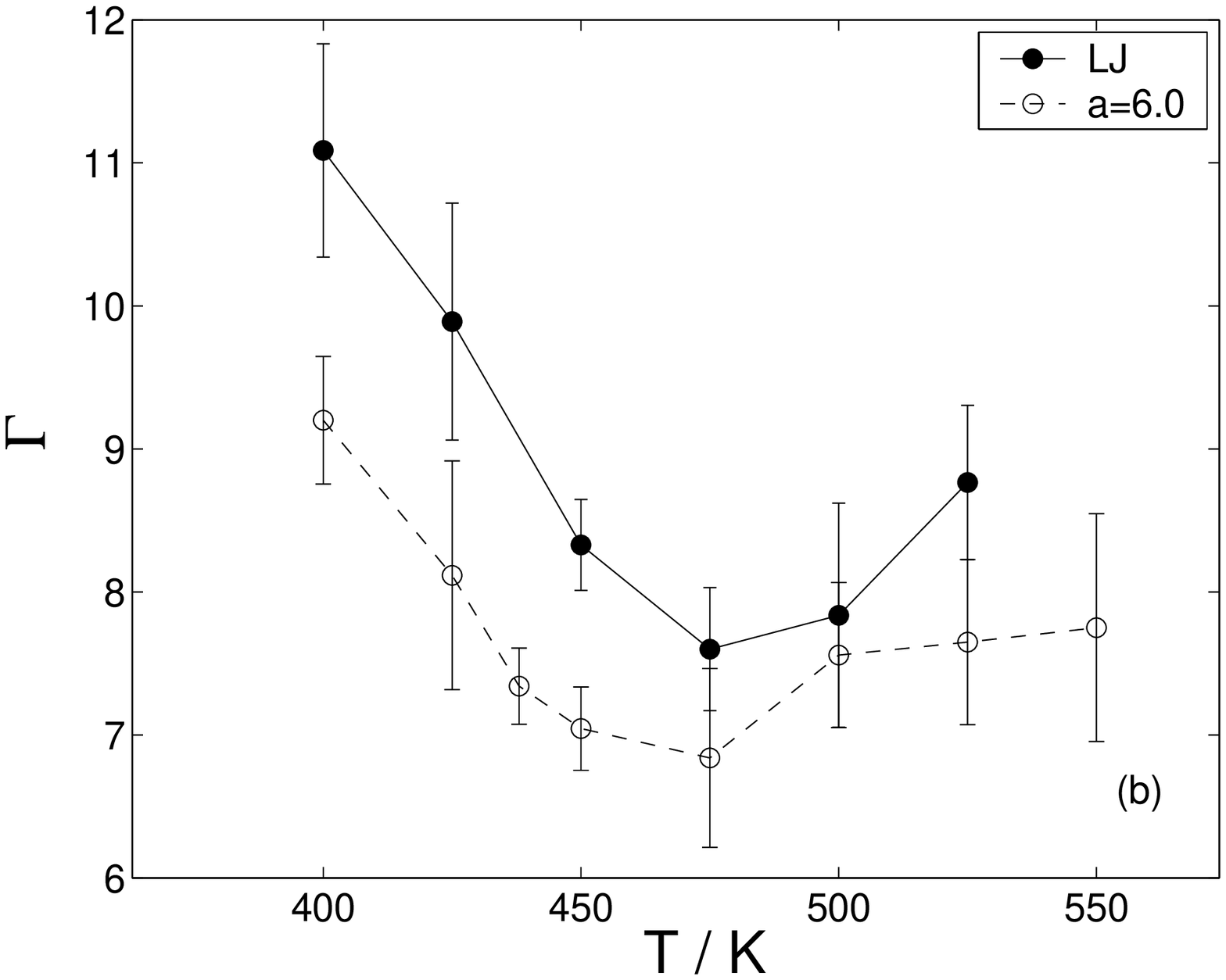}
\end{minipage}
\caption{$\Lambda$ (a) and $\Gamma$ (b) as functions of the temperature for the Lennard--Jones and 
Morse $a=6.0$ potential at $\varepsilon=5.0\%$, $\varepsilon=6.5\%$ respectively.}
\label{LGvsT}
\end{figure}
The enhanced mobility of the particles causes a more distinct separation of
the two particle types, resulting in more regular B stripes. 
As figure \ref{LGvsT}(a) shows the width of the B stripes approaches a constant value for the high temperature 
region. 
Further note, that in the region of high temperatures 
- like in the equilibrium simulations - 
nearly all B clusters are aligned in the $<11>$ directions in order to
achieve the energetically most favorable arrangement of particles.
\section{Lattice gas simulations}
\begin{figure}
\begin{minipage}{0.45 \textwidth}
  \epsfxsize= 0.99\textwidth
  \epsffile{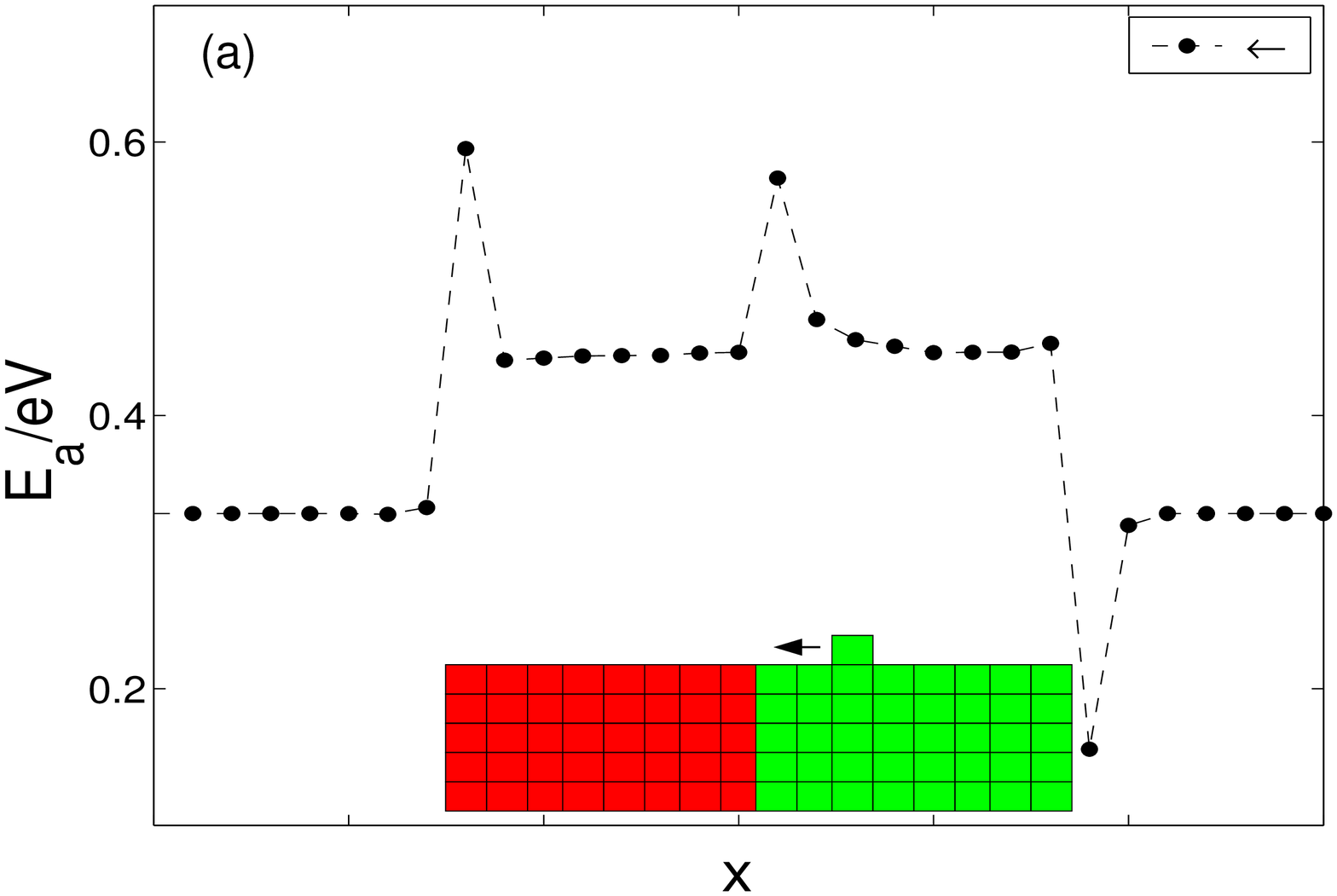}
\end{minipage}
\hfill
\begin{minipage}{0.45 \textwidth}
  \epsfxsize= 0.99\textwidth
  \epsffile{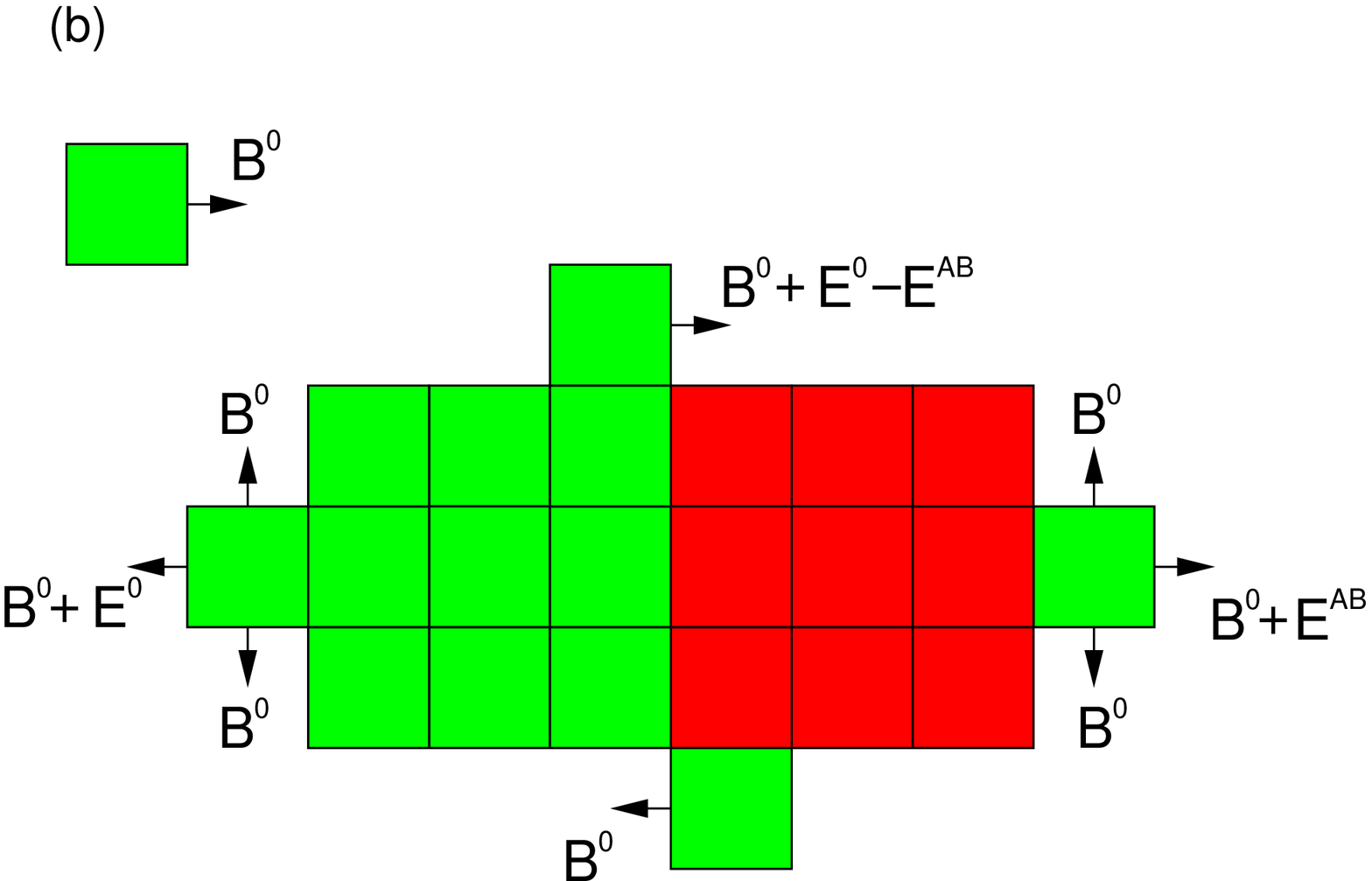}
\end{minipage}
\caption{(a) Barriers for diffusion of a B particle (light gray) from the right to the left, near an A--B interface. 
The values are given for the Lennard--Jones potential at a 
misfit $\varepsilon=4\%$. Note the increased barrier for a diffusion 
jump towards the A region. (b) Diffusion barriers for the diffusion process of a B particle in the lattice gas 
model. Note that the particle has to overcome an additional barrier $E^0-E^{AB}>0$ for a jump to the A region.}
\label{RGOTV}
\end{figure}
The question now is in which way the observed branches are related to the stripe structures found in the equilibrium
simulations.
Figure \ref{RGOTV}(a) 
shows the activation barrier for a B particle diffusing near an A--B interface for $\varepsilon=4\%$ 
and $E_{AB}=0.6E_A$. The weaker A--B interaction causes an extra step edge diffusion barrier for the jump from 
the B  to the A region. As already mentioned this diffusion barrier is believed to favor the formation of 
alternating stripes. We discuss now by means of a latticed based method 
how such a diffusion barrier influences the multi--component growth. Of special interest is here if the stripe formation
and the island morphology, as observed in our off--lattice simulations, can be explained within such a model.

\subsection{Simulation Model}
We consider now a lattice model with the two adsorbate species growing on a different 
planar substrate which provides a square lattice with $150 \times 150$ adsorption sites $(x,y)$. 
Unlike in the off--lattice simulation where a particle interacts with all particles within the 
range of the potential, A and B particles interact now only 
with their lateral nearest neighbors through attractive two--particle 
interactions $E^{AA}$, $E^{BB}$ and $E^{AB}$. 
The diffusion of adatoms on the surface
is described by thermally activated nearest neighbor hopping processes with Arrhenius rates 
$R_i=\nu_0 \exp(-E_{a,i}/kT)$, 
where we again 
choose $\nu_0=10^{12}$Hz as value for the attempt frequency. Unless otherwise
indicated the temperature $T$ is again set to $500K$.
For the activation energy $E_{a,i}$ of the diffusion event $i$ we use Kawasaki type energy barriers \cite{Newman:1999:MCM}:
\begin{equation}
\label{kawasaki_barriers}
E_{a,i} = \max \{B^X, B^X+ \Delta E^X\}. 
\end{equation}
Here $B^X$ gives the diffusion barrier for a free particle of type $X=A$ or B on the substrate. $\Delta E^X$ denotes
the change in the total energy caused by the event  which is determined 
by a simple bond--counting scheme from the NN particle--interactions. For example, for a diffusion event of 
an A particle we obtain
\begin{equation}
\label{bond_counting}
\Delta E^A = \Delta n^{AA} E^{AA}+ \Delta n^{AB} E^{AB},
\end{equation}
where $\Delta n^{AA}$ and $\Delta n^{AB}$ count the difference between the number of A--A and A--B bonds 
before and after the diffusion step.
To keep the number of parameters manageable we assume equal diffusion barriers for free A and B particles,
$B^A=B^B=B^0$, and also the strength of A--A and B--B bonds shall be the same: 
$E^{AA}=E^{BB}=E^0$. The interaction between particles of different types is assumed to be weaker than between those 
of the same type: $E^{AB} < E^0$. 

Figure \ref{RGOTV}(b) shows activation energies for exemplary diffusion processes of a free B particle and
B particles at the boundary of a small A--B cluster.
For A particles the picture is essentially the same, except that the activation energies for crossing 
the A--B interface have to be exchanged, as well as those for detachment from A and B step edges.
Since $E^{AB} < E^0$ one reads from figure \ref{RGOTV}(b) that the difference between $E^{AB}$ and $E^0$ has two
main consequences. First, an A particle diffusing along a step edge made up of A particles faces an 
enhanced diffusion barrier $B^0+E^0-E^{AB} > B^0$ when it attempts to cross an A--B interface. 
The same happens to a B particle coming from the other side. Thus, A and B particles diffusing along step edges
are likely to be reflected at A--B interfaces. Second, the barrier for detachment of an A particle from a step edge 
made up of B particles is lower than that for detachment from an A step edge and vice versa. 
This reproduces basically the influence of a weaker A--B interaction in the off--lattice simulations, 
disregarding though all influences of strain and long range interactions.

\subsection{Influence of the step edge barrier}
We first investigate the influence of the binding energy $E^{AB}$ between A and B 
particles on the morphology of growing films. Therefore we fix $B^0=0.37eV$ and $E^0=0.51eV$ - this 
reproduces roughly the homoepitaxy ($\varepsilon=0$) barriers for diffusion on planar substrate and 
detachment from an island edge as measured in the off--lattice simulations.
$E^{AB}$ is varied between $0.31 E^0$ and $0.71 E^0$.

\subsubsection{Island geometry}
Following the off--lattice simulation for 
all simulation runs the deposition rate for both types of particles is set to 
$0.005ML/s$ resulting in a overall deposition rate of $R_d=0.01ML/s$.
When the total adsorbate coverage has reached $0.5ML$ the particle fluxes are switched off and the simulation is halted. 

Figure \ref{LATI} shows snapshots of example configurations obtained after the end of simulation runs 
for four different values of the binding energy $E^{AB}$.

\begin{figure}[hbt]
\begin{minipage}{0.24 \textwidth}
  \epsfxsize= 0.99\textwidth
  \epsffile{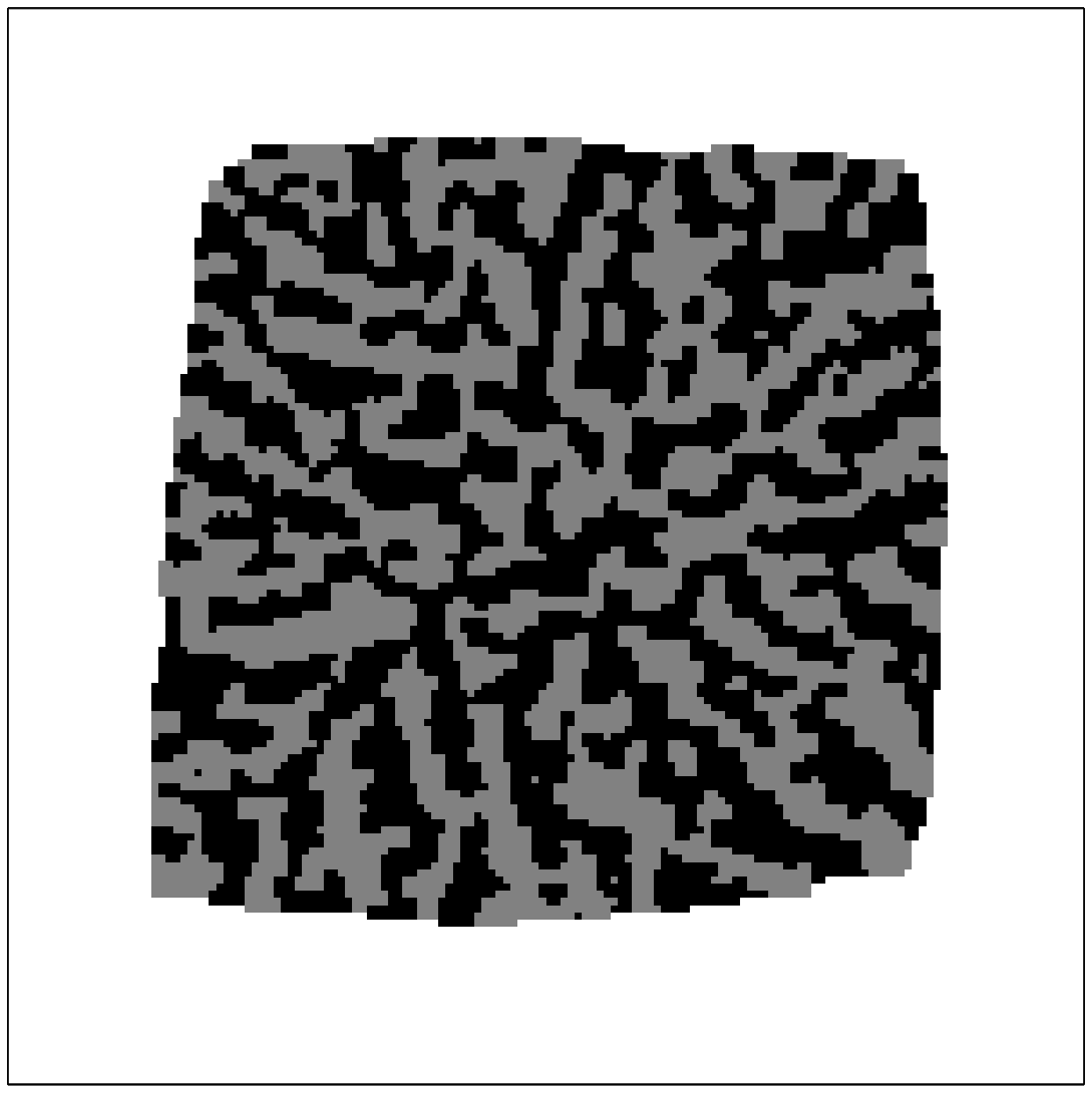}
\end{minipage}
\begin{minipage}{0.24 \textwidth}
  \epsfxsize= 0.99\textwidth
  \epsffile{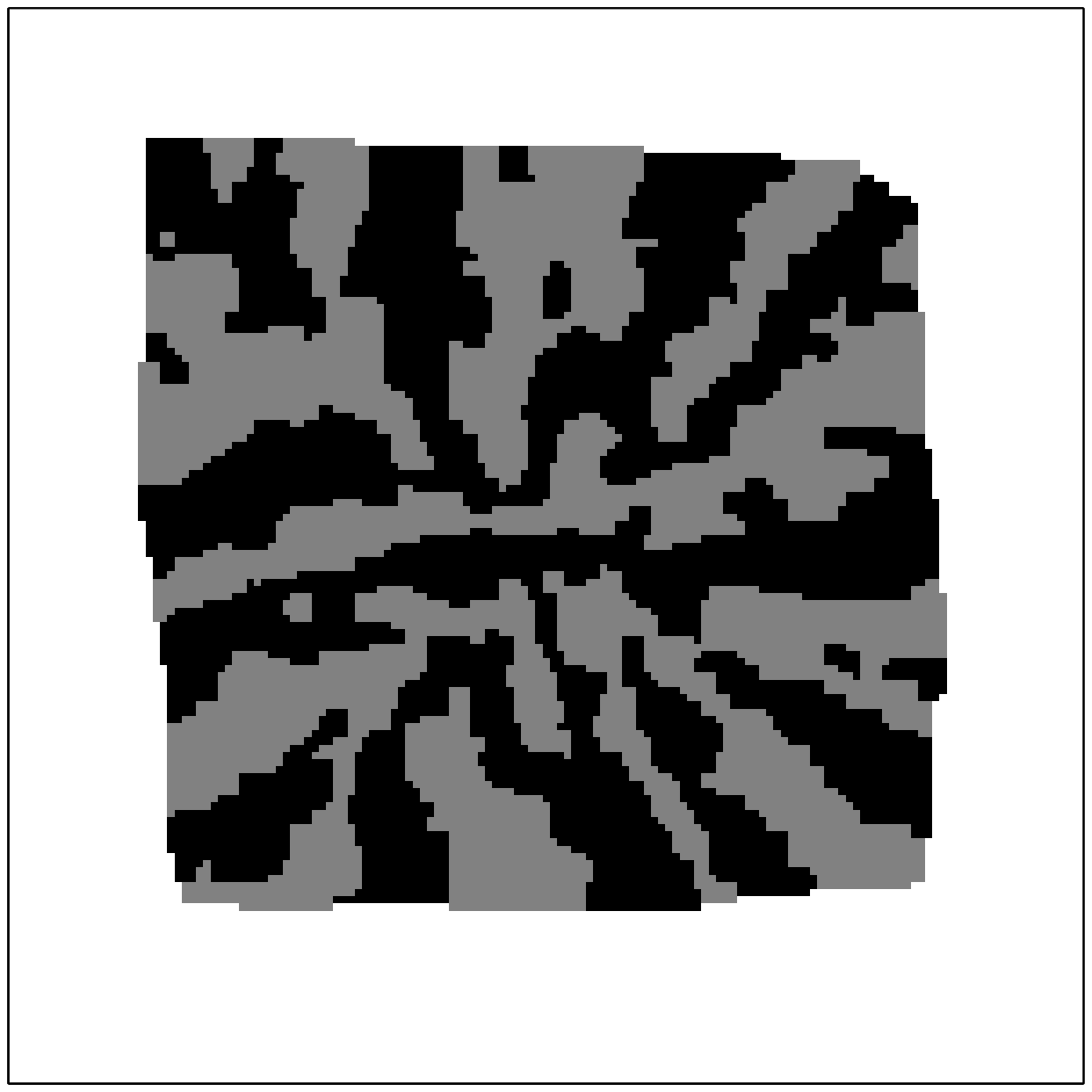}
\end{minipage}
\begin{minipage}{0.24 \textwidth}
  \epsfxsize= 0.99\textwidth
  \epsffile{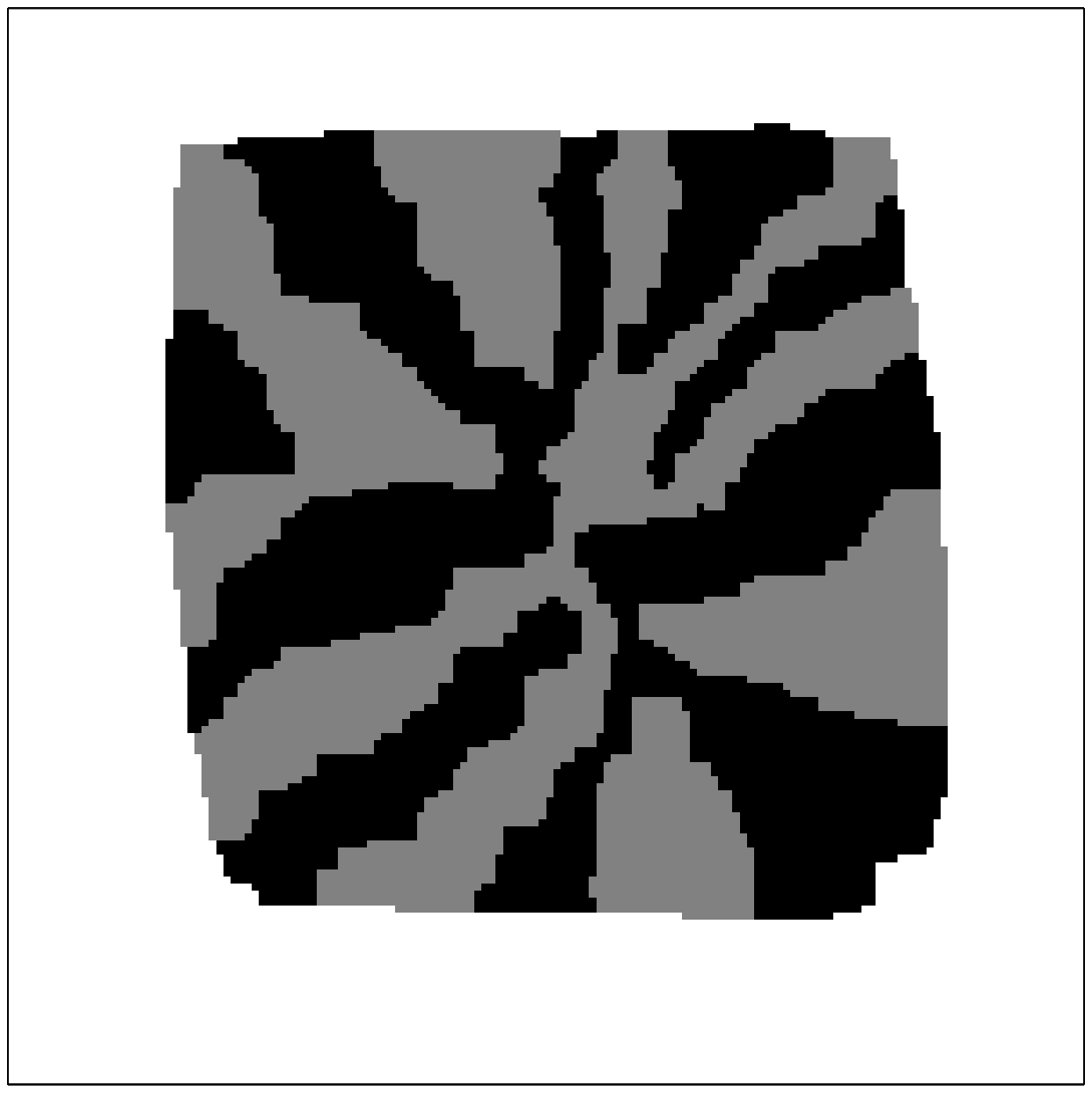}
\end{minipage}
\begin{minipage}{0.24 \textwidth}
  \epsfxsize= 0.99\textwidth
  \epsffile{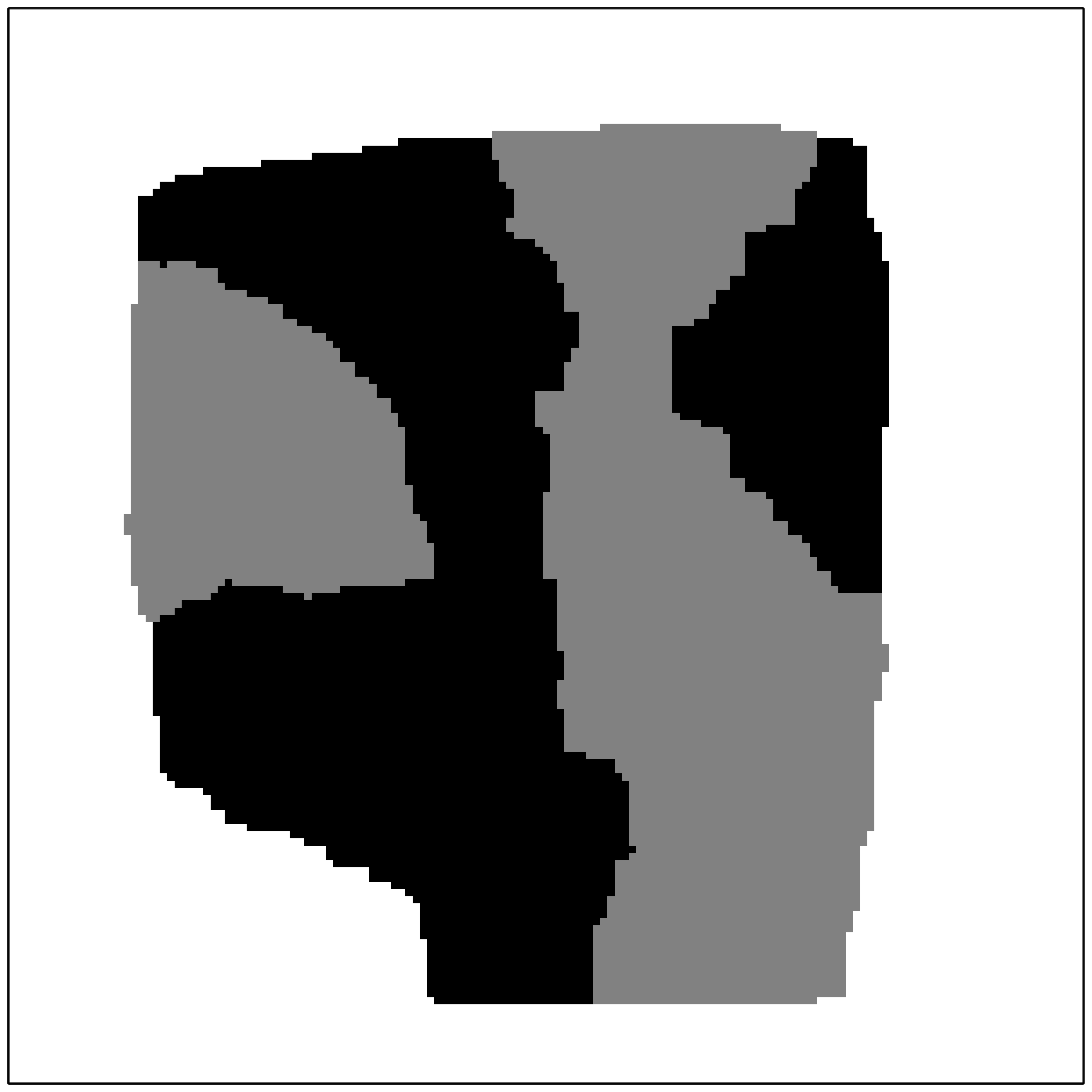}
\end{minipage}
\caption{Islands obtained by simultaneous deposition of A and B particles for 
different binding energies between A and B
particles (from left to the right): $E^{AB}=0.71E^0$, $E^{AB}=0.59E^0$, $E^{AB}=0.51 E^0$ 
and  $E^{AB}=0.39E^0$. The system size
is $150 \times 150$ and the total coverage is $\theta=0.5ML$. The figures are by courtesy of Th. Volkmann.}
\label{LATI}
\end{figure}

For all values of $E^{AB}$ one observes compact island shapes with the island boundaries roughly parallel to
the lattice directions. The weaker binding energy between A and B particles leads to an aggregation of
particles of the same type in clusters which can be characterized as stripes. 
While for the higher value of $E^{AB}$  these stripes are rather thin 
and show a considerable degree of irregular intermixing for lower values of $E^{AB}$ the 
stripes are both much thicker and there is a tendency for them to stretch outwards. One also sees that at 
a certain stage of the island growth a stripe of one particle type may become wide enough for particles of 
the other type to form a stable nucleus within this stripe, thus leading to a branch--like structure. 
The occurrence of the stripe--like structures is here a pure kinetic effect.

\subsubsection{Step geometry}
Due to the island topology chosen we have a fourfold symmetry of the stripe--structures. In the following we focus
on the stripe formation in only one growth direction. Therefore we choose a different growth topology for our simulations
where the growth proceeds from a step edge. 
We assume here that our system represents 
one particular terrace of a vicinal surface with elongated terraces of constant width. 
Therefore we use periodic boundary conditions in the direction parallel to the step edges ($x$--direction). 
In the perpendicular $y$--direction the system is bounded by the lower part
of a step edge and the upper part of the next step edge. While adsorbate particles of both types 
may attach to the lower part of the step edge with the same binding energy $E^0$ as to next neighbors
of the same type, they become reflected at the upper part of the step edge, again representing an infinite 
Ehrlich--Schwoebel barrier for interlayer jumps.

We mostly studied systems with $256 \times 100$ and $512 \times 100$ lattice sites. The combination of the 
specific boundary conditions in the $y$--direction together with the values used for the activation energies 
guarantees that for the given growth temperature $T=500K$ and system sizes adsorbate particles deposited on 
the substrate will reach the step edge at $y=0$. Thus terrace nucleation is suppressed and growth proceeds
from the step edge. All other simulation parameters remain unchanged.

Figure \ref{STUFEN} (left panel) shows snapshots of system configurations after the end of simulation runs 
for various values of $E^{AB}$.
\begin{figure}
\begin{minipage}{0.49 \textwidth}
\begin{minipage}{0.99 \textwidth}
  \epsfxsize= 0.99\textwidth
  \epsffile{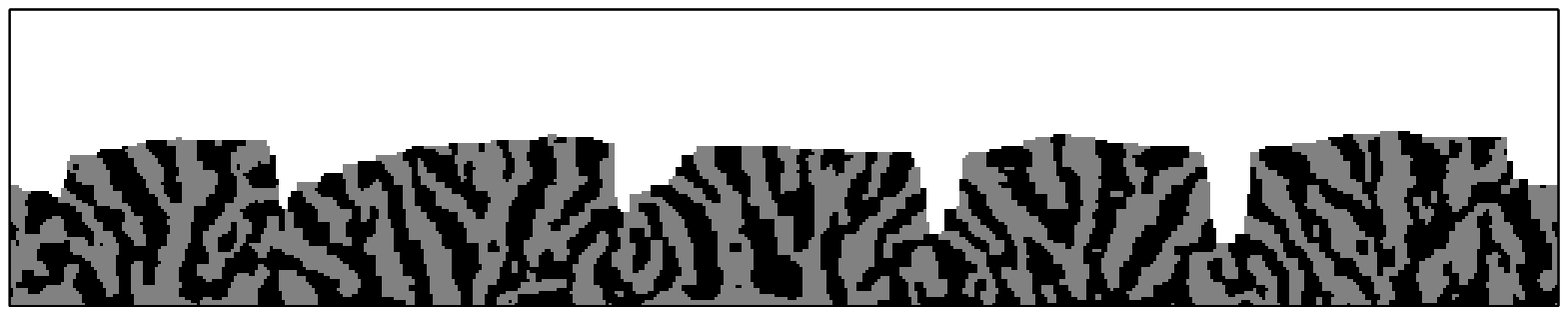}
\end{minipage}

\begin{minipage}{0.99 \textwidth}
  \epsfxsize= 0.99\textwidth
  \epsffile{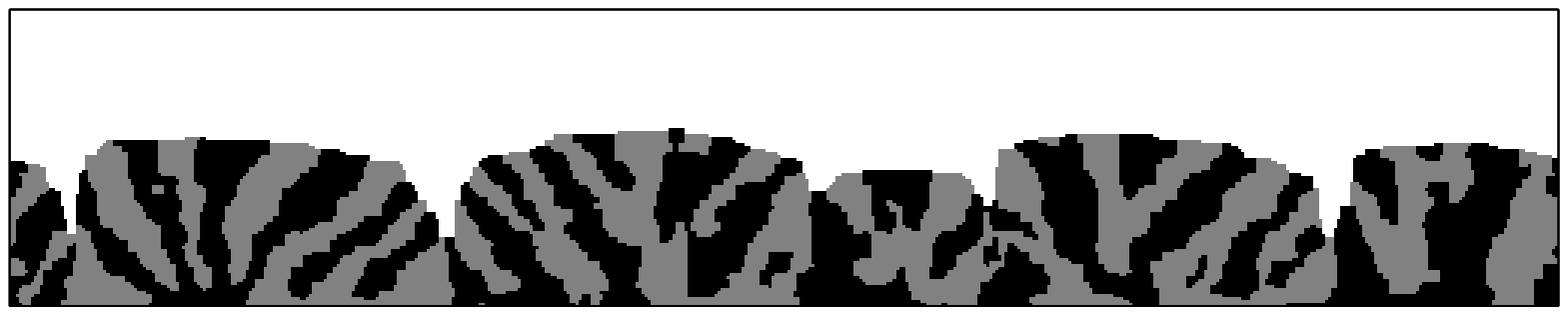} 
\end{minipage}

\begin{minipage}{0.99 \textwidth}
  \epsfxsize= 0.99\textwidth
  \epsffile{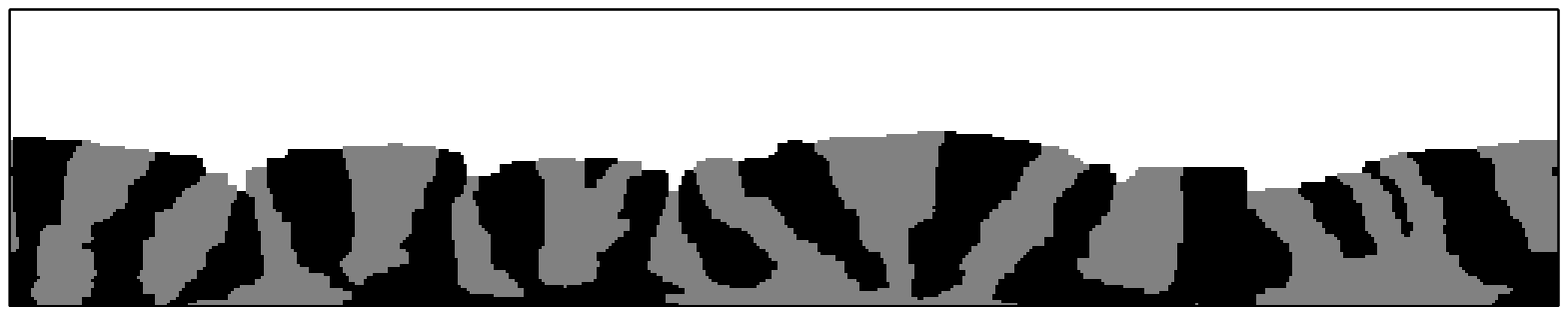} 
\end{minipage}

\begin{minipage}{0.99 \textwidth}
  \epsfxsize= 0.99\textwidth
  \epsffile{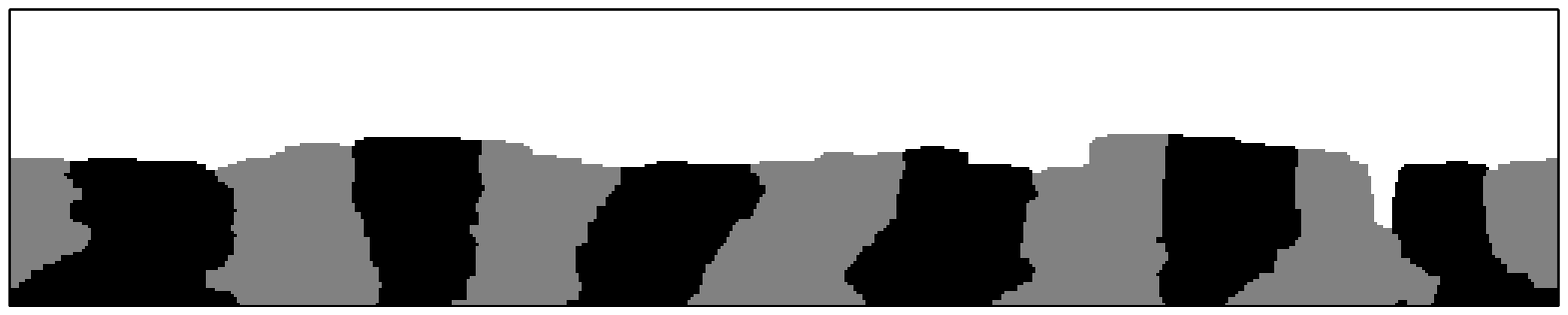} 
\end{minipage}
\begin{minipage}{0.99 \textwidth}
  \epsfxsize= 0.99\textwidth
  \epsffile{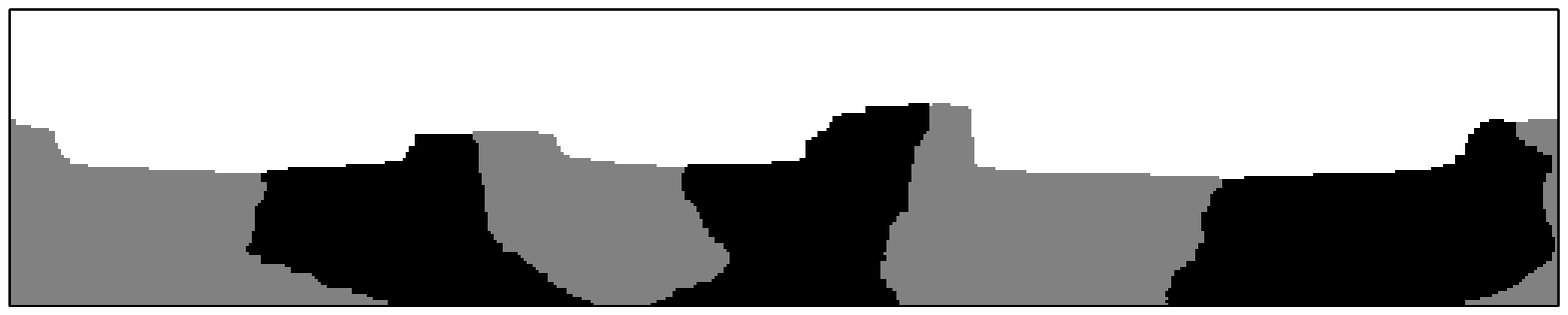} 
\end{minipage}
\end{minipage}
\hfill
\begin{minipage}{0.49 \textwidth}
 \epsfxsize= 0.99\textwidth
  \epsffile{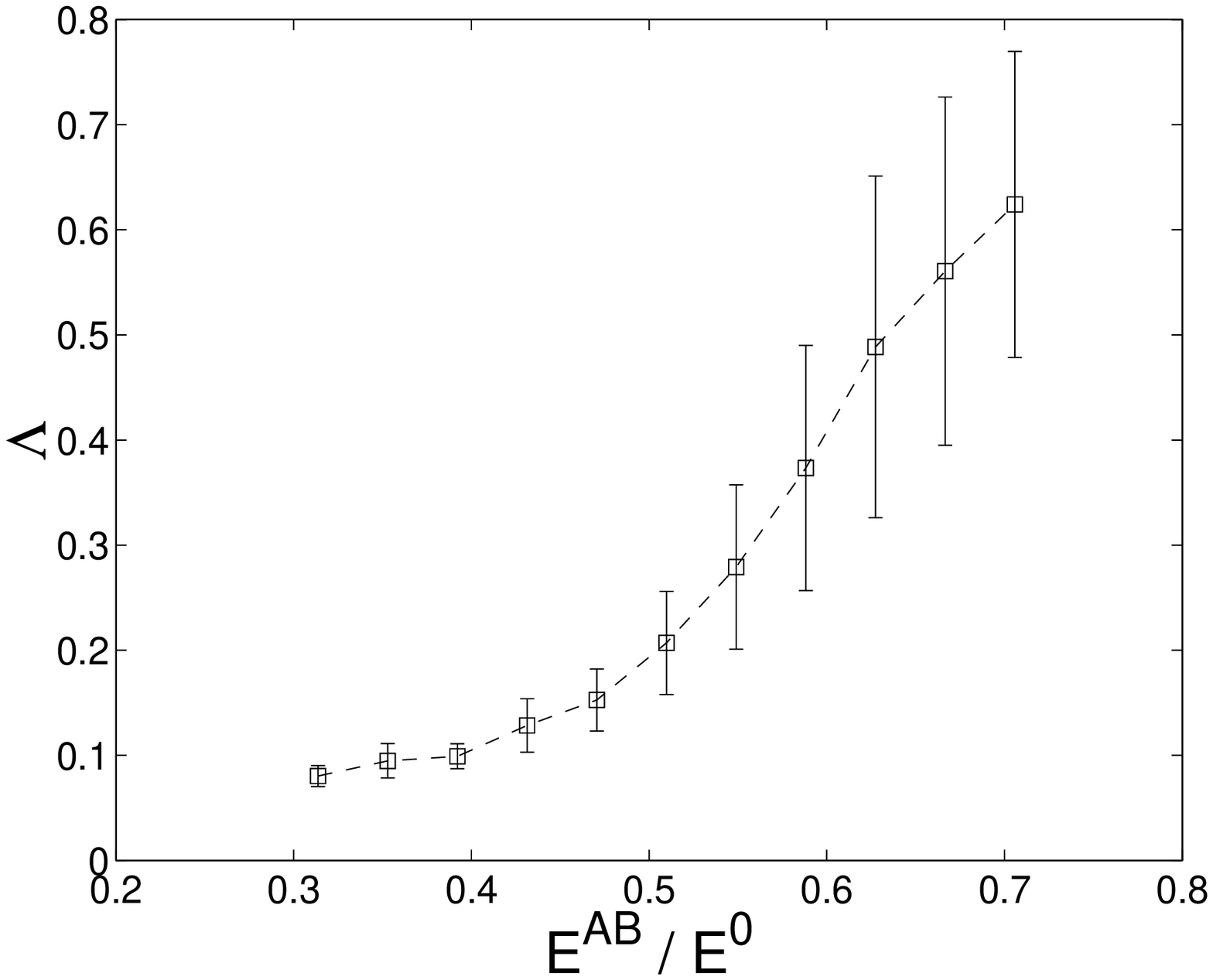}
\end{minipage}
\caption{Left panel: simulation results for $E^{AB}/E^0=0.63,0.55,0.47,0.39,0.31$ (from top to bottom).
A particles appear in dark gray, B particles in light gray. 
Right panel: dependence of $\Lambda$ on $E^{AB}/E^0$. Each value is obtained by
averaging over the A clusters of $10$ independent simulation runs. The errorbars are given by the standard
deviation. The shown results are by courtesy of Th. Volkmann.}
\label{STUFEN}
\end{figure}
The difference between $E^{AB}$ and $E^0$ leads to a separation of A and B particles in stripe--like clusters. 
Figure \ref{STUFEN} (left panel) shows that for a decreasing value of $E^{AB}$ the average thickness of the stripes 
tends to increase. While for a low value of $E^{AB}$ A and B cluster are well separated they become more and
more intertwined with increasing $E^{AB}$. Both observations can be explained by the fact that with 
decreasing $E^{AB}$ it is more favorable for A and B particles to attach to particles of the same type.
We notice also that there is no clear orientation of the stripes, for example parallel to the $y$--direction. 
In contrast stripes of one particle type may very well grow sideways as can best be seen in the downmost sample.

We now have a closer look at the influence of $E^{AB}$ on the stripe width. Therefore we determine for 
each connected cluster of A particles the ratio $\Lambda$ between its perimeter length and its volume. 
For each value of $E^{AB}$ we did 10 independent simulation runs and averaged over the occurring 
A clusters. To get a better statistics all clusters were initially sorted by their size and very small clusters
($\leq 0.2 \times$ the mean cluster size), 
for which $\Lambda \approx 1$ holds, were omitted from the averaging.

Figure \ref{STUFEN} (right panel) shows the dependence of $\Lambda$ on the binding energy $E^{AB}$. As one sees $\Lambda$
increases monotonously with increasing $E^{AB}$ confirming that the stripe width decreases with the difference
in binding energies becoming smaller. For $E^{AB}/E^0=0.31$ we obtain $\Lambda \approx 0.08$ which is already 
close to the value which we would expect for the limit 
$E^{AB}/E^0 \rightarrow 0$ when only one A cluster with straight edges was present.
When $E^{AB}/E^0$ approaches $1$, $\Lambda$ also should tend to $1$ because A and B particles should then
be perfectly mixed, since there is no longer a preference for a particle to stick to its own kind. This means
that most of the cluster contain $\mathcal{O}(1)$ particles which then results in $\Lambda \approx 1$.

We conclude from our lattice gas simulations, that the step edge barrier indeed gives reason for stripe 
formation. Similar to the equilibrium simulations the width of the stripes can be controlled by adjusting the  
binding energy between A  and B particles. However as figure \ref{LATI} shows neither asymmetries between 
A  and B clusters nor a ramification of the islands is observed here.
Certainly this is not surprising since in the lattice gas simulations A and B particles are treated in a 
symmetric way, whereas in the off--lattice simulations the different sign of the misfits causes 
different diffusion barriers for A and B particles, 
respectively.

\subsection{Enhanced lattice gas model}
For that reason we propose now an enhanced lattice model taking basic differences of both particle types
into account.
This is done in order to determine whether a simple misfit dependence of the diffusion barriers could
lead to the observed results within a lattice model. For example in the off--lattice case for $\varepsilon>0$ 
the substrate diffusion for the bigger B particles is always faster than for the smaller A particles
(cf. chapter \ref{KAP-1}).

Furthermore in the off--lattice method 
the barriers for edge diffusion are higher than the substrate diffusion barriers.
This could also give rise to a ramified island morphology.
Therefore we extract the barriers for free diffusion on the substrate  
as well as averaged values for edge diffusion and detachment for a {\it fixed} island size (see also fig. \ref{SCAN})
as a function of the misfit. Theses barriers are then used in the lattice model as parameters.
The thus modified lattice model incorporates the basic misfit dependence of the diffusion barriers. 
But effects of the long--range interaction, like the reduced barriers for jumps towards an island still have to be 
neglected here.
\begin{figure}[h]
\begin{minipage}{0.50 \textwidth}
\epsfxsize= 0.99\textwidth
\epsffile{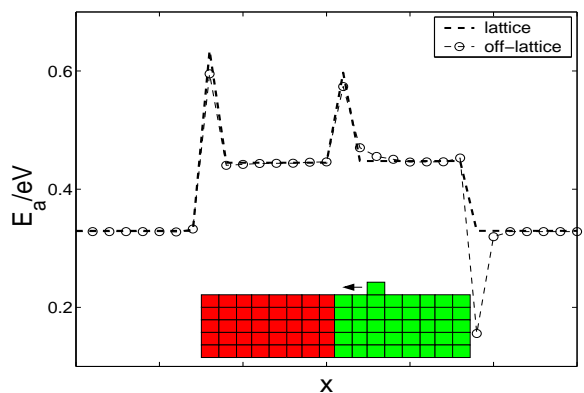}
\end{minipage}
\hfill
\begin{minipage}{0.48 \textwidth}
\caption{Barriers for in--plane diffusion 
of a B particle (light gray) from the right to the left, near an A--B interface. 
The values are given
for the Lennard--Jones potential at a misfit $\varepsilon=4\%$. 
Note that for the off--lattice barriers the particle {\it feels} the influence of the island due to the 
range of the potential before attaching to it. This results in a rather low barrier for the jump towards the island.}
\label{SCAN}
\end{minipage}
\end{figure} 

Figure \ref{COMP_O_L} shows a comparison between the lattice model and the off--lattice simulation for the 
Lennard--Jones potential. 
Simular results are obtained by fitting the barriers for the Morse potentials to the lattice model.
\begin{figure}[hbt]
\begin{minipage}{0.24 \textwidth}
  \epsfxsize= 0.99\textwidth
  \epsffile{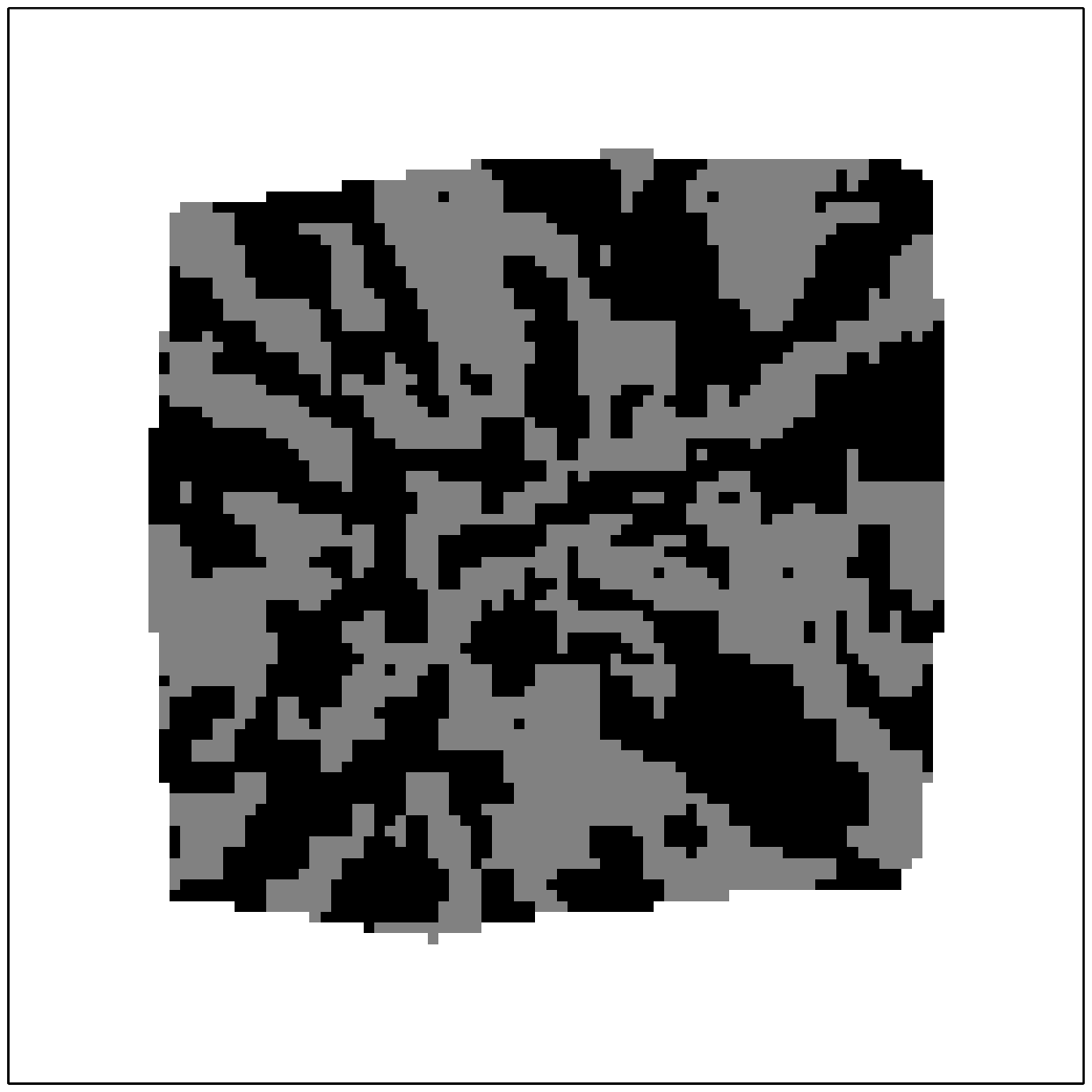}
\end{minipage}
\hfill
\begin{minipage}{0.24 \textwidth}
  \epsfxsize= 0.99\textwidth
  \epsffile{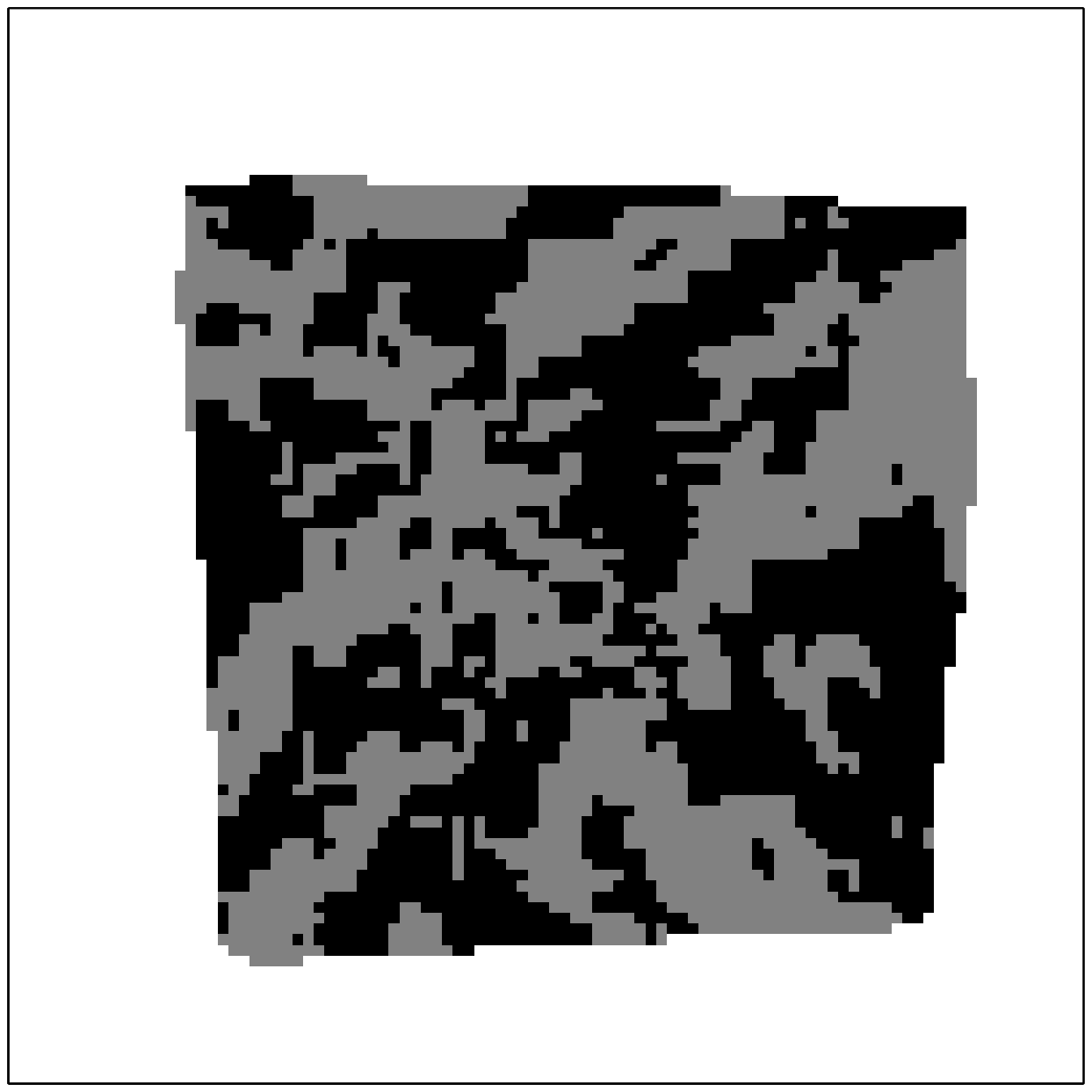} 
\end{minipage}
\hfill
\begin{minipage}{0.24 \textwidth}
  \epsfxsize= 0.99\textwidth
  \epsffile{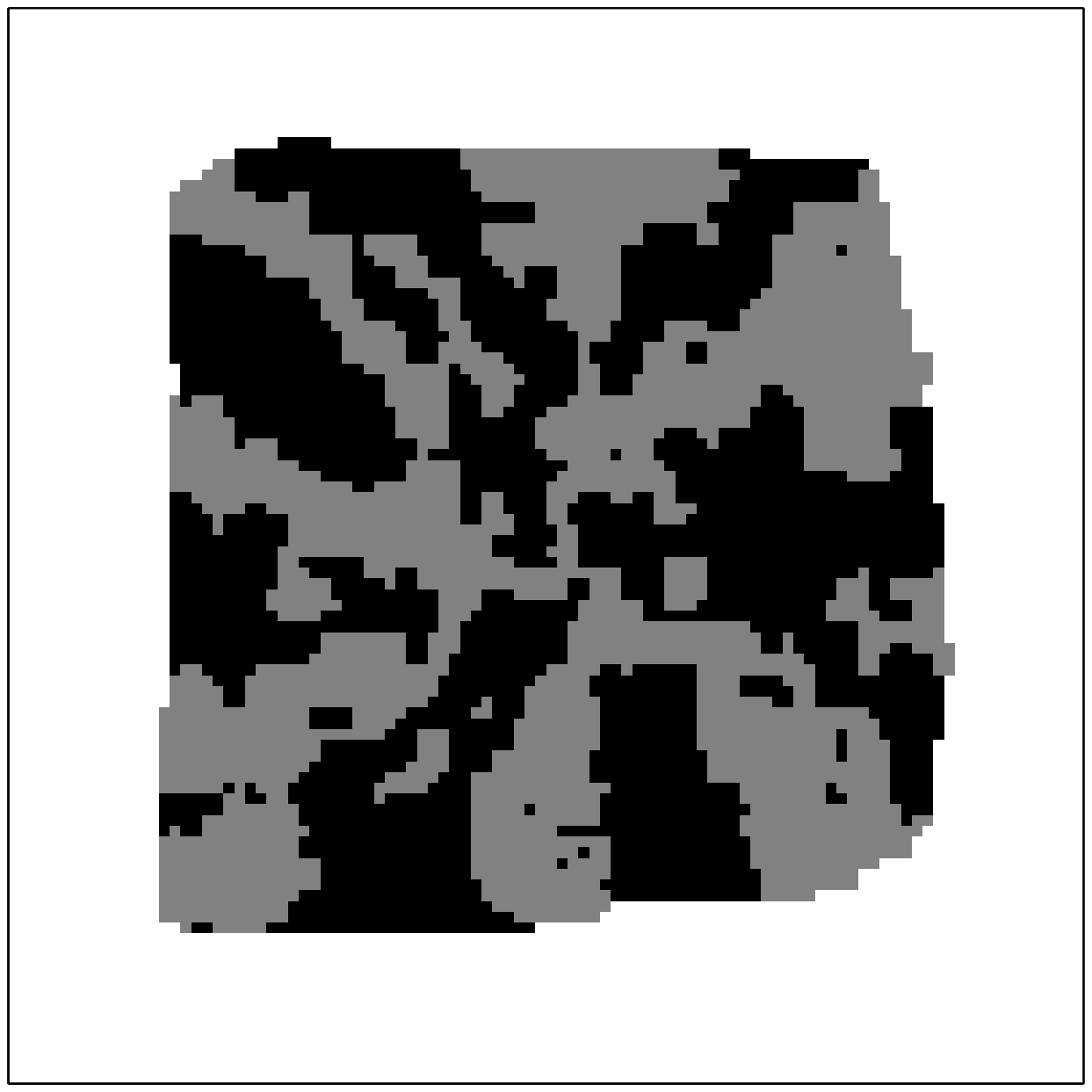} 
\end{minipage}
\hfill
\begin{minipage}{0.24 \textwidth}
  \epsfxsize= 0.99\textwidth
  \epsffile{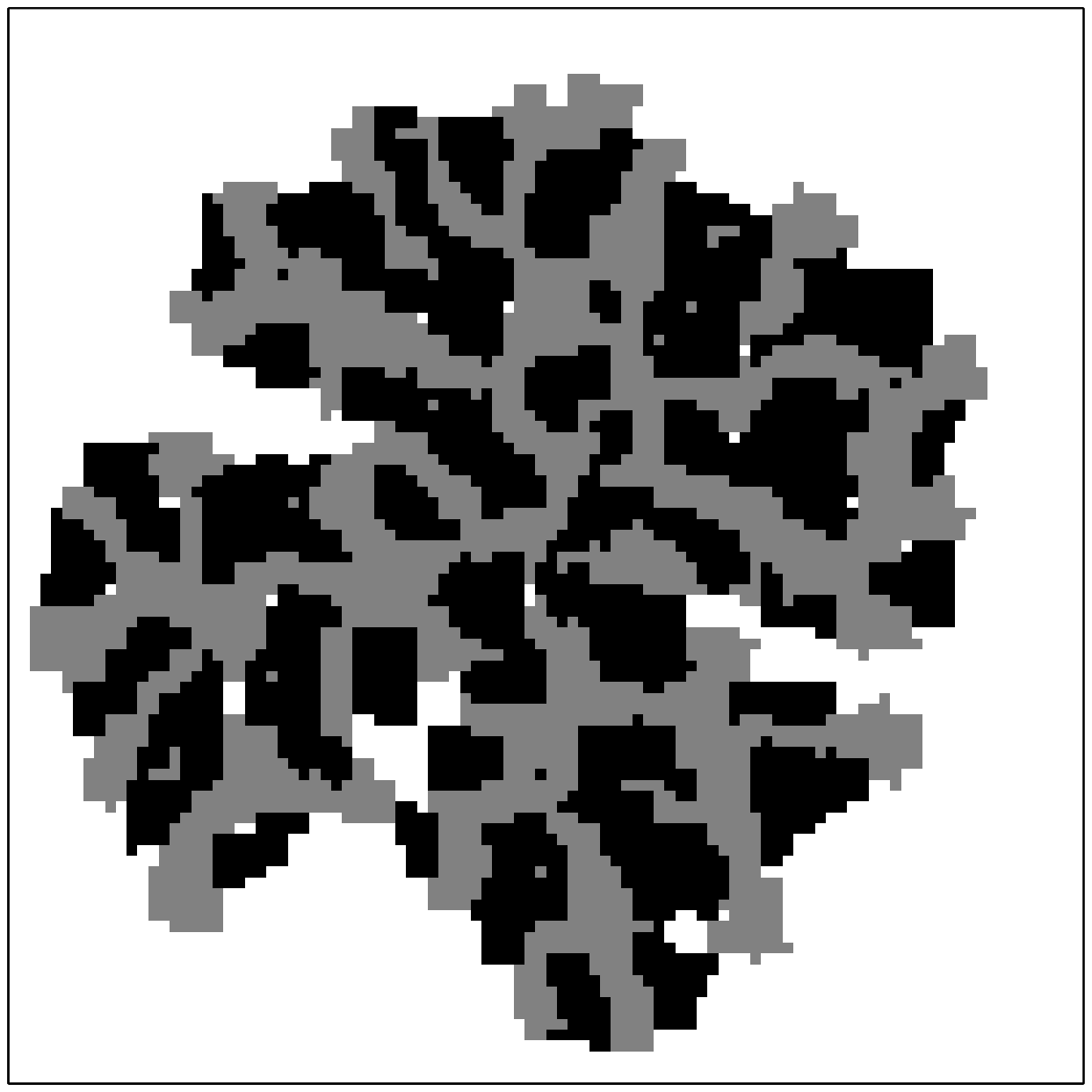} 
\end{minipage}
\caption{Snapshots for the Lennard--Jones potential for the enhanced lattice and the off--lattice model. 
The panels show (from left to the right) lattice, off--lattice results at $\varepsilon=0$ and lattice, off--lattice
results for  $\varepsilon=5\%$. The lattice gas results are by courtesy of Th. Volkmann.}
\label{COMP_O_L}
\end{figure}
As expected, the islands for both models look similar in the case of zero misfit. 
However for $\varepsilon=5\%$ the results for both models seem to have little in common. For the lattice model 
the separation of A and B regions is more pronounced as for the $\varepsilon=0$ parameters but no size 
limitation of the stripes and therefore no ramification is observable here.
To confirm this we have measured the ramification $\Gamma$ for the lattice model. Figure \ref{RAMCOMP} shows
this in comparison to the values obtained from the off--lattice simulations. Only at $\varepsilon=0$ both
curves collapse. 
\begin{figure}[hbt]
\begin{minipage}{0.50 \textwidth}
\epsfxsize= 0.99\textwidth
\epsffile{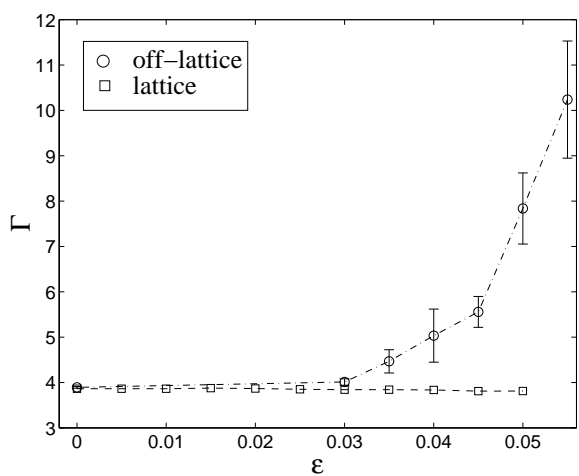}
\end{minipage}
\hfill
\begin{minipage}{0.48 \textwidth}
\caption{Comparison of the island--ramification in case of lattice and off--lattice simulations. 
Each value is obtained by averaging over $10$ independent simulation runs.  
In case of off--lattice simulations the errorbars are given by the standard deviation. For the lattice 
simulation errorbars are smaller than the symbols.}
\label{RAMCOMP}
\end{minipage}
\end{figure} 
From this examinations it is quite clear that the basic differences of the diffusion barriers of both particle types 
neither cause the width restriction of the B branches nor the ramification of the islands with growing misfit.
\begin{figure}[hbt]
\begin{minipage}{0.24 \textwidth}
  \epsfxsize= 0.99\textwidth
  \epsffile{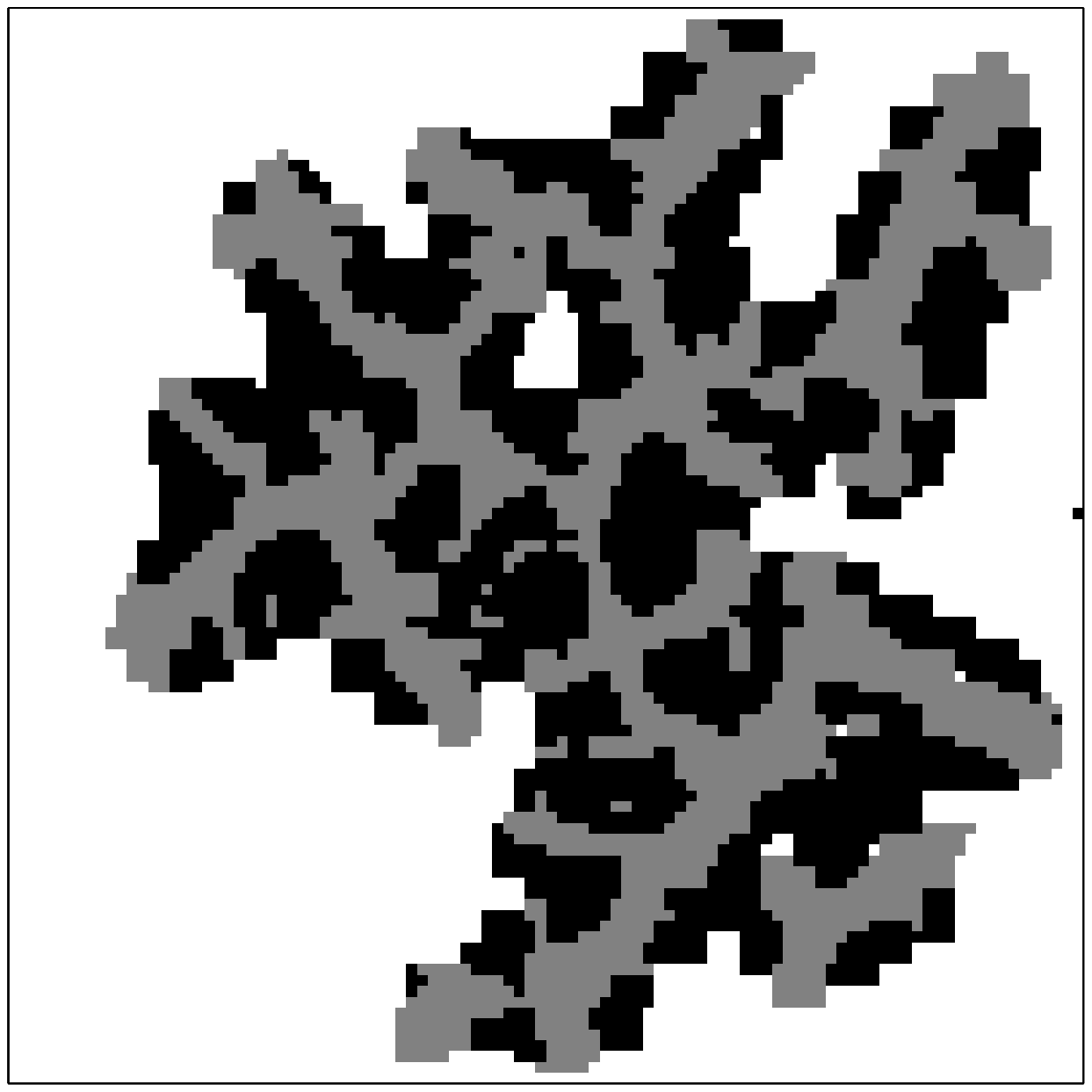}
\end{minipage}
\hfill
\begin{minipage}{0.24 \textwidth}
  \epsfxsize= 0.99\textwidth
  \epsffile{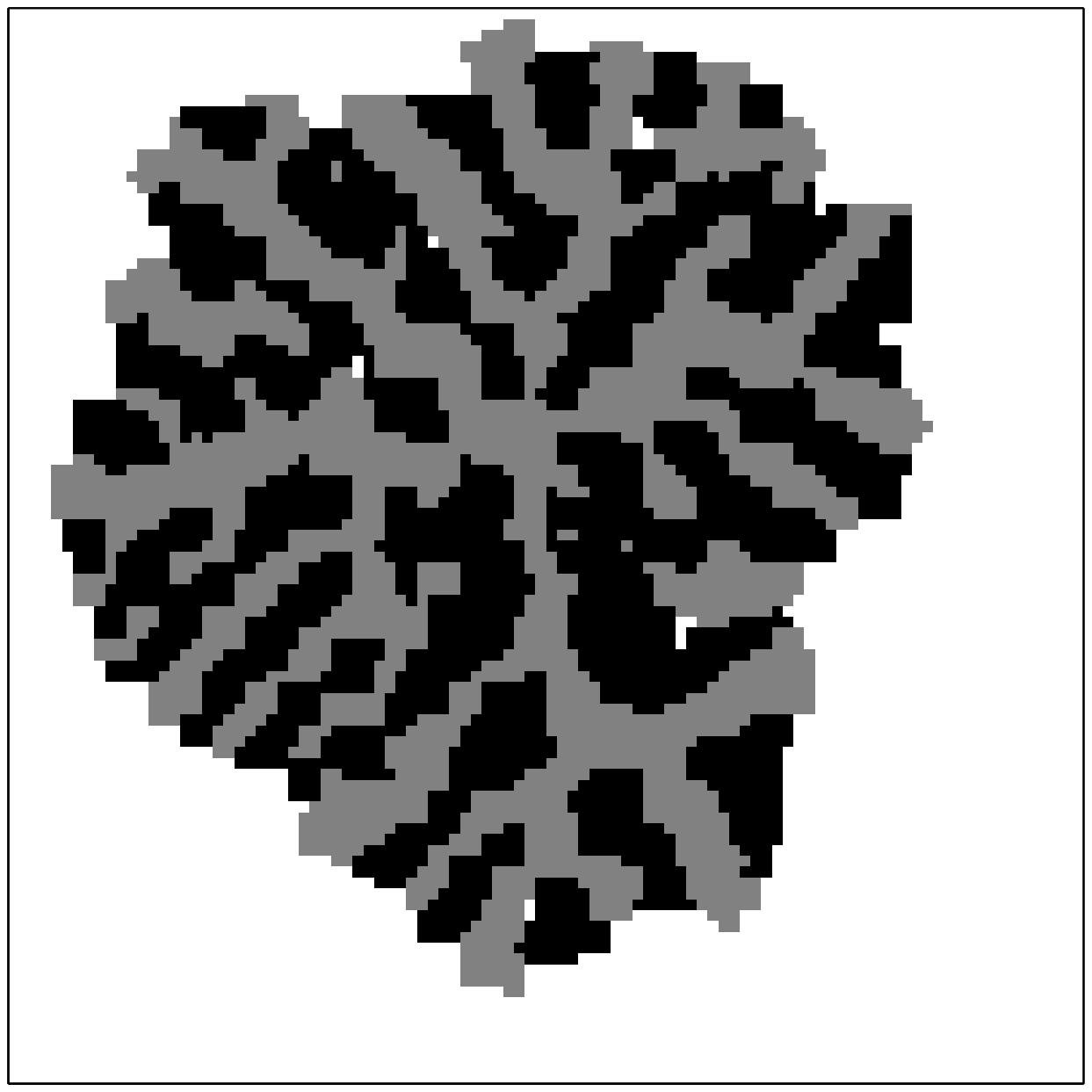}
\end{minipage}
\hfill
\begin{minipage}{0.48 \textwidth}
  \caption{Snapshots for simulation runs with (left panel) and without (right panel) the reduced barrier for 
jumps toward island. Both samples are given for the $a=6.0$ Morse potential at  $\varepsilon=6.5\%$.}
\label{KILLBIAS}
\end{minipage}
\end{figure}

From additional off--lattice simulations, where the reduced barrier for a jump towards an island (cf. fig. \ref{SCAN})
is suppressed we find that - though the resulting islands are less ramified - the width of the B branches stays
the same (see fig. \ref{KILLBIAS}). The reduction of the ramification is due to a higher mobility of the particles:
once a particle detaches from an island it has the same probability for jumps towards the island as away from it.
The capturing of diffusing particles by islands is less pronounced and therefore the particles are more uniformly 
distributed around the island.

In conclusion we learn from the enhanced lattice model with fitted diffusion barriers that neither the ramification 
nor the width restriction of the B stripes is due to the basic differences of the diffusion barriers of the both 
particle species. However, there has to be a kinetic reason for the obvious difference between both species.
For that reason we take in the following an closer look on the diffusion behavior of particle attached to island edges.

\section{Diffusion bias for the edge diffusion}
\begin{figure}
\begin{minipage}{0.45 \textwidth}
  \epsfxsize= 0.99\textwidth
  \epsffile{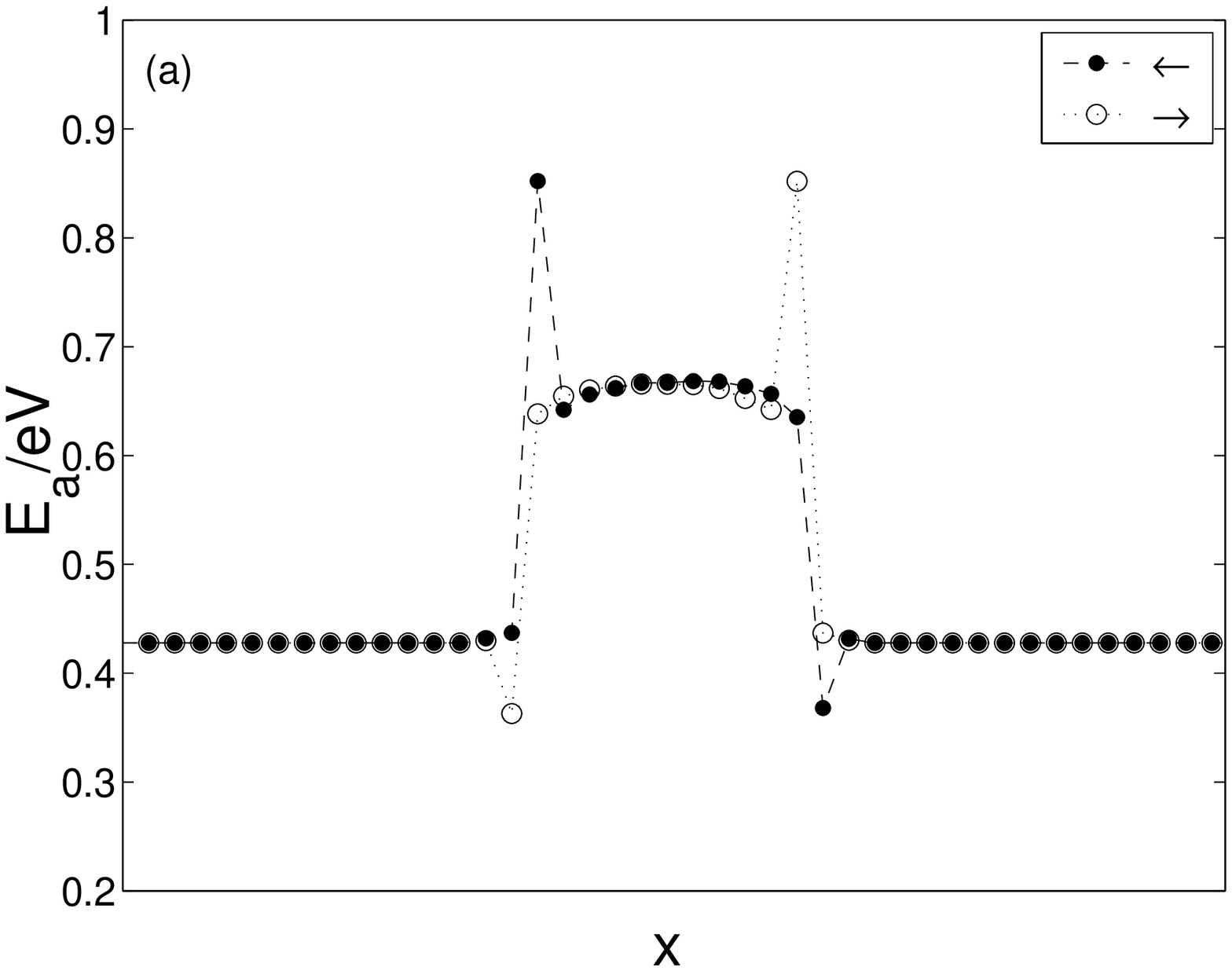}
\end{minipage}
\hfill
\begin{minipage}{0.45 \textwidth}
  \epsfxsize= 0.99\textwidth
  \epsffile{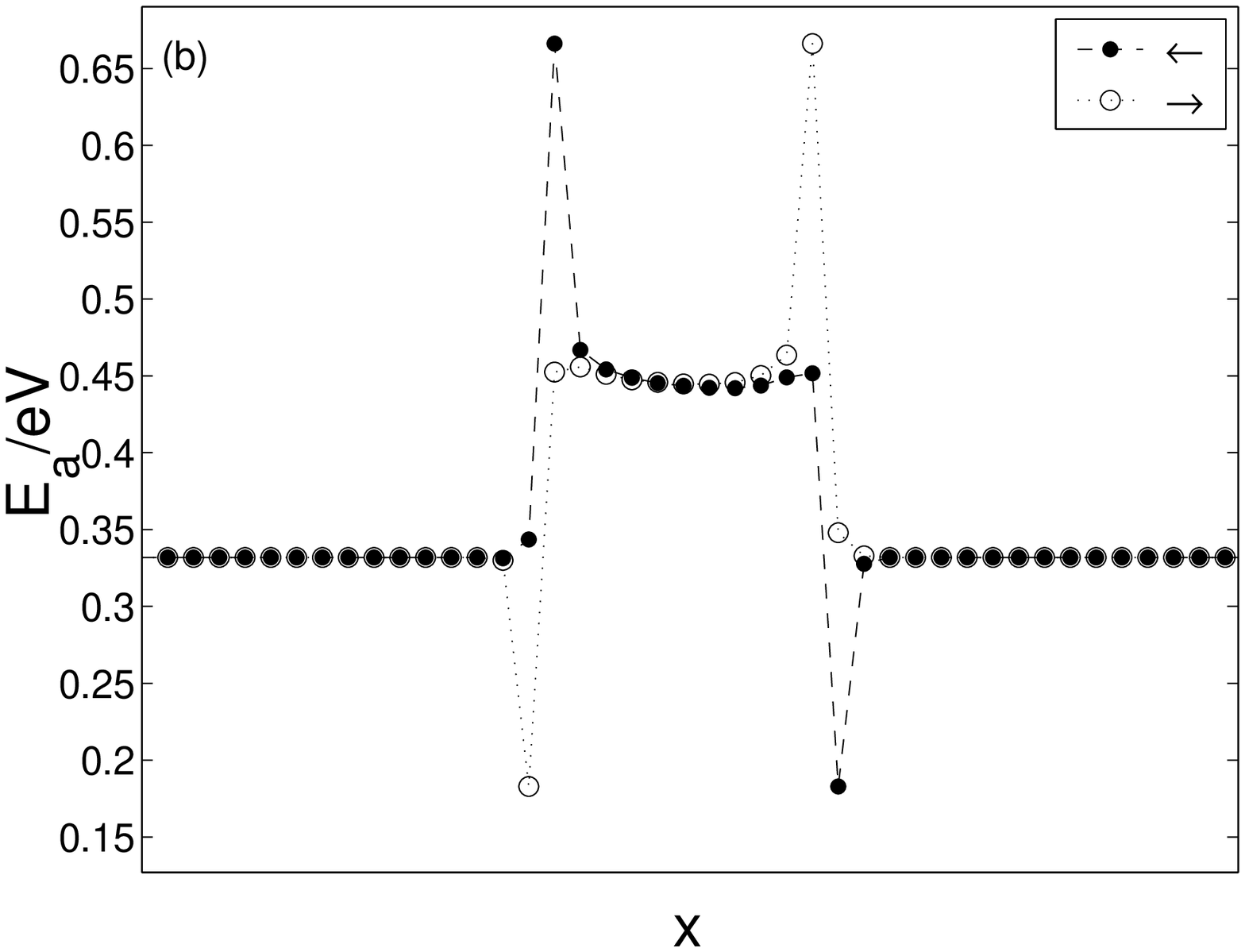}
\end{minipage}
\caption{Barriers for diffusion of an A particle near and on an A edge (a)  and for a B particle 
diffusing near and on a B edge (b). The barriers were calculated for the Morse $a=6.0$ potential, stripe 
width $l=11$ and misfit $\varepsilon=5\%$.}
\label{BIAS_GK}
\end{figure}
By taking in our lattice model only averaged diffusion barriers into account 
we neglected the diffusion bias for edge diffusion. 
This bias is thereby due to the same reason as explained in chapters \ref{KAP-1} and \ref{KAP-5} for the diffusion 
on an island: near the center of an edge the particles are bound with the substrate lattice constant, towards the
rims of the island they can approach their own lattice constant. 
Figure \ref{BIAS_GK} shows the resulting activation energy $E_a$ for 
a particle diffusing near and on an A ($\varepsilon<0$) or B ($\varepsilon>0$) island edge. 
Here, the diffusing particle is of the same type as the island. However, the direction of the diffusion 
bias only depends on the type of the island, whereas the value of the barriers changes
according to the interaction strength $E_{AB}$ between the particle types. 
\begin{figure}[h]
\begin{minipage}{0.50 \textwidth}
\epsfxsize= 0.99\textwidth
\epsffile{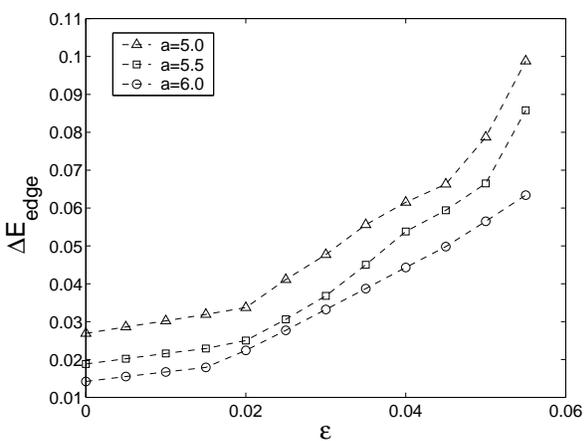}
\hfill
\end{minipage}
\begin{minipage}{0.45 \textwidth}
\caption{Relative difference $\Delta E_{edge}$ 
between the minimum and the maximum value of the edge diffusion barrier (width $l=11$) as a function 
of the misfit $\varepsilon$.}
\label{BIAS}
\end{minipage}
\end{figure} 

One reads from figure \ref{BIAS_GK}(a) that in case of an A edge the particle diffusion is biased towards the rims 
of the islands, whereas for B edges particles are more likely to jump towards the center of the edge.
This explains qualitatively the different morphology of A and B clusters in our simulations. The diffusion bias
towards the center of a B edge (see fig. \ref{BIAS_GK}(b)) 
assists the formation of width restricted B stripes. The increased probability for 
particles to be found near the rims of A edges causes the observed broader structures.

Within this framework also the dependency of the B stripes width on $\varepsilon$ and the used potential is understood. 
Figure \ref{BIAS} shows the relative difference $\Delta E_{edge}$ 
between the minimum and the maximum value of the barrier 
for diffusion in one direction of a B particle attached to a B edge. The values are taken for a fixed stripe width 
($l=11$). For the considered potentials increases  $\Delta E_{edge}$ with increasing misfit, i.e. the influence of 
the bias becomes stronger. It is also seen from figure \ref{BIAS} that at a given misfit the bias is the stronger the 
shallower the potential is. Both observations correspond well with the behavior of the stripe width as a function 
of misfit and potential (see fig. \ref{QANT}).  

\section{Conclusions and outlook}
In this chapter we have demonstrated that our off--lattice method is capable of 
simulating multi--component growth in $2+1$ dimensions. 
It was shown that
surface confined alloying of the two adsorbate species is indeed a possible 
strain relaxation mechanism. By means of equilibrium simulations the competition between 
strain and binding energy was found to yield regular stripe patterns, similar to the ones 
reported in experiments. 

But - as our lattice simulations proved - the edge diffusion barrier between regions of different 
particle types can already cause a stripe--like separation, without any consideration 
of an atomic mismatch. However, it seems clear that 
strain effects cause both, a restriction of the stripe width and a pronounced asymmetry in the behavior of the two
different particle types. Such an asymmetry, where one particle type forms the backbone of a ramified island and the 
other adsorbate species gathers inbetween these branches is also reported in experimental studies 
(e.g. \cite{Hwang:1996:CIS}). 
We have identified the different edge diffusion behavior of both adsorbate types as an important 
driving force for the observed island morphology.

Despite the success in describing multi--component growth within our off--lattice model, several interesting 
questions still remain. 
One important issue is the influence of the lattice structure on the made observations. 
Especially the island morphology
and the pattern formation in the equilibrium simulations are of interest here.
It will be necessary to extend the method to more realistic lattice symmetries (e.g. fcc).

Also, the use of more material specific interaction potentials is an interesting task. Along with a 
realistic lattice structure this could allow for the prediction of phenomena in real systems. 
The application of so--called EAM (embedded atom method) many body potentials would  be 
of particular relevance to metallic systems.
Since most of the experimental systems exhibit misfits $|\varepsilon| \geq 8\%$, dislocations are supposed to be of
some importance. The simulations should therefore be extended beyond the lattice based method. This would 
also facilitate investigations on the competition between misfit dislocations and surface alloying as possible 
strain relaxation mechanisms in the submonolayer regime.

  \cleardoublepage
\appendix 
\chapter{Pair--potentials used in this work}
\label{AP-1}
In this work pair--potentials $U_{ij}$ which depend on 
the distance $|\vec{r}_{ij}|$ between two particles $i$ and $j$ of the crystal 
are used to compute the binding and transition energies for surface diffusion.

The binding of two atoms - which is caused by
changes in the electron density - is modeled by an approximation $U_{ij}$ of the interactions
within the two--particle system. Many particle--interactions are neglected.
This becomes problematic for the case of metallic systems where the bonding
electrons are distributed over a large number of atoms and the binding--strength 
between two atoms is likely to depend highly on the local arrangement of the crystal
(see e.g. \cite{Nick:Wilson}).

However, pair--potentials are  numerically easy to handle and provide the possibility of 
real--time calculations in KMC simulations. Since the aim of this work is not to model 
specific materials realistically but to gain general insight into relevant mechanisms 
of heteroepitaxial growth the advantages of pair--potentials
outweigh their disadvantages.
\section{Lennard--Jones potential}
\label{AP-LJ}
The Lennard--Jones $n,m$ potential \cite{Lennard} for two particles $i$ and $j$ is given by
\begin{equation}
\label{LJnm}
 {U}_{ij} =\ 4 E_{ij} \left[\left(\frac{{\sigma}_{ij}}{r_{ij}}\right)^{n}
-\left(\frac{{\sigma}_{ij}}{r_{ij}}\right)^{m}\right], n>m. 
\end{equation}
It has an attractive tail for large particle distance $r_{ij}$ and reaches a minimum
at the equilibrium distance
\begin{equation}
\label{LJnm_minimum}
 r_{0}=\sqrt[n-m]{\frac{n}{m}}{\sigma}_{ij},
\end{equation}
where the depth of the potential is given by $-E_{ij}$ for the case $n=2m$.
The potential is zero at $r_{ij}={\sigma}_{ij}$ and becomes
strongly repulsive with decreasing particle distance \cite{Ercolessi:20xx:MDP}.
\begin{figure}[htb]
\centerline{
\begin{minipage}{0.5 \textwidth}
  \epsfxsize= 0.99\textwidth
  \epsffile{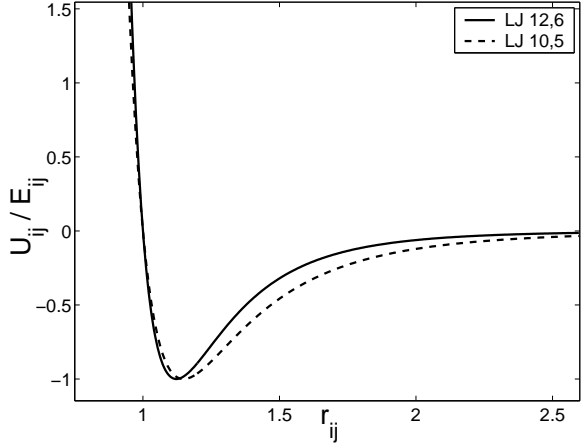}
\end{minipage}
\hfill
\begin{minipage}{0.45 \textwidth}
\caption{Lennard--Jones $12,6$ and $10,5$ potential for ${\sigma}_{ij}=1$.}
\label{LJ_FIG}
\end{minipage}}
\end{figure} 

The most common form of the Lennard--Jones potential is the $12,6$ form,
where the attractive part $\propto 1/{r}_{ij}^6$ is motivated by dipole--dipole
interactions due to fluctuating dipoles. These weak interactions dominate the bonding
character of rare gases.
The repulsive $\propto 1/{r}_{ij}^{12}$ term represents the repulsion between
the electron clouds of the two atoms. The exponent $n=12$ is chosen just for 
practical reasons since it allows a very {\it cheap} computation of the potential (see {\it e.g.}
\cite{Ercolessi:20xx:MDP}).
According to equation (\ref{LJnm_minimum}) $r_0 \propto \sigma_{ij}$ and the misfit $\varepsilon$ 
can be applied to the system by an appropriate choice of $\sigma_{ij}$: for substrate--substrate interaction we
choose $\sigma_{ij}=\sigma_{s}=1$, the adsorbate--adsorbate interaction is given according to 
$\sigma_{ij}=\sigma_{a}=1+\varepsilon$ and finally we set $\sigma_{ij}=\sigma_{as}=1+\varepsilon/2$ for
the  adsorbate--substrate interaction.
\section{Morse potential}
\label{AP-MORSE}
Another pair--potential used in this work is the Morse potential \cite{Morse:1929:DMA} given by
\begin{equation}
\label{Morse}
 {U}_{ij} =\ E_{ij} e^{a(\sigma_{ij}-r_{ij})} \left(e^{a(\sigma_{ij}-r_{ij})}-2\right).
\end{equation}
Like the Lennard--Jones potential it is attractive at large particle distances $r_{ij}$ 
and reaches a minimum at the equilibrium distance $r_0=\sigma_{ij}$,
where the depth of the potential is given by $-E_{ij}$.
\begin{figure}[hbt]
\centerline{
\begin{minipage}{0.5 \textwidth}
  \epsfxsize= 0.99\textwidth
  \epsffile{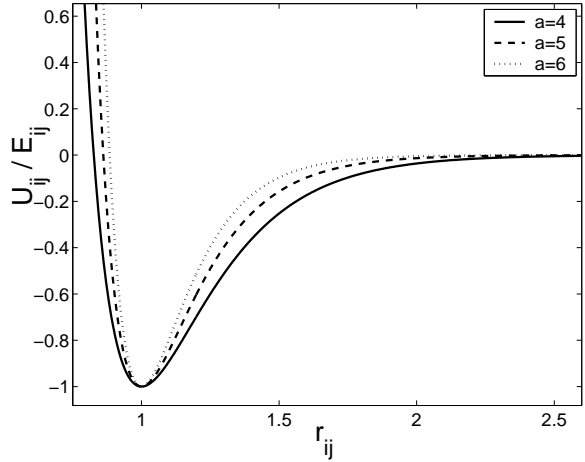}
\end{minipage}
\hfill
\begin{minipage}{0.45 \textwidth}
\caption{Morse potential for ${b}_{ij}=1$ and different values of $a$.}
\label{MORSE_FIG}
\end{minipage}}
\end{figure} 

In contrast to the Lennard--Jones potential the Morse potential remains finite in the repulsive
part  $r_{ij} \to 0$ although it may become very large provided $e^{a\sigma_{ij}}>>2$.
In this work the Morse potential is used for values of $a$ between $4.0$ and $7.0$. 
As figure \ref{MORSE_FIG} displays the potential becomes the steeper both for repulsive and
attractive part, the larger the value of $a$ is. In this way larger values of $a$ lead to a better 
localization of the particles in the crystal.

By using the experimental values for the energy of vaporization, the lattice constant and the compressibility, 
the Morse potential was fitted to a number of fcc and bcc metals \cite{Girifalco:1959:AMP}. This leads to 
values for $a$ between $1.0$ and $3.0$. However, as mentioned above the
application of pair--potentials to metallic systems is problematic.
Therefore, fitted potentials give only a rough estimate of the binding energies in the bulk lattice.

In order to introduce the misfit $\varepsilon$ into an heteroepitaxial system 
we choose $\sigma_{ij}=\sigma_{s}=1$, $\sigma_{ij}=\sigma_{a}=1+\varepsilon$ 
and $\sigma_{ij}=\sigma_{as}=1+\varepsilon/2$
for the substrate--substrate, adsorbate--adsorbate and adsorbate--substrate interaction, respectively.
\section{Simple cubic lattice}
\label{AP-CUBIC}
\begin{figure}[htb]
\centerline{
\begin{minipage}{0.50 \textwidth}
  \epsfxsize= 0.99\textwidth
  \epsffile{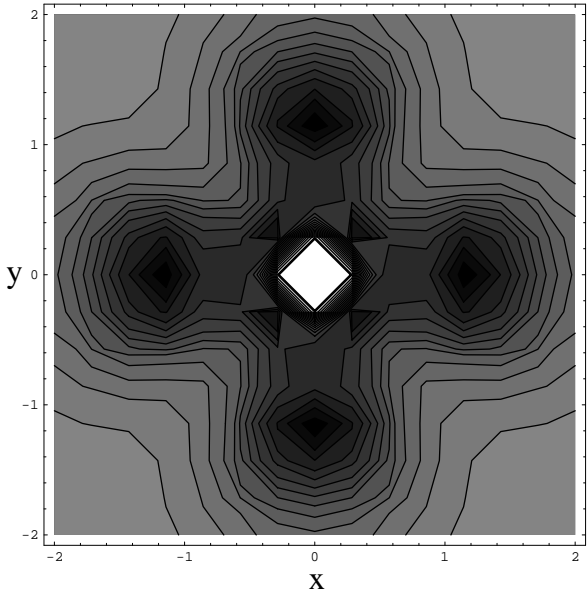}
\end{minipage}
\hfill
\begin{minipage}{0.45 \textwidth}
\caption{Contour plot of a cubic Lennard--Jones $12,6$ potential with $z=0.7$. Darker regions 
correspond to lower values of $U_{cub}$.}
\label{CUBIC_FIG}
\end{minipage}}
\end{figure} 
Due to the isotropy of the central force potentials (\ref{LJnm}) and (\ref{Morse}) particles
arrange into {\it triangular} or {\it fcc} lattices for the $2d$ and $3d$ case, respectively.
However, sometimes a simple cubic lattice structure is favorable for growth simulations. 
Although this lattice structure does not represent any relevant material it has some advantages
for basic examinations. Due to the lower coordination number less particles
have to be taken into account for energy calculations than in a closed--packed lattice. 
There are only four possible in--plane transition sites per 
particle, in comparison with six transition sites in case of, e.g., the fcc lattice. 
This makes faster calculations of activation energies possible and allows for larger system sizes and more 
simulation runs.

In order to stabilize a simple cubic lattice in our simulations an anisotropic potential,
\begin{equation}
\label{CUBIC}
 {U}_{cub} = \left(0.1 + 8\,\left( \frac{x_{ij}^2}{r_{ij}^2}-\frac{1}{2} \right) 
\left( \frac{y_{ij}^2}{r_{ij}^2}-\frac{1}{2} \right)
\left( \frac{z_{ij}^2}{r_{ij}^2}-\frac{1}{2} \right)\right)U_{ij}
\end{equation}
is used \cite{Schroeder:1997:DSS}. $U_{ij}$ denotes one of the central force potentials (\ref{LJnm}), 
(\ref{Morse}). The particle distance $\vec{r}_{ij}$ is given in Cartesian coordinates 
($x_{ij}$, $y_{ij}$, $z_{ij}$) where growth proceeds in $z$--direction.
Figure \ref{CUBIC_FIG} shows a contour plot for the Lennard--Jones $12,6$ type of the cubic 
potential (\ref{CUBIC})
with $z=0.7$.
The binding sites around the central particle are arranged in a cubic symmetry with the
local minima along the lattice directions.

\addcontentsline{toc}{chapter}{Bibliography}
\bibliography{bibliography}

\begin{thebibliography}{100}

\bibitem{Barabasi:1995:FCS}
A.-L. Barab\'{a}si and H.E. Stanley.
\newblock {\em Fractal concepts in surface growth}.
\newblock Cambridge University Press, Cambridge, Great Britain, 1995.

\bibitem{Zangwill:1992:PS}
A.~Zangwill.
\newblock {\em Physics at surfaces}.
\newblock Cambridge University Press, Cambridge, Great Britain, 1992.

\bibitem{Pimpinelli:1998:PCG}
A.~Pimpinelli and J.~Villain.
\newblock {\em Physics of Crystal Growth}.
\newblock Cambridge University Press, 1998.

\bibitem{Ahr:2002:SPE}
M.~Ahr.
\newblock {\em Surface properties of epitaxially grown crystals}.
\newblock PhD thesis, Bayerische Julius-Maximilians-Universit{\"a}t
  W{\"u}rzburg, W{\"u}rzburg, 2002.

\bibitem{Volkmann:Phd}
T.~Volkmann.
\newblock PhD thesis, Bayerische Julius-Maximilians-Universit{\"a}t
  W{\"u}rzburg, W{\"u}rzburg, 2004.

\bibitem{Spjut:1994:CSS}
H.~Spjut and D.A. Faux.
\newblock Computer simulation of strain-induced diffusion enhancement of {Si}
  adatoms on the {Si}(001) surface.
\newblock {\em Surf. Sci.}, 306:233, 1994.

\bibitem{Kratzer:2001:SKT}
P.~Kratzer and M.~Scheffler.
\newblock Surface knowledge toward a predictive theory of material.
\newblock {\em Comp. Sci. Eng.}, 3(6):16, 2001.

\bibitem{Jensen:1999:ICC}
F.~Jensen.
\newblock {\em Introduction to computational chemistry}.
\newblock John Wiley and Sons, New York, 1999.

\bibitem{Schwoebel:1966:SMC}
R.L. Schwoebel and E.J. Shipsey.
\newblock Step motion on crystal surfaces.
\newblock {\em J. Appl. Phys.}, 37(10):3682, 1966.

\bibitem{Schwoebel:1969:SMC}
R.L. Schwoebel.
\newblock Step motion on crystal surfaces.
\newblock {\em II. J. Appl. Phys.}, 40(2):614, 1969.

\bibitem{Schroeder:1997:DSS}
M.~Schroeder and D.E. Wolf.
\newblock Diffusion on strained surfaces.
\newblock {\em Surf. Sci.}, 375:129, 1997.

\bibitem{Cherepanov:2002:ISD}
V.~Cherepanov and B.~Voigtl{\"a}nder.
\newblock Influence of strain on diffusion at {Ge(111)} surfaces.
\newblock {\em Appl. Phys. Lett.}, 81(25):4745, 2002.

\bibitem{Penev:2001:ESS}
E.~Penev, P.~Kratzer, and M.~Scheffler.
\newblock Effect of strain on surface diffusion in semiconductor heteroepitaxy.
\newblock {\em Phys. Rev. B}, 64:085401, 2001.

\bibitem{Wrigley:1980:SDA}
J.D. Wrigley and G.~Ehrlich.
\newblock Surface diffusion by an atomic exchange mechanism.
\newblock {\em Phys. Rev. Lett.}, 44(10):661, 1980.

\bibitem{Politi:2000:ICG}
P.~Politi, G.~Grenet, A.~Marty, A.~Ponchet, and J.~Villain.
\newblock Instabilities in crystal growth by atomic or molecular beams.
\newblock {\em Physics Reports}, 324:271, 2000.

\bibitem{Hirth:1968:TD}
J.P. Hirth and J.~Lothe.
\newblock {\em Theory of Dislocations}.
\newblock McGraw-Hill Book Company, 1968.

\bibitem{Matthews:1974:DEM}
J.W. Matthews and A.E. Blakeslee.
\newblock Defects in epitaxial multilayers.
\newblock {\em J. Cryst. Growth}, 27:118, 1974.

\bibitem{Matthews:1975:DAA}
J.W. Matthews.
\newblock Defects associated with the accommodation of misfit between crystals.
\newblock {\em J. Vac. Sci. Technol.}, 12(1):116, 1975.

\bibitem{Trushin:2003:EAM}
O.~Trushin, E.~Granato, S.C. Ying, P.~Salo, and T.~Ala-Nissila.
\newblock Energetics and atomic mechanisms of dislocation nucleation in
  strained epitaxial layers.
\newblock {\em cond-mat}, 0308186, 2003.

\bibitem{Bauer:1986:SGC}
E.~Bauer and J.H. van~der Merwe.
\newblock Structure and growth of crystalline superlattices: From monolayer to
  superlattice.
\newblock {\em Phys. Rev. B}, 33(6):3657, 1986.

\bibitem{Chambers:2000:EGP}
S.A. Chambers.
\newblock Epitaxial growth and properties of thin film oxides.
\newblock {\em Surf. Sci. Rep.}, 39:105, 2000.

\bibitem{Seifert:1996:SGQ}
W.~Seifert, N.~Carlsson, M.~Millera, M.-E. Pistol, L.~Samuelson, and L.R.
  Wallenberg.
\newblock In-situ growth of quantum dot structures by the
  {S}transki-{K}rastanow growth mode.
\newblock {\em Prog. Crystal Growth and Charact.}, 33:423, 1996.

\bibitem{Hull:2000:TFH}
R.~Hull and E.A. Stach.
\newblock {\em Thin Films: Heteroepitaxial Systems}, chapter~7, page 299.
\newblock World Scientific, Singapore, 2000.

\bibitem{Dutartre:1994:DFS}
D.~Dutartre, P.~Warren, F.~Chollet, F.~Gisbert, M.~B{\'e}renguer, and
  I.~Berb{\'e}zier.
\newblock Defect-free {S}transki-{K}rastanov growth of strained
  {Si}$_{1-x}${Ge}$_{x}$ layers on {Si}.
\newblock {\em J. Cryst. Growth}, 142:78, 1994.

\bibitem{Landin:1999:OSI}
L.~Landin.
\newblock {\em Optical studies of {InAs} quantum dots in {III}-{V}
  semiconductors}.
\newblock PhD thesis, School of Information Science, Computer and Electrical
  Engineering, 1999.

\bibitem{Stoyanov:1982:TET}
S.~Stoyanov and I.~Markov.
\newblock On the 2d-3d transition in epitaxial thin film growth.
\newblock {\em Surf. Sci.}, 116:313, 1982.

\bibitem{Korutcheva:2000:CSK}
E.~Korutcheva, A.M. Turiel, and I.~Markov.
\newblock Coherent {Stranski}--{Krastanov} growth in 1+1 dimensions with
  anharmonic interactions: An equilibrium study.
\newblock {\em Phys. Rev. B}, 61(24):16890, 2000.

\bibitem{Kuo:2002:FSA}
M.C. Kuo, C.S. Yang, P.Y. Tseng, J.~Lee, J.L. Shen, W.C. Chou, Y.T. Shih, C.T.
  Ku, M.C. Lee, and W.K. Chen.
\newblock Formation of self-assembled {ZnTe} quantum dots on {ZnSe} buffer
  layer grown on {GaAs} substrate by molecular beam epitaxy.
\newblock {\em J. Cryst. Growth}, 242:533, 2002.

\bibitem{Evans:1995:EGM}
M.M.R. Evans, J.C. Glueckstein, and J.~Nogami.
\newblock Epitaxial growth of manganese on silicon: {Volmer}-{Weber} growth on
  the {Si}(111) surface.
\newblock {\em Phys. Rev. B}, 53(7):4000, 1995.

\bibitem{Raviswaran:2001:ECI}
A.~Raviswaran, C.-P. Liu, J.~Kim, D.G. Cahill, and J.M. Gibson.
\newblock Evolution of coherent islands during strained-layer {Volmer}-{Weber}
  growth of {Si} on {Ge}(111).
\newblock {\em Phys. Rev. B}, 63:125314, 2001.

\bibitem{Tersoff:1995:SCA}
J.~Tersoff.
\newblock Surface--confined alloy formation in immiscible systems.
\newblock {\em Phys. Rev. Lett.}, 74(3):434, 1995.

\bibitem{Neugebauer:1993:MIF}
J.~Neugebauer and M.~Scheffler.
\newblock Mechanisms of island formation of alkali-metal adsorbates on
  {Al(111)}.
\newblock {\em Phys. Rev. Lett.}, 71(4):577--580, 1993.

\bibitem{Stampfl:1992:ISM}
C.~Stampfl, M.~Scheffler, H.~Over, J.~Burchhardt, M.~Nielsen, D.L. Adams, and
  W.~Morit.
\newblock Identification of stable and metastable adsorption sites of {K}
  adsorbed on {Al(111)}.
\newblock {\em Phys. Rev. Lett.}, 69(10):1532, 1992.

\bibitem{Nielsen:1993:IGA}
L.~Pleth Nielsen, F.~Besenbacher, I.~Stensgaard, E.~Laegsgaard, C.~Engdahl,
  P.~Stoltze, K.W. Jacobsen, and J.K. N{\o}rskov.
\newblock Initial growth of {Au} on {Ni(110)}: Surface alloying of immiscible
  metals.
\newblock {\em Phys. Rev. Lett.}, 71(754-757):5, 1993.

\bibitem{Roder:1993:MCM}
H.~R{\"o}der, R.~Roder, H.~Brune, and K.~Kern.
\newblock Monolayer-confined mixing at the {Ag-Pt(111)} interface.
\newblock {\em Phys. Rev. Lett.}, 71(13):2086, 1993.

\bibitem{Oppo:1993:TAS}
S.~Oppo, V.~Fiorentini, and M.~Scheffler.
\newblock Theory of adsorption and surfactant effect of {Sb} on {Ag(111)}.
\newblock {\em Phys. Rev. Lett.}, 71(15):2437, 1993.

\bibitem{Hwang:1996:CIS}
R.Q. Hwang.
\newblock Chemically induced step edge diffusion barriers: Dendritic growth in
  2d alloys.
\newblock {\em Phys. Rev. Lett.}, 76(25):4757, 1996.

\bibitem{Hwang:1997:STM}
R.Q. Hwang and M.C. Bartelt.
\newblock Scanning tunneling microscopy studies of metal on metal epitaxy.
\newblock {\em Chem. Rev.}, 97:1063, 1997.

\bibitem{Thayer:2001:RST}
G.E. Thayer, V.~Ozolins, A.K. Schmid, N.C. Bartelt, M.~Asta, J.J. Hoyt,
  S.~Chiang, and R.Q. Hwang.
\newblock Role of stress in thin film alloy thermodynamics: Competition between
  alloying and dislocation formation.
\newblock {\em Phys. Rev. Lett.}, 86(4):660, 2001.

\bibitem{Thayer:2002:LSS}
G.E. Thayer, N.C. Bartelt, V.~Ozolins, A.K. Schmid, S.~Chiang, and R.Q. Hwang.
\newblock Linking surface stress to surface structure: Measurement of atomic
  strain in a surface alloy using scanning tunneling microscopy.
\newblock {\em Phys. Rev. Lett.}, 89(3):036101, 2002.

\bibitem{Tober:1998:SAL}
E.D. Tober, R.F.C. Farrow, R.F. Marks, G.~Witte, K.~Kalki, and D.D. Chambliss.
\newblock Self--assembled lateral multilayers from thin film alloys of
  immiscible metals.
\newblock {\em Phys. Rev. Lett.}, 81(9):1897, 1998.

\bibitem{Stevens:1995:SSA}
J.L. Stevens and R.Q. Hwang.
\newblock Strain stabilized alloying of immiscible metals in thin films.
\newblock {\em Phys. Rev. Lett.}, 74(11):2078, 1995.

\bibitem{Sadigh:1999:SRO}
B.~Sadigh, M.~Asta, V.~Ozolins, N.C. Bartelt, A.A. Quong, and R.Q. Hwang.
\newblock Short--range order of phase stability of surface alloys: {PdAu} on
  {Ru(0001)}.
\newblock {\em Phys. Rev. Lett.}, 83(7):1379, 1999.

\bibitem{Scheffler:2000:ITS}
M.~Scheffler, P.~Kratzer, and L.G. Wang.
\newblock Ab initio thermodynamics and statistics of semiconductor growth and
  self--assembly of quantum dots.
\newblock In {\em Proceedings of the 4th Symposium an Atom--Scale Surface and
  Interface Dynamics}, Tsukuba, Japan, 2000.

\bibitem{Moll:1998:ISS}
N.~Moll, M.~Scheffler, and E.~Pehlke.
\newblock Influence of surface stress on the equilibrium shape of strained
  quantum dots.
\newblock {\em Phys. Rev. B}, 58(7):4566, 1998.

\bibitem{Abraham:1998:MOA}
Y.B. Abraham.
\newblock Multivariable optimization approach for the construction of many-body
  potentials.
\newblock http://www.wfu.edu/\symbol{126}abray00g/root1/root1.html, 1998.

\bibitem{Tersoff:1988:NEA}
J.~Tersoff.
\newblock New empirical approach for the structure and energy of covalent
  systems.
\newblock {\em Phys. Rev. B}, 37(12):6991, 1988.

\bibitem{Dong:1998:SRM}
L.~Dong, J.~Schnitker, R.W. Smith, and D.J. Srolovitz.
\newblock Stress relaxation and misfit dislocation nucleation in the growth of
  misfitting films: A molecular dynamics simulation study.
\newblock {\em J. Apll. Phys.}, 83(1):217, 1998.

\bibitem{Patriarca:2002:MMD}
M.~Patriarca, M.~Robles, and K.~Kaski.
\newblock Microscopic methods for dislocation tracking.
\newblock cond-mat/0212318, 2002.

\bibitem{Bailey:2000:DMT}
N.P. Bailey, J.P. Sethna, and C.R. Myers.
\newblock Dislocation mobility in two--dimensional {Lennard--Jones} material.
\newblock cond-mat/0001167, 2000.

\bibitem{Schiotz:1998:SMS}
The 19th Ris{\o} International Symposium on Materials Science: Modelling of
  Structure and Mechanics from Microstructure to Product.
\newblock {\em Simulations of mechanics and structures of nanomaterials - from
  nanoscale to coarser scales}, 1998.

\bibitem{Madhukar:1983:FEV}
A.~Madhukar.
\newblock Far from equilibrium vapor phase growth of lattice matched {III}-{V}
  compound semiconductor interfaces: some basic concepts and {Monte}--{Carlo}
  computer simulations.
\newblock {\em Surf. Sci.}, 132:344, 1983.

\bibitem{Ghaisas:1986:RSM}
S.V. Ghaisas and A.~Madhukar.
\newblock Role of surface molecular reactions influencing the growth mechanism
  and nature of nonequilibrium surfaces: a {Monte}--{Carlo} study of
  molecular--beam epitaxy.
\newblock {\em Phys. Rev. Lett.}, 56:1066, 1986.

\bibitem{Orr:1992:MSI}
B.G. Orr, D.~Kessler, C.W. Snyder, and L.M. Sander.
\newblock A model for strain-induced roughening and coherent island growth.
\newblock {\em Europhys. Lett.}, 19:33, 1992.

\bibitem{Barabasi:1997:SAI}
A.-L. Barab\'{a}si.
\newblock Self--assembled island formation in heteroepitaxial growth.
\newblock {\em Appl. Phys. Lett.}, 70(19):2566, 1997.

\bibitem{Khor:2000:QDS}
K.E. Khor and S.~Das Sarma.
\newblock Quantum dot self--assembly in growth of strained--layer thin films: a
  kinetic {Monte} {Carlo} study.
\newblock {\em Phys. Rev. B}, 62:16657, 2000.

\bibitem{Lam:2002:CRM}
C.H. Lam, C.K. Lee, and L.M. Sander.
\newblock Competing roughening mechanisms in strained heteroepitaxy: A fast
  kinetic {Monte Carlo} study.
\newblock {\em Phys. Rev. Lett.}, 89(21):216102, 2002.

\bibitem{Plotz:1992:MCS}
W.M. Plotz, K.~Hingerl, and H.~Sitter.
\newblock {Monte} {Carlo} simulation of growth.
\newblock {\em Phys. Rev. B.}, 44(20):12122, 1992.

\bibitem{Sitter:1995:MGM}
H.~Sitter.
\newblock {MBE} growth mechanisms; studies by {Monte}--{Carlo} simulation.
\newblock {\em Thin Solid Films}, 267:37, 1995.

\bibitem{Kew:1993:CSM}
J.~Kew, M.R. Wilby, and D.D. Vvedensky.
\newblock Continuous--space {Monte Carlo} simulations of epitaxial growth.
\newblock {\em J. Cryst. Growth}, 127:508, 1993.

\bibitem{Schindler:1999:TAG}
A.~Schindler.
\newblock {\em Theoretical aspects of growth on one and two dimensional
  strained crystal surfaces}.
\newblock PhD thesis, Gerhard-Mercator-Universit{\"a}t Duisburg, Duisburg,
  1999.

\bibitem{Trushin:1997:EBS}
O.S. Trushin, M.~Kotrla, and F.~M\'{a}ca.
\newblock Energy barriers on stepped {Ir}/{Ir}(111) surfaces: a molecular
  statics calculation.
\newblock {\em Surf. Sci}, 389:55, 1997.

\bibitem{Maca:1999:EBD}
F.~M\'{a}ca, M.~Kotrla, and O.S. Trushin.
\newblock Energy barriers for diffusion on stepped {Pt}(111) surface.
\newblock {\em Vacuum}, 54:113, 1999.

\bibitem{Maca:2000:EBD}
F.~M\'{a}ca, M.~Kotrla, and O.S. Trushin.
\newblock Energy barriers for diffusion on stepped {Rh}(111) surface.
\newblock {\em Surf. Sci.}, 454-456:579, 2000.

\bibitem{Liu:1993:DBS}
C.L. Liu and J.B. Adams.
\newblock Diffusion behavior of single adatoms near and at steps during growth
  of metallic thin films on {Ni} surfaces.
\newblock {\em Surf. Sci.}, 294:197, 1993.

\bibitem{Villarba:1994:DMR}
M.~Villarba and H.~J{\`o}nsson.
\newblock Diffusion mechanisms relevant to metal crystal growth: {Pt/Pt(111)}.
\newblock {\em Surf. Sci.}, 317:15, 1994.

\bibitem{Feibelman:1998:ISD}
P.J. Feibelman.
\newblock Interlayer self-diffusion on stepped {Pt}(111).
\newblock {\em Phys. Rev. Lett.}, 81(1):168, 1998.

\bibitem{Feibelman:1999:SDA}
P.J. Feibelman.
\newblock Self-diffusion along step bottoms on {Pt}(111).
\newblock {\em Phys. Rev. B}, 60(7):4972, 1999.

\bibitem{Press:1992:NRC}
W.H. Press, S.A.Teukolsky, W.T. Vetterling, and B.P. Flannery.
\newblock {\em Numerical Recipes in C}.
\newblock Cambridge University Press, Cambridge, second edition, 1992.

\bibitem{Vey:Diplom}
C.~Vey.
\newblock Master's thesis, Lehrstuhl f\"ur theoretische Physik {III},
  Universit\"at W\"urzburg, 2004.

\bibitem{Barkema:Mund}
G.T. Barkema.
\newblock private communication.

\bibitem{Barkema:1996:EBR}
G.T. Barkema and N.~Mousseau.
\newblock Event--based relaxation of continuous disordered systems.
\newblock {\em Phys. Rev. Lett.}, 77(21):4358, 1996.

\bibitem{Mousseau:1998:TTP}
N.~Mousseau and G.T. Barkema.
\newblock Traveling through potential energy landscapes of disordered
  materials: the activation--relaxation technique.
\newblock {\em Phys. Rev. E}, 57(2):2419, 1998.

\bibitem{Malek:2000:DLJ}
R.~Malek and N.~Mousseau.
\newblock Dynamics of {L}ennard--{J}ones clusters: A characterization of the
  activation-relaxation technique.
\newblock {\em Phys. Rev. E}, 62(6):7723, 2000.

\bibitem{Shiang:1994:MDSa}
K.D. Shiang, C.M. Wei, and T.T. Tsong.
\newblock A molecular dynamics study of self--diffusion on metal surfaces.
\newblock {\em Surf. Sci.}, 30:136, 1994.

\bibitem{Voter:1984:TST}
A.F. Voter and J.D. Doll.
\newblock Transition state theory description of surface self--diffusion:
  Comparison with classical trajectory results.
\newblock {\em J. Chem. Phys.}, 80:5832, 1984.

\bibitem{Shiang:1994:MDSb}
K.D. Shiang and T.T. Tsong.
\newblock Molecular--dynamics study of self--diffusion: Iridium dimers on
  iridium surfaces.
\newblock {\em Phys. Rev. B}, 49(1):7670, 1994.

\bibitem{Fu:1996:SDD}
T.Y. Fu, Y.R. Tzeng, and T.T. Tsong.
\newblock Self--diffusion and dynamic behavior of atoms at step edges of
  iridium surfaces.
\newblock {\em Phys. Rev. B}, 54(8):5932, 1996.

\bibitem{Iijima:1997:MDS}
T.~Iijima and O.~Sugino.
\newblock Molecular dynamics study of adatom diffusion on {Si(100)} surface -
  importance of the exchange mechanism.
\newblock {\em Surf. Sci.}, 391:1199, 1997.

\bibitem{Newman:1999:MCM}
M.E.J. Newman and G.T. Barkema.
\newblock {\em {Monte Carlo} methods in statistical physics.}
\newblock Clarendon Press, Oxford, 1999.

\bibitem{Pinardi:1998:CTS}
K.~Pinardi, U.~Jain, S.C. Jain, H.E. Maes, R.~Van Overstraeten, and
  M.~Willander.
\newblock Critical thickness and strain relaxation in lattice mismatched
  {II-VI} semiconductor layers.
\newblock {\em J. Appl. Phys.}, 83(9):4724, 1998.

\bibitem{Sabiryanov:2003:SDG}
R.F. Sabiryanov, M.I. Larsson, K.J. Cho, W.D. Nix, and B.M. Clemens.
\newblock Surface diffusion and growth of patterned nanostructures on strained
  surfaces.
\newblock {\em Phys. Rev. B}, 67:125412, 2003.

\bibitem{Ashcroft:1976:SSP}
N.W. Ashcroft and N.D. Mermin.
\newblock {\em Solid State Physics}.
\newblock Saunders College Publishing, Philadelphia, 1976.

\bibitem{Strandburg:1988:TDM}
K.J. Strandburg.
\newblock Two-dimensional melting.
\newblock {\em Rev. Mod. Phys.}, 60:161, 1988.

\bibitem{Frank:1949:ODDa}
F.C. Frank and J.H. van~der Merwe.
\newblock One-dimensional dislocations. 1. static theory.
\newblock {\em Proc. Roy. Soc A}, 198:205, 1949.

\bibitem{Frank:1949:ODDb}
F.C. Frank and J.H. van~der Merwe.
\newblock One-dimensional dislocations. 2. misfitting monolayers and oriented
  overgrowth.
\newblock {\em Proc. Roy. Soc A}, 198:216, 1949.

\bibitem{Liu:1999:FMD}
X.W. Liu, A.A. Hopgood, B.F. Usher, H.~Wang, and N.St.J. Braithwaite.
\newblock Formation of misfit dislocations during growth of
  {In$_x$Ga$_{1-x}$/GaAs} strained--layer heterostructures.
\newblock {\em Semicond. Sci. Technol.}, 14:1154, 1999.

\bibitem{Cohen-Solal:1994:CTH}
G.~Cohen-Solal, F.~Bailly, and M.~Barb\'{e}.
\newblock Critical thickness in heteroepitaxial growth of zinc-blende
  semiconductor compounds.
\newblock {\em J. Cryst. Growth}, 138:68, 1994.

\bibitem{Tsao:1987:CSS}
J.Y. Tsao, B.W. Dodson, S.T. Picraux, and D.M. Cornelison.
\newblock Critical stresses for {Si$_x$Ge$_{1-x}$} strained-layer plasticity.
\newblock {\em Phys. Rev. Lett.}, 59(21):2455, 1987.

\bibitem{Zou:1996:TDG}
J.~Zou, D.J.H. Cockayne, and B.F. Usher.
\newblock Temperature-dependent generation of misfit dislocations in
  {In$_{0.2}$Ga$_{0.8}$/GaAs} single heterostructures.
\newblock {\em Appl. Phys. Lett.}, 68(5):673, 1996.

\bibitem{Bailly:1995:SMD}
F.~Bailly, M.~Barb\'{e}, and G.~Cohen-Solal.
\newblock Setting up of misfit dislocations in heteroepitaxial growth and
  critical thickness.
\newblock {\em J. Cryst. Growth}, 153:115, 1995.

\bibitem{Bader:2003:RTS}
A.S. Bader, W.~Faschinger, C.~Schumacher, J.~Guerts, and L.W. Molenkamp.
\newblock Real-time in situ {X}-ray diffraction as a method to control
  epitaxial growth.
\newblock {\em Appl. Phys. Lett.}, 82(26):4684, 2003.

\bibitem{Faschinger:2003:IVH}
W.~Faschinger, R.~Neder, and J.~Guerts.
\newblock {\em {II-VI} Halbleiter Wachstumsmechanismen niedrig dimensionaler
  Strukturen und Grenzfl{\"a}chen}, chapter~B2, pages B2--1.
\newblock Sonderforschungsbereich 410, Universit{\"a}t W{\"u}rzburg, 2003.

\bibitem{Evans:1991:FMS}
J.W. Evans.
\newblock Factors mediating smoothness in epitaxial thin-film growth.
\newblock {\em Phys. Rev. B}, 43(5):3897, 1991.

\bibitem{Evans:1990:LTE}
J.W. Evans, D.E. Sanders, P.A. Thiel, and A.E. DePristo.
\newblock Low-temperature epitaxial growth of thin metal films.
\newblock {\em Phys. Rev. B}, 41(8):5410, 1990.

\bibitem{Gilmore:1991:MDS}
C.M. Gilmore and J.A. Sprague.
\newblock Molecular-dynamics simulation of the energetic deposition of {Ag}
  thin films.
\newblock {\em Phys. Rev. B}, 44(16):8950, 1991.

\bibitem{Yue:1998:MDS}
Y.~Yue, Y.K. Ho, and Z.Y. Pan.
\newblock Molecular-dynamics study of transient-diffusion mechanisms in
  low-temperature epitaxial growth.
\newblock {\em Phys. Rev. B}, 57(11):6685, 1998.

\bibitem{Heyn:2001:CCSa}
Ch. Heyn.
\newblock Critical coverage for strain-induced formation of {InAs} quantum
  dots.
\newblock {\em Phys. Rev. B}, 64:165306, 2001.

\bibitem{Cullis:2002:SKT}
A.G. Cullis, D.J. Norris, T.~Walther, M.A. Migliorato, and M.~Hopkinson.
\newblock {Stranski-Krastanow} transition and epitaxial island growth.
\newblock {\em Phys. Rev. B}, 66:081305, 2002.

\bibitem{Chkoda:2003:TDM}
L.~Chkoda, M.~Schneider, V.~Shklover, L.~Kilian, M.~Sokolowski, C.~Heske, and
  E.~Umbach.
\newblock Temperature-dependent morphology and structure of ordered
  {3,4,9,10-perylene-tetracarboxylicacid-dianhydride} ({PTCDA}) thin films on
  {Ag(111)}.
\newblock {\em Chem. Phys. Lett.}, 371:548, 2003.

\bibitem{Sutter:2000:NTD}
P.~Sutter and M.G. Lagally.
\newblock Nucleationless three--dimensional island formation in low--misfit
  heteroepitaxy.
\newblock {\em Phys. Rev. Lett.}, 84(20):4637, 2000.

\bibitem{Kastner:1999:KSL}
M.~K{\"a}stner and B.~Voigtl{\"a}nder.
\newblock Kinetically self--limiting growth of {Ge} islands on {Si(001)}.
\newblock {\em Phys. Rev. Lett.}, 82(13):2745, 1999.

\bibitem{Roland:1993:GGF}
C.~Roland and G.H. Gilmer.
\newblock Growth of germanium films on {Si(001)} substrates.
\newblock {\em Phys. Rev. B}, 47(24):16286, 1993.

\bibitem{Osipov:2002:SDN}
A.V. Osipov, F.~Schmitt, S.A. Kukushkin, and P.~Hess.
\newblock Stress-driven nucleation of coherent islands: theory and experiment.
\newblock {\em Appl. Surf. Sci.}, 188:156, 2002.

\bibitem{Raiteri:2002:SSE}
P.~Raiteri, F.~Valentinotti, and L.~Miglio.
\newblock Stress, strain and elastic energy at nanometric {Ge} dots on
  {Si(001)}.
\newblock {\em Appl. Surf. Sci.}, 188:4, 2002.

\bibitem{Schittenhelm:1998:SAG}
P.~Schittenhelm, C.~Engel, F.~Findeis, G.~Abstreiter, A.A. Darhuber, G.~Bauer,
  A.O. Kosogov, and P.~Werner.
\newblock Self--assembled {Ge} dots: Growth, characterization, ordering and
  applications.
\newblock {\em J. Vac. Sci. Technol. B.}, 16(3):1575, 1998.

\bibitem{Thanh:1999:FSA}
V.L. Thanh, P.~Boucaud, Y.~Zheng, A.~Younsi, D.~D{\'e}barre, D.~Bouchier, and
  J.-M. Lourtioz.
\newblock On the formation of self--assembled {Ge}/{Si(001)} quantum dots.
\newblock {\em J. Cryst. Growth}, 201:1212, 1999.

\bibitem{Johansson:2002:KSAa}
J.~Johansson and W.~Seifert.
\newblock Kinetics of self--assembled island formation: part {II}--island size.
\newblock {\em J. Cryst. Growth}, 234:139, 2002.

\bibitem{Johansson:1998:MSD}
J.~Johansson, N.~Carlsson, and W.~Seifert.
\newblock Manipulations of size and density of self--assembled quantum dots
  grown by {MOVPE}.
\newblock {\em Physica E}, 2:667, 1998.

\bibitem{Johansson:2002:KSAb}
J.~Johansson and W.~Seifert.
\newblock Kinetics of self--assembled island formation: part {I}--island
  density.
\newblock {\em J. Cryst. Growth}, 234:132, 2002.

\bibitem{Carlsson:1994:STD}
N.~Carlsson, W.~Seifert, A.~Petersson, P.~Castrillo, M.-E. Pistol, and
  L.~Samuelson.
\newblock Study of the two--dimensional--three--dimensional growth mode
  transition in metalorganic vapor phase epitaxy of {GaInP/InP} quantum--sized
  structures.
\newblock {\em Appl. Phys. Lett.}, 65(24):3093, 1994.

\bibitem{Seifert:1997:SGN}
W.~Seifert, N.~Carlsson, J.~Johansson, M.-E. Pistol, and L.~Samuelson.
\newblock In situ growth of nano--structures by metal--organic vapour phase
  epitaxy.
\newblock {\em J. Cryst. Growth}, 170:39, 1997.

\bibitem{Snyder:1991:ESS}
C.W. Snyder, B.G. Orr, D.~Kessler, and L.M. Sander.
\newblock Effect of strain on surface morphology in highly strained {InAsGa}
  films.
\newblock {\em Phys. Rev. Lett.}, 66(23):3032, 1991.

\bibitem{Leonard:1994:CLT}
D.~Leonard, K.~Pond, and P.M. Petroff.
\newblock Critical layer thickness for self--assembled {InAs} islands on
  {GaAs}.
\newblock {\em Phys. Rev. B}, 50(16):11687, 1994.

\bibitem{Snyder:1992:KCC}
C.W. Snyder, J.F. Mansfield, and B.G. Orr.
\newblock Kinetically controlled critical thickness for coherent islanding and
  thick highly strained pseudomorphic films of {In$_x$Ga$_{1-x}$As} on
  {GaAs(100)}.
\newblock {\em Phys. Rev. B}, 46(15):9551, 1992.

\bibitem{Heyn:2001:CCSb}
Ch. Heyn.
\newblock Critical coverage for strain--induced formation of {InAs} quantum
  dots.
\newblock {\em Phys. Rev. B}, 64:165306, 2001.

\bibitem{Heyn:2001:FSE}
Ch. Heyn and C.~Dumat.
\newblock Formation and size evolution of self--assembled quantum dots.
\newblock {\em J. Cryst. Growth}, 227:990, 2001.

\bibitem{Heyn:2000:FDI}
Ch. Heyn, D.~Endler, K.~Zhang, and W.~Hansen.
\newblock Formation and dissolution of {InAs} quantum dots on {GaAs}.
\newblock {\em J. Cryst. Growth}, 210:421, 2000.

\bibitem{Geiger:1997:OGM}
M.~Geiger, A.~Bauknecht, F.~Adler, H.~Schweizer, and F.~Scholz.
\newblock Observation of the 2d-3d growth mode transition in the {InAs}/{GaA}
  system.
\newblock {\em J. Cryst. Growth}, 170:558, 1997.

\bibitem{Ruvimov:1995:SCG}
S.~Ruvimov, P.~Werner, K.~Scheerschmidt, U.~G{\"o}sele, J.~Heydenreich,
  U.~Richter, N.N. Ledentsov, M.~Grundmann, D.~Bimberg, V.~M. Ustinov, A.~Yu.
  Egorov, P.S. Kop{\'e}v, and Zh.I. Alferov.
\newblock Structural characterization of {(In,Ga)As} quantum dots in a {GaAs}
  matrix.
\newblock {\em Phys. Rev. B}, 51(20):14766, 1995.

\bibitem{Schikora:2000:ISK}
D.~Schikora, S.~Schwedhelm, D.J. As, K.~Lischka, D.~Litvinov, A.~Rosenauer,
  D.~Gerthsen, M.~Strassburg, A.~Hoffmann, and D.~Bimberg.
\newblock Investigations on the {Stranski--Krastanow} growth of {CdSe} quantum
  dots.
\newblock {\em Appl. Phys. Lett.}, 76(4):418, 2000.

\bibitem{Alchalabi:2003:SAS}
K.~Alchalabi, D.~Zimin, G.~Kostorz, and H.~Zogg.
\newblock Self--assembled semiconductor quantum dots with nearly uniform size.
\newblock {\em Phys. Rev. Lett.}, 90(2):026104, 2003.

\bibitem{Schikora:2000:IFK}
D.~Schikora, S.~Schwedhelm, I.~Kudryashov, K.~Lischka, D.~Litvinov,
  A.~Rosenauer, D.~Gerthsen, M.~Strassburg, A.~Hoffmann, and D.~Bimberg.
\newblock Investigations on the formation kinetics of {CdSe} quantum dots.
\newblock {\em J. Cryst. Growth}, 214:698, 2000.

\bibitem{InAs_web}
http://www.trnmag.com/Stories/101100/Measuring\_Quantum\_Dots\_101100.htm.

\bibitem{Cherepanov:2002:IMS}
V.~Cherepanov and B.~Voigtl{\"a}nder.
\newblock Influence of material, surface reconstruction and strain on diffusion
  at a {Ge(111)} surface.
\newblock {\em preprint}, 2002.

\bibitem{Dobbs:1997:MFT}
H.T. Dobbs, D.D. Vvedensky, A.~Zangwill, J.~Johansson, N.~Carlsson, and
  W.~Seifert.
\newblock Mean-field theory of quantum dot formation.
\newblock {\em Phys. Rev. Lett.}, 79(5):897, 1997.

\bibitem{Daruka:2003:TCF}
I.~Daruka and J.C. Hamilton.
\newblock A two--component {Frenkel}--{Kontorowa} model for surface alloy
  formation.
\newblock {\em J. Phys.: Condens. Matter}, 15:1827, 2003.

\bibitem{Krack:2002:DSB}
B.D. Krack, V.~Ozolins, M.~Asta, and I.~Daruka.
\newblock Devil`s staircases in bulk--immiscible ultrathin alloy films.
\newblock {\em Phys. Rev. Lett.}, 88(18):186101, 2002.

\bibitem{Yu:1997:POE}
B.D. Yu and M.~Scheffler.
\newblock Physical origin of exchange diffusion on fcc(100) metal surfaces.
\newblock {\em Phys. Rev. B}, 56(24):15569, 1997.

\bibitem{Kellog:1990:SSD}
G.L. Kellog and P.J. Feibelman.
\newblock Surface self--diffusion on {Pt(001)} by atomic exchange mechanism.
\newblock {\em Phys. Rev. Lett.}, 64(26):3143, 1990.

\bibitem{Ahr:2002:FSI}
M.~Ahr and M.~Biehl.
\newblock Flat {(001)} surfaces of {II-VI} semiconductors: a lattice gas model.
\newblock {\em Surf. Sci.}, 505:124, 2002.

\bibitem{Nick:Wilson}
N.~Wilson.
\newblock http://www.tc.bham.ac.uk/\symbol{126}nickw/research/bham.ac.uk/PhD/.

\bibitem{Lennard}
J.E. Lennard-Jones.
\newblock {\em Proc. Roy. Soc.}, A106:463, 1924.

\bibitem{Ercolessi:20xx:MDP}
F.~Ercolessi.
\newblock A molecular dynamics primer.
\newblock http://www.fisica.uniud.it/\symbol{126}er\-co\-les\-si/md/.

\bibitem{Morse:1929:DMA}
P.M. Morse.
\newblock Diatomic molecules according to the wave mechanics. {II.} vibrational
  levels.
\newblock {\em Phys. Rev.}, 34:57, 1929.

\bibitem{Girifalco:1959:AMP}
L.A. Girifalco and V.G. Weizer.
\newblock Application of the {Morse} potential function to cubic metals.
\newblock {\em Phys. Rev.}, 114(3):687, 1959.

\end{thebibliography}
\bibliographystyle{unsrt}
\newpage
\thispagestyle{empty}
\selectlanguage{ngerman}
\chapter*{Danksagungen}
Allen, die zum Gelingen dieser Arbeit beigetragen haben m\"ochte ich auf diesem Weg meinen
herzlichen Dank aussprechen. Insbesondere bedanke ich mich bei
\begin{itemize}
\item Priv. Doz. Dr. Michael Biehl f\"ur die ausgezeichnete Betreuung dieser Arbeit, die stete 
Bereitschaft auf Fragen einzugehen, sowie f\"ur zahlreiche Anregungen und Ideen.
\item Prof. Dr. Wolfgang Kinzel f\"ur viele Anregungen und 
die M\"oglichkeit diese Arbeit an seinem Lehrstuhl anzufertigen.
\item Dr. Martin Ahr, Priv. Doz. Dr. Michael Biehl, Dr. Miroslav Kotrla und Thorsten Volkmann f\"ur
die fruchtbare Zusammenarbeit, deren Ergebnisse wir gemeinsam ver\"offentlicht haben.
\item Christian Vey und Thorsten Volkmann f\"ur das gr\"undliche Korrekturlesen dieser Arbeit. 
\item Ansgar Freking, Andreas Ruttor, Christian Vey und Thorsten Volkmann f\"ur viele Anregungen und 
Diskussionen.
\item den Systembetreuern Ansgar Freking, Andreas Klein, Georg Reents, Alexander Wagner und Andreas Vetter f\"ur den
str\"orungsfreien Betrieb des Computersystems.
\item allen Mitgliedern unserer Arbeitsgruppe f\"ur das freundliche und kreative Ar\-beits\-klima.
\item der deutschen Forschungsgemeinschaft f\"ur die Finanzierung dieser Arbeit.
\item meiner Familie f\"ur die Unterst\"utzung w\"ahrend meines Studiums.
\end{itemize}
\selectlanguage{USenglish}

\newpage
\end{document}